# A Catalog of Galaxies behind the Southern Milky Way. – I. The Hydra/Antlia Extension ($l \approx 266° - 296°$) *

**Renée C. Kraan-Korteweg**

Departamento of Astronomía, Universidad de Guanajuato, Apartado Postal 144, Guanajuato Gto 36000, Mexico

**Abstract.** A deep optical galaxy search in the southern Milky Way – aimed at reducing the width of the Zone of Avoidance – revealed 3279 galaxy candidates above the diameter limit of D $\gtrsim 0\farcm2$, of which only 112 (3.4%) were previously catalogued. The surveyed region ($266° \lesssim \ell \lesssim 296°$ and $-10° \lesssim b \lesssim +8°$) lies in the extension of the Hydra and Antlia clusters – where a supercluster is suspected – *and* in the approximate direction of the dipole anisotropy in the Cosmic Microwave Background radiation.

Here we present the optical properties of the unveiled galaxies such as positions, diameters, magnitudes, morphological types, including a detailed discussion on the quality of these data and the completeness limits as a function of the foreground dust extinction. For 127 of the 227 positional matches in the IRAS PSC, a reliable cross-identification could be found.

Several distinct overdensities and filaments of galaxies can be identified that are apparently uncorrelated with the Galactic foreground extinction hence the probable signature of extragalactic large-scale structures.

This catalog constitutes the first part in a series of five equally conducted optical searches for galaxies in the southern Milky Way ($245° \lesssim \ell \lesssim 350°$). With these surveys, the entire Zone of Avoidance will have been covered by means of visual inspection. The catalogs build the basis for various spectroscopic and photometric follow-up programs which eventually will allow a thorough analyse of the galaxy distribution in redshift space and the peculiar velocity fields within the Zone of Avoidance, as well an an improved understanding of the Galactic foreground extinction.

## 1. Introduction

Due to the foreground extinction of the Milky Way, galaxies become increasingly fainter, smaller and are of lower surface brightness as the dust extinction increases. Although most of them are not intrinsically faint or small, galaxies close to the Galactic Plane fail to meet the criteria for inclusion in magnitude or diameter-limited catalogs and only few galaxies are known below Galactic latitudes of $|b| \lesssim 10°$. Added to this are the enormous numbers of foreground stars that frequently fall on the galaxy images and crowd the field of view. Because of this, most extragalactic studies are done in regions which are free of the effects of this 'foreground pollution' – they avoid the so-called Zone of Avoidance (ZOA). Various questions with regard to the dynamics in the local Universe, however, require knowledge of the galaxy distribution in the ZOA:

(1) To explain the peculiar velocity of the Local Group (LG) with the irregular mass distribution in the local Universe and compare this motion to the dipole in the Cosmic Microwave Background (CMB) requires "whole-sky" coverage. Kolatt et al. (1995) have shown that the gravitational acceleration of the LG is strongly affected by the mass distribution in the ZOA. The dipole direction determined from the visible mass distribution changes significantly (by 31°) if the mass distribution in the ZOA is not accounted for.

(2) For our understanding of velocity flow fields – such as the flow induced by the Great Attractor centered in the Galactic Plane – we need to know the galaxy distribution in the ZOA. The comparison of the actual galaxy distribution with dynamically implicated mass excesses will tell us whether galaxies are fair tracers of mass.

(3) This not only concerns large-scale structures such as clusters, voids and walls. Hidden nearby massive galaxies will influence the internal dynamics of the LG, its mass derivation and the present density determination of the Universe from timing arguments (Peebles 1994). Moreover, the gravitational attraction of the nearest galaxies ($v < 300\,\mathrm{km\,s^{-1}}$) generates 20% of the total dipole moment (Kraan-Korteweg 1989), and six of the nine apparent

*Send offprint requests to*: Renée C. Kraan-Korteweg

* Tables 1 and 2 are available in electronic form. See the Editorial in A&AS 1994, Vol. 103, No.1

brightest galaxies are located in the ZOA. Others might still remain uncovered.

For these reasons, various groups have initiated projects in recent years to unveil the galaxy distribution behind our Milky Way such as galaxy searches on optical sky surveys, near-infrared surveys (e.g., DENIS and 2MASS), far-infrared surveys (e.g., IRAS), and systematic blind H I searches. Although all are subjected to different limitations and selection effects, they have the advantage that they are complementary in the galaxies they unveil and the latitude ranges they are optimal for (see Kraan-Korteweg & Woudt 1999, for a review).

### *1.1. Optical galaxy searches in the southern Milky Way*

We here report on the first of a series of five deep optical galaxy searches. One of the methods to reduce the width of the ZOA is to identify galaxies on sky surveys to fainter magnitude limits and lower diameter limits compared to existing catalogs. Here, examination by eye is still the best technique. A separation of galaxy and star images cannot as yet be done by automated measuring machines such as COSMOS or APM on a viable basis below $|b| \lesssim {}^\circ - 15^\circ$ (see Sect. 3.4) though surveys by eye are clearly both very trying and time consuming, and maybe not as objective.

Using the IIIaJ film copies of the ESO/SRC sky survey, we have pursued this approach and systematically surveyed five contiguous areas in the southern Zone of Avoidance (see Fig. 1). The southern Milky Way is especially exciting as many suspected large-scale structures are bisected or hidden by the Milky Way: we see traces of the nearby Puppis filament, a possible extension of the Hydra/Antlia clusters across the ZOA, the crossing of the Supergalactic Plane, and it hides a large fraction of the Great Attractor overdensity which is centered in the ZOA at $(\ell, b, v) \sim (320^\circ, 0^\circ, 4500\,\mathrm{km\ s^{-1}}$, Kolatt et al. 1995).

Fig. 1 shows a diameter-coded distribution of all galaxies with $D \geq 1\farcm0$ in the southern sky for declinations $\delta \leq -17\fdg5$ as taken from the Lauberts Catalog of Galaxies (Lauberts 1982). Part of the above discussed suspected features can be identified in this distribution. The most important clusters are indicated as well as the dipole direction of the CMB ($\ell = 280^\circ$, $b = 27^\circ$, Kogut et al. 1993) and the predicted center of the Great Attractor region. Most conspicuous in this distribution is the very broad, nearly empty band of about 20° wide that stretches across this equal area projection i.e., the Zone of Avoidance. This is the area in which we performed our galaxy searches. The five survey regions are marked with solid lines.

The thick contour marks the search area discussed in this paper. It lies in the extension of the Hydra/Antlia clusters towards the Galactic Plane and covers about 400 □° on the sky from $266^\circ \lesssim \ell \lesssim 296^\circ$ between $-10^\circ \lesssim b \lesssim +8^\circ$). This area was chosen because of the suspicion that the Hydra and Antlia clusters are part of a much larger structure, a possible supercluster that stretches across the Milky Way. At a mean redshift distance of $v \sim 2500\,\mathrm{km\ s^{-1}}$ this could contribute significantly to the peculiar motoin of the LG, and – seen its vicinity to the CMB dipole direction – bring the gravitationally determined peculiar motion vector in better agreement with the observed anisotropy in the CMB radiation.

The subsequent regions are the Crux region ($287^\circ \lesssim \ell \lesssim 318^\circ$, $|b| \lesssim 10^\circ$, Woudt 1998, Woudt & Kraan-Korteweg, in prep.) and the Great Attractor region ($318^\circ \lesssim \ell \lesssim 340^\circ$, $|b| \lesssim 10^\circ$, Woudt 1998, Woudt & Kraan-Korteweg in prep.), paper II and III of these series. The extensions from the Hydra/Antlia region to Puppis, as well as the Great Attractor region towards the Galactic Bulge, i.e., the Scorpius region, have already been surveyed and the respective catalogs are in preparation (Salem and Kraan-Korteweg, respectively, Fairall and Kraan-Korteweg).

The five survey regions will connect to optical searches done by other groups. The other optical surveys performed in the southern sky are the Puppis region ($230^\circ \lesssim \ell \lesssim 250^\circ$, $|b| \lesssim \pm 10^\circ$) by Saito et al. (1991) on the right hand side of our survey (dashed line in Fig. 1), the Ophiuchus Supercluster Region by Wakamatsu et al. (1994) and its extension (Wakamatsu et al., in preparation) as well as the Sagittarius region below the Galactic Center ($-7^\circ \lesssim \ell \lesssim 16^\circ$, $-19^\circ \lesssim b \lesssim -1^\circ$) by Roman & Saito (1997) and the Aquila/Sagittarius survey ($8^\circ \lesssim \ell \lesssim 43^\circ$, $|b| \lesssim \pm 15^\circ$) by Roman et al. (1996) on the lefthand side of Fig. 1).

We have allowed for a small overlap of our search regions with the surveys connecting to our search areas. The interleaving of the Ophiuchus area with our Scorpius region (the narrow gap visible in Fig. 1) is currently being done by Wakamatsu et al. This will allow a homogenisation of the data given in the different catalogs into one coherent catalog of galaxies in the southern ZOA, complete to well-defined identification criteria.

The galaxy catalogs build the basis for various distinct redshift and photometric observational follow-up programs. Optical spectroscopy (individual) of all the brightest (extinction-corrected) galaxies with high central surface brightness has been obtained at the 1.9 m telescope of the SAAO for the here presented galaxy catalog (Kraan-Korteweg et al. 1995), for the Crux region (Fairall et al. 1998), and for the GA region (Woudt et al. 1999). Low-surface brightness galaxies have been observed in H I with the 64 m radio telescope at Parkes (see Kraan-Korteweg et al. 1997), and multifiber spectroscopy has been applied to high-density areas at the 3.6m telescope of ESO with Optopus and Mefos. The latter data have been reduced and will soon be submitted for publication. Some of the results have been presented earlier (Felenbok et al. 1997, Woudt 1998).

In this first paper of the catalog series of deep optical galaxy searches in the southern ZOA, a description of the search method is given in Sect. 2 including a discussion on the uncovered galaxy distribution. This is followed by the catalog of the 3279 galaxy candidates in Sect. 3, and a detailed discussion of the quality of the listed galaxy properties in Sect. 4, including cross-identifications in the IRAS Point Source Catalog (Joint IRAS Science Working Group 1988, IRAS PSC). In the last section, the completeness of the galaxy catalog is analysed as a function of extinction, leading to a new, complete diameter-limited southern sky distribution of extinction-corrected galaxies, with the gap in the ZOA in the Hydra-Antlia region filled in to its completeness level at $A_B = 3$ mag.

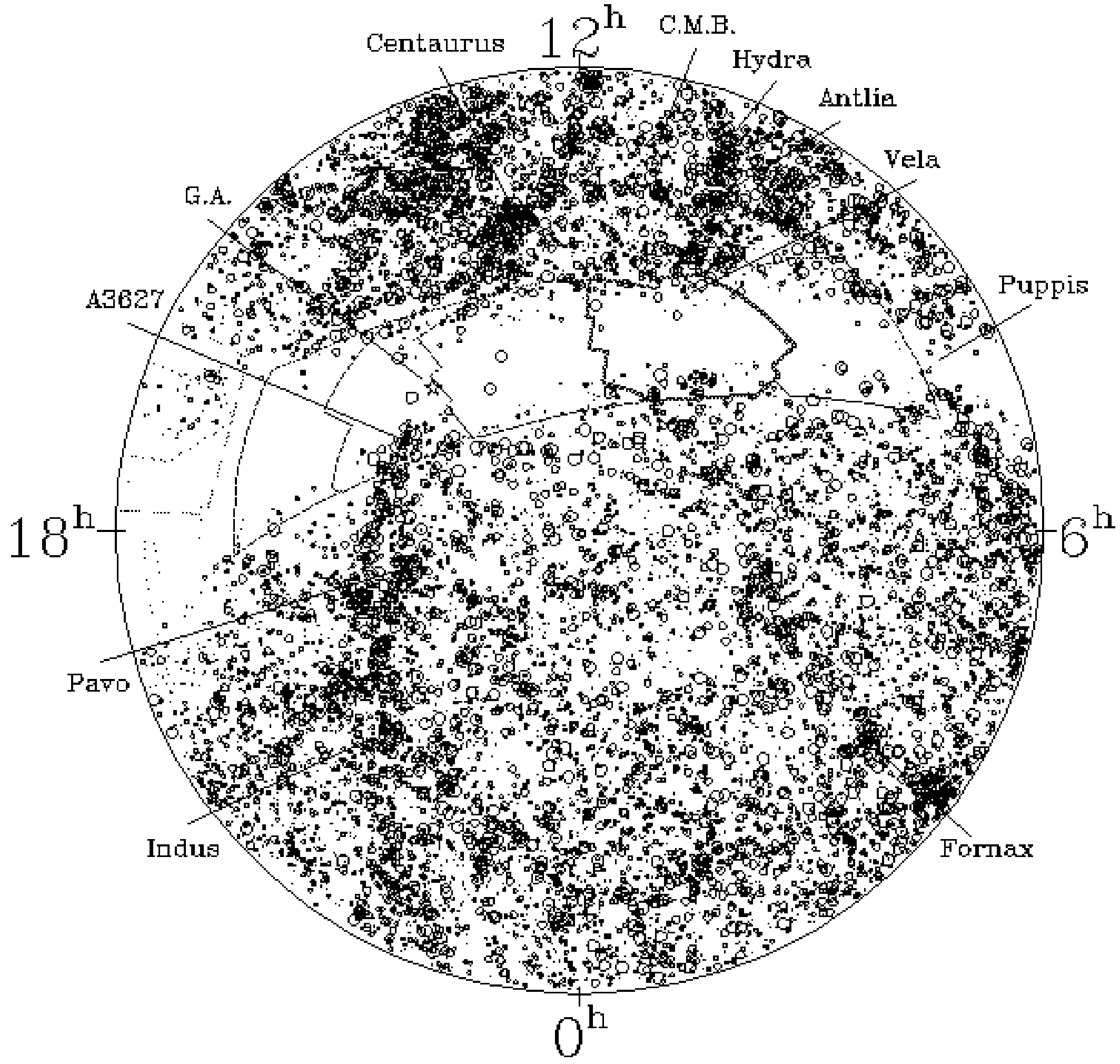

**Fig. 1.** Equal area projection of all Lauberts galaxies (D≥ 1.′0) in the southern sky ($\delta \leq -17.^\circ5$) in equatorial coordinates. The galaxies are diameter-coded: the galaxies are displayed as points for $1.'0 \leq D < 1.'3$, as small circles for $1.'3 \leq D < 2'$, as larger circles for $2' \leq D < 3'$, and as big circles for $D \geq 3'$. The most important clusters are labelled, as well as the CMB-dipole direction and the center of the Great Attractor. Within the ZOA only few galaxies are catalogued. This region has been covered by our five deep optical surveys (outlined by the solid lines). The thick contour marks the search area discussed in this paper. The dashed areas mark optical surveys performed by other groups.

## 2. The Galaxy Search

The tools for this galaxy search are simple. They comprise a viewer with the ability to magnify 50 times (a prototype blinking machine on semi-permanent loan from the Astronomisches Institut der Universität Basel) and the IIIaJ film copies of the ESO/SRC sky survey. The viewer projects an area of 3.′5 x 4.′0 on a screen which is viewed in a darkened room making the visual systematic scanning of the plates straightforward and comfortable.

The ZOA is the only part of the sky where it remains more efficient to scan the sky surveys by eye rather than by using modern plate measuring machines or sophisticated galaxy identification algorithms, as for instance with COSMOS, SUPERCOSMOS and MAMA. Automatic searches all fail close to the Galactic Plane, due to the crowding effects when blended stars are mistakingly identified as galaxies or when star subtraction of superimposed stars break the galaxies up in various 'small' galaxies (see Sect. 3.4).

Even though Galactic extinction effects are stronger in the blue, the IIIaJ films are chosen over their red counterparts. A careful inspection between the various surveys demonstrated that the hypersensitized and fine grained emulsion of the IIIaJ films go deeper and show more resolution. Even in the deepest extinction layers of the ZOA, the red films were found to have no advantage over the IIIaJ films.

The success of optically identifying extragalactic objects at very low latitudes is proven by the fact that less than 3% of the over 10% spectroscopically observed galaxy candidates of the catalog (Kraan-Korteweg et al. 1994, Felenbok et al. 1997) have a star-like signature, hence are either foreground stars or are overshadowed by foreground stars, and only a few were found to be Galactic nebulae.

We imposed a diameter limit of $D \gtrsim 0.2$ arcminutes for our search. Below this diameter the refraction crosses of the stars disappear, making it hard to differentiate consistently between stars, blended stars and faint round galaxies. In a few cases of clear clustering, smaller galaxies – mainly early-type galaxies – are retained in the list.

For every galaxy, we recorded the major and minor diameter, an estimate of the average surface brightness and the morphological type of the galaxy. From the diameters and the average surface brightness a magnitude estimate is derived. The reliability of the recorded diameters and the apparent magnitude are discussed in detail in Sect. 3. A surprisingly good relation is found for the estimated magnitudes, with no deviations from linearity even for the faintest galaxies, and a scatter of only $\sigma = 0\overset{m}{.}5$.

The positions of all the galaxies were subsequently measured with the Optronics machine at ESO in Garching. The accuracy of the positions is about $1''$.

Due to the locally varying extinction it is difficult to give a homegeneous galaxy classification. The distinction between, for instance, the bulge of a spiral galaxy and an early-type galaxy remains ambiguous in very obscured regions. The details of the morphological classification thus depends on the identifiable details on the enlarged survey image.

In this way, 3279 galaxy candidates have been discovered in the Hydra/Antlia search region, a region of approximately 400 $\square^\circ$ that encompasses 18 fields of the ESO/SRC survey (F91-F93, F125-F129, F165-F170, F211-F214) within Galactic latitudes of $-10^\circ \lesssim b \lesssim +8^\circ$ and longitudes of $266^\circ \lesssim \ell \lesssim 296^\circ$. Of these 3279 galaxy candidates, 2818 are certain galaxies and 453 are likely galaxy candidates. 8 unlikely candidates were retained in the catalog as well. Among the 3279 identifications, only 97 galaxies were previously recorded by Lauberts (1982) – of which 4 Lauberts objects turned out to be two close galaxies. Eleven further galaxies were listed in other catalogs, leading to 3167 (96.6%) newly identified galaxies.

The galaxies found in our search are entered as small dots in Fig. 2, the larger dots identify the known Lauberts galaxies in the illustrated region. It is obvious from the distribution of the newly identified galaxies that this method is quite succesful: the Zone of Avoidance has been narrowed down to $-4^\circ \lesssim b \lesssim 2\overset{\circ}{.}5$.

The asymmetry of the galaxy distribution with respect to the Galactic equator reflects the asymmetry of the dustlayer. The latter can be traced by the contours in Fig. 2 which mark absorption levels of $A_B = 1\overset{m}{.}0, 3\overset{m}{.}0$ (thick line) and $5\overset{m}{.}0$. These values are based on the DIRBE/IRAS extinction maps by Schlegel et al. (1998). The offset to the south was already established by Kerr & Westerhout in (1965) for the longitude range $\ell = 200^\circ - 330^\circ$ from the hydrogen column densities.

Galaxies remain visible through obscuration layers of 3 magnitudes of extinction; a few galaxies still are recognisable up to extinction levels of $A_B = 5\overset{m}{.}0$. Overall, the mean number density follows the dust distribution remarkably well at low Galactic latitudes. The contour level of $A_B = 5\overset{m}{.}0$, for instance, is nearly indistinguishable from the galaxy density contour at 0.5 galaxies per square degree. At intermediate extinction levels, distinct under- and overdensities are noticeable in the unveiled galaxy distribution that are uncorrelated with the foreground obscuration. They must be the signature of large-scale structures. Strong clustering is evident around $\ell = 280^\circ, b = +6^\circ$ (the Vela overdensity), $\ell = 275^\circ, b = -9^\circ$ and $\ell = 292^\circ, b = +8^\circ$. A conspicous underdensity in the longitude range $285^\circ - 290^\circ$ above the Galactic Plane remains unexplained by the dust distribution, as well the distinct decrease in galaxy density below the Galactic Plane from the right-hand side to the left.

Although a handful of very small galaxy candidates have been found at high extinction levels, these galaxies most likely indicate holes in the dust layer. Overall, the Milky Way remains optically opaque for extinction levels above $A_B \gtrsim 5\overset{m}{.}0$. This intransparent part of the ZOA regions is currently being filled in through the systematic blind H I searches the Dwingeloo Obscured Galaxy Survey (DOGS) in the north (Henning et al. 1998, Rivers et al. 1999), and the Multibeam ZOA-survey in the south (Kraan-Korteweg et al. 1999, Henning et al. 1999).

A discussion on structures in redshift-space based on the galaxies identified in the here presented search area can be found in Kraan-Korteweg (1992), Kraan-Korteweg & Woudt (1993,1994), Kraan-Korteweg et al. (1994) and Kraan-Korteweg et al. (1996). Our redshift follow-up programs have proven that the prominent overdensity visible at $(\ell, b) = (280^\circ, +6^\circ)$ is actually a superposition of three structures: a filamentary thin structure that can indeed be traced from the Hydra and Antlia clusters to $(\ell, b) = (280^\circ, -7^\circ)$ to the oposite side of the ZOA at a mean recession velocity of 2500 km s$^{-1}$ , a shallow but very extended supercluster, the Vela SCL (Kraan-Korteweg & Woudt 1993) centered at $(\ell, b, v) = (280°, +6°, \sim 6000\,\mathrm{km\ s^{-1}})$ as well as a number of clusters at about 16000 km s$^{-1}$. The overdense clumps below the Plane at $\ell = (268° - 286°)$ are due to clusters in the same high redshift range. This gave rise to the suspicion that these clusters mark a possible connection between the Horologium clusters below the ZOA and the Shapley clusters above the ZOA. Given its approximate distance ($\sim$ 16000 km s$^{-1}$) and its extent on the sky (roughly 100°), this would imply the largest known structures to date.

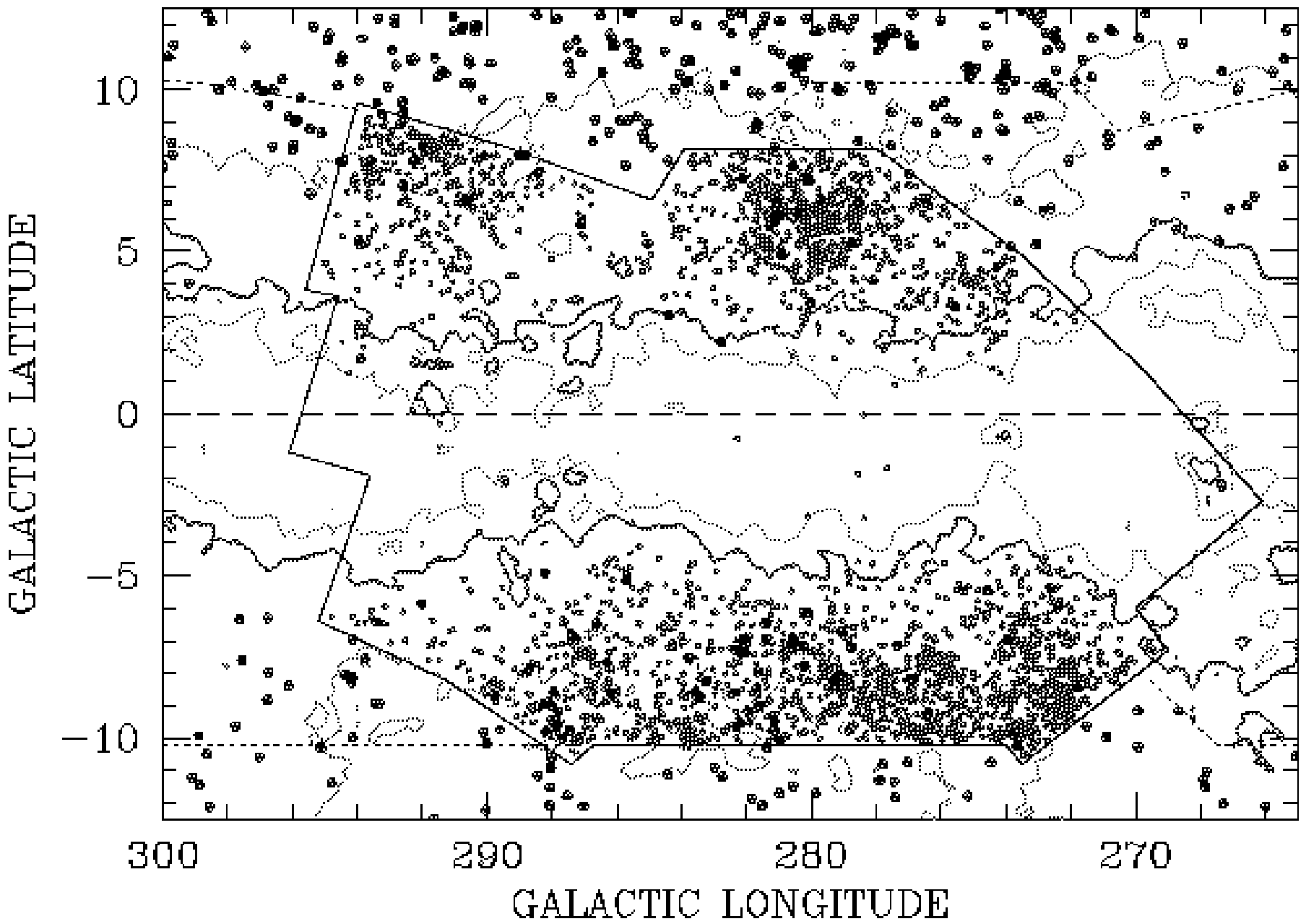

**Fig. 2.** Distribution of galaxies in the Hydra/Antlia extension. The outlined area marks the search region in the ZOA where the dashed lines indicate the adjacent areas covered by us. The 3279 unveiled galaxy candidates (D$\gtrsim$ 0.′2) are shown as small dots. In the surrounding area the Lauberts galaxies are displayed (large dots, D≥ 1.′0). The contours mark the dust extinction as determined from the 100$\mu$m DIRBE maps (Schlegel et al. 1998) at the levels $A_B = 1\overset{m}{.}0$, $3\overset{m}{.}0$ (thick line) and $5\overset{m}{.}0$.

## 3. The Catalog

### 3.1. Short description

The catalog itself (Table 1) is listed at the end of the paper. A short description of the entries in the catalog is given below. Detailed information on the listed properties of the detected galaxies and/or an assessment of the quality of the presented data is presented in the following subsections. A table listing the possible IRAS galaxies is given in Sect. 3.3.

*Column 1:* RKK running number of the detected 3279 galaxies and galaxy candidates, ordered in Right Ascension (1950.0).

*Column 2:* Second name.

*Column 3:* Codes for identifications with objects in the IRAS PSC within a radius of 2 arcmin. The entries signify certain identification (I), possible identification (P), questionable identification (Q) and no credible cross-identification (N).

*Column 4 and 5:* Right Ascension and Declination (1950.0).

*Column 6 and 7:* Galactic longitude and latitude.

*Column 8:* Field number of the ESO/SRC Survey on which the galaxy was detected.

*Column 9 and 10:* X- and Y-coordinate in mm as measured from the center of the field listed in column 8.

*Column 11:* Large and small diameter D x d in arcsec.

*Column 12:* Apparent magnitude $B_J$.

*Column 13:* Galactic reddening based on the DIRBE/IRAS extinction maps (Schlegel et al. 1998).

*Column 14:* Morphological type including codes with regard to uncertainty of the identification, the orientation and superimposed stars.

Column 15: Descriptive remarks.

### 3.2. Second names

Most of the second identifications given in Column 2 of Table 1 originate from the ESO/Uppsala Survey of the ESO(B) Atlas (Lauberts, 1982), recognisable as 'L' plus the respective field and running number. After closer inspection, a number of Lauberts galaxies turned out to be two individual galaxies. The Lauberts identification is then given in both case. L126-9 = RKK1251 & RKK1253, corresponding to ESO-LV 126-0090 and ESO-LV 126-0091 in the ESO-LV Catalog (Lauberts & Valentijn 1989). Four galaxies abbreviated as FGCE# are listed in the Flat Galaxy Catalog (Karenchentsev et al. 1993), two in the Arp Madore (1987) Catalog (marked as AM), two in the Parkes-MIT-NRAO 5GHz Radio Survey (Wright et al. 1994, code PMN), and one in the Catalog of Southern Ring Galaxies (Buta 1995).

97 galaxies (3%) were previously identified by Lauberts (1982), of which four each have two counterparts in the here presented catalog. Including the overlaps in the above mentioned catalogs 112 have entries in earlier optical catalogs (3.4%) and 3167 are newly identified galaxies.

### *3.3. IRAS identifications*

The IRAS PSC (Joint *IRAS* Science Working Group 1988) has been used extensively for studies of the large-scale structures in the Universe. It is therefore of interest to identify IRAS counterparts of the here uncovered galaxies.

IRAS sources were searched in a radius of 2 arcmin around the optical galaxies. This led to 227 coincidences in the IRAS PSC. In a second step, the possible matches were investigated individually, taking into account
(a) the positional offset between galaxy and IRAS source in combination with the IRAS position uncertainty ellipses
(b) the colors as deduced from the different IRAS wavebands, such as $col_1 = f_{12} \cdot f_{25}/(f_{60})^2$ and $col_2 = f_{100}/f_{60}$,
(c) coincidences with other astronomical objects and their respective colors or color expectation values.

We more or less adopted the selection criteria used by Yamada et al. (1993) in their search for IRAS galaxy candidates in the ZOA as being characteristic of IRAS galaxies, i.e., $col_1 < 1$ and $0.8 < col_2 < 5$. However, neither a lower limit for the $60\mu$ flux was imposed (0.6 Jy in Yamada et al. 1993), nor a flux quality restriction.

Depending on the probability of the cross-identification being correct, the following categories were defined (denoted as such in column 3 of Table 1):

**I** : high-certainty identification with IRAS PSC object

**P** : possible match in the IRAS PSC, but either the $f_{60}$ flux was only a lower limit, the separation with regard to the uncertainty ellipse relatively large, or the colors $col_1$ or $col_2$ atypical for galaxies

**U** : an unlikely cross-identification because of large positional offset and/or unlikely IRAS-colors

**N** : the cross-identification is not accepted as credible because of the large positional offset and the unlikely colors and, in addition, that the identification with another object is more likely. If this concerns another galaxy, it is marked as **NG** (20 cases) in Table 1, if a star as **NS** (10 cases).

If a number is added to the code I, P or Q (e.g., **Q2, I3**), this number indicates the equally possible galaxy counterparts for a given IRAS source.

Overall, 135 certain IRAS galaxies were identified. For 6 of these sources, two galaxies are equally likely to be the counterpart of the IRAS source (marked as I2), and in 2 cases three galaxies (I3) could match the optical identification, leading to a total of 145 galaxy counterparts for the 135 certain IRAS sources.

The number of galaxies with certain IRAS PSC cross-identification is reduced to 87 (91 galaxies, 87 IRAS sources), if – as in most IRAS color-selected galaxy searches – a strict lower limit for the flux density at $60\mu$m of $f_{60} = 0.6$ Jy is demanded, as well as high flux quality at $60\mu$, i.e., $Q_{60} = 3$.

Surprisingly, only 66 IRAS sources were identified by Yamada et al. (1993) in their ZOA IRAS galaxy survey, leaving 22 IRAS galaxies (25%) unaccounted for. Of these, 7 comply with all selection criteria set out by Yamada, hence it is not clear why these IRAS galaxies were missed. The remaining 15 have a high ratio of $f_{100}/f_{60} > 5$, though they are bonafida galaxies. Yamada et al. noted already that color-selected, 'blind' IRAS galaxy samples are incomplete because of the $f_{100}/f_{60}$ upper limit restriction which was implemented to avoid the contamination by Galactic cirrus. Overall, the IRAS PSC traces a population of large and bright galaxies (see Woudt 1998, for further details). Nevertheless, 'blind' IRAS galaxy searches apparently miss a significant fraction of nearby galaxies.

There are 21 possible cross-identifications in the IRAS PSC (one with two possible galaxy counterparts) and 27 questionable cross-identifications (3 with two likely galaxy counterparts).

In Table 2, given at the end of the article, the certain (I and I2 or I3), the possible (P or P2), and questionable (Q and Q2) IRAS sources in the Hydra/Antlia deep optical galaxy catalog are listed with their optical and IRAS properties. The unlikely cross-identifications (NG and NS) are marked in Table 1 but not entered in the IRAS table.

The entries in Table 2 are as follows:
*Column 1:* Identification in the IRAS Point Source Catalog. If the IRAS name is followed by 'Y', it is also in the IRAS galaxy list of Yamada et al. (1993), if followed by a '*', it satisfies all the Yamada et al. selection criteria, but is not listed there.
*Column 2:* Quality parameter of the IRAS PSC cross-identification, I, P and Q as explained above. If followed by a number, the latter indicates the equally likely galaxy counterparts for a given IRAS source.
*Column 3:* The RKK identification number as in the optical galaxy catalog (Table 1). The 'L' signifies whether this is also a Lauberts galaxy.
*Column 4 – 10:* As column 4 –7, 11, 12 and 14 of Table 1.
*Column 11:* Angular separation in arcsec between the optical position (Column 4 and 5) and the position given in the IRAS PSC.
*Column 12 – 15*: The flux densities at $12\mu$m, $25\mu$m, $60\mu$m, respectively $100\mu$m.
*Column 16:* The IRAS flux qualities at $12\mu$m, $25\mu$m,

60$\mu$m, and 100$\mu$m, where '1' indicates a lower limit, '2' an uncertain flux, and '3' a good flux quality.
*Column 17 – 18:* The IRAS colors $col_1 = f_{12} \cdot f_{25}/(f_{60})^2$ and $col_2 = f_{100}/f_{60}$.

### *3.4. Quality of optical parameters*

To assess the quality of the listed positions, diameters and magnitudes, whose derivations are discussed in the following sections, comparisons with three independent samples were made.

The first is an internal consistency check. A generous overlap on the borders of adjacent fields led to independent values for parameters of a given galaxy. In the Hydra/Antlia ZOA survey region, 146 galaxies were found on the borders of two up to four different ESO/SRC survey fields.

A further comparison was made for galaxies in common with "The Surface Photometry Catalogue of the ESO-Uppsala Galaxies" (Lauberts & Valentijn 1989), henceforth the ESO–LV catalog. Although the ESO–LV catalog generally avoided the ESO/SRC fields close to the Galactic Plane ($|b| \lesssim 15°$), some exceptions were made. This resulted in an overlap with our survey of the fields F91, F92 and F126 on which we have 49 (12, 14, and 23) of the brighter galaxies in common.

For the field F213 centered at RA = $10^h00^m$, Dec = $-50°$ (1950.0), $\ell = 277\overset{\circ}{.}3, b = 4\overset{\circ}{.}0$, MacGillivray used COSMOS to extract the galaxy parameters at the positions of the galaxies identified by us. 330 galaxies are listed in the optical catalog for F213. However, a third of these lie outside the central 5° x 5° area of the field, hence have no COSMOS parameters. For 20% of the galaxies within the 5° x 5° boundaries, COSMOS extracted two galaxy candidates at the position of the visually identified galaxy. In general, the parameters of only 'one' of the candidates – sometimes none – matched the actual galaxy. For another 8% of the galaxies, the parameters such as diameters and magnitudes diverged strongly from the visually determined values. After careful inspection on the sky survey plate, it was clear that the superposition of stars on the galaxy images generally caused the confusion: a star superimposed on the border of a galaxy image resulted in a smaller galaxy, stars superimposed more centrally on a galaxy could result in the breaking up of one galaxy into various "COSMOS galaxies". Quite often the fainter outer parts of LSB spiral were not recognised as being part of a galaxy, and parts of a spiral arms sometimes were lost. About 4% of the galaxies were – for no obvious reasons – not recovered at all by COSMOS.

The above comparison stresses the inherent difficulties in achieving a high success rate of the galaxy identification procedure from automated extraction algorithms at low Galactic latitudes and in obtaining reliable data. Still, the parameters of 186 galaxies could be used for our comparative purposes.

### *3.5. Positional accuracy of coordinates*

The X and Y measurements listed in Column 9 and 10 are offsets with respect to the center of the field identified in Column 8, i.e., the field on which the galaxy was first identified in the course of the galaxy search. This therefore is not necessarily identical to the field on which that galaxy would belong based on its coordinates and the optimal survey fields. Positive X-values indicate increasing RA, negative X-values decreasing RA. Positive Y-values point north, negative values south of the field center.

The positions of the galaxies as well as up to 30 standard stars per field were measured with the measuring machine Optronics at the ESO in Garching (one advantage of working in the ZOA is the availability of numerous standard stars). Fitting a polynomial to the X- and Y-measurements of the standards stars resulted in an rms of the star positions of 0.3 – 0.5 arcsec. For galaxies this precision can not be achieved due to the uncertainties in the determination (by eye) of the center of the extended galaxy images. As this is straightforward for small galaxies, the positions of smaller galaxies generally are of better precision.

Comparing the positions of galaxies derived from the borders of neighboring fields (generally the lowest precision cases) did reveal minor systematic offsets of typically $0 - 1.5''$ in RA or Dec with a dispersion of $\sigma = 2''$. A comparison of our positions with COSMOS positions for the 186 galaxies in common on field 213 revealed similar trends (offsets in RA and Dec of $2''$, $\sigma = 1\overset{''}{.}5$). Positions of ESO–LV galaxies show no offset but a dispersion of $4''$ compared to the Optronics positions. Taking the quoted error of $3''$ for the positions in the ESO–LV catalog into account, the $1''$ for COSMOS, as well as our internal consistency, together indicate that the in Table 1 listed galaxy positions have an accuracy of about 1 arcsec.

The positions of a few galaxies are of lower precision as they were not derived with the measuring machine Optronics. These are marked with a colon following the equatorial coordinates in column 4 and 5 of Table 1.

### *3.6. Large and small diameters*

A comparison of galaxy diameters based on the COSMOS and ESO–LV overlaps, shows a linear correlation with decreasing scatter towards smaller galaxies. On average the present diameters are 10% smaller compared to the COSMOS and ESO–LV values suggesting that the listed diameters correspond approximately to the isophote of 24.5 mag/arcsec$^2$ (compared to 25.0 mag/arcsec$^2$ of COSMOS and ESO–LV).

An internal consistency check of diameters measured on different plates reveals an error of the order of $1''$, while the deviations for the galaxies in common with COSMOS are lower and with ESO–LV slightly higher ($\varepsilon = 0\farcs3$, respectively $\varepsilon = 1\farcs5$).

In the surveyed region, 103 galaxies have – according to our determinations – a major diameter of $D \geq 60''$, i.e., the Lauberts (1982) diameter limit. Of the 97 galaxies identified by Lauberts in the Hydra/Antlia survey region (of which 4 are double systems), 25 actually are smaller than $60''$, leaving 76 galaxies above the Lauberts diameter limit of $D = 60''$. A comparison with the here identified 103 galaxies with $D \geq 60''$ hence indicates that 27 galaxies (26%) were missed by Lauberts. These statistics improve somewhat in favor of the Lauberts catalog for galaxies larger than $1\farcm35$, the diameter limit for which the Lauberts catalog is claimed to be complete (Hudson and Lynden-Bell 1991). Still, 5 (of the 49) galaxies larger than $1\farcm35$ have not been identified by Lauberts.

### *3.7. Apparent magnitude $B_J$*

Using a KODAK Photographic Step Tablet (exposed and processed acetate photographic silver density film) of 21 steps as a comparison scale (from 0.05 to 3.05 in increments of 0.15 in density), the average surface density (blackness) of each galaxy was determined. The surface densities SD of the two 7-step wedges on the survey fields were determined as well. Applying the log relative intensity scale as given in the UK Schmidt Telescope Handbook (1980) to the step wedges leads to surface density vs. log intensity calibrations of the form:

$$\log(\mathrm{I}) = \mathrm{C}_1 \cdot \mathrm{SD} + \mathrm{C}_2. \tag{1}$$

This relation was found to be linear in the surface density range of the galaxies (density steps 10 – 16.5, cf., Fig. 1.1 in Woudt 1998, Vol. II).

Combining the dimensions of the galaxy (in arcsec) with the relative intensity scale leads to a relative magnitude $B_J$ estimate via the equation

$$\mathrm{B_J} = \mathrm{C}_3 - 2.5 \cdot (\log(\pi \cdot (\mathrm{D}/2) \cdot (\mathrm{d}/2)) + \log(\mathrm{I})). \tag{2}$$

The resulting *isophotal* magnitudes were then calibrated using the 49 galaxies in common with the ESO–LV catalog (Lauberts & Valentijn 1989), leading to a value of the constant $C_3$ of

$$\mathrm{C}_3 = 26\overset{m}{.}4. \tag{3}$$

The here derived magnitude estimates compare best with the $B_{25}$ of the ESO–LV: as illustrated in the left panel of Fig. 3, no deviation from linearity is observed over the common magnitude range ($13\overset{m}{.}0 - 17\overset{m}{.}0$), and a surprisingly low dispersion of $1\sigma = 0\overset{m}{.}46$.

A comparison with the COSMOS magnitudes for the 186 galaxies on field 213 is displayed in the middle panel of Fig. 3). As the COSMOS magnitudes were not calibrated a zeropoint correction of $\Delta B_J^{COS} = +0\overset{m}{.}25$ had to be applied. With the zeropoint set, a linear relationship up to the faintest magnitudes of $B_J = 19\overset{m}{.}5$ was found with a dispersion of $1\sigma = 0\overset{m}{.}47$

To check the internal consistency, a comparison was made for galaxies found on borders of adjacent fields. Even though measurements on the plate borders are the least reliable, the magnitudes scatter with a dispersion of $\sigma = 0\overset{m}{.}33$ only and show no systematic offsets (right panel of Fig. 3).

Hence, we conclude that our eye-estimates yield magnitudes with no deviations from linearity from the brightest to the faintest galaxies, with a $1\sigma$ dispersion of $0\overset{m}{.}5$.

### *3.8. Column 13: Galactic reddening*

Column 13 lists the Galactic reddening at the position of the galaxy as given by the DIRBE/IRAS extinction maps (Schlegel et al. 1998). Based on CCD photometry and measurements of the $Mg_2$-index of 18 early type galaxies at low Galactic latitudes, Woudt (1998) has tested the calibration of the DIRBE/IRAS extinction maps. He found that the extinction for moderate to high DIRBE/IRAS reddenings is systematically underestimated by a factor of f=0.86. As his new calibration so far is based on a small sample of galaxies only which do not cover a substantial fraction of the southern ZOA, we have at this point of time not adopted his correction for this suggested underestimation.

### *3.9. Column 14: Morphological classification*

The morphological types are coded similarly to the precepts of the RC2 (de Vaucouleurs et al. 1976). Due to the varying foreground extinction a homogenous and detailed type classification could not always be accomplished and some codes were added. In the first column, a question mark denotes the uncertainty about the galaxian nature of the object. In the second column, the code F for the Hubble type E-S0 (T= $-3$ in the RC2 classification) was added to the normal designations of E, L, S and I. In the fourth column the subtypes E, M and L are introduced next to the general subtypes 0 to 9. They stand for early spiral (S0/a-Sab), middle spiral (Sb-Sd) and late spiral or irregular (Sdm-Im). The cruder subtypes are a direct indication of the fewer details visible on the obscured galaxy image. The questionmark at the end marks uncertainty of the main type, the colon uncertainty in the subtype. The third column (o) marks the orientation of the galaxy: E, N, F, V stands for edge-on, nearly edge-on, face-on and nearly face-on. The fourth subcolumn (*) indicates the superposition of a single star (1) or multiple stars (S) on the galaxy.

The mixture of galaxy types in our survey – (E–SO: S–I : unclassified) = (11% : 60% : 29%) – is consistent with most optical surveys.

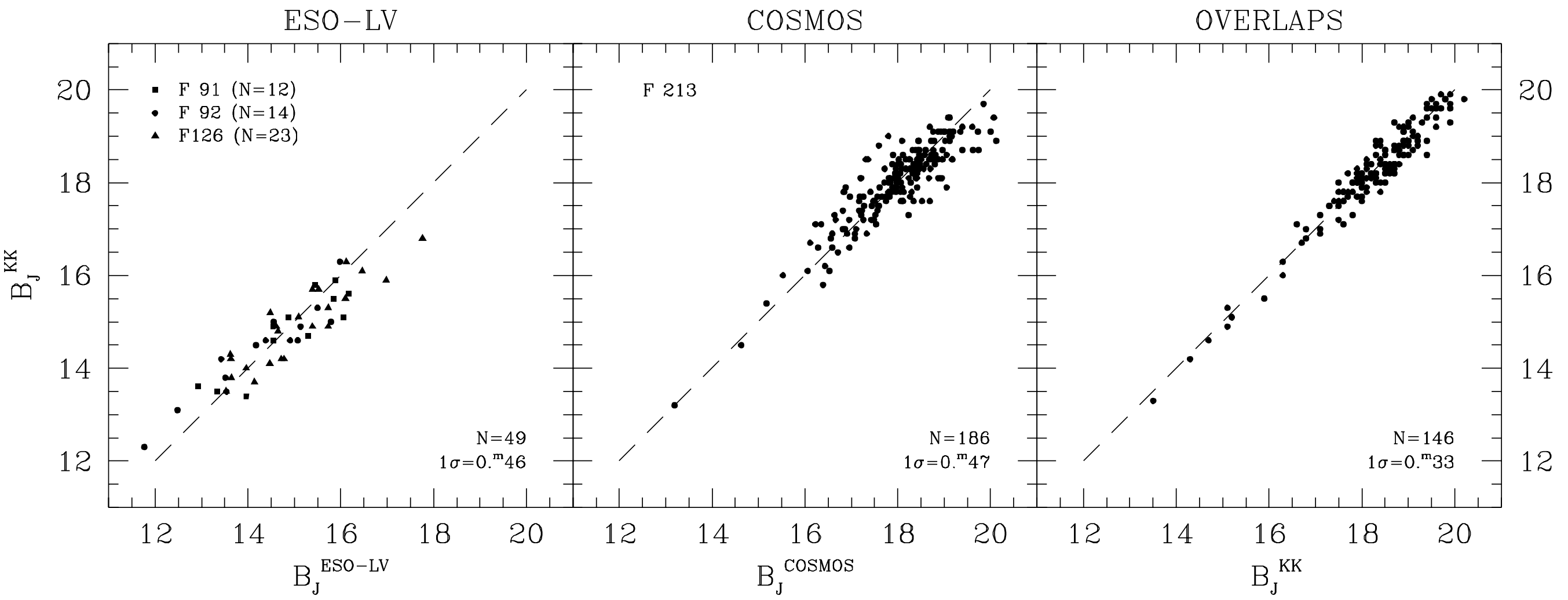

**Fig. 3.** Comparison of the estimated magnitudes $B_J^{KK}$ with the 49 galaxies in common with the ESO-LV catalog (left), the 186 galaxies on field 213 in common with COSMOS (middle), and the 146 independent magnitude estimates from the borders of neighbouring fields. For the uncalibrated COSMOS magnitudes a zeropoint correction of $\Delta B_J^{COS} = +0\overset{m}{.}25$ was applied.

### *3.10. Column 15: Descriptive remarks*

In general, the remarks and abbreviations of column 15 are selfexplanatory. The most common abbreviations are:

| | |
|---|---|
| * or ** | star or stars |
| asym. | asymmetric |
| blg | bulge |
| br. | bright; to indicate bright bulges, nearby or superimposed bright stars or bright bulges |
| cl.to br. * | close to bright star |
| dbl | double |
| diff. | diffuse |
| dist. | distinct |
| HSB | extremely high surface brightness |
| i.a.w. | interacting with; close pair of galaxies which display disturbances in the image or light bridges between the galaxies. |
| LSB | low surface brightness galaxy. Also vLSB for very low surface brightness, and vvLSB for galaxies just barely visible against the sky background |
| neb? | nebula? |
| neighb. of | neighbor of a listed galaxy. In general, these galaxies are below the diameter limit of D = 13″ but close to a catalogued galaxy and possibly associated with the named neighbor. |
| PN | Planetary Nebula |
| p.cov.by * | partly covered by star |
| poss.larger | possibly larger |
| s.p. | superimposed (generally used to note superimposed star or stars on the galaxy image |
| sp. | spiral (e.g. spiral arm) |
| struct. | structure |
| in st.diff.pat. | galaxy lies within a stellar diffraction |

## 4. Properties of the Galaxies

### *4.1. Magnitude and diameter distribution*

The top panel of Fig. 4 shows the distribution of the observed magnitudes (left) and diameters (right) of the 3279 galaxy candidates identified in the Hydra/Antlia ZOA galaxy search. On average the galaxies are quite small ($<\mathrm{D}> = 21\farcs8$) and faint ($<\mathrm{B_J}> = 18\fm2$).

However, even the galaxies at the highest latitudes are viewed through an obscuration layer of $\mathrm{A_B} \approx 1\fm0$ (see Fig. 2) which thickens as we approach the Galactic equator. The observed diameters and magnitudes are heavily influenced by the obscuring effects of the Milky Way. The extinction dims the magnitudes by the amount $\mathrm{A_B}$ *plus* an additional dimming $\Delta$ ($\mathrm{B_J^o} = \mathrm{B_J} - \mathrm{A_B} - \Delta$) because the observed diameters are reduced, hence also the surface area of a galaxy within the defined isophotal limit.

These obscuration effects on the intrinsic properties of galaxies have been studied in detail by Cameron (1990) who artificially absorbed the intensity profiles of various Virgo galaxies. This led to analytical descriptions of the diameter and isophotal magnitude corrections for early-type and spiral galaxies. For example, a spiral galaxy, seen through an extinction of $\mathrm{A_B} = 1^m$, is reduced to $\sim 80\%$ of its unobscured size. Only $\sim 22\%$ of a (spiral) galaxy's original dimension is seen when it is observed through $\mathrm{A_B} = 3^m$. The *additional* magnitude correction in this case amounts to $1\fm1$, a non-negligable amount.

In earlier papers, we used the neutral hydrogen (H I) content in the Milky Way with a constant gas-to-dust ratio as indicator of the foreground extinction. However, the gas-to-dust ratio does vary (e.g., Burstein et al. 1987). Moreover, close to the Galactic plane the Galactic H I line might be saturated, leading to an underestimation of the true extinction. With the recent availability of the 100 micron extinction maps from the DIRBE experiment (Schlegel et al. 1998), we have started implementing these values as they provide a direct measure of the dust column density and the maps have better angular resolution ($6\farcm1$ compared to $\sim 20-30'$ of the H I maps). Following Cardelli et al. (1989), the Galactic foreground in the blue was determined as

$$\mathrm{A_B} = 4.14 \cdot \mathrm{E(B-V)}. \tag{4}$$

Applying the Cameron corrections to the observed magnitudes and diameters of the galaxies identified in the ZOA result in a considerable shift of the respective means to $<\mathrm{B_J^o}> = 16\fm8$ and $<\mathrm{D}^o> = 34\farcs7$ (cf., lower panel of Fig. 4). We have avoided unrealistically large extinction-corrections for galaxy candidates in the deepest extinction layers by limiting the maximum correction factors to $\mathrm{A_B} = 6^m$.

A total of 277 galaxies have extinction-corrected diameters larger or equal than $60''$, i.e., the Lauberts (1982) diameter limit. This means that in the absence of the obscuration by the Milky Way, Lauberts would have detected 277 galaxies in the ZOA search region instead of the recorded 97 galaxies in his catalog, respectively the 76 galaxies that really have a diameter above $60''$. Comparing this to the diameter limit of $1\farcm35$ for which the Lauberts catalog is claimed to be complete (Hudson and Lynden-Bell 1991), 178 galaxies larger than $1\farcm35$ are identified, compared to the 49 galaxies in the Lauberts catalog. These numbers demonstrate the incompleteness in the Lauberts catalog near the plane of the Milky Way. More importantly, it shows the effectiveness of deep optical surveys in retrieving these galaxies.

### *4.2. Dependence on foreground extinction*

The effects of the absorption on the observed parameters of these low-latitude galaxies is reflected clearly in Fig 5. Here, the magnitudes and major diameters are plotted against the Galactic extinction E(B-V) derived from the 100 micron DIRBE/IRAS dust maps. The top panels show the observed magnitudes (left) and diameters (right) and the bottom panels the for extinction corrected parameters.

The distribution of both the observed magnitudes and diameters show a distinct cut-off as a function of extinction – all the galaxies lie in the lower right triangle of the diagram leaving the upper left triangle empty of points. At low extinction values, bright to faint galaxies, respectively large to small galaxies can be identified, whereas only apparently fainter and smaller galaxies enter our catalog for higher extinction values. The division in the diagram defines an upper envelope of the intrinsically brightest and largest galaxies. This fiducial line, i.e. the shift $\Delta m$ to fainter apparent magnitudes of the intrinisically brightest galaxies, is a direct measure of the absorption $\mathrm{A_B}$. In fact, this shift in magnitude is tightly correlated with the absorption in the blue $\mathrm{A_B} = 4.14 \cdot \mathrm{E(B-V)}$.

In the lower panels of Fig 5, the for extinction-corrected magnitudes and diameters are plotted as a function of the foreground extinction. Clearly, the faintest galaxies ($\mathrm{B_J^o} \lesssim 19\fm0$) are only uncovered at the high latitude borders of our survey, whereas the brightest galaxies can still be identified through obscuration layers of $\mathrm{A_B} \approx 4\fm0$. This distribution also has a very well-defined upper envelope which can be used to assess the completeness of the survey as a function of extinction. The distribution indicates that at extinction levels of $\mathrm{A_B} = 4\fm0$, the survey is complete only for the brighest and largest galaxies ($\mathrm{B_J^o} \lesssim 12\fm5$, $\mathrm{D^o} \gtrsim 160''$), however at extinction levels of $\mathrm{A_B} = 3\fm0$ we are still complete for galaxies with $\mathrm{B_J^o} \lesssim 15\fm5$ and $\mathrm{D^o} \gtrsim 60''$, and at extinction levels of $\mathrm{A_B} = 2\fm0$, for galaxies with $\mathrm{B_J^o} \lesssim 18\fm0$ or $\mathrm{D^o} \gtrsim 20''$.

### *4.3. Completeness of the optical survey*

A more qualitative assessment of the completeness of our deep optical galaxy survey in the ZOA can be achieved by analysing the cumulative diameter and magnitude distributions (observed and extinction-corrected) as displayed for four different extinction intervals in Fig. 6. The cumulative distribution has not been normalised by the area corresponding to each different interval of galactic extinction. The respective number of galaxies in the extinction intervals are 1089 for $0\overset{m}{.}59 \leq A_B \leq 1\overset{m}{.}0$, 1921 for $1\overset{m}{.}0 < A_B \leq 2\overset{m}{.}0$, 218 for $2\overset{m}{.}0 < A_B \leq 3\overset{m}{.}0$ and 37 for $3\overset{m}{.}0 < A_B \leq 4\overset{m}{.}0$.

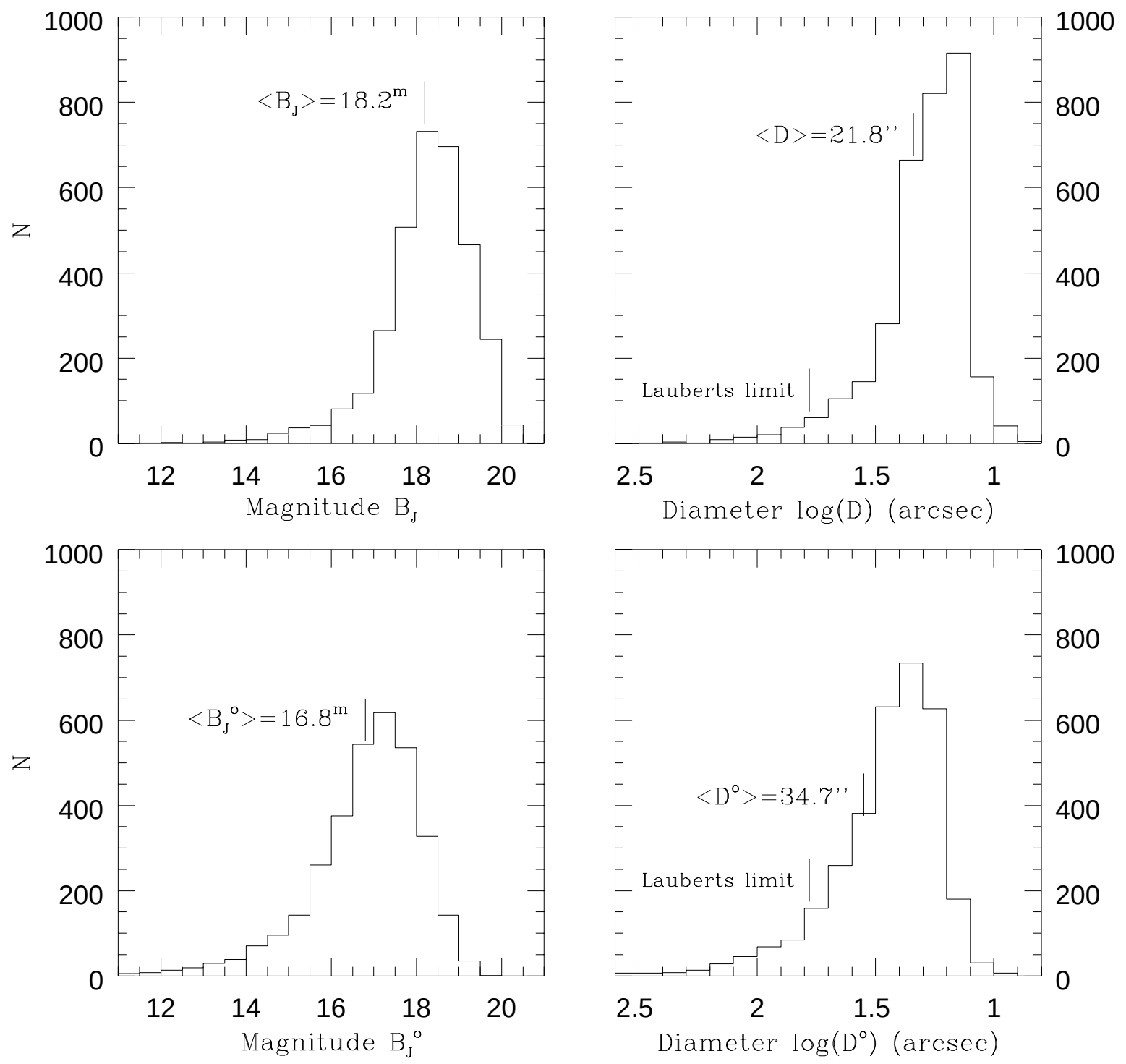

**Fig. 4.** The distribution of the observed (top panels) and extinction-corrected (bottom) magnitudes (left) and diameters (right) of the 3279 galaxy candidates discovered in the Hydra/Antlia region.

The slopes of the $B_J - \log N$ and $\log D - \log N$ distributions are slightly lower compared to unobscured regions. With the exception of the bright and large galaxy end of the cumulative distributions and for the highest extinction bin where number counts are low, we find a linear increase of the cumulative curves up to magnitudes of $B_J = 18\overset{m}{.}5$ and diameters of $\log D = 1.15$ ($D = 14''$). Then the curves start to flatten. These values hence indicate the completeness limits for the apparent (obscured) parameters of the galaxies of our survey.

The bottom panels of Figure 6 show the same distributions, but for extinction-corrected magnitudes and diameters. Here, the point at which the curves start to flatten out obviously depends on the amount of foreground extinction. We find that our deep optical galaxy search becomes seriously incomplete only in the interval $3^m < A_B \leq 4^m$ (filled triangles). A detailed analysis based on various extinction bins (not plotted here) finds that we are complete for galaxies $B_J^o \lesssim 15\overset{m}{.}5$ and diameters of $\log D^o \gtrsim 1.78$ ($D^o = 60''$) up to extinction levels of $A_B \leq 3\overset{m}{.}0$ (the open triangles in Fig. 6).

At $A_B = 3\overset{m}{.}0$, a spiral galaxy with $D^o = 60''$ will be visible with $D = 14''$ only, and an elliptical with $D = 17''$. Vice-versa, an obscured spiral or an elliptical galaxy at our *apparent* completeness limit of $D = 14''$ would have an intrinsic diameter of $D^o \approx 60''$, respectively $D^o \approx 50''$. At extinction levels higher than $A_B = 3\overset{m}{.}0$, an elliptical galaxy with $D^o = 60''$ would appear smaller than the completeness limit $D = 14''$ of this catalog and might have gone unnoticed. The here presented galaxy catalog should thus be complete for all galaxy types with $D^o \geq 60''$ down to extinction levels of $A_B = 3\overset{m}{.}0$. Only intrinsically very large and bright galaxies – particularly galaxies with high surface brightness – will be recovered in deeper extinction layers.

With the above relations between foreground extinction and completeness limit for extinction-corrected galaxies, the first step in arriving at a complete whole-sky survey can be undertaken.

According to Hudson and Lynden-Bell (1991), the Lauberts catalog is complete for galaxies larger than $1\overset{\prime}{.}35$. The optical ZOA-survey is complete $D^o = 1\overset{\prime}{.}0$ at extinction levels of $A_B \leq 3\overset{m}{.}0$. Fig. 7 combines the two catalogs and shows in an equal-area projection of equatorial coordinates all galaxies with extinction-corrected diameters larger than $D^o \geq 1\overset{\prime}{.}3$. The Hydra/Antlia ZOA survey region is now filled to Galactic latitudes of $-4° \lesssim b \lesssim 2\overset{\circ}{.}5$ (i.e., extinction levels $A_B \leq 3\overset{m}{.}0$, cf., Fig. 2). A com-parison of Fig. 7 with Fig. 1 demonstrates convincingly how the deep optical galaxy search realizes a considerable reduction of the ZOA. Moreover, the display of the extinction-corrected, diameter-coded galaxy distribution with its well-defined completeness limit clearly reveals the dynamically important large-scale structures of the nearby Universe.

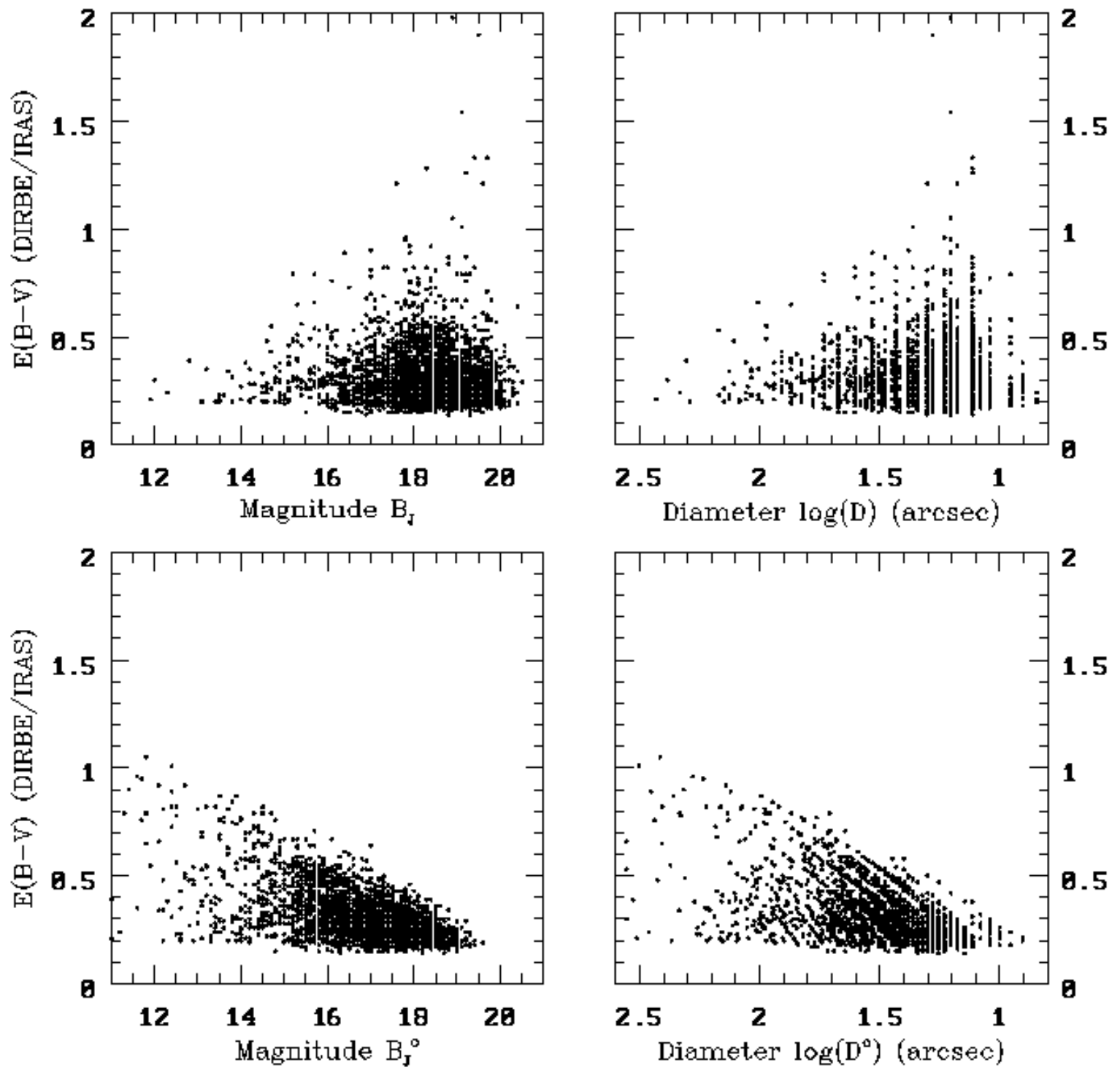

**Fig. 5.** The observed (top panels) and extinction-corrected (bottom) magnitudes (left) and diameters (right) of the 3279 galaxy candidates discovered in the Hydra/Antlia region as a function of the foreground extinction E(B − V). Some extinction-corrected values fall outside the magnitude/diameter range displayed here.

With the other forthcoming optical galaxy searches, we soon will have a much improved consensus about the most important galaxy overdensities in the southern sky.

## 5. Summary

A deep optical galaxy search for obscured galaxies behind the dust layer and crowded star fields of the Milky Way revealed 3279 galaxy candidates of which only 112 (3.4%) were previously catalogued and of which 127 (of the 227 positional matches) have a reliable counterpart in the IRAS PSC.

A comparison with the foreground extinction levels (Schlegel et al. 1998) indicates that galaxies remain easily detectable through obscuration layers of 3 magnitudes of extinction and recognisable up to levels of $A_B = 5\overset{m}{.}0$. At higher extinction levels, the Milky Way remains mostly opaque. Overall, the mean number density follows the dust distribution remarkably well.

An analysis of the completeness of this visual search indicates that we are complete to a diameter limit of $D = 14''$. Correcting the observed parameters for the foreground obscuration (DIRBE extinction values with Cameron extinction corrections) we find that our ZOA survey is complete to $D^o \geq 60''$ and – depending on the surface brightness of the galaxy – to $B_J^o \lesssim 15\overset{m}{.}5$ foreground obscuration levels of $A_B \leq 3\overset{m}{.}0$. 277 galaxies were identified above this diameter limit compared to the 76 in this area by Lauberts. These numbers demonstrate the success of deep visual galaxy searches at low Galactic latitudes.

With the understanding of our completeness limit for extinction-corrected galaxies, the first step in arriving at a complete whole-sky survey with a considerably reduced ZOA could be undertaken (cf., Fig. 7, i.e. a southern-sky galaxy distribution complete to $D^o \geq 1\overset{\prime}{.}3$ for extinction levels of $A_B \leq 3\overset{m}{.}0$).

Several distinct overdensities and filaments of galaxies can be identified that are apparently uncorrelated with the Galactic foreground extinction hence the probable signature of extragalactic large-scale structures. The catalog (the first in a series of five) build the basis for various spectroscopic and photometric follow-up programs.

*Acknowledgements.* Foremost, I would like to thank P.A. Woudt for many stimulating discussions. Without the prestation of the old blinking machine from the Astronomical Institute of Basel and the donation of the surveyed film copies of the ESO/SRC survey by ESO, this project could not have been pursued. The derivations of positions, diameters and mag-nitudes for the visually detected galaxies on F213 from COSMOS scans by H. MacGillivray were extremely valuable for the analysis. The enthousiastic collaborations with my colleagues C. Balkowski, V. Cayatte, A.P. Fairall, P.A. Henning and P.A. Woudt in the various redshift follow-ups programs is greatly appreciated.

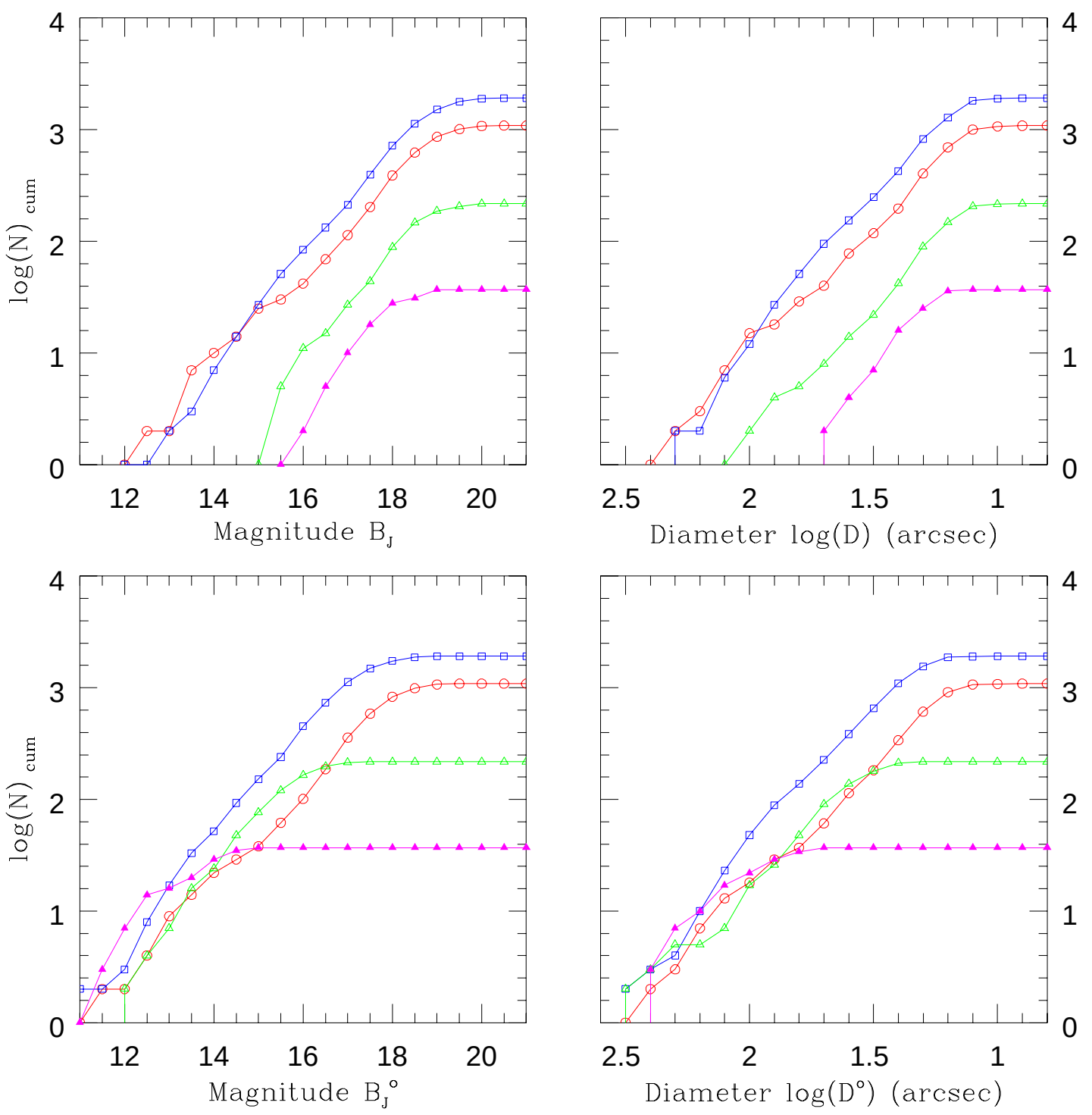

**Fig. 6.** The cumulative distribution of observed (top panels) and extinction-corrected (bottom) magnitudes (left) and diameters (right) for four different intervals of galactic foreground extinction. The open circles display the galaxies with $A_B \leq 1^m$, the squares are galaxies with $1^m < A_B \leq 2^m$, the triangles correspond to galaxies with $2^m < A_B \leq 3^m$, and the filled triangles are galaxies with $3^m < A_B \leq 4^m$.

## References

Arp H.C., Madore B.F. 1987, A Catalog of Southern Peculiar Galaxies and Associations, Cambridge: Cambridge University Press

Burstein, D., Davies, R.L., Dressler, A., et al. 1987, ApJS 64, 601

Buta R. 1995, The Catalog of Southern Ringed Galaxies, ApJS 96, 39

Cameron L.M. 1990, A&A 233, 16

Cardelli J.A., Clayton G.C., Mathis J.S. 1989, ApJ 345, 245

Fairall A.P., Woudt P.A., Kraan-Korteweg R.C. 1998, A&ASS 127, 463

Felenbok P., Guérin J., Fernandez A., et al. 1997, Experimental Astronomy 7, 65

Henning P.A., Kraan-Korteweg R.C., Rivers A.J. et al. 1998, AJ 115, 584

Henning P.A., Staveley-Smith L., Kraan-Korteweg R.C., Sadler E.M. 1999, Proc. Astron. Soc. Aust. 16, 35

Hudson, M.J., Lynden-Bell, D. 1991, MNRAS 252, 219

Joint *IRAS* Science Working Group 1988, *IRAS* Point Source Catalog, Version 2 (Washington: US Govt. Printing Office) (IRAS PSC)

Karachentsev I.D., Karachentseva V.E., Parnovsky S.L. 1993, Flat Galaxy Catalog, ANac 314, 97

Kerr F.J., Westerhout G. 1965 in Galactic Structure, Chicago: Univ. of Chicago, p186

Kogut A., Lineweaver C., Smoot G.F. et al. 1993, ApJ 419, 1

Kolatt T., Dekel A., Lahav O. 1995, MNRAS 275, 797

Kraan-Korteweg R.C. 1989, in Reviews in Modern Astronomy 2, ed. G. Klare, Springer: Berlin, p119

Kraan-Korteweg, R.C. 1992, in Variable Stars and Galaxies, ed. B. Warner, ASP 15, p235

Kraan-Korteweg R.C., Woudt P.A. 1993, in $9^{th}$ IAP Astrophysics Meeting on "Cosmic Velocity Fields", eds. F. Bouchet & M. Lachièze-Rey, Ed. Frontières, Gif-sur-Yvette, p557

Kraan-Korteweg R.C., Woudt P.A. 1994, in Unveiling Large-Scale Structures behind the Milky Way, eds. C. Balkowski & R.C. Kraan-Korteweg, ASP 67, p89

Kraan-Korteweg R.C., Woudt P.A. 1999, Proc. Astron. Soc. Aust. 16, 53

Kraan-Korteweg R.C., Cayatte V., Fairall A.P. et al. 1994a, in Unveiling Large-Scale Structures behind the Milky Way, eds. C. Balkowski & R.C. Kraan-Korteweg, ASP 67, p99

Kraan-Korteweg R.C., Fairall A.P., Balkowski C. 1995, A&A 297, 617

Kraan-Korteweg R.C., Woudt P.A., Fairall A.P. et al. 1996 in XVth Moriond Astrophysics Meeting on "Clustering in the Universe", eds. J. Trân Thanh Vân et al. p71

Kraan-Korteweg, R.C., Woudt, P.A., Henning, P.A. 1997 Proc. Astron. Soc. Aust. 14, 15

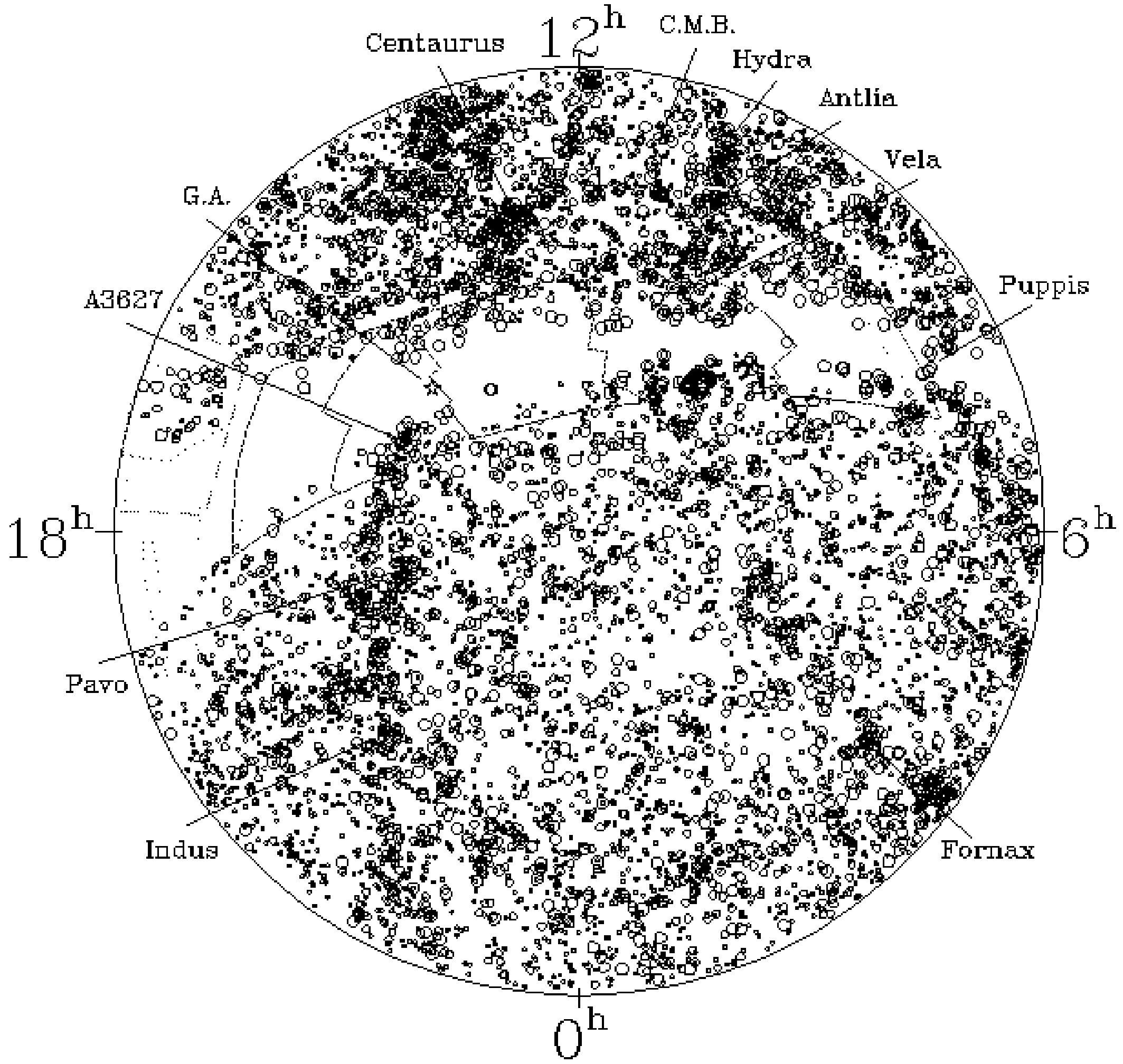

**Fig. 7.** An equal area distribution of all the Lauberts galaxies with extinction-corrected diameters $D^o \geq 1\farcm3$ in the southern sky ($\delta \leq -17.5°$), supplemented with galaxies from our the optical ZOA galaxy search following the same selection criterion. The galaxies are diameter-coded: the galaxies with $1\farcm3 \leq D^o < 2'$ are displayed as small, with $2' \leq D^o < 3'$ as middle and the galaxies with $D^o \geq 3'$ as large circles.

Kraan-Korteweg R.C., Koribalski B., Juraszek S. 1999, in ESO/ATNF Workshop on Looking Deep in the Southern Sky, eds. R. Morganti & W. Couch, Springer, p. 23

Lauberts A. 1982, The ESO/Uppsala Survey of the ESO (B) Atlas, ESO: Garching

Lauberts A., Valentijn E. 1989, The Surface Photometry Catalogue of the ESO/Uppsala Galaxies, Garching: ESO

Peebles P.J.E. 1994, ApJ 429, 43

Rivers, A., Henning, P.A., Kraan-Korteweg, R.C., 1999, Proc. Astron. Soc. Aust. 16, 48

Roman A.T., Saito M. 1998, Contributions from the Dept. of Astron., Kyoto University, No. 570

Roman A.T., Nakanishi K, Tomita A. et al. 1996, PASJ 48, 679

Saito M., Ohtani A., Baba A. et al. 1991, PASJ 43, 449

Schlegel D.J., Finkbeiner D.P., Davis M. 1998, ApJ 500, 525

Vaucouleurs G. de, Vaucouleurs A. de, Corwin H.G. 1976, $2^{nd}$ Reference Catalogue of Bright Galaxies (RC2), University of Texas Press, Austin

Wakamatsu K., Hasegawa T., Karoji H., et al. 1994, in Unveiling Large-Scale Structures behind the Milky Way, eds. C. Balkowski & R.C. Kraan-Korteweg, ASP 67, p131

Woudt P.A. 1998, PhD Thesis, University of Cape Town

Woudt P.A., Kraan-Korteweg R.C., Fairall A.P. 1999, A&A, in press

Wright A.E., Griffith M.R., Burke B.F., Ekers R.D. 1994, ApJS 91, 111

Yamada, T., Takata, T., Djamaluddin, T., et al. 1993, ApJS 89, 57

**Table 1.** Galaxies in the southern ZOA – I. The Hydra/Antlia Extension

| RKK Ident (1) | Other Ident (2) | IR (3) | R.A. (h m s) (4) | Dec. (° ′ ″) (5) | gal $\ell$ (°) (6) | gal $b$ (°) (7) | SRC (8) | X (mm) (9) | Y (mm) (10) | D x d (″) (11) | $B_J$ ($^m$) (12) | $N_{HI}$ (13) | u | Type class. (14) | o * | Remarks (15) |
|---|---|---|---|---|---|---|---|---|---|---|---|---|---|---|---|---|
| 1 | | | 08 30 42.4 | -57 28 39 | 273.40 | -10.55 | 165 | -128.7 | -135.4 | 23x 4 | 18.6 | .20 | | S 5 | E | |
| 2 | | | 08 30 44.7 | -57 25 27 | 273.36 | -10.51 | 165 | -128.6 | -132.5 | 23x 8 | 17.9 | .20 | | S 3 : | 1 | |
| 3 | | | 08 30 51.7 | -56 40 48 | 272.75 | -10.07 | 165 | -129.9 | -92.7 | 26x 3 | 19.3 | .21 | | S 7 | E | |
| 4 | | | 08 31 02.8 | -57 10 05 | 273.17 | -10.33 | 165 | -127.2 | -118.7 | 20x 4 | 19.0 | .19 | | S | E | |
| 5 | | | 08 31 04.5 | -55 56 28 | 272.16 | -9.62 | 165 | -129.7 | -53.6 | 23x 3 | 19.8 | .25 | | S 5 | E | |
| 6 | | | 08 31 04.9 | -57 28 35 | 273.43 | -10.51 | 165 | -126.0 | -135.2 | 34x 17 | 16.9 | .20 | | S 4 | S | |
| 7 | | | 08 31 07.9 | -57 30 56 | 273.47 | -10.53 | 165 | -125.6 | -137.3 | 15x 11 | 18.2 | .20 | | S | S | |
| 8 | | | 08 31 19.7 | -55 55 40 | 272.17 | -9.58 | 165 | -127.8 | -52.8 | 13x 13 | 18.2 | .24 | | S ? | F | ext. round |
| 9 | | | 08 31 31.2 | -55 39 54 | 271.97 | -9.40 | 165 | -127.2 | -38.6 | 13x 8 | 18.7 | .25 | | | 1 | |
| 10 | | | 08 31 37.3 | -56 58 46 | 273.06 | -10.16 | 165 | -123.5 | -108.4 | 13x 8 | 18.6 | .21 | ? | | | |
| 11 | | | 08 31 40.0 | -55 40 31 | 271.99 | -9.39 | 165 | -126.0 | -39.1 | 31x 11 | 17.3 | .25 | | E 3 | | |
| 12 | | | 08 31 44.2 | -55 45 08 | 272.06 | -9.43 | 165 | -125.3 | -43.2 | 13x 12 | 18.8 | .23 | | | | |
| 13 | | | 08 31 45.1 | -56 17 10 | 272.50 | -9.74 | 165 | -123.5 | -71.8 | 42x 4 | 18.1 | .27 | | | 1 | |
| 14 | | | 08 31 46.9 | -56 20 56 | 272.56 | -9.77 | 165 | -123.1 | -75.1 | 19x 8 | 18.3 | .27 | | E ? | | |
| 15 | | | 08 31 47.5 | -56 53 49 | 273.01 | -10.09 | 165 | -122.5 | -103.9 | 15x 4 | 19.2 | .22 | | S | | |
| 16 | | | 08 31 48.1 | -57 03 38 | 273.14 | -10.19 | 165 | -122.0 | -112.7 | 17x 8 | 18.4 | .20 | | L | | |
| 17 | | | 08 31 51.0 | -56 08 04 | 272.39 | -9.64 | 165 | -123.3 | -63.6 | 17x 12 | 18.2 | .25 | | S M | | |
| 18 | | | 08 31 52.3 | -53 54 37 | 270.56 | -8.33 | 165 | -129.7 | 55.5 | 17x 4 | 19.6 | .32 | | S ? | | |
| 19 | | | 08 31 52.8 | -57 09 38 | 273.23 | -10.24 | 165 | -121.2 | -118.0 | 17x 4 | 19.2 | .19 | | S | E | |
| 20 | | | 08 31 58.9 | -55 38 52 | 272.00 | -9.34 | 165 | -123.7 | -37.5 | 20x 12 | 17.9 | .24 | ? | S ? | S | |
| 21 | | | 08 32 01.5 | -54 59 27 | 271.46 | -8.95 | 165 | -125.3 | -2.3 | 13x 8 | 18.8 | .39 | ? | | | smooth |
| 22 | | | 08 32 06.9 | -55 45 11 | 272.09 | -9.39 | 165 | -122.4 | -43.1 | 40x 11 | 17.2 | .22 | | S E | | |
| 23 | | | 08 32 08.4 | -57 12 04 | 273.29 | -10.23 | 165 | -119.2 | -120.1 | 27x 27 | 16.4 | .19 | | | | |
| 24 | | | 08 32 09.9 | -53 55 46 | 270.60 | -8.31 | 165 | -127.3 | 54.6 | 17x 13 | 17.7 | .34 | | E ? | | |
| 25 | | | 08 32 09.9 | -55 45 05 | 272.10 | -9.38 | 165 | -122.8 | -42.4 | 13x 5 | 19.1 | .22 | | | | |
| 26 | | | 08 32 11.8 | -57 01 27 | 273.15 | -10.12 | 165 | -119.2 | -110.6 | 17x 5 | 19.1 | .21 | | S | | |
| 27 | | | 08 32 14.2 | -53 56 35 | 270.62 | -8.31 | 165 | -126.7 | 53.9 | 13x 5 | 19.2 | .34 | | S ? | | |
| 28 | | | 08 32 14.6 | -55 34 23 | 271.96 | -9.27 | 165 | -122.7 | -32.8 | 13x 7 | 19.0 | .24 | | | | |
| 29 | | | 08 32 15.5 | -55 36 17 | 271.98 | -9.28 | 165 | -121.4 | -39.6 | 16x 5 | 19.1 | .24 | | S | | |
| 30 | | | 08 32 23.0 | -55 31 21 | 271.93 | -9.22 | 165 | -121.0 | -30.6 | 13x 5 | 19.4 | .24 | | | | |
| 31 | | | 08 32 29.4 | -57 27 24 | 273.53 | -10.34 | 165 | -116.0 | -133.6 | 13x 3 | 19.5 | .20 | | S | E | |
| 32 | | | 08 32 32.9 | -55 38 46 | 272.04 | -9.28 | 165 | -120.2 | -36.6 | 11x 5 | 19.5 | .23 | | S | E1 | |
| 33 | | | 08 32 36.5 | -55 49 36 | 272.20 | -9.38 | 165 | -118.5 | -46.8 | 15x 4 | 19.0 | .22 | | | | |
| 34 | | | 08 32 37.4 | -55 32 19 | 271.96 | -9.20 | 165 | -119.2 | -31.4 | 15x 12 | 18.0 | .23 | | E | | |
| 35 | | | 08 32 37.8 | -55 01 18 | 271.54 | -8.90 | 165 | -120.6 | -3.7 | 13x 3 | 20.3 | .36 | | S | E | |
| 36 | | | 08 32 40.5 | -55 39 46 | 272.07 | -9.27 | 165 | -119.2 | -37.5 | 8x 7 | 19.4 | .22 | | E ? | | |
| 37 | | | 08 32 42.2 | -54 00 30 | 270.72 | -8.29 | 165 | -122.8 | 50.6 | 15x 11 | 18.1 | .44 | ? | | | |
| 38 | | P | 08 32 43.2 | -55 35 50 | 272.02 | -9.23 | 165 | -118.3 | -34.5 | 23x 7 | 18.3 | .22 | | S M | | |
| 39 | | | 08 32 43.7 | -55 21 30 | 271.82 | -9.09 | 165 | -118.9 | -21.7 | 17x 15 | 17.6 | .26 | | S M: | | |
| 40 | | | 08 32 48.7 | -55 44 16 | 272.14 | -9.30 | 165 | -117.2 | -42.0 | 20x 3 | 19.2 | .22 | | S | E | |
| 41 | | | 08 32 59.4 | -55 16 45 | 271.78 | -9.01 | 165 | -117.1 | -17.3 | 13x 7 | 18.9 | .27 | | | | |
| 42 | | | 08 33 00.5 | -56 15 52 | 272.59 | -9.59 | 165 | -114.3 | -70.1 | 17x 5 | 19.2 | .22 | | | | blg or s.p.* |
| 43 | | | 08 33 02.9 | -53 29 31 | 270.33 | -7.95 | 165 | -121.5 | 78.4 | 15x 7 | 18.8 | .30 | ? | | S | |
| 44 | | | 08 33 08.3 | -54 56 36 | 271.52 | -8.79 | 165 | -116.9 | .7 | 23x 20 | 17.2 | .38 | ? | S ? | | |
| 45 | | | 08 33 08.5 | -55 35 59 | 272.05 | -9.18 | 165 | -115.1 | -34.5 | 17x 3 | 19.9 | .22 | | S ? | | |
| 46 | | | 08 33 09.1 | -55 26 20 | 271.92 | -9.09 | 165 | -115.4 | -25.8 | 16x 16 | 18.1 | .26 | ? | S 5 | F | |
| 47 | | | 08 33 10.3 | -53 47 43 | 270.58 | -8.11 | 165 | -119.7 | 62.2 | 16x 5 | 19.3 | .32 | | | 1 | |
| 48 | | | 08 33 11.1 | -55 34 00 | 272.03 | -9.16 | 165 | -114.9 | -32.7 | 16x 11 | 18.2 | .22 | ? | | | |
| 49 | | | 08 33 17.5 | -56 45 04 | 273.01 | -9.84 | 165 | -110.9 | -96.1 | 17x 15 | 17.8 | .21 | | S 5 : | F | |
| 50 | | | 08 33 18.3 | -54 00 56 | 270.78 | -8.23 | 165 | -119.9 | 46.0 | 16x 7 | 18.6 | .51 | ? | S M | | |
| 51 | | | 08 33 18.7 | -55 38 56 | 272.11 | -9.19 | 165 | -113.7 | -37.0 | 20x 16 | 17.5 | .22 | | E ? | | |
| 52 | | | 08 33 21.6 | -55 05 40 | 271.66 | -8.86 | 165 | -114.8 | -7.3 | 16x 13 | 18.1 | .32 | | S | F | |
| 53 | | | 08 33 24.4 | -52 22 07 | 269.45 | -7.24 | 165 | -121.6 | 138.7 | 19x 16 | 18.2 | .65 | | S | F | LSB |
| 54 | | | 08 33 26.5 | -55 40 02 | 272.13 | -9.19 | 165 | -112.6 | -38.0 | 17x 12 | 18.0 | .22 | | S E: | | w.comp. |
| 55 | | | 08 33 28.5 | -53 04 42 | 270.03 | -7.65 | 165 | -119.2 | 100.7 | 22x 5 | 18.9 | .48 | | S | E | |
| 56 | | | 08 33 29.8 | -55 44 16 | 272.20 | -9.22 | 165 | -112.0 | -41.7 | 13x 9 | 18.6 | .22 | | S | V | |
| 57 | | | 08 33 32.7 | -53 37 48 | 270.48 | -7.97 | 165 | -117.2 | 71.2 | 15x 8 | 18.8 | .30 | | | | |
| 58 | | | 08 33 35.2 | -56 27 07 | 272.79 | -9.63 | 165 | -109.5 | -79.9 | 22x 20 | 16.8 | .22 | | S 3 | | |
| 59 | L165-001 | | 08 33 36.1 | -57 28 22 | 273.63 | -10.23 | 165 | -107.9 | -134.1 | 141x 34 | 14.8 | .20 | | S 2 | S | br.* s.p. |
| 60 | | I | 08 33 37.1 | -55 53 08 | 272.33 | -9.30 | 165 | -110.7 | -49.6 | 16x 16 | 17.4 | .22 | ? | E ? | | between ** |
| 61 | | | 08 33 37.7 | -56 59 22 | 273.23 | -9.94 | 165 | -107.7 | -108.7 | 20x 7 | 18.6 | .20 | | | 1 | |
| 62 | | | 08 33 38.0 | -56 18 00 | 272.67 | -9.54 | 165 | -109.5 | -71.8 | 22x 3 | 19.2 | .22 | | S 5 | E | |
| 63 | | | 08 33 39.9 | -53 23 42 | 270.30 | -7.82 | 165 | -116.8 | 83.8 | 20x 4 | 19.4 | .36 | ? | S | E | |
| 64 | | | 08 33 43.8 | -56 04 07 | 272.49 | -9.39 | 165 | -109.4 | -59.4 | 17x 9 | 18.4 | .22 | ? | | | |
| 65 | | | 08 33 44.3 | -55 33 03 | 272.06 | -9.08 | 165 | -110.7 | -31.6 | 20x 4 | 18.8 | .23 | | S | 1 | |
| 66 | | | 08 33 45.9 | -53 26 27 | 270.35 | -7.83 | 165 | -115.9 | 81.4 | 15x 7 | 18.7 | .35 | | | | |
| 67 | | I | 08 33 46.0 | -55 00 22 | 271.62 | -8.76 | 165 | -111.9 | -2.4 | 50x 34 | 15.7 | .33 | | S 0 | S | |
| 68 | | | 08 33 47.8 | -56 49 13 | 273.11 | -9.83 | 165 | -107.0 | -99.6 | 47x 7 | 17.4 | .20 | | S M | S | |
| 69 | | | 08 33 53.9 | -57 34 16 | 273.73 | -10.26 | 165 | -105.5 | -139.3 | 30x 4 | 18.6 | .19 | | S M | E | |
| 70 | | | 08 33 55.5 | -57 02 03 | 273.29 | -9.94 | 165 | -105.5 | -111.0 | 16x 5 | 19.3 | .19 | | S ? | | |
| 71 | | | 08 33 58.7 | -55 20 54 | 271.92 | -8.94 | 165 | -109.4 | -20.7 | 13x 7 | 18.7 | .28 | | | | |
| 72 | | | 08 34 00.5 | -56 46 55 | 273.09 | -9.78 | 165 | -105.5 | -97.5 | 24x 4 | 18.7 | .20 | | S | 1 | |
| 73 | | | 08 34 01.8 | -57 21 43 | 273.57 | -10.12 | 165 | -105.1 | -128.1 | 17x 16 | 18.2 | .19 | | S | F | |
| 74 | | | 08 34 02.0 | -55 56 33 | 272.41 | -9.28 | 165 | -107.5 | -52.5 | 17x 9 | 18.2 | .22 | | SX M | | |
| 75 | | | 08 34 02.0 | -57 25 24 | 273.62 | -10.15 | 165 | -104.9 | -131.4 | 13x 8 | 18.6 | .19 | | | | |
| 76 | | | 08 34 02.7 | -55 32 15 | 272.08 | -9.04 | 165 | -108.4 | -30.8 | 15x 12 | 18.0 | .24 | ? | | | |
| 77 | | | 08 34 04.8 | -53 42 17 | 270.59 | -7.95 | 165 | -112.8 | 67.4 | 23x 7 | 18.1 | .34 | | S L: | ES | |
| 78 | | | 08 34 07.5 | -55 57 49 | 272.43 | -9.28 | 165 | -106.7 | -53.6 | 15x 4 | 19.1 | .22 | | S ? | | |
| 79 | | NG | 08 34 11.1 | -54 33 08 | 271.29 | -8.44 | 165 | -109.8 | 22.0 | 13x 5 | 19.6 | .40 | | | | |
| 80 | | | 08 34 14.8 | -55 23 16 | 271.97 | -8.93 | 165 | -107.3 | -22.7 | 13x 5 | 19.6 | .27 | | S ? | | |
| 81 | | | 08 34 15.6 | -55 29 51 | 272.07 | -8.99 | 165 | -106.9 | -28.6 | 40x 16 | 16.8 | .25 | | S E: | | |
| 82 | | | 08 34 15.9 | -53 39 11 | 270.57 | -7.90 | 165 | -111.4 | 70.2 | 17x 16 | 17.6 | .34 | | E 1 | | |
| 83 | | | 08 34 16.5 | -53 25 06 | 270.38 | -7.76 | 165 | -111.9 | 82.8 | 20x 13 | 18.2 | .39 | ? | | | |
| 84 | | | 08 34 16.6 | -54 26 27 | 271.21 | -8.36 | 165 | -109.4 | 28.0 | 30x 7 | 18.2 | .44 | | S L | | |
| 85 | | | 08 34 19.0 | -53 38 47 | 270.56 | -7.89 | 165 | -111.0 | 70.6 | 12x 11 | 18.6 | .34 | | | | |
| 86 | | | 08 34 19.6 | -53 17 19 | 270.28 | -7.67 | 165 | -111.8 | 89.8 | 13x 5 | 19.5 | .44 | | S | E | |
| 87 | | I | 08 34 19.6 | -54 32 39 | 271.29 | -8.42 | 165 | -108.7 | 22.5 | 47x 23 | 16.2 | .40 | | F | | |
| 88 | | | 08 34 20.0 | -53 18 55 | 270.30 | -7.69 | 165 | -111.7 | 88.4 | 17x 8 | 18.6 | .43 | | S M: | | |
| 89 | | | 08 34 22.2 | -57 23 40 | 273.63 | -10.10 | 165 | -102.5 | -129.7 | 16x 11 | 17.9 | .19 | | S M | | cl.to br.* |
| 90 | | | 08 34 22.8 | -52 35 28 | 269.72 | -7.25 | 165 | -113.1 | 127.2 | 16x 7 | 19.3 | .80 | ? | | | |
| 91 | | | 08 34 24.7 | -53 45 49 | 270.67 | -7.95 | 165 | -110.0 | 64.4 | 20x 8 | 18.5 | .34 | | | | |
| 92 | | | 08 34 25.2 | -53 05 56 | 270.13 | -7.55 | 165 | -111.5 | 100.0 | 22x 8 | 18.2 | .47 | | S M | | |
| 93 | | | 08 34 26.3 | -56 43 52 | 273.09 | -9.70 | 165 | -102.5 | -94.6 | 16x 15 | 17.7 | .19 | | E 1 ? | | |
| 94 | | | 08 34 26.6 | -55 50 32 | 272.36 | -9.18 | 165 | -104.6 | -47.0 | 17x 7 | 18.1 | .20 | | S | | |
| 95 | | | 08 34 26.8 | -57 17 28 | 273.55 | -10.03 | 165 | -102.2 | -124.2 | 13x 5 | 19.1 | .18 | | | | |
| 96 | | | 08 34 28.5 | -55 35 28 | 272.16 | -9.02 | 165 | -105.0 | -33.5 | 22x 12 | 17.7 | .22 | | S ? | | dust, w.comp. |
| 97 | | I | 08 34 29.2 | -57 08 18 | 273.43 | -9.94 | 165 | -101.1 | -116.4 | 38x 23 | 16.1 | .18 | | | | i.a.dbl syst. |
| 98 | | | 08 34 29.6 | -55 50 00 | 272.36 | -9.17 | 165 | -104.3 | -46.5 | 34x 11 | 17.1 | .20 | | E ? | | |
| 99 | | | 08 34 30.2 | -53 11 27 | 270.21 | -7.59 | 165 | -110.6 | 95.1 | 13x 9 | 19.0 | .45 | | E ? | | |
| 100 | | | 08 34 34.5 | -56 47 44 | 273.15 | -9.73 | 165 | -101.3 | -98.0 | 20x 7 | 18.2 | .19 | | | 1 | |

**Table 1.** Galaxies in the southern ZOA – I. The Hydra/Antlia Extension (continued)

| RKK Ident (1) | Other Ident (2) | IR (3) | R.A. (h m s) (4) | Dec. (° ′ ″) (5) | gal ℓ (°) (6) | gal $b$ (°) (7) | SRC (8) | X (mm) (9) | Y (mm) (10) | D x d (″) (11) | $B_J$ ($^m$) (12) | $N_{HI}$ (13) | u | Type class. (14) | o | * | Remarks (15) |
|---|---|---|---|---|---|---|---|---|---|---|---|---|---|---|---|---|---|
| 101 | | | 08 34 40.5 | -54 15 12 | 271.09 | -8.21 | 165 | -106.7 | 38.2 | 20x 16 | 17.5 | .59 | | E ? | | | w.2 comp. |
| 102 | | | 08 34 40.8 | -55 37 04 | 272.20 | -9.02 | 165 | -103.4 | -34.9 | 16x 4 | 19.1 | .22 | | S | E | 1 | |
| 103 | | | 08 34 44.2 | -55 55 56 | 272.46 | -9.20 | 165 | -102.2 | -51.7 | 13x 11 | 18.5 | .22 | | | | 1 | |
| 104 | | | 08 34 44.3 | -55 36 41 | 272.20 | -9.01 | 165 | -103.0 | -34.5 | 20x 11 | 17.9 | .22 | | | | | |
| 105 | | | 08 34 45.9 | -57 16 39 | 273.56 | -9.99 | 165 | -100.0 | -123.3 | 22x 5 | 18.5 | .18 | | S | | | |
| 106 | | | 08 34 48.9 | -52 18 12 | 269.52 | -7.02 | 165 | -110.2 | 142.7 | 15x 12 | 18.3 | .67 | | S ? | F | | |
| 107 | | Q | 08 34 49.9 | -55 30 38 | 272.12 | -8.94 | 165 | -102.5 | -29.1 | 19x 9 | 18.1 | .24 | ? | | | | |
| 108 | | | 08 34 50.8 | -57 23 38 | 273.66 | -10.05 | 125 | -125.7 | 137.5 | 16x 11 | 18.4 | .18 | | | | 1 | |
| 109 | | | 08 34 52.6 | -53 59 30 | 270.89 | -8.03 | 165 | -105.8 | 52.3 | 27x 16 | 17.2 | .47 | | S E: | | | |
| 110 | | | 08 34 53. | -57 23.4 : | 273.66 | -10.04 | 165 | -98.8 | -129.5 | 13x 5 | 19.1 | .18 | | | | | |
| 111 | | | 08 34 53.4 | -56 53 58 | 273.26 | -9.75 | 165 | -98.7 | -103.5 | 15x 15 | 17.9 | .19 | | E ? | | | |
| 112 | | | 08 34 57.7 | -55 55 10 | 272.47 | -9.16 | 165 | -100.6 | -50.9 | 13x 8 | 18.4 | .21 | ? | S ? | | | |
| 113 | | | 08 34 59. | -57 09.7 : | 273.49 | -9.90 | 165 | -97.4 | -117.5 | 16x 5 | 18.9 | .18 | | | | | |
| 114 | | | 08 34 59.2 | -55 35 42 | 272.21 | -8.97 | 165 | -101.1 | -33.5 | 24x 13 | 17.7 | .22 | | S | | 1 | |
| 115 | | | 08 35 02.4 | -55 39 22 | 272.26 | -9.00 | 165 | -100.6 | -36.8 | 15x 11 | 18.2 | .21 | | E ? | | | w.sev.comp. |
| 116 | | | 08 35 02.7 | -56 19 54 | 272.81 | -9.40 | 165 | -99.0 | -73.0 | 19x 7 | 18.3 | .20 | | | | | |
| 117 | | | 08 35 03.1 | -56 47 44 | 273.19 | -9.67 | 165 | -97.8 | -97.9 | 24x 3 | 19.0 | .19 | | S | E | | |
| 118 | | | 08 35 03.1 | -57 08 53 | 273.48 | -9.88 | 165 | -97.0 | -116.7 | 11x 9 | 18.7 | .18 | | E ? | | | |
| 119 | | | 08 35 05.1 | -57 05 27 | 273.44 | -9.84 | 165 | -96.9 | -113.7 | 13x 13 | 18.6 | .18 | | | | | |
| 120 | | | 08 35 06.1 | -55 14 46 | 271.93 | -8.75 | 165 | -101.1 | -14.8 | 13x 7 | 19.0 | .27 | | | | | |
| 121 | | | 08 35 07.7 | -55 49 59 | 272.41 | -9.09 | 165 | -99.5 | -46.3 | 27x 20 | 16.9 | .20 | | S | F | S | |
| 122 | | | 08 35 07.7 | -57 09 07 | 273.49 | -9.88 | 165 | -96.4 | -116.9 | 20x 5 | 18.6 | .19 | | S | E | | in halo of 124 |
| 123 | | | 08 35 08.7 | -55 44 43 | 272.34 | -9.04 | 165 | -99.6 | -41.6 | 11x 8 | 18.9 | .20 | | E ? | | S | |
| 124 | | | 08 35 09.7 | -57 09 03 | 273.49 | -9.87 | 165 | -96.2 | -116.9 | 38x 31 | 16.0 | .19 | | E | | | |
| 125 | | | 08 35 11.7 | -55 37 31 | 272.25 | -8.96 | 165 | -99.5 | -35.1 | 8x 5 | 19.9 | .22 | | | | | |
| 126 | | | 08 35 13.6 | -55 38 39 | 272.27 | -8.97 | 165 | -99.2 | -36.1 | 13x 9 | 18.5 | .21 | | S ? | | | |
| 127 | | | 08 35 14.7 | -55 39 22 | 272.28 | -8.98 | 165 | -99.0 | -36.7 | 16x 13 | 18.1 | .21 | | S | | | |
| 128 | | | 08 35 16.2 | -55 44 16 | 272.35 | -9.02 | 165 | -98.7 | -41.1 | 16x 9 | 18.4 | .20 | | | | | |
| 129 | | | 08 35 16.2 | -57 02 00 | 273.40 | -9.79 | 165 | -95.7 | -110.5 | 27x 9 | 17.7 | .18 | | S E | | | |
| 130 | | | 08 35 16.4 | -55 42 30 | 272.32 | -9.00 | 165 | -98.7 | -39.5 | 15x 11 | 18.2 | .20 | | S ? | | S | |
| 131 | | | 08 35 18.5 | -55 45 45 | 272.37 | -9.03 | 165 | -98.3 | -42.4 | 13x 9 | 18.6 | .20 | | E | | | |
| 132 | | | 08 35 19.4 | -55 38 36 | 272.27 | -8.96 | 165 | -98.5 | -36.0 | 17x 7 | 18.7 | .21 | | S ? | | | |
| 133 | | | 08 35 19.6 | -54 18 28 | 271.19 | -8.16 | 165 | -101.5 | 35.5 | 13x 5 | 19.2 | .54 | | | | | |
| 134 | | | 08 35 24.9 | -57 09 33 | 273.52 | -9.85 | 165 | -94.3 | -117.2 | 16x 11 | 18.3 | .19 | | | F | | |
| 135 | | | 08 35 26.4 | -55 30 03 | 272.17 | -8.86 | 165 | -97.9 | -28.4 | 20x 20 | 17.1 | .24 | | E 0 | | S | |
| 136 | | | 08 35 26.5 | -55 37 32 | 272.27 | -8.93 | 165 | -97.6 | -35.0 | 7x 7 | 19.4 | .21 | | | | | |
| 137 | | | 08 35 26.7 | -55 37 52 | 272.27 | -8.94 | 165 | -97.6 | -35.3 | 13x 12 | 18.0 | .21 | | L | | | |
| 138 | | | 08 35 27.1 | -54 41 15 | 271.51 | -8.37 | 165 | -99.7 | 15.2 | 22x 11 | 18.1 | .31 | | S | | S | |
| 139 | | | 08 35 27.4 | -55 37 21 | 272.27 | -8.93 | 165 | -97.5 | -34.9 | 11x 9 | 18.7 | .21 | | | | | |
| 140 | | | 08 35 29.5 | -56 55 55 | 273.34 | -9.71 | 165 | -94.3 | -105.0 | 13x 5 | 18.8 | .19 | | S | | | |
| 141 | | | 08 35 31.4 | -56 10 45 | 272.73 | -9.25 | 165 | -95.8 | -64.7 | 13x 7 | 18.9 | .20 | | S ? | | | |
| 142 | | | 08 35 32.6 | -56 18 24 | 272.83 | -9.33 | 165 | -95.3 | -71.5 | 40x 27 | 16.4 | .19 | | SB 5 | | 1 | |
| 143 | | | 08 35 35.4 | -55 38 36 | 272.30 | -8.93 | 165 | -96.5 | -35.9 | 9x 5 | 19.8 | .21 | | | | | |
| 144 | | | 08 35 36.7 | -57 00 00 | 273.40 | -9.73 | 165 | -93.2 | -108.6 | 13x 3 | 19.8 | .19 | | S | E | | |
| 145 | | | 08 35 37.7 | -55 47 46 | 272.42 | -9.02 | 165 | -95.8 | -44.1 | 13x 7 | 18.9 | .21 | | | | | |
| 146 | | | 08 35 37.8 | -55 37 16 | 272.28 | -8.91 | 165 | -96.2 | -34.7 | 17x 11 | 17.8 | .21 | | L | | | or S..E |
| 147 | | | 08 35 38.9 | -55 38 46 | 272.30 | -8.92 | 165 | -96.0 | -36.1 | 15x 4 | 19.3 | .21 | | S M | E | | |
| 148 | | | 08 35 39.7 | -57 14 27 | 273.61 | -9.87 | 165 | -93.5 | -121.1 | 12x 7 | 19.1 | .20 | | S ? | | | |
| 149 | | | 08 35 39.8 | -56 59 08 | 273.40 | -9.72 | 165 | -92.9 | -107.8 | 13x 11 | 18.1 | .19 | | E | | | w.comp. |
| 150 | | | 08 35 40.7 | -55 18 43 | 272.04 | -8.72 | 165 | -96.5 | -18.2 | 13x 5 | 19.0 | .27 | ? | | | 1 | |
| 151 | | | 08 35 40.7 | -55 41 52 | 272.35 | -8.95 | 165 | -95.2 | -42.6 | 31x 8 | 17.5 | .21 | | S 1 | | | cl.to br.* |
| 152 | | | 08 35 41.7 | -55 37 53 | 272.30 | -8.91 | 165 | -95.7 | -35.3 | 16x 5 | 18.9 | .21 | | S E | | | |
| 153 | | | 08 35 41.9 | -55 24 24 | 272.11 | -8.78 | 165 | -96.2 | -23.2 | 16x 13 | 18.0 | .26 | | S 3 | | | |
| 154 | | | 08 35 42.0 | -57 04 46 | 273.48 | -9.77 | 165 | -92.4 | -112.9 | 24x 8 | 17.8 | .19 | | S | E | | |
| 155 | | | 08 35 44.5 | -55 42 52 | 272.37 | -8.95 | 165 | -94.7 | -43.5 | 15x 9 | 18.6 | .21 | | E | | | w.comp. |
| 156 | | | 08 35 48.2 | -56 03 14 | 272.65 | -9.15 | 165 | -93.9 | -57.9 | 16x 13 | 18.0 | .20 | | E ? | | | |
| 157 | | | 08 35 51.2 | -55 21 04 | 272.08 | -8.72 | 165 | -95.1 | -20.2 | 13x 7 | 19.4 | .26 | | S | | | |
| 158 | | | 08 35 53.8 | -55 18 02 | 272.04 | -8.69 | 165 | -94.9 | -17.5 | 16x 7 | 18.6 | .27 | | S | | | |
| 159 | | | 08 35 59.6 | -57 03 48 | 273.49 | -9.73 | 165 | -90.3 | -111.9 | 23x 16 | 17.6 | .19 | | | | | |
| 160 | | | 08 36 00.8 | -57 04 51 | 273.50 | -9.74 | 165 | -90.1 | -112.8 | 20x 13 | 18.2 | .19 | | S L ? | | | |
| 161 | | | 08 36 04.1 | -55 44 20 | 272.42 | -8.93 | 165 | -92.6 | -40.9 | 27x 15 | 17.2 | .21 | | | | | |
| 162 | L165-002 | I | 08 36 05.0 | -54 56 51 | 271.77 | -8.46 | 165 | -94.2 | 1.5 | 242x215 | 12.0 | .30 | | L | | S | |
| 163 | | | 08 36 05.6 | -55 46 57 | 272.45 | -8.95 | 165 | -92.4 | -43.3 | 40x 22 | 16.4 | .21 | | E | | | p.cov.by *, w.comp. |
| 164 | | | 08 36 06.0 | -55 42 09 | 272.39 | -8.91 | 165 | -92.5 | -39.0 | 11x 8 | 19.0 | .21 | | S ? | | | |
| 165 | | | 08 36 08.0 | -55 24 47 | 272.16 | -8.73 | 165 | -92.9 | -23.4 | 31x 24 | 16.6 | .26 | | F | | | |
| 166 | L165-003 | Q | 08 36 09.7 | -57 22 41 | 273.76 | -9.90 | 165 | -88.4 | -128.7 | 74x 16 | 16.1 | .19 | | S 3 | | | w.2 comp. |
| 167 | | | 08 36 11.2 | -55 26 21 | 272.18 | -8.74 | 165 | -92.4 | -24.8 | 16x 11 | 18.2 | .26 | | E | | | |
| 168 | | | 08 36 11.4 | -57 17 01 | 273.68 | -9.84 | 165 | -89.6 | -123.3 | 11x 4 | 19.9 | .20 | | S ? | | | |
| 169 | | | 08 36 11.5 | -55 56 50 | 272.60 | -9.04 | 165 | -91.3 | -52.1 | 17x 8 | 18.3 | .21 | | E ? | | | |
| 170 | | | 08 36 14.2 | -56 36 51 | 273.14 | -9.43 | 165 | -89.5 | -87.8 | 24x 17 | 17.0 | .19 | | F | | S | |
| 171 | | P | 08 36 15.0 | -56 56 14 | 273.41 | -9.63 | 165 | -88.7 | -105.1 | 13x 7 | 18.9 | .19 | | S ? | | | |
| 172 | | | 08 36 15.1 | -54 36 24 | 271.51 | -8.23 | 165 | -93.6 | 19.8 | 13x 7 | 19.2 | .34 | | | | | |
| 173 | | | 08 36 15.4 | -55 25 26 | 272.18 | -8.72 | 165 | -91.9 | -24.0 | 15x 5 | 19.1 | .26 | | S | | | |
| 174 | | | 08 36 16.4 | -55 42 05 | 272.40 | -8.89 | 165 | -91.2 | -38.9 | 13x 11 | 18.4 | .20 | | S ? | | | dist.blg |
| 175 | | | 08 36 16.4 | -57 07 50 | 273.57 | -9.74 | 165 | -88.1 | -115.4 | 20x 8 | 18.1 | .20 | | | | | |
| 176 | | | 08 36 18.8 | -57 07 34 | 273.56 | -9.73 | 165 | -87.9 | -115.2 | 11x 8 | 18.8 | .19 | | | | | neighb.of 175 |
| 177 | | | 08 36 19.1 | -54 36 48 | 271.52 | -8.23 | 165 | -93.1 | 19.5 | 13x 7 | 19.2 | .34 | | | | | |
| 178 | | | 08 36 20.0 | -57 22 50 | 273.77 | -9.88 | 165 | -87.2 | -128.8 | 15x 7 | 18.6 | .19 | | I ? | | | |
| 179 | | | 08 36 20.9 | -57 12 31 | 273.64 | -9.78 | 165 | -87.4 | -119.6 | 13x 13 | 18.2 | .20 | | | | | w.2 comp. |
| 180 | | | 08 36 21.5 | -57 13 52 | 273.65 | -9.79 | 165 | -88.5 | -120.4 | 11x 4 | 19.9 | .20 | | S | | | dist.blg |
| 181 | | | 08 36 24.4 | -57 22 47 | 273.78 | -9.87 | 165 | -86.6 | -128.7 | 15x 4 | 19.3 | .19 | | S | E | | |
| 182 | | | 08 36 26.6 | -57 23 42 | 273.79 | -9.88 | 125 | -114.2 | 138.0 | 13x 11 | 18.9 | .19 | | | | | |
| 183 | | | 08 36 29.8 | -56 07 26 | 272.76 | -9.11 | 165 | -88.6 | -61.4 | 16x 5 | 19.1 | .21 | | I ? | | | |
| 184 | | | 08 36 30.4 | -55 12 07 | 272.02 | -8.56 | 165 | -90.4 | -12.0 | 34x 11 | 17.2 | .28 | | S E | | | larger LSBdisk? |
| 185 | | | 08 36 30.9 | -53 49 03 | 270.90 | -7.73 | 165 | -93.2 | 62.2 | 27x 17 | 17.6 | .33 | ? | | | | LSB |
| 186 | | | 08 36 32.9 | -57 19 32 | 273.75 | -9.82 | 165 | -86.9 | -125.4 | 12x 5 | 19.6 | .20 | | S M | | | |
| 187 | | | 08 36 34.1 | -55 36 20 | 272.35 | -8.80 | 165 | -89.2 | -33.6 | 16x 7 | 18.6 | .21 | | | | | |
| 188 | | | 08 36 34.5 | -55 34 02 | 272.32 | -8.77 | 165 | -89.2 | -31.6 | 13x 11 | 18.6 | .22 | | S ? | F | | |
| 189 | | | 08 36 35.6 | -57 15 31 | 273.70 | -9.78 | 165 | -85.5 | -122.2 | 27x 20 | 16.9 | .20 | | | | S | w.2 comp. |
| 190 | | | 08 36 36.3 | -53 18 07 | 270.49 | -7.41 | 165 | -93.5 | 89.8 | 13x 5 | 19.4 | .50 | | | | | |
| 191 | | | 08 36 36.9 | -55 12 57 | 272.04 | -8.56 | 165 | -89.6 | -12.7 | 36x 8 | 17.4 | .27 | | S E | | 1 | |
| 192 | | | 08 36 37.3 | -56 48 47 | 273.34 | -9.51 | 165 | -86.3 | -98.3 | 38x 28 | 16.3 | .20 | | S 3 | | S | |
| 193 | | | 08 36 38.1 | -55 25 10 | 272.20 | -8.68 | 165 | -89.0 | -23.6 | 31x 11 | 17.4 | .26 | | S | | | |
| 194 | | | 08 36 41.3 | -55 58 11 | 272.66 | -9.00 | 165 | -87.5 | -53.1 | 34x 8 | 17.7 | .22 | | S M | | | |
| 195 | | | 08 36 42.3 | -54 23 52 | 271.38 | -8.06 | 165 | -90.6 | 31.1 | 13x 5 | 19.5 | .43 | ? | S ? | | | |
| 196 | | | 08 36 43.8 | -57 29 52 | 273.90 | -9.91 | 165 | -84.1 | -135.0 | 16x 7 | 18.3 | .19 | | S | | 1 | |
| 197 | | | 08 36 45.7 | -57 47 49 | 274.15 | -10.08 | 125 | -110.8 | 116.6 | 34x 17 | 16.9 | .17 | | E | | 1 | |
| 198 | | | 08 36 50.1 | -57 06 02 | 273.59 | -9.66 | 165 | -85.3 | -113.3 | 11x 7 | 19.3 | .20 | | | | | |
| 199 | | | 08 36 50.4 | -56 31 08 | 273.11 | -9.31 | 165 | -85.3 | -82.5 | 20x 13 | 17.6 | .20 | | | | S | |
| 200 | | | 08 36 51.8 | -55 43 55 | 272.48 | -8.84 | 165 | -86.7 | -40.3 | 16x 11 | 18.0 | .20 | | S | | | |

**Table 1.** Galaxies in the southern ZOA – I. The Hydra/Antlia Extension (continued)

| RKK Ident (1) | Other Ident (2) | IR (3) | R.A. (h m s) (4) | Dec. (° ′ ″) (5) | gal ℓ (°) (6) | gal $b$ (°) (7) | SRC (8) | X (mm) (9) | Y (mm) (10) | D x d (″) (11) | $B_J$ ($^m$) (12) | $N_{HI}$ (13) | u | Type class. (14) | | o | * | Remarks (15) |
|---|---|---|---|---|---|---|---|---|---|---|---|---|---|---|---|---|---|---|
| 201 | | | 08 36 51.8 | -57 11 16 | 273.66 | -9.71 | 165 | -84.9 | -118.0 | 7x 5 | 19.7 | .20 | | | | | | neighb.of 206 |
| 202 | | | 08 36 52.1 | -56 53 02 | 273.41 | -9.53 | 165 | -84.3 | -102.1 | 16x 8 | 18.5 | .20 | | | | | | |
| 203 | | | 08 36 53.0 | -55 29 30 | 272.28 | -8.69 | 165 | -87.7 | -27.1 | 11x 9 | 18.5 | .25 | | | | | | dist.blg, larger LSBdisk? |
| 204 | | | 08 36 54.8 | -55 41 39 | 272.45 | -8.81 | 165 | -86.4 | -38.3 | 16x 7 | 18.4 | .20 | | S | | | | |
| 205 | | | 08 36 56.8 | -55 40 43 | 272.44 | -8.80 | 165 | -86.1 | -37.4 | 13x 5 | 19.1 | .20 | | S | | | | |
| 206 | | | 08 36 56.9 | -57 11 28 | 273.67 | -9.70 | 165 | -80.6 | -121.9 | 15x 3 | 20.1 | .20 | | S | M | | E | | LSB |
| 207 | | | 08 36 57.3 | -55 17 26 | 272.13 | -8.56 | 165 | -87.5 | -16.3 | 9x 7 | 19.4 | .25 | | S | ? | | | | |
| 208 | | | 08 36 57.6 | -54 42 41 | 271.66 | -8.21 | 165 | -87.9 | 14.4 | 27x 24 | 17.1 | .31 | | S | M | | F | | |
| 209 | | | 08 36 58.8 | -55 29 29 | 272.29 | -8.68 | 165 | -87.0 | -27.0 | 15x 5 | 19.2 | .25 | | S | | | | |
| 210 | | | 08 36 59.3 | -57 10 16 | 273.66 | -9.68 | 165 | -84.0 | -117.1 | 12x 9 | 18.6 | .20 | | | | | | dbl syst, cl.to br.* |
| 211 | | | 08 37 01.2 | -57 06 04 | 273.60 | -9.64 | 165 | -83.9 | -113.3 | 11x 7 | 19.5 | .20 | | S | | E | | v.obs, LSB |
| 212 | | | 08 37 02.8 | -55 39 16 | 272.43 | -8.77 | 165 | -85.4 | -36.1 | 30x 4 | 18.7 | .21 | | S | M | E | | |
| 213 | | | 08 37 03.4 | -55 00 04 | 271.90 | -8.38 | 165 | -86.6 | -1.1 | 38x 24 | 16.4 | .28 | | F | | | S | |
| 214 | | | 08 37 03.9 | -54 36 47 | 271.59 | -8.14 | 165 | -87.3 | 19.7 | 13x 4 | 19.6 | .35 | | | | | | |
| 215 | | | 08 37 04.0 | -55 08 22 | 272.02 | -8.46 | 165 | -86.3 | -8.5 | 19x 8 | 18.4 | .28 | | S | | | | |
| 216 | | | 08 37 05.3 | -57 22 08 | 273.83 | -9.79 | 165 | -81.7 | -128.0 | 13x 9 | 18.6 | .20 | | S | M | | | br.blg, vLSBdisk |
| 217 | | | 08 37 06.6 | -57 01 22 | 273.55 | -9.58 | 165 | -82.3 | -109.4 | 15x 13 | 18.3 | .20 | | | | | | |
| 218 | | | 08 37 07.4 | -56 58 27 | 273.51 | -9.55 | 165 | -82.3 | -106.8 | 20x 3 | 19.4 | .20 | | S | | E | | cl.to br.* |
| 219 | | | 08 37 07.4 | -57 19 30 | 273.79 | -9.76 | 165 | -81.6 | -125.6 | 23x 11 | 18.0 | .20 | | S | ? | | | smooth |
| 220 | | | 08 37 08.0 | -55 42 41 | 272.48 | -8.79 | 165 | -84.7 | -39.1 | 12x 9 | 18.8 | .21 | | E | ? | | 1 | w.comp. |
| 221 | | | 08 37 08.6 | -57 22 22 | 273.83 | -9.79 | 165 | -81.3 | -128.2 | 17x 11 | 18.2 | .20 | | S | ? | | | pair w.222 |
| 222 | | | 08 37 10.6 | -57 22 14 | 273.83 | -9.78 | 165 | -81.1 | -128.0 | 9x 8 | 18.8 | .20 | | E | 1 | | | pair w.221 |
| 223 | | | 08 37 17.0 | -57 41 23 | 274.10 | -9.96 | 125 | -107.4 | 122.5 | 28x 7 | 18.0 | .17 | | S | M | | | |
| 224 | | | 08 37 17.7 | -57 22 39 | 273.85 | -9.77 | 165 | -80.2 | -128.4 | 13x 4 | 19.6 | .20 | | | | | | |
| 225 | | | 08 37 19.2 | -55 17 44 | 272.16 | -8.52 | 165 | -84.0 | -16.8 | 19x 5 | 18.9 | .26 | | S | | E | 1 | |
| 226 | | | 08 37 20.1 | -57 22 41 | 273.85 | -9.77 | 165 | -79.9 | -128.4 | 16x 5 | 19.1 | .20 | | | | | | |
| 227 | | | 08 37 20.7 | -54 17 43 | 271.36 | -7.92 | 165 | -85.7 | 36.8 | 13x 13 | 18.1 | .46 | | S | ? | | | |
| 228 | | | 08 37 21.0 | -53 57 04 | 271.08 | -7.71 | 165 | -86.4 | 55.3 | 27x 27 | 17.0 | .44 | ? | | | | S | LSB |
| 229 | | NS | 08 37 23.8 | -53 18 52 | 270.57 | -7.32 | 165 | -87.2 | 89.4 | 17x 15 | 17.8 | .52 | | S | | | | |
| 230 | | | 08 37 24.2 | -57 53 14 | 274.27 | -10.07 | 125 | -106.0 | 112.0 | 51x 27 | 16.2 | .17 | | S | M | | | dist.blg, vLSBdisk |
| 231 | | | 08 37 28.3 | -57 14 43 | 273.76 | -9.68 | 165 | -79.2 | -121.2 | 27x 11 | 17.6 | .20 | | S | | | | |
| 232 | | | 08 37 29.6 | -55 23 23 | 272.25 | -8.56 | 165 | -82.6 | -21.8 | 13x 13 | 18.7 | .26 | ? | | | | S | smooth |
| 233 | | | 08 37 29.6 | -56 16 19 | 272.97 | -9.09 | 165 | -80.9 | -69.1 | 23x 16 | 17.5 | .22 | | S | 5 | | | |
| 234 | | | 08 37 31.1 | -54 47 50 | 271.78 | -8.20 | 165 | -83.5 | 9.9 | 20x 20 | 17.7 | .29 | ? | D | ? | | | LSB |
| 235 | | | 08 37 35.1 | -55 29 59 | 272.35 | -8.62 | 165 | -81.7 | -27.7 | 16x 3 | 19.8 | .25 | | S | | E | | neighb.of 238 |
| 236 | | | 08 37 35.3 | -56 05 50 | 272.84 | -8.98 | 165 | -80.5 | -59.7 | 19x 7 | 18.5 | .23 | | S | ? | | | |
| 237 | | | 08 37 36.6 | -53 54 40 | 271.07 | -7.66 | 165 | -84.4 | 57.5 | 27x 20 | 17.4 | .42 | | S | ? | | S | LSB |
| 238 | | | 08 37 38.6 | -55 29 45 | 272.35 | -8.61 | 165 | -81.2 | -27.5 | 15x 13 | 17.9 | .25 | | E | ? | | | |
| 239 | | | 08 37 41.8 | -56 38 23 | 273.28 | -9.29 | 165 | -78.7 | -88.7 | 16x 3 | 19.7 | .23 | | S | | E | | |
| 240 | | | 08 37 43.6 | -57 48 09 | 274.23 | -9.98 | 125 | -103.9 | 116.6 | 13x 7 | 18.7 | .17 | | | | | | |
| 241 | | | 08 37 44.3 | -55 41 15 | 272.52 | -8.71 | 165 | -80.1 | -37.7 | 15x 4 | 19.2 | .23 | | S | | | | |
| 242 | | | 08 37 45.1 | -55 41 48 | 272.52 | -8.72 | 165 | -80.0 | -38.2 | 17x 5 | 18.7 | .23 | | S | ? | | | |
| 243 | | | 08 37 49.3 | -54 56 21 | 271.92 | -8.25 | 165 | -80.9 | 2.4 | 27x 9 | 18.0 | .29 | | S | M | | 1 | w.comp. |
| 244 | | | 08 37 50.4 | -56 58 19 | 273.57 | -9.47 | 165 | -77.0 | -106.5 | 16x 13 | 17.7 | .20 | | S | | | S | |
| 245 | | | 08 37 51.3 | -55 41 22 | 272.53 | -8.70 | 165 | -79.3 | -37.8 | 13x 11 | 18.3 | .24 | | S | E? | | | |
| 246 | | | 08 37 53.3 | -57 23 02 | 273.90 | -9.71 | 165 | -75.9 | -128.6 | 16x 11 | 18.2 | .20 | | S | | | | |
| 247 | | | 08 37 55.3 | -57 18 24 | 273.84 | -9.66 | 165 | -75.8 | -124.4 | 17x 8 | 18.3 | .20 | | E | ? | | | |
| 248 | | | 08 37 56.1 | -54 51 40 | 271.87 | -8.19 | 165 | -80.1 | 6.6 | 17x 11 | 18.0 | .30 | | S | ? | | S | |
| 249 | | | 08 37 56.4 | -54 51 42 | 271.87 | -8.19 | 165 | -80.8 | 3.2 | 24x 24 | 17.5 | .30 | ? | D | ? | | | LSB |
| 250 | | | 08 37 58.8 | -55 10 42 | 272.13 | -8.38 | 165 | -79.2 | -10.3 | 17x 5 | 19.2 | .30 | | S | ? | | | cl.to br.* |
| 251 | | | 08 37 59.3 | -55 25 25 | 272.32 | -8.53 | 165 | -79.4 | -23.2 | 12x 5 | 19.4 | .27 | | S | ? | | | |
| 252 | | | 08 38 00.2 | -55 31 07 | 272.40 | -8.58 | 165 | -78.4 | -28.6 | 20x 16 | 17.9 | .25 | | S | ? | F | | |
| 253 | | | 08 38 00.3 | -53 54 31 | 271.11 | -7.61 | 165 | -81.3 | 57.7 | 16x 4 | 19.6 | .47 | | | | | | |
| 254 | | | 08 38 02.6 | -57 17 54 | 273.85 | -9.65 | 165 | -75.0 | -124.0 | 16x 4 | 19.1 | .20 | | S | | E | | |
| 255 | | | 08 38 03.8 | -55 27 40 | 272.36 | -8.54 | 165 | -78.1 | -25.5 | 13x 9 | 18.6 | .26 | | S | | | | w.sev.comp. |
| 256 | | | 08 38 06.3 | -56 17 12 | 273.03 | -9.03 | 165 | -76.3 | -69.7 | 19x 9 | 18.3 | .23 | | S | ? | | | |
| 257 | | | 08 38 07.0 | -54 42 02 | 271.75 | -8.07 | 165 | -79.0 | 15.3 | 16x 5 | 19.0 | .33 | | | | | | |
| 258 | | | 08 38 07.7 | -55 26 54 | 272.36 | -8.52 | 165 | -78.3 | -24.5 | 12x 7 | 18.8 | .27 | | S | E | | | neighb.of 255 |
| 259 | | | 08 38 09.1 | -57 36 56 | 274.11 | -9.82 | 125 | -101.3 | 126.8 | 15x 4 | 19.8 | .17 | | S | | E | | |
| 260 | | | 08 38 09.3 | -57 09 10 | 273.74 | -9.55 | 165 | -75.6 | -115.8 | 12x 4 | 19.6 | .19 | | S | | E | | |
| 261 | | | 08 38 12.9 | -57 18 16 | 273.87 | -9.63 | 165 | -73.7 | -124.2 | 17x 11 | 18.0 | .20 | | F | | | | w.sev.comp. |
| 262 | | | 08 38 13.2 | -58 00 44 | 274.44 | -10.05 | 125 | -99.9 | 105.6 | 36x 7 | 17.8 | .17 | | S | M | E | | |
| 263 | | | 08 38 17.9 | -56 00 10 | 272.82 | -8.84 | 165 | -75.4 | -54.5 | 23x 16 | 17.5 | .22 | | | | | | w.comp. |
| 264 | | | 08 38 20.9 | -57 23 12 | 273.94 | -9.67 | 165 | -72.6 | -128.6 | 31x 15 | 17.3 | .20 | | S | | | 1 | |
| 265 | | | 08 38 22.6 | -57 37 18 | 274.14 | -9.80 | 125 | -99.7 | 126.5 | 13x 4 | 19.5 | .17 | | S | | E | | |
| 266 | | | 08 38 22.7 | -54 52 06 | 271.91 | -8.15 | 165 | -76.7 | 6.4 | 16x 13 | 17.6 | .30 | | E | ? | | | |
| 267 | | | 08 38 25.2 | -55 31 21 | 272.44 | -8.54 | 165 | -75.3 | -28.7 | 19x 17 | 17.5 | .26 | | S | | F | 1 | |
| 268 | | | 08 38 26.6 | -55 21 35 | 272.31 | -8.44 | 165 | -75.4 | -20.0 | 13x 8 | 18.6 | .29 | | | | | | |
| 269 | | | 08 38 26.8 | -57 50 11 | 274.32 | -9.92 | 125 | -98.7 | 115.1 | 26x 5 | 18.4 | .17 | | S | M | | | |
| 270 | | | 08 38 30.0 | -55 29 43 | 272.43 | -8.51 | 165 | -74.7 | -27.2 | 20x 5 | 18.9 | .27 | | S | | N | | |
| 271 | | | 08 38 30.4 | -55 16 17 | 272.25 | -8.37 | 165 | -75.0 | -15.2 | 47x 40 | 15.6 | .31 | ? | E | 0 | | | |
| 272 | | | 08 38 30.9 | -54 44 53 | 271.83 | -8.06 | 165 | -75.8 | 12.8 | 24x 8 | 17.7 | .31 | | S | E: | | 1 | |
| 273 | | | 08 38 32.6 | -55 16 56 | 272.26 | -8.38 | 165 | -74.7 | -15.8 | 16x 8 | 18.7 | .31 | | S | ? | | | neighb.of 271 |
| 274 | | | 08 38 36.3 | -53 49 39 | 271.10 | -7.49 | 165 | -76.7 | 62.2 | 20x 9 | 18.3 | .46 | | | | | | smooth |
| 275 | | | 08 38 37.3 | -55 51 30 | 272.73 | -8.72 | 165 | -73.2 | -46.6 | 13x 5 | 19.1 | .23 | | | | | | |
| 276 | | | 08 38 46.0 | -57 08 38 | 273.78 | -9.47 | 165 | -70.0 | -115.5 | 13x 8 | 18.8 | .19 | | S | M | | | |
| 277 | | | 08 38 53.8 | -54 57 34 | 272.03 | -8.14 | 165 | -72.6 | 1.6 | 17x 11 | 17.8 | .29 | | E | | | | |
| 278 | | | 08 38 55.4 | -56 16 05 | 273.09 | -8.93 | 165 | -70.3 | -68.5 | 15x 11 | 18.1 | .24 | | | | | | |
| 279 | | | 08 38 55.6 | -57 46 25 | 274.30 | -9.84 | 125 | -95.4 | 118.6 | 17x 11 | 18.1 | .18 | | S | | | S | |
| 280 | | | 08 38 57.5 | -53 46 36 | 271.09 | -7.42 | 165 | -74.0 | 65.0 | 13x 13 | 18.4 | .48 | ? | S | ? | F | | |
| 281 | | | 08 38 59.4 | -53 00 10 | 270.47 | -6.94 | 165 | -74.9 | 106.5 | 15x 8 | 18.7 | .52 | ? | | | | | |
| 282 | | | 08 39 01.7 | -56 13 31 | 273.06 | -8.89 | 165 | -69.5 | -66.2 | 13x 8 | 18.8 | .23 | ? | | | | | |
| 283 | | | 08 39 02.4 | -56 34 08 | 273.34 | -9.10 | 165 | -68.9 | -84.6 | 20x 7 | 18.3 | .26 | | S | | | | |
| 284 | | | 08 39 03.8 | -56 12 29 | 273.05 | -8.88 | 165 | -69.3 | -65.3 | 15x 11 | 18.7 | .23 | ? | | | | | LSB |
| 285 | | | 08 39 04.3 | -54 13 38 | 271.46 | -7.68 | 165 | -72.4 | 40.9 | 13x 9 | 18.7 | .45 | ? | | | | | |
| 286 | | | 08 39 08.5 | -56 15 36 | 273.10 | -8.90 | 165 | -68.6 | -68.1 | 13x 5 | 19.0 | .24 | | S | | | | |
| 287 | | | 08 39 09.3 | -57 36 51 | 274.19 | -9.72 | 125 | -94.1 | 127.2 | 22x 5 | 18.9 | .18 | | S | | E | | |
| 288 | | | 08 39 11.7 | -57 35 37 | 274.18 | -9.70 | 125 | -93.9 | 128.3 | 17x 13 | 18.0 | .19 | | | | | | |
| 289 | | | 08 39 11.8 | -56 15 38 | 273.10 | -8.89 | 165 | -68.2 | -68.1 | 12x 12 | 17.9 | .24 | | E | 0 | | | |
| 290 | | | 08 39 13.4 | -54 53 38 | 272.01 | -8.06 | 165 | -70.1 | 5.2 | 15x 8 | 18.3 | .27 | | | | | | w.sev.comp. |
| 291 | | | 08 39 20.0 | -53 44 02 | 271.09 | -7.35 | 165 | -71.0 | 67.4 | 11x 7 | 19.2 | .49 | | | | | | |
| 292 | | P | 08 39 29.3 | -55 22 16 | 272.41 | -8.32 | 165 | -67.4 | -20.3 | 13x 4 | 19.6 | .31 | | | | | | |
| 293 | | | 08 39 29.4 | -57 52 01 | 274.43 | -9.83 | 125 | -91.2 | 113.7 | 16x 5 | 19.0 | .18 | | S | | E | | |
| 294 | | | 08 39 30.7 | -55 20 56 | 272.40 | -8.31 | 165 | -67.3 | -19.1 | 34x 7 | 17.8 | .31 | | S | | | S | |
| 295 | | | 08 39 30.8 | -55 28 34 | 272.50 | -8.38 | 165 | -67.0 | -26.0 | 12x 7 | 18.7 | .29 | | | | | | |
| 296 | | | 08 39 39.1 | -54 45 09 | 271.93 | -7.93 | 165 | -67.1 | 12.9 | 16x 13 | 17.4 | .28 | | E | ? | | | w.comp. |
| 297 | | | 08 39 39.3 | -56 18 40 | 273.18 | -8.87 | 165 | -64.7 | -70.7 | 13x 7 | 18.9 | .25 | | | | | | |
| 298 | | | 08 39 40.3 | -54 10 23 | 271.47 | -7.57 | 165 | -67.7 | 43.9 | 40x 9 | 17.1 | .51 | | S | 1 | | | |
| 299 | | | 08 39 42.9 | -55 30 59 | 272.55 | -8.39 | 165 | -65.5 | -28.1 | 19x 12 | 17.8 | .28 | | S | E | F | | |
| 300 | | | 08 39 44.1 | -54 41 10 | 271.88 | -7.88 | 165 | -66.5 | 16.4 | 54x 13 | 16.5 | .31 | | S | E | E | | |

**Table 1.** Galaxies in the southern ZOA – I. The Hydra/Antlia Extension (continued)

| RKK Ident (1) | Other Ident (2) | IR (3) | R.A. (h m s) (4) | Dec. (° ′ ″) (5) | gal ℓ (°) (6) | gal b (°) (7) | SRC (8) | X (mm) (9) | Y (mm) (10) | D x d (″) (11) | $B_J$ ($^m$) (12) | $N_{HI}$ (13) | u (14) | Type class. (14) | o (14) | * (14) | Remarks (15) |
|---|---|---|---|---|---|---|---|---|---|---|---|---|---|---|---|---|---|
| 301 | | | 08 39 44.1 | -58 06 33 | 274.64 | -9.95 | 125 | -89.0 | 100.8 | 13x 7 | 19.1 | .18 | | | | | |
| 302 | | | 08 39 45.9 | -55 45 44 | 272.75 | -8.53 | 165 | -64.7 | -41.2 | 19x 19 | 17.7 | .26 | ? | | | S | LSB |
| 303 | | | 08 39 50.1 | -54 45 36 | 271.95 | -7.91 | 165 | -65.6 | 12.5 | 23x 19 | 17.2 | .28 | | S ? | | S | |
| 304 | | | 08 39 50.5 | -54 53 31 | 272.06 | -7.99 | 165 | -65.4 | 5.4 | 13x 9 | 18.5 | .26 | | | | S | |
| 305 | | | 08 39 50.5 | -56 01 10 | 272.96 | -8.68 | 165 | -63.8 | -55.0 | 16x 4 | 19.1 | .22 | | S | E | | |
| 306 | | | 08 39 55.7 | -54 52 23 | 272.05 | -7.97 | 165 | -64.7 | 6.5 | 19x 8 | 18.1 | .26 | ? | E | | | w.2 comp. |
| 307 | | NG | 08 39 58.5 | -54 39 52 | 271.89 | -7.84 | 165 | -64.7 | 17.6 | 32x 20 | 16.6 | .31 | | F | | | |
| 308 | | | 08 40 02.3 | -56 18 12 | 273.21 | -8.83 | 165 | -61.9 | -70.2 | 13x 9 | 18.7 | .25 | ? | | | | |
| 309 | | | 08 40 05.1 | -54 34 29 | 271.83 | -7.77 | 165 | -63.9 | 22.5 | 17x 9 | 18.3 | .34 | | | | S | |
| 310 | | I | 08 40 07.9 | -54 39 35 | 271.90 | -7.82 | 165 | -63.5 | 17.9 | 27x 9 | 17.7 | .31 | | S L | | S | |
| 311 | | | 08 40 09.2 | -58 10 56 | 274.74 | -9.95 | 125 | -85.9 | 97.0 | 19x 4 | 19.3 | .18 | | S | | | |
| 312 | | | 08 40 15.3 | -55 50 20 | 272.85 | -8.52 | 165 | -60.4 | -48.2 | 34x 7 | 17.9 | .24 | | S M | E | | |
| 313 | | | 08 40 16.5 | -55 40 05 | 272.72 | -8.42 | 165 | -61.0 | -36.1 | 20x 7 | 18.6 | .28 | | S | E | | |
| 314 | | | 08 40 17.4 | -57 11 51 | 273.95 | -9.34 | 165 | -58.8 | -118.1 | 15x 13 | 17.7 | .24 | | E | | | |
| 315 | | | 08 40 18.0 | -54 02 20 | 271.42 | -7.42 | 165 | -63.0 | 51.2 | 20x 9 | 18.1 | .58 | | S | | | |
| 316 | | | 08 40 18.9 | -57 12 12 | 273.96 | -9.34 | 165 | -58.6 | -118.4 | 13x 9 | 18.2 | .25 | | S 1 | | | |
| 317 | | | 08 40 20.6 | -54 47 48 | 272.03 | -7.88 | 165 | -61.6 | 10.6 | 13x 8 | 18.4 | .26 | | E ? | | | |
| 318 | | | 08 40 22.1 | -54 32 21 | 271.82 | -7.72 | 165 | -61.8 | 24.4 | 17x 13 | 18.0 | .34 | ? | | | | smooth |
| 319 | | | 08 40 25.0 | -54 50 59 | 272.08 | -7.90 | 165 | -61.0 | 7.8 | 22x 4 | 18.9 | .26 | | S | E | | |
| 320 | | | 08 40 25.1 | -54 41 18 | 271.95 | -7.80 | 165 | -61.2 | 16.4 | 47x 27 | 15.9 | .30 | | SB 4 | | S | |
| 321 | | | 08 40 26.1 | -54 40 18 | 271.94 | -7.79 | 165 | -61.1 | 17.4 | 16x 9 | 18.4 | .30 | | S ? | | | |
| 322 | | | 08 40 27.1 | -55 04 02 | 272.25 | -8.03 | 165 | -61.0 | -3.6 | 24x 7 | 18.4 | .28 | | S | E | | |
| 323 | | | 08 40 27.3 | -55 05 42 | 272.28 | -8.05 | 165 | -60.4 | -5.3 | 27x 17 | 17.2 | .28 | ? | S | F | S | |
| 324 | | | 08 40 30.1 | -57 15 46 | 274.02 | -9.36 | 165 | -57.2 | -121.5 | 13x 11 | 18.4 | .25 | ? | | | | |
| 325 | | | 08 40 34.0 | -54 19 06 | 271.66 | -7.56 | 165 | -60.5 | 36.3 | 20x 20 | 17.5 | .41 | | S ? | F | | LSB |
| 326 | | | 08 40 42.4 | -54 47 02 | 272.05 | -7.83 | 165 | -58.9 | 11.4 | 13x 7 | 18.9 | .27 | | | | | mult.syst. |
| 327 | | | 08 40 44.4 | -55 20 22 | 272.50 | -8.16 | 165 | -57.9 | -18.4 | 20x 13 | 17.7 | .28 | | | | | |
| 328 | | | 08 40 45.1 | -57 04 30 | 273.89 | -9.22 | 165 | -55.6 | -111.4 | 13x 9 | 18.7 | .24 | | | | S | |
| 329 | | | 08 40 45.4 | -55 00 06 | 272.23 | -7.96 | 165 | -58.2 | -.3 | 13x 3 | 19.4 | .28 | | S | E | | |
| 330 | | | 08 40 45.7 | -55 47 33 | 272.86 | -8.44 | 165 | -57.2 | -42.7 | 13x 12 | 18.3 | .25 | | S ? | F | | |
| 331 | | | 08 40 46.8 | -54 35 36 | 271.90 | -7.70 | 165 | -58.5 | 21.6 | 20x 13 | 18.0 | .33 | | S ? | | | |
| 332 | | | 08 40 47.6 | -57 47 35 | 274.47 | -9.65 | 125 | -82.1 | 118.0 | 19x 3 | 19.5 | .19 | | S | E | | |
| 333 | | | 08 40 47.7 | -56 19 00 | 273.28 | -8.75 | 165 | -56.3 | -70.8 | 13x 7 | 19.0 | .23 | ? | | | | |
| 334 | | Q | 08 40 48.8 | -57 19 38 | 274.10 | -9.36 | 165 | -54.9 | -124.9 | 13x 8 | 18.4 | .24 | | E ? | | | |
| 335 | | | 08 40 50.8 | -56 20 16 | 273.31 | -8.76 | 165 | -55.9 | -71.9 | 34x 8 | 17.8 | .23 | | S | E | S | |
| 336 | | Q | 08 40 51.2 | -55 30 08 | 272.64 | -8.25 | 165 | -56.8 | -27.1 | 16x 11 | 18.3 | .28 | | | | S | |
| 337 | | | 08 40 52.8 | -54 52 41 | 272.14 | -7.87 | 165 | -57.4 | 6.4 | 16x 7 | 18.5 | .27 | | S | | | |
| 338 | | | 08 40 55.1 | -57 02 59 | 273.88 | -9.18 | 165 | -54.4 | -110.0 | 16x 4 | 19.5 | .24 | | S | | | |
| 339 | | | 08 40 58.4 | -55 53 10 | 272.95 | -8.47 | 165 | -55.5 | -47.7 | 13x 9 | 18.5 | .24 | | | | | |
| 340 | | | 08 40 58.8 | -55 22 27 | 272.55 | -8.16 | 165 | -56.0 | -20.2 | 19x 11 | 18.2 | .28 | | | | | |
| 341 | | | 08 41 05.6 | -56 03 00 | 273.10 | -8.56 | 165 | -54.4 | -56.4 | 47x 20 | 16.8 | .22 | | | | | LSB |
| 342 | | | 08 41 06.1 | -57 05 38 | 273.94 | -9.19 | 165 | -53.1 | -112.4 | 15x 3 | 20.0 | .23 | | S | E | | dist.nucl. |
| 343 | | | 08 41 06.7 | -55 09 10 | 272.38 | -8.01 | 165 | -55.3 | -8.3 | 31x 20 | 17.0 | .28 | | S ? | F | S | |
| 344 | | | 08 41 10.0 | -55 37 45 | 272.77 | -8.29 | 165 | -54.3 | -33.8 | 16x 5 | 19.1 | .28 | | S ? | | 1 | |
| 345 | | | 08 41 13.0 | -55 02 40 | 272.30 | -7.93 | 165 | -54.6 | -2.5 | 20x 17 | 17.3 | .29 | | E | | | |
| 346 | | | 08 41 14.9 | -55 57 18 | 273.03 | -8.48 | 165 | -53.3 | -51.3 | 20x 7 | 18.5 | .23 | ? | S ? | | | |
| 347 | FGCE0717 | | 08 41 16.5 | -55 12 43 | 272.44 | -8.03 | 165 | -54.0 | -11.5 | 60x 7 | 17.2 | .28 | | S | | S | |
| 348 | | | 08 41 16.5 | -55 14 21 | 272.46 | -8.04 | 165 | -53.9 | -12.9 | 24x 16 | 17.4 | .28 | | | | S | |
| 349 | | | 08 41 18.5 | -53 44 35 | 271.27 | -7.12 | 165 | -55.4 | 67.3 | 67x 7 | 17.3 | .38 | | S M | E | | |
| 350 | | | 08 41 18.5 | -56 14 31 | 273.27 | -8.65 | 165 | -52.5 | -66.7 | 13x 8 | 18.5 | .22 | ? | E ? | | S | |
| 351 | | | 08 41 19.3 | -56 13 56 | 273.26 | -8.64 | 165 | -52.5 | -66.2 | 17x 13 | 17.7 | .21 | ? | E ? | | S | |
| 352 | | | 08 41 21.3 | -55 21 11 | 272.56 | -8.10 | 165 | -53.2 | -19.0 | 40x 5 | 18.3 | .28 | ? | S | E | | p.cov.by br.** |
| 353 | | | 08 41 21.4 | -55 34 38 | 272.74 | -8.24 | 165 | -52.9 | -31.0 | 13x 5 | 19.0 | .28 | | S ? | | | |
| 354 | | | 08 41 25.3 | -54 43 32 | 272.07 | -7.71 | 165 | -53.4 | 14.6 | 23x 4 | 18.7 | .29 | | S | E | | |
| 355 | | | 08 41 28.0 | -54 35 35 | 271.96 | -7.62 | 165 | -53.2 | 21.4 | 13x 11 | 18.0 | .34 | ? | | | 1 | |
| 356 | | | 08 41 33.8 | -56 08 47 | 273.21 | -8.56 | 165 | -50.7 | -61.5 | 15x 9 | 18.2 | .21 | ? | | | | |
| 357 | | | 08 41 37.2 | -56 10 09 | 273.24 | -8.57 | 165 | -50.3 | -62.7 | 17x 5 | 18.7 | .21 | | S | | | |
| 358 | | | 08 41 42.3 | -56 19 27 | 273.37 | -8.66 | 165 | -49.5 | -71.0 | 13x 11 | 18.1 | .21 | ? | E ? | | 1 | |
| 359 | | | 08 41 47.2 | -54 33 02 | 271.96 | -7.56 | 165 | -50.8 | 24.1 | 20x 7 | 17.8 | .36 | | S E | | | in st.diffr.pat. |
| 360 | | | 08 41 47.3 | -58 05 05 | 274.79 | -9.72 | 125 | -74.5 | 102.7 | 22x 15 | 17.7 | .18 | | | | | |
| 361 | | | 08 41 47.6 | -56 55 30 | 273.86 | -9.01 | 165 | -48.2 | -103.2 | 24x 5 | 18.7 | .22 | | S | | | |
| 362 | | | 08 41 49.8 | -58 05 10 | 274.79 | -9.72 | 125 | -74.2 | 102.6 | 17x 15 | 18.5 | .18 | | S ? | F | | vLSB |
| 363 | | | 08 41 51.0 | -58 03 25 | 274.77 | -9.70 | 125 | -74.1 | 104.2 | 9x 7 | 19.5 | .18 | | | | | |
| 364 | | | 08 41 51.6 | -58 05 44 | 274.80 | -9.72 | 125 | -73.9 | 102.1 | 13x 3 | 20.0 | .18 | | S | E | | |
| 365 | | | 08 41 56.4 | -58 23 48 | 275.05 | -9.90 | 125 | -72.9 | 86.0 | 15x 7 | 18.8 | .17 | | S E: | | | |
| 366 | | | 08 41 59.5 | -56 44 00 | 273.72 | -8.88 | 165 | -47.0 | -92.9 | 13x 11 | 18.5 | .23 | ? | | | | |
| 367 | | | 08 42 00.5 | -58 41 09 | 275.29 | -10.06 | 125 | -71.9 | 70.5 | 20x 4 | 19.4 | .18 | | S | E | 1 | |
| 368 | | | 08 42 01.4 | -53 30 51 | 271.16 | -6.90 | 165 | -49.9 | 79.7 | 70x 20 | 16.0 | .44 | | S E | | S | |
| 369 | | | 08 42 02.1 | -54 58 18 | 272.32 | -7.79 | 165 | -48.4 | 1.5 | 17x 9 | 18.5 | .29 | | | | | |
| 370 | | | 08 42 02.8 | -58 32 22 | 275.18 | -9.97 | 125 | -71.9 | 78.3 | 15x 4 | 19.2 | .18 | | S | | | |
| 371 | | | 08 42 05.4 | -56 54 30 | 273.87 | -8.97 | 165 | -46.0 | -102.3 | 20x 5 | 18.9 | .22 | | S | | | |
| 372 | | | 08 42 07.3 | -54 41 29 | 272.10 | -7.61 | 165 | -48.0 | 16.6 | 17x 13 | 17.8 | .34 | ? | | | | in st.diffr.pat. |
| 373 | | | 08 42 09.3 | -56 01 34 | 273.17 | -8.43 | 165 | -46.4 | -55.0 | 67x 47 | 15.0 | .23 | | E 2 | | S | |
| 374 | | | 08 42 09.7 | -55 09 37 | 272.48 | -7.89 | 165 | -47.2 | -8.6 | 20x 16 | 18.0 | .29 | | | | | smooth, LSB |
| 375 | | | 08 42 15.7 | -53 50 59 | 271.45 | -7.08 | 165 | -47.7 | 61.7 | 16x 16 | 17.9 | .42 | | S ? | F | | |
| 376 | | | 08 42 16.3 | -56 01 54 | 273.18 | -8.42 | 165 | -45.6 | -55.3 | 15x 8 | 18.6 | .23 | | | | S | |
| 377 | | | 08 42 19.7 | -57 33 16 | 274.41 | -9.34 | 165 | -43.7 | -136.9 | 40x 11 | 17.0 | .19 | | S 1 | | | w.comp. |
| 378 | | | 08 42 21.2 | -55 11 16 | 272.52 | -7.89 | 165 | -45.7 | -10.0 | 34x 9 | 17.8 | .30 | | | | | |
| 379 | | | 08 42 21.4 | -56 36 05 | 273.65 | -8.76 | 165 | -44.4 | -85.8 | 19x 13 | 18.0 | .23 | | | | | |
| 380 | | | 08 42 21.4 | -57 32 50 | 274.40 | -9.33 | 125 | -71.3 | 131.6 | 12x 8 | 18.6 | .19 | | E | | | |
| 381 | | | 08 42 23.8 | -53 00 46 | 270.79 | -6.54 | 165 | -47.4 | 106.6 | 15x 9 | 18.4 | .61 | | | | | |
| 382 | | P | 08 42 25.8 | -54 51 15 | 272.26 | -7.67 | 165 | -45.5 | 7.9 | 13x 8 | 18.6 | .32 | | S ? | | | |
| 383 | | | 08 42 27.9 | -54 58 02 | 272.35 | -7.74 | 165 | -45.1 | 1.8 | 16x 7 | 19.1 | .30 | | | | 1 | LSB |
| 384 | | | 08 42 34.6 | -58 04 04 | 274.84 | -9.63 | 125 | -68.9 | 103.7 | 13x 4 | 19.6 | .19 | | | | | |
| 385 | | | 08 42 36.0 | -56 23 36 | 273.50 | -8.60 | 165 | -42.8 | -74.6 | 40x 40 | 16.0 | .20 | | SB 5 : | | S | |
| 386 | | | 08 42 42.3 | -58 51 52 | 275.49 | -10.10 | 125 | -66.8 | 61.1 | 17x 3 | 19.6 | .18 | | S | E | | |
| 387 | | | 08 42 42.6 | -54 52 43 | 272.30 | -7.66 | 165 | -43.2 | 6.4 | 13x 3 | 19.9 | .32 | | S | E | | |
| 388 | | | 08 42 42.8 | -58 12 34 | 274.97 | -9.70 | 125 | -67.7 | 96.2 | 20x 13 | 18.1 | .18 | | S ? | | 1 | |
| 389 | | | 08 42 44.6 | -56 05 16 | 273.27 | -8.40 | 165 | -42.0 | -58.2 | 15x 11 | 18.3 | .22 | | S ? | | S | |
| 390 | | | 08 42 46.0 | -54 48 06 | 272.25 | -7.60 | 165 | -42.9 | 10.7 | 12x 7 | 18.8 | .34 | ? | | | | |
| 391 | | | 08 42 46.9 | -56 41 02 | 273.75 | -8.76 | 165 | -41.2 | -90.2 | 27x 4 | 19.2 | .21 | | S | E | | dbl syst. |
| 392 | | | 08 42 47.2 | -55 17 53 | 272.64 | -7.91 | 165 | -42.3 | -15.9 | 17x 5 | 19.1 | .29 | | | | 1 | |
| 393 | | | 08 42 47.4 | -54 51 31 | 272.29 | -7.64 | 165 | -42.6 | 7.4 | 8x 8 | 19.0 | .33 | | | | | blended w.* |
| 394 | | | 08 42 51.2 | -54 48 44 | 272.26 | -7.60 | 165 | -42.2 | 10.2 | 13x 7 | 18.7 | .34 | ? | | | | |
| 395 | | | 08 42 51.5 | -58 41 32 | 275.37 | -9.98 | 125 | -66.0 | 70.3 | 31x 4 | 18.9 | .18 | | S | E | | |
| 396 | | | 08 42 57.4 | -57 20 23 | 274.29 | -9.14 | 165 | -39.3 | -125.3 | 40x 8 | 17.4 | .22 | | S E: | | | |
| 397 | | | 08 43 16.2 | -55 33 40 | 272.90 | -8.02 | 165 | -38.5 | -29.9 | 24x 4 | 18.9 | .29 | | S | E | | |
| 398 | | | 08 43 18.2 | -57 02 32 | 274.08 | -8.92 | 165 | -37.1 | -109.3 | 40x 16 | 16.9 | .21 | | S ? | | S | |
| 399 | | | 08 43 23.2 | -53 15 34 | 271.08 | -6.58 | 165 | -39.3 | 93.5 | 22x 16 | 17.9 | .52 | | S | | | dist.nucl. |
| 400 | | | 08 43 23.7 | -56 50 39 | 273.93 | -8.79 | 165 | -36.5 | -98.7 | 17x 8 | 18.3 | .20 | | | | | |

**Table 1.** Galaxies in the southern ZOA – I. The Hydra/Antlia Extension (continued)

| RKK Ident (1) | Other Ident (2) | IR (3) | R.A. (h m s) (4) | Dec. (° ′ ″) (5) | gal ℓ (°) (6) | gal $b$ (°) (7) | SRC (8) | X (mm) (9) | Y (mm) (10) | D x d (″) (11) | $B_J$ ($^m$) (12) | $N_{HI}$ (13) | u | Type class. (14) | o * | Remarks (15) |
|---|---|---|---|---|---|---|---|---|---|---|---|---|---|---|---|---|
| 401 | | | 08 43 25.9 | -56 53 39 | 273.97 | -8.82 | 165 | -36.2 | -101.4 | 34x 27 | 16.2 | .20 | | L | S | |
| 402 | | P | 08 43 27.0 | -53 08 29 | 270.99 | -6.50 | 165 | -38.8 | 99.8 | 13x 7 | 18.9 | .60 | ? | | | |
| 403 | | | 08 43 28.2 | -58 02 11 | 274.89 | -9.52 | 125 | -62.6 | 105.6 | 24x 5 | 18.7 | .20 | | S M | S | w.comp. |
| 404 | | | 08 43 38.9 | -56 54 11 | 274.00 | -8.80 | 165 | -34.6 | -101.8 | 13x 8 | 18.6 | .21 | ? | | | |
| 405 | | | 08 43 39.4 | -56 31 38 | 273.70 | -8.57 | 165 | -34.8 | -81.7 | 15x 4 | 19.3 | .22 | | S ? | | |
| 406 | | | 08 43 40.4 | -58 47 00 | 275.50 | -9.95 | 125 | -60.2 | 65.6 | 15x 3 | 19.6 | .18 | | S | E | |
| 407 | | | 08 43 42.4 | -56 59 50 | 274.08 | -8.85 | 165 | -34.2 | -106.9 | 16x 5 | 19.1 | .21 | | S | | |
| 408 | | | 08 43 42.5 | -58 49 53 | 275.55 | -9.98 | 125 | -59.9 | 63.0 | 13x 5 | 19.2 | .18 | | S | | |
| 409 | | | 08 43 43.1 | -57 38 55 | 274.60 | -9.25 | 125 | -61.4 | 126.4 | 15x 8 | 18.6 | .18 | | | | |
| 410 | | | 08 43 43.9 | -57 57 54 | 274.85 | -9.44 | 125 | -60.9 | 109.4 | 16x 9 | 18.6 | .20 | | S | | dist.blg, vLSBdisk |
| 411 | | | 08 43 47.3 | -57 47 39 | 274.72 | -9.33 | 125 | -60.7 | 118.6 | 20x 15 | 17.6 | .18 | | | | dbl syst. |
| 412 | | | 08 43 49.3 | -56 31 44 | 273.71 | -8.55 | 165 | -33.6 | -81.7 | 13x 13 | 18.6 | .22 | | S ? | F | |
| 413 | | | 08 43 50.9 | -57 45 11 | 274.69 | -9.30 | 125 | -60.3 | 120.8 | 20x 5 | 18.9 | .18 | | S | | br.blg, vLSBdisk |
| 414 | | | 08 43 51.5 | -57 15 07 | 274.29 | -8.99 | 165 | -32.9 | -120.5 | 20x 11 | 17.4 | .25 | | E ? | | |
| 415 | | | 08 43 52.0 | -58 05 15 | 274.96 | -9.51 | 125 | -59.8 | 102.9 | 23x 5 | 19.0 | .21 | | I | | |
| 416 | | | 08 43 52.5 | -56 59 53 | 274.09 | -8.84 | 165 | -32.9 | -106.9 | 19x 9 | 18.1 | .21 | | | | |
| 417 | | | 08 43 54.7 | -57 42 40 | 274.66 | -9.27 | 125 | -59.9 | 123.1 | 15x 4 | 19.3 | .18 | | S | | |
| 418 | | | 08 43 55.0 | -58 35 14 | 275.37 | -9.81 | 125 | -58.8 | 76.1 | 15x 11 | 18.1 | .18 | | E ? | 1 | |
| 419 | | | 08 44 02.6 | -54 13 16 | 271.90 | -7.10 | 165 | -33.4 | 42.0 | 27x 19 | 16.9 | .37 | | S 5 | | |
| 420 | | | 08 44 05.2 | -58 54 20 | 275.64 | -9.98 | 125 | -57.2 | 59.1 | 17x 11 | 18.2 | .20 | | S | | |
| 421 | | | 08 44 07.2 | -56 54 41 | 274.04 | -8.76 | 165 | -31.2 | -102.2 | 40x 34 | 16.1 | .21 | | S | S | |
| 422 | | | 08 44 07.7 | -55 26 17 | 272.87 | -7.84 | 165 | -32.0 | -23.2 | 13x 5 | 19.6 | .28 | | S ? | | |
| 423 | | | 08 44 09.2 | -58 37 46 | 275.42 | -9.81 | 125 | -57.1 | 73.9 | 12x 8 | 18.8 | .18 | | | | |
| 424 | | | 08 44 09.9 | -56 33 14 | 273.76 | -8.53 | 165 | -31.1 | -83.1 | 17x 9 | 18.1 | .23 | | S ? | 1 | |
| 425 | | | 08 44 13.2 | -56 51 17 | 274.01 | -8.71 | 165 | -30.5 | -99.2 | 20x 13 | 17.9 | .21 | | | | cl.to br.* |
| 426 | | | 08 44 13.2 | -57 46 31 | 274.74 | -9.28 | 125 | -57.6 | 119.7 | 17x 8 | 18.8 | .18 | | | | |
| 427 | | | 08 44 13.3 | -57 06 22 | 274.21 | -8.86 | 165 | -30.3 | -112.7 | 16x 8 | 18.4 | .23 | | | S | |
| 428 | | | 08 44 14.2 | -56 09 22 | 273.45 | -8.28 | 165 | -30.8 | -61.7 | 15x 4 | 19.6 | .24 | | | | |
| 429 | | | 08 44 20.5 | -54 12 38 | 271.92 | -7.06 | 165 | -31.1 | 42.6 | 17x 13 | 17.8 | .36 | | E ? | | |
| 430 | | | 08 44 20.5 | -56 46 15 | 273.95 | -8.64 | 165 | -29.6 | -94.7 | 20x 12 | 17.7 | .21 | | | 1 | |
| 431 | | | 08 44 21.4 | -54 21 02 | 272.03 | -7.14 | 165 | -30.9 | 35.1 | 22x 13 | 17.7 | .38 | | S M | 1 | |
| 432 | | | 08 44 22.8 | -54 24 56 | 272.09 | -7.18 | 165 | -30.7 | 31.6 | 15x 4 | 19.0 | .39 | | S | | in st.diffr.pat. |
| 433 | | | 08 44 26.4 | -58 45 40 | 275.55 | -9.86 | 125 | -54.9 | 66.9 | 23x 9 | 18.2 | .19 | | | 1 | |
| 434 | | | 08 44 29.2 | -56 00 46 | 273.36 | -8.16 | 165 | -29.0 | -54.0 | 13x 3 | 20.3 | .27 | | S | E | |
| 435 | | | 08 44 29.4 | -56 18 15 | 273.59 | -8.34 | 165 | -28.8 | -69.6 | 13x 9 | 18.8 | .24 | | | | |
| 436 | | | 08 44 30.5 | -55 06 15 | 272.64 | -7.59 | 165 | -29.3 | -5.3 | 24x 4 | 18.9 | .30 | | S ? | | |
| 437 | | | 08 44 35.7 | -55 03 36 | 272.62 | -7.56 | 165 | -28.7 | -2.9 | 20x 8 | 18.6 | .30 | ? | | | |
| 438 | | | 08 44 36.1 | -58 47 14 | 275.58 | -9.86 | 125 | -53.8 | 65.5 | 36x 17 | 16.8 | .19 | | S 3 | | dist.blg, dist.sp.arms |
| 439 | | | 08 44 45.0 | -54 27 12 | 272.15 | -7.16 | 165 | -27.8 | 29.6 | 13x 13 | 18.1 | .37 | | | | dbl nucl, in st.diffr.pat. |
| 440 | | | 08 44 47.6 | -58 54 35 | 275.70 | -9.92 | 125 | -52.3 | 59.0 | 23x 4 | 19.0 | .20 | | S | E | |
| 441 | | | 08 44 47.7 | -56 10 53 | 273.52 | -8.23 | 165 | -26.6 | -63.0 | 20x 11 | 18.0 | .25 | | | 1 | |
| 442 | | | 08 44 48.4 | -55 35 13 | 273.05 | -7.86 | 165 | -26.8 | -31.2 | 13x 4 | 19.7 | .27 | | S | E | |
| 443 | | | 08 44 50.8 | -58 51 40 | 275.66 | -9.88 | 125 | -52.0 | 61.6 | 13x 7 | 18.9 | .20 | | S | | |
| 444 | | | 08 44 51.0 | -58 48 44 | 275.62 | -9.85 | 125 | -52.0 | 64.2 | 17x 16 | 17.7 | .20 | | E 1 | | |
| 445 | | | 08 44 52.7 | -54 57 02 | 272.55 | -7.46 | 165 | -26.5 | 3.0 | 36x 36 | 16.1 | .32 | | E 0 | | |
| 446 | | | 08 44 53.7 | -53 59 00 | 271.79 | -6.85 | 165 | -26.8 | 54.8 | 27x 11 | 17.8 | .41 | | S ? | | smooth |
| 447 | | | 08 44 54.0 | -54 02 52 | 271.84 | -6.89 | 165 | -26.8 | 51.4 | 17x 13 | 17.9 | .40 | | | | |
| 448 | | | 08 44 55.1 | -56 07 49 | 273.49 | -8.19 | 165 | -25.7 | -60.3 | 34x 11 | 17.5 | .26 | | S | 1 | |
| 449 | | | 08 44 58.2 | -57 22 18 | 274.48 | -8.95 | 165 | -24.8 | -126.8 | 15x 3 | 20.0 | .23 | | S | E | |
| 450 | | | 08 45 00.5 | -57 22 08 | 274.48 | -8.94 | 165 | -24.5 | -126.7 | 15x 8 | 18.2 | .23 | | E ? | | |
| 451 | | | 08 45 01.7 | -58 48 01 | 275.63 | -9.82 | 125 | -50.8 | 64.9 | 15x 8 | 18.6 | .20 | | S | | |
| 452 | | | 08 45 04.0 | -58 56 05 | 275.74 | -9.90 | 125 | -50.4 | 57.7 | 20x 17 | 17.3 | .20 | | E | | |
| 453 | | | 08 45 04.1 | -57 03 42 | 274.24 | -8.75 | 165 | -24.2 | -110.2 | 34x 7 | 17.8 | .23 | | S E | S | |
| 454 | | | 08 45 08.7 | -59 13 09 | 275.97 | -10.07 | 125 | -49.6 | 42.4 | 13x 8 | 18.7 | .19 | ? | | | |
| 455 | | | 08 45 11.4 | -55 37 26 | 273.11 | -7.84 | 165 | -23.9 | -33.1 | 20x 20 | 17.5 | .27 | ? | | | or PN |
| 456 | | | 08 45 12.9 | -58 25 42 | 275.34 | -9.58 | 125 | -49.9 | 84.8 | 13x 9 | 18.7 | .19 | | | 1 | |
| 457 | | | 08 45 15.1 | -58 40 49 | 275.55 | -9.73 | 125 | -49.4 | 71.3 | 12x 11 | 18.7 | .20 | | | | |
| 458 | | | 08 45 15.9 | -54 57 23 | 272.59 | -7.42 | 165 | -23.5 | 2.7 | 20x 17 | 17.6 | .31 | | | | |
| 459 | | | 08 45 16.2 | -58 55 20 | 275.74 | -9.87 | 125 | -49.0 | 58.4 | 34x 23 | 16.5 | .20 | | S 0 | 1 | |
| 460 | | | 08 45 17.2 | -58 58 42 | 275.79 | -9.91 | 125 | -48.8 | 55.4 | 32x 13 | 17.3 | .20 | | S M | S | |
| 461 | | | 08 45 22.6 | -58 47 39 | 275.65 | -9.78 | 125 | -48.4 | 65.3 | 20x 8 | 18.3 | .21 | | S | 1 | |
| 462 | | | 08 45 23.4 | -55 00 26 | 272.65 | -7.44 | 165 | -22.6 | -.1 | 30x 8 | 18.1 | .30 | | S | | |
| 463 | | | 08 45 23.5 | -56 54 56 | 274.16 | -8.62 | 165 | -21.9 | -102.4 | 27x 7 | 18.7 | .23 | ? | | S | |
| 464 | | | 08 45 24.9 | -56 13 21 | 273.61 | -8.19 | 165 | -21.9 | -65.2 | 13x 13 | 18.7 | .25 | ? | D ? | | neighb.of 466 |
| 465 | | | 08 45 24.9 | -57 06 29 | 274.31 | -8.74 | 165 | -21.6 | -112.7 | 20x 13 | 18.0 | .24 | | S E: | | br.blg, LSBdisk |
| 466 | | | 08 45 27.9 | -56 12 50 | 273.61 | -8.18 | 165 | -21.6 | -64.7 | 20x 4 | 19.4 | .25 | | S ? | ES | |
| 467 | | | 08 45 32.9 | -56 25 42 | 273.78 | -8.30 | 165 | -20.9 | -76.2 | 22x 13 | 17.3 | .25 | | S | | |
| 468 | | | 08 45 36.9 | -59 10 45 | 275.98 | -10.00 | 125 | -46.4 | 44.6 | 12x 5 | 19.2 | .20 | | S | | |
| 469 | | | 08 45 37.8 | -54 27 52 | 272.24 | -7.07 | 165 | -20.9 | 29.1 | 15x 15 | 18.0 | .33 | ? | S ? | S | |
| 470 | | | 08 45 38.7 | -58 50 55 | 275.72 | -9.79 | 125 | -46.5 | 62.4 | 15x 15 | 17.9 | .21 | | E ? | | |
| 471 | | | 08 45 42.3 | -58 03 47 | 275.09 | -9.30 | 125 | -46.7 | 104.5 | 13x 7 | 18.9 | .21 | | | | |
| 472 | | | 08 45 42.9 | -55 51 30 | 273.35 | -7.93 | 165 | -19.8 | -45.7 | 20x 17 | 17.4 | .28 | | S E | | |
| 473 | | | 08 45 46.4 | -55 43 42 | 273.25 | -7.84 | 165 | -19.4 | -38.7 | 30x 8 | 17.7 | .26 | | S | | |
| 474 | | | 08 45 47.1 | -58 04 48 | 275.11 | -9.30 | 125 | -46.2 | 103.6 | 20x 17 | 17.4 | .21 | | S | S | |
| 475 | | | 08 45 51.7 | -53 34 35 | 271.56 | -6.49 | 165 | -19.3 | 76.7 | 13x 7 | 18.9 | .49 | | S ? | | |
| 476 | | | 08 45 53.0 | -54 15 19 | 272.10 | -6.91 | 165 | -19.0 | 40.3 | 34x 34 | 16.4 | .37 | | S ? | F | or E? |
| 477 | | | 08 45 54.9 | -57 47 52 | 274.90 | -9.11 | 125 | -45.5 | 118.7 | 16x 8 | 18.6 | .20 | | S E | | |
| 478 | | | 08 45 55.3 | -54 13 59 | 272.08 | -6.89 | 165 | -18.7 | 41.5 | 20x 16 | 17.7 | .37 | | S ? | F | or E? |
| 479 | | | 08 45 57.3 | -57 15 18 | 274.47 | -8.77 | 165 | -17.7 | -120.6 | 15x 11 | 18.7 | .25 | | S | | LSB |
| 480 | | | 08 46 04.5 | -57 39 47 | 274.81 | -9.01 | 125 | -44.4 | 125.9 | 20x 7 | 18.5 | .20 | | S | | |
| 481 | | | 08 46 08.7 | -55 30 46 | 273.11 | -7.67 | 165 | -16.7 | -27.1 | 27x 4 | 18.9 | .26 | | S L | E1 | |
| 482 | | | 08 46 09.5 | -58 37 15 | 275.58 | -9.60 | 125 | -43.1 | 74.6 | 17x 7 | 18.5 | .22 | | S E | | |
| 483 | | | 08 46 11.2 | -58 48 55 | 275.73 | -9.72 | 125 | -42.8 | 64.2 | 20x 12 | 18.1 | .23 | | S ? | | |
| 484 | | | 08 46 13.3 | -57 59 44 | 275.08 | -9.20 | 125 | -43.1 | 108.2 | 19x 16 | 17.7 | .21 | | L ? | 1 | |
| 485 | | | 08 46 13.9 | -54 48 51 | 272.57 | -7.22 | 165 | -16.1 | 10.3 | 20x 8 | 18.1 | .32 | | S E: | S | |
| 486 | | | 08 46 14.6 | -56 56 45 | 274.25 | -8.55 | 165 | -13.6 | -105.7 | 13x 4 | 19.7 | .24 | | | | |
| 487 | | | 08 46 15.7 | -58 34 30 | 275.55 | -9.56 | 125 | -42.4 | 77.1 | 17x 7 | 19.1 | .21 | | S ? | 1 | |
| 488 | | | 08 46 23.0 | -55 31 53 | 273.15 | -7.65 | 165 | -14.8 | -28.1 | 22x 16 | 17.4 | .26 | | S ? | S | |
| 489 | | | 08 46 28.9 | -57 57 20 | 275.07 | -9.15 | 125 | -41.3 | 110.3 | 20x 9 | 18.3 | .21 | | S | | w.comp. |
| 490 | | | 08 46 31.8 | -58 51 54 | 275.80 | -9.71 | 125 | -40.3 | 61.6 | 16x 7 | 18.8 | .24 | | S | | |
| 491 | | | 08 46 34.2 | -58 27 46 | 275.48 | -9.46 | 125 | -40.3 | 83.1 | 20x 8 | 18.3 | .19 | | | | |
| 492 | | | 08 46 35.8 | -53 09 03 | 271.30 | -6.14 | 165 | -13.5 | 99.5 | 30x 20 | 16.9 | .54 | ? | | | smooth E? |
| 493 | | | 08 46 44.5 | -58 23 13 | 275.44 | -9.39 | 125 | -39.2 | 87.2 | 20x 8 | 18.1 | .20 | | | | neighb.of 491 |
| 494 | | | 08 46 44.6 | -54 46 10 | 272.58 | -7.14 | 165 | -12.2 | 12.7 | 27x 20 | 17.2 | .30 | | S M | | |
| 495 | | | 08 46 45.1 | -59 32 24 | 276.36 | -10.11 | 125 | -38.3 | 25.4 | 16x 11 | 18.4 | .19 | | | | smooth |
| 496 | | | 08 46 45.9 | -58 40 15 | 275.66 | -9.57 | 125 | -38.8 | 72.0 | 40x 8 | 17.7 | .23 | | S M | N | |
| 497 | | | 08 46 48.0 | -52 18 57 | 270.67 | -5.59 | 211 | -108.3 | -125.7 | 15x 9 | 18.0 | .68 | | S 0 ? | | |
| 498 | | | 08 46 52.6 | -54 20 36 | 272.26 | -6.85 | 165 | -11.2 | 35.6 | 16x 4 | 19.6 | .34 | | S | | |
| 499 | | | 08 46 56.2 | -55 27 38 | 273.14 | -7.55 | 165 | -10.7 | -24.3 | 16x 7 | 18.9 | .25 | | S | ES | |
| 500 | | | 08 46 58.6 | -53 47 21 | 271.83 | -6.49 | 165 | -10.4 | 65.3 | 16x 8 | 18.1 | .37 | ? | | | or blended ** |

**Table 1.** Galaxies in the southern ZOA – I. The Hydra/Antlia Extension (continued)

| RKK Ident (1) | Other Ident (2) | IR (3) | R.A. (h m s) (4) | Dec. (° ′ ″) (5) | gal $\ell$ (°) (6) | gal $b$ (°) (7) | SRC (8) | X (mm) (9) | Y (mm) (10) | D x d (″) (11) | $B_J$ ($^m$) (12) | $N_{HI}$ (13) | u | Type class. (14) | o * | Remarks (15) |
|---|---|---|---|---|---|---|---|---|---|---|---|---|---|---|---|---|
| 501 | | | 08 46 59.0 | -55 45 23 | 273.38 | -7.73 | 165 | -10.3 | -40.2 | 23x 4 | 19.1 | .26 | | S L | E | |
| 502 | | | 08 46 59.5 | -59 08 52 | 276.06 | -9.84 | 125 | -37.0 | 46.5 | 19x 8 | 18.2 | .23 | | S | | neighb.of 507 |
| 503 | | | 08 47 01.2 | -54 08 02 | 272.11 | -6.71 | 165 | -10.1 | 46.8 | 13x 13 | 18.3 | .35 | | E ? | | |
| 504 | | | 08 47 02.8 | -56 06 11 | 273.66 | -7.94 | 165 | -9.8 | -58.8 | 16x 4 | 19.3 | .27 | | S | | |
| 505 | | | 08 47 02.9 | -54 54 58 | 272.72 | -7.19 | 165 | -9.8 | 4.9 | 22x 13 | 17.6 | .29 | | S | S | larger LSBdisk? |
| 506 | | | 08 47 03.2 | -58 21 40 | 275.44 | -9.34 | 125 | -37.0 | 88.6 | 27x 5 | 18.7 | .20 | | S | 1 | dist.blg |
| 507 | | | 08 47 05.2 | -59 08 45 | 276.07 | -9.83 | 125 | -36.3 | 46.6 | 17x 9 | 18.3 | .24 | | | | dbl syst? cl.to br.* |
| 508 | | | 08 47 06.4 | -55 24 28 | 273.12 | -7.49 | 165 | -9.4 | -21.5 | 16x 5 | 18.9 | .25 | | S | E | |
| 509 | | | 08 47 06.5 | -57 00 31 | 274.38 | -8.50 | 165 | -9.3 | -107.3 | 16x 12 | 17.9 | .25 | ? | | | |
| 510 | | | 08 47 11.9 | -53 23 36 | 271.55 | -6.22 | 165 | -8.7 | 86.5 | 20x 12 | 17.7 | .55 | | S ? | | |
| 511 | | | 08 47 13.6 | -54 19 38 | 272.28 | -6.80 | 165 | -8.5 | 36.5 | 20x 8 | 18.1 | .35 | | | | w.comp. |
| 512 | | | 08 47 14.7 | -55 17 50 | 273.04 | -7.41 | 165 | -8.3 | -15.6 | 13x 4 | 19.8 | .26 | | S ? | E | |
| 513 | | | 08 47 15.5 | -58 14 29 | 275.36 | -9.25 | 125 | -35.6 | 95.1 | 20x 8 | 18.4 | .20 | | | | |
| 514 | | | 08 47 16.2 | -55 53 38 | 273.51 | -7.78 | 165 | -8.1 | -47.6 | 20x 13 | 17.9 | .28 | ? | S ? | S | smooth |
| 515 | | | 08 47 16.4 | -58 35 53 | 275.65 | -9.47 | 125 | -35.3 | 75.9 | 19x 8 | 18.4 | .22 | | S M | 1 | |
| 516 | | | 08 47 31.0 | -59 38 01 | 276.49 | -10.09 | 125 | -33.1 | 20.5 | 15x 7 | 18.8 | .18 | | S | | |
| 517 | | | 08 47 33.4 | -57 32 04 | 274.83 | -8.78 | 165 | -6.1 | -135.5 | 17x 8 | 18.3 | .22 | | | 1 | |
| 518 | | | 08 47 34.2 | -57 40 33 | 274.94 | -8.86 | 125 | -33.7 | 125.4 | 26x 11 | 17.5 | .21 | | S M | S | |
| 519 | | | 08 47 35.3 | -54 23 21 | 272.36 | -6.80 | 165 | -5.6 | 33.1 | 23x 13 | 18.1 | .33 | | S | F S | LSB |
| 520 | | | 08 47 36.9 | -52 18 52 | 270.74 | -5.49 | 211 | -101.7 | -125.4 | 16x 9 | 18.1 | .76 | | S 1 : | S | |
| 521 | | | 08 47 38.2 | -57 39 27 | 274.93 | -8.85 | 125 | -33.3 | 126.4 | 23x 7 | 18.1 | .21 | | S | ES | |
| 522 | | | 08 47 38.7 | -53 48 33 | 271.91 | -6.43 | 165 | -5.1 | 64.2 | 23x 7 | 18.4 | .36 | | S M | E | |
| 523 | | | 08 47 40.9 | -58 01 04 | 275.22 | -9.07 | 125 | -32.8 | 107.1 | 22x 15 | 18.2 | .21 | | S | F | v.obs. |
| 524 | | | 08 47 46.6 | -55 38 40 | 273.36 | -7.57 | 165 | -4.3 | -34.2 | 13x 4 | 19.7 | .24 | | S | E | |
| 525 | | | 08 47 51.8 | -55 25 48 | 273.20 | -7.43 | 165 | -3.6 | -22.7 | 17x 5 | 18.9 | .25 | | S ? | | |
| 526 | | | 08 47 52.6 | -55 32 14 | 273.29 | -7.49 | 165 | -3.5 | -28.4 | 17x 5 | 19.1 | .24 | | S | | |
| 527 | | | 08 47 53.8 | -55 34 54 | 273.32 | -7.52 | 165 | -3.4 | -30.8 | 23x 20 | 16.7 | .24 | | E ? | | |
| 528 | | | 08 47 53.8 | -55 39 41 | 273.39 | -7.57 | 165 | -3.4 | -35.1 | 20x 8 | 18.3 | .24 | | S M | | |
| 529 | | | 08 48 05.9 | -58 57 53 | 276.01 | -9.61 | 125 | -29.4 | 56.3 | 16x 4 | 19.5 | .29 | | S M | | |
| 530 | | | 08 48 10.8 | -56 54 44 | 274.39 | -8.32 | 165 | -1.5 | -102.2 | 17x 3 | 19.6 | .27 | | S | E | |
| 531 | | | 08 48 14.2 | -59 45 24 | 276.65 | -10.09 | 125 | -28.2 | 13.9 | 12x 9 | 19.1 | .18 | | | | smooth |
| 532 | | | 08 48 19.1 | -56 13 28 | 273.86 | -7.88 | 165 | -.3 | -65.3 | 15x 7 | 18.4 | .29 | | S | | |
| 533 | | | 08 48 20.1 | -54 33 49 | 272.57 | -6.83 | 165 | .2 | 23.8 | 16x 13 | 18.0 | .26 | ? | | | neighb.of 537 |
| 534 | | | 08 48 20.5 | -54 44 00 | 272.70 | -6.93 | 165 | .2 | 14.7 | 13x 8 | 18.5 | .27 | | S E | | |
| 535 | | | 08 48 23.7 | -54 17 48 | 272.36 | -6.65 | 165 | .7 | 38.1 | 13x 11 | 18.6 | .33 | | S | F | |
| 536 | | | 08 48 31.8 | -56 40 25 | 274.24 | -8.13 | 165 | 1.2 | -89.4 | 20x 13 | 17.9 | .28 | | E ? | | |
| 537 | | | 08 48 32.1 | -54 33 00 | 272.57 | -6.80 | 165 | 1.7 | 24.5 | 20x 3 | 19.3 | .26 | | S | E1 | |
| 538 | | | 08 48 33. | -59 17.1 : | 276.30 | -9.77 | 125 | -26.2 | 39.2 | 13x 8 | 19.0 | .18 | | S | | vLSBdisk |
| 539 | | | 08 48 43.5 | -59 37 18 | 276.58 | -9.96 | 125 | -24.9 | 21.2 | 13x 8 | 19.1 | .18 | | | | w.comp. |
| 540 | | | 08 48 44.6 | -55 08 05 | 273.05 | -7.14 | 165 | 3.2 | -6.9 | 12x 5 | 19.4 | .29 | | | | neighb.of 542 |
| 541 | | | 08 48 45.6 | -54 33 41 | 272.60 | -6.78 | 165 | 3.5 | 23.9 | 16x 5 | 19.1 | .26 | | S | E | |
| 542 | | | 08 48 46.7 | -55 07 55 | 273.05 | -7.14 | 165 | 3.4 | -6.7 | 23x 7 | 18.1 | .29 | | S | ES | w.comp. |
| 543 | | | 08 48 47.1 | -59 39 46 | 276.62 | -9.98 | 125 | -24.5 | 19.0 | 12x 8 | 19.1 | .18 | | S ? | | cl.to br.* |
| 544 | | | 08 48 47.3 | -59 20 24 | 276.36 | -9.78 | 125 | -24.6 | 36.3 | 13x 12 | 18.4 | .17 | | S ? | | |
| 545 | | I2 | 08 48 48.6 | -58 05 43 | 275.38 | -9.00 | 125 | -24.7 | 103.0 | 31x 7 | 17.8 | .21 | | S | 1 | i.a.w.547 |
| 546 | | | 08 48 50.6 | -53 59 53 | 272.17 | -6.41 | 165 | 4.3 | 54.1 | 15x 5 | 19.1 | .34 | | S ? | | |
| 547 | | I2 | 08 48 50.6 | -58 05 41 | 275.38 | -8.99 | 125 | -24.5 | 103.0 | 16x 8 | 18.4 | .21 | | | | i.a.w.545 |
| 548 | | NS | 08 48 51.1 | -56 56 30 | 274.47 | -8.27 | 165 | 3.4 | -103.8 | 13x 4 | 19.8 | .28 | | S | E | |
| 549 | | | 08 48 54.0 | -59 25 48 | 276.44 | -9.82 | 125 | -23.8 | 31.4 | 13x 5 | 19.2 | .18 | | S | | |
| 550 | | | 08 49 01.0 | -57 54 33 | 275.25 | -8.86 | 125 | -23.3 | 113.0 | 17x 4 | 19.7 | .20 | | S | E1 | |
| 551 | | | 08 49 01.7 | -55 06 57 | 273.06 | -7.10 | 165 | 5.4 | -5.9 | 17x 16 | 17.9 | .29 | ? | | | LSB |
| 552 | | | 08 49 03.0 | -55 25 07 | 273.30 | -7.29 | 165 | 5.4 | -22.1 | 40x 8 | 17.6 | .25 | | S M | E | in st.diffr.pat. |
| 553 | | | 08 49 04.9 | -57 34 18 | 274.99 | -8.64 | 165 | 5.0 | -137.6 | 13x 12 | 18.3 | .23 | | S E? | | |
| 554 | | | 08 49 06.0 | -58 15 38 | 275.53 | -9.07 | 125 | -22.6 | 94.1 | 22x 15 | 17.8 | .21 | | S ? | | |
| 555 | | Q | 08 49 08.0 | -57 07 49 | 274.65 | -8.36 | 165 | 5.4 | -113.9 | 17x 8 | 18.9 | .34 | | S | | dist.nucl, LSBdisk |
| 556 | | | 08 49 11.7 | -58 35 13 | 275.80 | -9.27 | 125 | -21.9 | 76.6 | 13x 4 | 19.8 | .28 | | | | |
| 557 | | | 08 49 13.9 | -55 29 11 | 273.37 | -7.31 | 165 | 6.8 | -25.8 | 9x 5 | 19.8 | .25 | | | | dbl syst.w.558 |
| 558 | | | 08 49 15.6 | -55 28 54 | 273.37 | -7.30 | 165 | 7.0 | -25.5 | 13x 7 | 18.9 | .25 | | | | dbl syst.w.557 |
| 559 | | | 08 49 16.4 | -54 56 42 | 272.95 | -6.96 | 165 | 7.3 | 3.3 | 17x 5 | 19.2 | .29 | | S | E | |
| 560 | | | 08 49 17.2 | -57 29 50 | 274.95 | -8.57 | 125 | -21.4 | 135.0 | 13x 9 | 19.0 | .27 | | | | smooth |
| 561 | | | 08 49 21.2 | -55 36 17 | 273.47 | -7.37 | 165 | 6.9 | -32.2 | 23x 9 | 18.2 | .26 | | | | on st.diffr.ring |
| 562 | | | 08 49 21.3 | -56 35 59 | 274.25 | -8.00 | 165 | 7.3 | -85.4 | 16x 5 | 19.5 | .30 | | | | |
| 563 | | | 08 49 22.9 | -55 17 24 | 273.23 | -7.17 | 165 | 8.0 | -15.2 | 13x 9 | 18.6 | .26 | | S | | |
| 564 | | | 08 49 24.1 | -55 17 55 | 273.24 | -7.17 | 165 | 8.1 | -15.7 | 13x 13 | 18.3 | .26 | | S | | i.a.w.565,in st.diffr.pat. |
| 565 | | | 08 49 24.6 | -55 17 44 | 273.24 | -7.17 | 165 | 8.2 | -15.5 | 13x 13 | 18.3 | .26 | | S | | i.a.w.564,in st.diffr.pat. |
| 566 | | | 08 49 25.1 | -59 47 37 | 276.77 | -10.00 | 125 | -16.8 | 10.8 | 60x 13 | 16.9 | .18 | | S M | 1 | |
| 567 | | | 08 49 26.8 | -59 00 58 | 276.16 | -9.51 | 125 | -20.1 | 53.7 | 13x 9 | 18.7 | .24 | | S | | w.comp. |
| 568 | | | 08 49 29.2 | -58 26 55 | 275.71 | -9.15 | 125 | -19.9 | 84.1 | 17x 8 | 18.7 | .27 | | S | | |
| 569 | | | 08 49 34.0 | -55 33 22 | 273.45 | -7.32 | 165 | 9.3 | -29.5 | 16x 5 | 19.0 | .26 | | S | | in st.diffr.pat. |
| 570 | | | 08 49 35.4 | -55 34 16 | 273.47 | -7.32 | 165 | 9.5 | -30.3 | 16x 3 | 19.6 | .26 | ? | S | E | neighb.of 569 |
| 571 | | | 08 49 35.9 | -59 38 10 | 276.66 | -9.88 | 125 | -19.0 | 20.4 | 13x 8 | 18.5 | .18 | | | 1 | |
| 572 | | | 08 49 36.8 | -56 43 14 | 274.37 | -8.05 | 165 | 9.1 | -91.9 | 13x 8 | 19.0 | .31 | ? | | | |
| 573 | | | 08 49 38.3 | -59 11 20 | 276.31 | -9.60 | 125 | -18.8 | 44.4 | 19x 5 | 18.8 | .20 | | | | |
| 574 | | | 08 49 39.4 | -53 44 05 | 272.04 | -6.15 | 165 | 10.8 | 68.1 | 8x 8 | 19.1 | .38 | ? | | | larger LSBdisk? |
| 575 | | | 08 49 42.2 | -58 06 20 | 275.46 | -8.91 | 125 | -18.4 | 102.4 | 17x 11 | 18.0 | .20 | | | S | |
| 576 | | | 08 49 42.8 | -57 26 21 | 274.94 | -8.49 | 165 | 9.5 | -130.5 | 40x 5 | 18.1 | .31 | | S L | E | |
| 577 | | | 08 49 48.7 | -59 43 08 | 276.74 | -9.92 | 125 | -14.1 | 14.9 | 22x 9 | 18.0 | .18 | | S E | | |
| 578 | | | 08 49 49.8 | -58 55 19 | 276.12 | -9.41 | 125 | -17.5 | 58.7 | 15x 11 | 18.1 | .26 | | E ? | | |
| 579 | | | 08 49 50.1 | -55 19 55 | 273.30 | -7.15 | 165 | 11.4 | -17.5 | 13x 8 | 18.7 | .25 | | | | |
| 580 | | | 08 49 50.5 | -57 29 22 | 274.99 | -8.51 | 165 | 10.4 | -133.2 | 17x 4 | 19.4 | .28 | | S | E | |
| 581 | | | 08 49 50.8 | -59 26 03 | 276.52 | -9.73 | 125 | -17.3 | 31.3 | 16x 9 | 18.4 | .19 | | | | |
| 582 | | | 08 49 51.8 | -59 37 32 | 276.67 | -9.85 | 125 | -17.2 | 21.0 | 15x 5 | 18.8 | .18 | | S E | | |
| 583 | | | 08 49 51.9 | -58 50 33 | 276.06 | -9.36 | 125 | -17.2 | 63.0 | 8x 7 | 19.6 | .28 | | | | |
| 584 | | | 08 49 52.3 | -55 20 30 | 273.31 | -7.15 | 165 | 11.7 | -18.0 | 12x 9 | 18.7 | .25 | | | | |
| 585 | | | 08 49 52.7 | -55 21 00 | 273.32 | -7.15 | 165 | 11.8 | -18.5 | 9x 5 | 19.6 | .25 | | | | |
| 586 | | | 08 49 54.9 | -53 36 01 | 271.96 | -6.04 | 165 | 12.9 | 75.3 | 16x 8 | 18.6 | .38 | | S | | |
| 587 | | | 08 49 57.3 | -59 19 40 | 276.45 | -9.66 | 125 | -16.6 | 36.9 | 15x 8 | 18.4 | .20 | | S E: | | w.comp. |
| 588 | | | 08 49 57.4 | -55 21 14 | 273.33 | -7.15 | 165 | 12.3 | -18.7 | 9x 5 | 19.6 | .25 | | | | in st.diffr.pat. |
| 589 | | | 08 49 59.2 | -58 41 26 | 275.95 | -9.25 | 125 | -16.4 | 71.1 | 27x 11 | 17.6 | .28 | | S | | dbl syst? |
| 590 | | | 08 49 59.7 | -58 56 20 | 276.14 | -9.41 | 125 | -16.3 | 57.8 | 15x 11 | 18.2 | .26 | | E ? | 1 | |
| 591 | | | 08 50 00.0 | -58 50 41 | 276.07 | -9.35 | 125 | -16.3 | 62.8 | 20x 12 | 17.7 | .28 | | S E: | 1 | |
| 592 | | | 08 50 03.2 | -54 47 36 | 272.90 | -6.78 | 165 | 13.4 | 11.4 | 13x 4 | 19.7 | .28 | | S | E | |
| 593 | | | 08 50 03.3 | -57 56 29 | 275.36 | -8.77 | 125 | -15.9 | 111.2 | 23x 11 | 17.9 | .19 | | S ? | | |
| 594 | | | 08 50 04.5 | -58 15 14 | 275.61 | -8.97 | 125 | -15.8 | 94.5 | 16x 7 | 18.8 | .21 | | S M | S | |
| 595 | | | 08 50 07.0 | -58 56 21 | 276.15 | -9.40 | 125 | -15.5 | 57.8 | 12x 5 | 19.6 | .26 | | S | | |
| 596 | | | 08 50 07.5 | -58 50 48 | 276.08 | -9.34 | 125 | -15.4 | 62.7 | 13x 5 | 19.5 | .28 | | S | | |
| 597 | | | 08 50 20.1 | -59 08 51 | 276.33 | -9.51 | 125 | -14.0 | 46.6 | 40x 9 | 17.6 | .21 | | S M | 1 | |
| 598 | | | 08 50 21.7 | -55 51 33 | 273.76 | -7.42 | 165 | 15.1 | -45.8 | 13x 9 | 18.7 | .29 | | | | |
| 599 | | | 08 50 24.5 | -52 48 40 | 271.40 | -5.48 | 165 | 17.3 | 117.6 | 20x 13 | 18.0 | .61 | | S | | |
| 600 | | | 08 50 25.4 | -57 28 54 | 275.03 | -8.44 | 125 | -13.3 | 135.9 | 22x 5 | 18.9 | .29 | | S M | | i.a.w.602,cl.to br.* |

**Table 1.** Galaxies in the southern ZOA – I. The Hydra/Antlia Extension (continued)

| RKK Ident (1) | Other Ident (2) | IR (3) | R.A. (h m s) (4) | Dec. (° ′ ″) (5) | gal ℓ (°) (6) | gal b (°) (7) | SRC (8) | X (mm) (9) | Y (mm) (10) | D x d (″) (11) | $B_J$ ($^m$) (12) | $N_{HI}$ (13) | u | Type class. (14) | o * | Remarks (15) |
|---|---|---|---|---|---|---|---|---|---|---|---|---|---|---|---|---|
| 601 | | | 08 50 25.7 | -59 40 21 | 276.76 | -9.83 | 125 | -13.4 | 18.5 | 15x 3 | 20.1 | .18 | | S | E | |
| 602 | | | 08 50 27.6 | -57 29 07 | 275.04 | -8.44 | 125 | -16.9 | 135.7 | 22x 7 | 18.4 | .28 | | S M | | i.a.w.600,cl.to br.* |
| 603 | | | 08 50 30.7 | -55 29 29 | 273.49 | -7.17 | 165 | 16.5 | -26.1 | 26x 12 | 17.4 | .26 | | | 1 | in st.diffr.pat. |
| 604 | | | 08 50 30.7 | -59 41 08 | 276.77 | -9.83 | 125 | -12.8 | 17.8 | 13x 3 | 20.2 | .18 | | S | E | |
| 605 | | | 08 50 31.9 | -59 40 37 | 276.77 | -9.82 | 125 | -12.7 | 18.2 | 9x 9 | 19.1 | .18 | | E | | neighb.of 601 |
| 606 | | I | 08 50 32.6 | -55 11 07 | 273.25 | -6.98 | 165 | 16.9 | -9.7 | 16x 7 | 18.6 | .27 | | | | in st.diffr.pat. |
| 607 | | | 08 50 37.2 | -55 31 34 | 273.52 | -7.18 | 165 | 17.3 | -28.0 | 13x 4 | 19.1 | .26 | | | | in st.diffr.pat. |
| 608 | | | 08 50 40.3 | -56 02 32 | 273.93 | -7.51 | 165 | 18.1 | -56.6 | 17x 9 | 18.3 | .31 | | S | | in st.diffr.pat. |
| 609 | | | 08 50 43.0 | -59 47 04 | 276.87 | -9.87 | 125 | -11.4 | 12.5 | 13x 7 | 18.7 | .18 | | S E | | |
| 610 | | | 08 50 43.2 | -59 41 22 | 276.79 | -9.81 | 125 | -11.4 | 17.6 | 12x 9 | 18.4 | .18 | | E | | or blg |
| 611 | | | 08 50 45.0 | -56 03 11 | 273.95 | -7.50 | 165 | 17.9 | -56.3 | 13x 11 | 18.6 | .31 | ? | | | smooth |
| 612 | | | 08 50 45.7 | -56 19 27 | 274.16 | -7.68 | 165 | 17.8 | -70.8 | 20x 7 | 18.3 | .31 | | S | S | |
| 613 | | | 08 50 48.1 | -58 17 01 | 275.69 | -8.91 | 125 | -10.6 | 92.9 | 13x 5 | 19.4 | .22 | | S | | br.blg, vLSBdisk |
| 614 | | | 08 50 51.0 | -56 09 44 | 274.04 | -7.56 | 165 | 18.6 | -62.1 | 13x 9 | 18.8 | .31 | ? | | | smooth |
| 615 | | | 08 50 53.9 | -54 59 17 | 273.13 | -6.81 | 165 | 19.8 | .8 | 20x 17 | 17.4 | .26 | ? | | S | dbl syst? in st.diffr.pat. |
| 616 | | | 08 50 55.0 | -55 23 16 | 273.44 | -7.06 | 165 | 19.6 | -20.6 | 13x 4 | 19.1 | .25 | | S | | in st.diffr.pat. |
| 617 | | | 08 50 55.2 | -58 48 02 | 276.11 | -9.23 | 125 | -9.9 | 65.2 | 20x 11 | 17.8 | .27 | | S E | | |
| 618 | | Q | 08 50 56.9 | -57 32 37 | 275.13 | -8.43 | 165 | 18.3 | -136.2 | 13x 3 | 19.9 | .26 | | S | | |
| 619 | | | 08 50 59.6 | -55 32 20 | 273.57 | -7.15 | 165 | 20.1 | -28.7 | 28x 13 | 17.7 | .26 | | S M: | | in st.diffr.pat. |
| 620 | | | 08 51 00.1 | -59 58 00 | 277.03 | -9.96 | 125 | -9.6 | 2.7 | 15x 9 | 18.6 | .18 | | | | |
| 621 | | | 08 51 03.3 | -53 18 23 | 271.84 | -5.72 | 165 | 22.2 | 91.0 | 13x 8 | 18.7 | .46 | | E ? | | |
| 622 | | | 08 51 03.9 | -53 35 28 | 272.07 | -5.90 | 165 | 22.1 | 75.7 | 13x 4 | 19.5 | .37 | | | 1 | |
| 623 | | | 08 51 03.9 | -57 49 34 | 275.36 | -8.60 | 125 | -8.7 | 117.4 | 20x 15 | 17.5 | .22 | | | 1 | |
| 624 | | I | 08 51 04.8 | -54 22 03 | 272.67 | -6.39 | 165 | 21.6 | 34.1 | 16x 11 | 18.3 | .29 | | S ? | | |
| 625 | | | 08 51 05.0 | -57 51 36 | 275.39 | -8.62 | 125 | -8.6 | 115.6 | 17x 13 | 17.8 | .21 | | | S | |
| 626 | | | 08 51 11.5 | -57 50 54 | 275.39 | -8.60 | 125 | -7.8 | 116.2 | 16x 12 | 17.9 | .22 | | | 1 | |
| 627 | | | 08 51 16.5 | -54 50 32 | 273.05 | -6.68 | 165 | 22.8 | 8.6 | 15x 8 | 18.6 | .27 | | | | cl.to br.* |
| 628 | | | 08 51 16.6 | -59 07 21 | 276.39 | -9.40 | 125 | -7.5 | 47.9 | 13x 5 | 19.4 | .24 | | S | | |
| 629 | | | 08 51 17.3 | -59 30 53 | 276.70 | -9.64 | 125 | -7.5 | 26.9 | 12x 11 | 18.7 | .20 | | | | w.comp. |
| 630 | | | 08 51 19.8 | -55 19 40 | 273.44 | -6.98 | 165 | 22.8 | -17.4 | 20x 5 | 18.9 | .25 | | S | | in st.diffr.pat. |
| 631 | | I | 08 51 22.8 | -54 27 24 | 272.77 | -6.42 | 165 | 23.9 | 29.3 | 13x 7 | 18.6 | .31 | | S E: | | |
| 632 | | | 08 51 24.0 | -59 55 54 | 277.04 | -9.89 | 125 | -6.9 | 4.6 | 19x 15 | 17.4 | .18 | | E | | |
| 633 | | | 08 51 24.7 | -57 36 23 | 275.22 | -8.42 | 125 | -6.2 | 129.2 | 19x 13 | 17.8 | .25 | | | S | |
| 634 | | | 08 51 25.1 | -59 43 27 | 276.88 | -9.76 | 125 | -6.7 | 15.7 | 15x 11 | 18.2 | .18 | | F | | |
| 635 | | | 08 51 27.0 | -58 18 18 | 275.77 | -8.86 | 125 | -6.1 | 91.7 | 31x 7 | 17.9 | .24 | | S M | | |
| 636 | | | 08 51 27.6 | -59 32 59 | 276.74 | -9.65 | 125 | -6.4 | 25.0 | 20x 7 | 18.2 | .20 | | S E | | |
| 637 | | | 08 51 27.9 | -55 31 29 | 273.60 | -7.09 | 165 | 23.7 | -28.0 | 13x 4 | 19.8 | .26 | ? | | | in st.diffr.pat. |
| 638 | | | 08 51 28.6 | -56 03 15 | 274.01 | -7.43 | 165 | 23.4 | -56.4 | 13x 7 | 19.1 | .33 | ? | | | smooth |
| 639 | | | 08 51 32.9 | -56 12 54 | 274.14 | -7.52 | 165 | 23.8 | -65.0 | 16x 5 | 19.3 | .32 | ? | | S | |
| 640 | | | 08 51 34.4 | -55 00 22 | 273.21 | -6.75 | 165 | 25.0 | -.2 | 23x 13 | 17.5 | .27 | | E ? | S | in st.diffr.pat. |
| 641 | | | 08 51 36.4 | -55 46 53 | 273.81 | -7.24 | 165 | 24.6 | -41.8 | 13x 7 | 18.6 | .29 | ? | | | in st.diffr.pat. |
| 642 | | | 08 51 40.6 | -57 08 19 | 274.87 | -8.10 | 165 | 23.9 | -114.5 | 27x 3 | 18.9 | .30 | | S | E | |
| 643 | | | 08 51 40.8 | -59 41 50 | 276.88 | -9.72 | 125 | -4.9 | 17.1 | 24x 13 | 17.5 | .18 | | S E | | |
| 644 | | | 08 51 41.9 | -59 42 23 | 276.89 | -9.72 | 125 | -4.8 | 16.6 | 15x 9 | 18.4 | .18 | | | | |
| 645 | | | 08 51 48.3 | -58 12 30 | 275.72 | -8.76 | 125 | -3.5 | 96.9 | 40x 17 | 16.8 | .23 | | S E | | |
| 646 | | | 08 51 50.0 | -57 14 30 | 274.97 | -8.15 | 165 | 25.0 | -120.1 | 20x 4 | 19.5 | .31 | | S ? | E | LSB |
| 647 | | | 08 51 52.6 | -59 42 15 | 276.90 | -9.70 | 125 | -3.6 | 16.7 | 12x 8 | 18.7 | .18 | | | | trpl syst.w.648,649 |
| 648 | | | 08 51 53.6 | -59 42 30 | 276.90 | -9.71 | 125 | -3.5 | 16.5 | 17x 13 | 17.8 | .18 | | E | | trpl syst.w.647,648 |
| 649 | | | 08 51 54.8 | -59 42 16 | 276.90 | -9.70 | 125 | -3.3 | 16.7 | 13x 13 | 18.1 | .18 | | E | | trpl syst.w.647,648 |
| 650 | | | 08 51 55.8 | -53 43 47 | 272.26 | -5.89 | 165 | 28.8 | 68.2 | 13x 11 | 18.4 | .36 | | | | |
| 651 | | | 08 51 56.0 | -59 41 57 | 276.90 | -9.70 | 125 | -3.2 | 17.0 | 8x 7 | 19.3 | .18 | | | | |
| 652 | | | 08 51 57.5 | -59 42 15 | 276.90 | -9.70 | 125 | -3.0 | 16.7 | 8x 5 | 19.6 | .18 | | | | |
| 653 | | | 08 52 01.1 | -54 47 04 | 273.08 | -6.56 | 165 | 28.6 | 11.6 | 17x 15 | 17.9 | .29 | ? | | S | w.2 E comp. |
| 654 | L165-006 | | 08 52 09.6 | -53 53 41 | 272.40 | -5.97 | 165 | 30.5 | 59.3 | 134x121 | 13.7 | .34 | ? | D ? | | |
| 655 | | | 08 52 10.7 | -58 13 51 | 275.77 | -8.74 | 125 | -.9 | 95.7 | 16x 4 | 19.5 | .25 | | S M | | |
| 656 | | | 08 52 10.9 | -58 56 16 | 276.32 | -9.19 | 125 | -1.2 | 57.8 | 54x 20 | 16.2 | .28 | | L | S | |
| 657 | | | 08 52 13.2 | -56 03 29 | 274.08 | -7.35 | 165 | 28.9 | -56.7 | 13x 4 | 19.6 | .34 | | S | | |
| 658 | | | 08 52 17.3 | -59 43 21 | 276.95 | -9.68 | 125 | -.8 | 15.7 | 17x 7 | 18.8 | .18 | | S | 1 | dist.blg |
| 659 | | | 08 52 17.6 | -59 01 35 | 276.40 | -9.23 | 125 | -.5 | 53.0 | 91x 89 | 14.0 | .28 | | E 1 | S | |
| 660 | | | 08 52 20.0 | -56 39 13 | 274.55 | -7.72 | 165 | 29.2 | -88.6 | 34x 8 | 17.7 | .28 | | S | E | |
| 661 | | | 08 52 24.3 | -60 14 12 | 277.36 | -9.99 | 125 | -.3 | -11.8 | 54x 15 | 16.5 | .19 | | S M | | |
| 662 | | | 08 52 25.1 | -59 37 49 | 276.88 | -9.60 | 125 | .1 | 20.7 | 17x 8 | 18.9 | .19 | | S | | vLSB |
| 663 | | | 08 52 25.4 | -54 10 17 | 272.64 | -6.12 | 165 | 32.3 | 44.4 | 15x 5 | 19.1 | .32 | | | S | |
| 664 | | | 08 52 30.2 | -55 41 15 | 273.82 | -7.08 | 165 | 31.4 | -36.9 | 50x 15 | 16.6 | .29 | | S M: | S | in st.diffr.pat. |
| 665 | | | 08 52 30.4 | -58 11 15 | 275.76 | -8.68 | 125 | 1.4 | 98.0 | 20x 20 | 17.3 | .25 | | E 0 | | |
| 666 | | | 08 52 30.7 | -54 17 04 | 272.74 | -6.18 | 165 | 32.9 | 38.4 | 20x 8 | 18.4 | .33 | | | | |
| 667 | | | 08 52 33.9 | -55 34 10 | 273.74 | -7.00 | 165 | 32.0 | -30.5 | 15x 4 | 19.4 | .30 | | S | E S | in st.diffr.pat. |
| 668 | | | 08 52 37.1 | -55 28 33 | 273.67 | -6.93 | 165 | 32.5 | -25.5 | 24x 19 | 17.5 | .32 | | S | F | in st.diffr.pat. |
| 669 | | | 08 52 37.6 | -55 44 33 | 273.87 | -7.10 | 165 | 32.3 | -39.8 | 32x 24 | 16.8 | .29 | | S M | F | in st.diffr.pat. |
| 670 | | | 08 52 37.7 | -59 11 19 | 276.56 | -9.30 | 125 | 1.8 | 44.3 | 19x 5 | 18.9 | .25 | | S | S | dist.blg, vLSBdisk |
| 671 | | | 08 52 38.5 | -56 21 33 | 274.35 | -7.50 | 165 | 31.8 | -72.9 | 13x 3 | 20.3 | .29 | | S | | |
| 672 | | | 08 52 40.0 | -54 50 53 | 273.19 | -6.53 | 165 | 33.5 | 8.1 | 13x 7 | 19.2 | .29 | ? | S ? | S | |
| 673 | | | 08 52 40.2 | -56 20 06 | 274.34 | -7.48 | 165 | 32.0 | -71.6 | 16x 16 | 18.0 | .29 | ? | | | blg or s.p.* |
| 674 | | | 08 52 43.4 | -55 27 51 | 273.67 | -6.91 | 165 | 33.3 | -24.9 | 27x 4 | 18.9 | .33 | | S | E | in st.diffr.pat. |
| 675 | | | 08 52 43.7 | -59 40 53 | 276.95 | -9.61 | 125 | 2.2 | 17.9 | 20x 7 | 18.2 | .19 | | S | S | |
| 676 | | | 08 52 46.5 | -59 10 32 | 276.56 | -9.28 | 125 | 2.8 | 45.0 | 12x 9 | 19.0 | .25 | | | | |
| 677 | | | 08 52 51.2 | -58 09 27 | 275.77 | -8.62 | 125 | 3.9 | 99.6 | 34x 28 | 16.4 | .25 | | S 5 | S | |
| 678 | | | 08 52 52.7 | -59 46 54 | 277.04 | -9.66 | 125 | 3.1 | 12.5 | 17x 5 | 19.1 | .18 | | S | N | |
| 679 | | | 08 52 55.8 | -58 22 46 | 275.95 | -8.76 | 125 | 4.3 | 87.7 | 13x 4 | 19.8 | .28 | | S | | |
| 680 | | | 08 52 57.7 | -59 59 10 | 277.21 | -9.78 | 125 | 3.6 | 1.6 | 15x 7 | 18.8 | .19 | | | 1 | cl.to br.* |
| 681 | | | 08 53 03.2 | -57 10 30 | 275.02 | -7.98 | 165 | 33.9 | -116.7 | 20x 16 | 17.6 | .33 | ? | | | triangle-shaped |
| 682 | | | 08 53 07.1 | -54 35 17 | 273.03 | -6.31 | 165 | 37.3 | 22.0 | 34x 20 | 17.2 | .34 | | | | |
| 683 | | | 08 53 07.1 | -57 36 43 | 275.37 | -8.25 | 125 | 6.1 | 128.8 | 20x 12 | 17.6 | .26 | | S E | 1 | or blg |
| 684 | | | 08 53 09.3 | -53 41 38 | 272.34 | -5.73 | 165 | 38.6 | 69.9 | 13x 8 | 18.7 | .34 | | | | |
| 685 | | | 08 53 09.3 | -60 06 56 | 277.32 | -9.84 | 125 | 4.8 | -5.4 | 12x 5 | 19.6 | .20 | ? | | | |
| 686 | | P | 08 53 13.4 | -55 36 33 | 273.83 | -6.95 | 165 | 37.0 | -32.8 | 58x 40 | 15.4 | .31 | | S L | S | in st.diffr.pat. |
| 687 | | | 08 53 15.4 | -59 51 22 | 277.13 | -9.67 | 125 | 5.6 | 8.5 | 12x 7 | 19.2 | .18 | ? | | | |
| 688 | | | 08 53 15.8 | -54 33 26 | 273.02 | -6.27 | 165 | 38.5 | 23.6 | 27x 24 | 17.1 | .35 | | S M | S | |
| 689 | | | 08 53 18.0 | -54 37 16 | 273.07 | -6.31 | 165 | 38.7 | 20.2 | 17x 17 | 18.1 | .34 | | | S | |
| 690 | | | 08 53 23.5 | -56 13 13 | 274.31 | -7.33 | 165 | 37.5 | -65.6 | 20x 7 | 18.8 | .29 | ? | | S | |
| 691 | | | 08 53 30.0 | -58 25 50 | 276.04 | -8.73 | 125 | 8.3 | 84.9 | 12x 5 | 19.3 | .28 | | | | |
| 692 | | | 08 53 30.1 | -59 42 02 | 277.03 | -9.54 | 125 | 7.4 | 16.8 | 16x 13 | 17.9 | .20 | | E | | |
| 693 | | | 08 53 31.7 | -55 59 43 | 274.15 | -7.17 | 165 | 38.8 | -53.5 | 13x 13 | 18.6 | .33 | ? | | | |
| 694 | L165-007 | I | 08 53 32.8 | -55 41 58 | 273.93 | -6.98 | 165 | 39.3 | -37.7 | 67x 13 | 16.3 | .32 | | S ? | | in st.diffr.pat. |
| 695 | | | 08 53 33.5 | -54 46 06 | 273.21 | -6.38 | 165 | 40.5 | 12.3 | 23x 13 | 17.9 | .32 | ? | | | |
| 696 | | | 08 53 36.9 | -59 41 46 | 277.03 | -9.53 | 125 | 8.2 | 17.0 | 9x 7 | 19.0 | .20 | | E | | neighb.of 307 |
| 697 | | | 08 53 42.0 | -57 38 28 | 275.44 | -8.21 | 125 | 10.2 | 127.2 | 17x 13 | 17.8 | .25 | | | S | i.a.w.699 |
| 698 | | | 08 53 43.5 | -58 43 36 | 276.29 | -8.90 | 125 | 9.6 | 69.0 | 67x 50 | 15.2 | .31 | | E | S | |
| 699 | | | 08 53 44.0 | -57 38 50 | 275.45 | -8.21 | 125 | 10.5 | 126.8 | 20x 17 | 17.8 | .25 | | S | S | i.a.w.697 |
| 700 | | | 08 53 48.9 | -55 57 20 | 274.15 | -7.11 | 165 | 41.0 | -51.4 | 27x 11 | 18.0 | .33 | ? | | S | |

**Table 1.** Galaxies in the southern ZOA – I. The Hydra/Antlia Extension (continued)

| RKK Ident (1) | Other Ident (2) | IR (3) | R.A. (h m s) (4) | Dec. (° ′ ″) (5) | gal ℓ (°) (6) | gal $b$ (°) (7) | SRC (8) | X (mm) (9) | Y (mm) (10) | D x d (″) (11) | $B_J$ ($^m$) (12) | $N_{HI}$ (13) | u | Type class. (14) | | | o | * | Remarks (15) |
|---|---|---|---|---|---|---|---|---|---|---|---|---|---|---|---|---|---|---|---|
| 701 | | | 08 53 51.8 | -53 55 58 | 272.60 | -5.80 | 165 | 43.9 | 57.0 | 20x 5 | 19.0 | .34 | | | | | | 1 | |
| 702 | | | 08 53 51.8 | -55 26 06 | 273.75 | -6.77 | 165 | 42.0 | -23.5 | 12x 7 | 19.1 | .32 | | | | | | | in st.diffr.pat. |
| 703 | | | 08 53 52.1 | -59 18 35 | 276.75 | -9.26 | 125 | 10.2 | 37.7 | 16x 11 | 18.2 | .21 | ? | S | | ? | | | br.blg, vLSBdisk |
| 704 | | | 08 53 54.0 | -57 27 42 | 275.32 | -8.07 | 165 | 39.7 | -132.2 | 13x 8 | 19.0 | .27 | | S | | | | | |
| 705 | | | 08 53 58.0 | -55 44 00 | 273.99 | -6.95 | 165 | 42.4 | -39.5 | 15x 9 | 18.6 | .34 | | | | | | | in st.diffr.pat. |
| 706 | | | 08 53 58.2 | -59 40 59 | 277.05 | -9.49 | 125 | 10.6 | 17.7 | 16x 16 | 17.5 | .21 | | E | 0 | | | | |
| 707 | | | 08 54 00.6 | -55 02 28 | 273.46 | -6.50 | 165 | 43.6 | -2.5 | 22x 11 | 17.8 | .29 | | S | | | | S | in st.diffr.pat. |
| 708 | | | 08 54 03.6 | -55 46 11 | 274.03 | -6.97 | 165 | 43.1 | -41.5 | 17x 7 | 18.6 | .33 | | | | | | | in st.diffr.pat. |
| 709 | | | 08 54 05.6 | -55 43 04 | 273.99 | -6.93 | 165 | 43.4 | -38.7 | 15x 8 | 18.7 | .34 | | | | | | | in st.diffr.pat. |
| 710 | | | 08 54 06.0 | -60 14 38 | 277.50 | -9.83 | 125 | 11.0 | -12.3 | 13x 3 | 20.4 | .19 | | S | | | E | | v.obs. |
| 711 | | | 08 54 17.6 | -57 53 53 | 275.69 | -8.31 | 125 | 14.3 | 113.3 | 12x 7 | 19.1 | .26 | | S | | | | | vLSBdisk |
| 712 | | | 08 54 19.6 | -60 38 32 | 277.83 | -10.07 | 125 | 12.2 | -33.7 | 16x 5 | 19.0 | .22 | | S | | | | | dist.blg |
| 713 | | | 08 54 19.8 | -53 52 13 | 272.59 | -5.71 | 165 | 47.7 | 60.2 | 34x 5 | 18.6 | .35 | | S | | | E | | |
| 714 | | | 08 54 22.5 | -55 26 42 | 273.81 | -6.72 | 165 | 45.9 | -24.2 | 20x 9 | 18.3 | .31 | | | | | | 1 | in st.diffr.pat. |
| 715 | | | 08 54 30.4 | -54 51 47 | 273.37 | -6.33 | 165 | 47.7 | 7.0 | 17x 13 | 18.3 | .37 | | | | | | S | |
| 716 | | | 08 54 31.2 | -60 39 28 | 277.86 | -10.06 | 125 | 13.4 | -34.6 | 17x 15 | 17.6 | .22 | | | | | | | |
| 717 | | | 08 54 33.3 | -54 46 12 | 273.30 | -6.27 | 165 | 48.2 | 12.0 | 16x 7 | 19.0 | .35 | | | | | | S | |
| 718 | | | 08 54 39.5 | -54 52 28 | 273.39 | -6.32 | 165 | 48.9 | 6.4 | 67x 60 | 15.1 | .38 | | S | | ? | F | S | |
| 719 | | | 08 54 43.5 | -57 34 33 | 275.48 | -8.06 | 125 | 17.7 | 130.5 | 13x 5 | 19.7 | .25 | ? | S | | | | 1 | |
| 720 | | NS | 08 54 45.4 | -55 18 10 | 273.73 | -6.59 | 165 | 49.0 | -16.6 | 17x 9 | 18.3 | .29 | | | | | | S | in st.diffr.pat. |
| 721 | | | 08 54 46.4 | -57 19 25 | 275.29 | -7.89 | 165 | 46.2 | -124.9 | 40x 5 | 17.9 | .26 | | S | | | E | | |
| 722 | | | 08 54 47.0 | -60 36 59 | 277.85 | -10.01 | 125 | 15.2 | -32.4 | 13x 7 | 19.0 | .22 | ? | | | | | | |
| 723 | | | 08 54 49.1 | -59 33 19 | 277.02 | -9.32 | 125 | 16.4 | 24.5 | 13x 5 | 19.8 | .22 | ? | | | | | | |
| 724 | | | 08 54 50.0 | -59 51 51 | 277.26 | -9.52 | 125 | 16.3 | 7.9 | 12x 8 | 19.1 | .19 | | S | | | | | dist.blg, vLSBdisk |
| 725 | | | 08 54 51.1 | -59 51 54 | 277.27 | -9.52 | 125 | 16.4 | 7.9 | 13x 5 | 19.5 | .19 | | S | | | | | dist.blg, vLSBdisk |
| 726 | | | 08 54 52.9 | -57 42 40 | 275.60 | -8.13 | 125 | 18.6 | 123.3 | 28x 22 | 17.0 | .25 | | S | | | | | dist.blg, v.obs. |
| 727 | | | 08 54 52.9 | -59 52 16 | 277.27 | -9.52 | 125 | 16.6 | 7.5 | 13x 5 | 19.5 | .19 | | S | | | | | neighb.of 725, cl.to br.* |
| 728 | | | 08 54 56.6 | -55 05 18 | 273.58 | -6.43 | 165 | 50.7 | -5.2 | 20x 16 | 17.4 | .34 | | E | | ? | | | w.LSBcomp, in st.diffr.pat |
| 729 | | | 08 54 56.8 | -58 30 11 | 276.22 | -8.63 | 125 | 18.3 | 80.8 | 12x 5 | 19.4 | .33 | | | | | | | |
| 730 | | | 08 54 58.5 | -58 18 29 | 276.07 | -8.51 | 125 | 18.7 | 91.3 | 11x 9 | 18.4 | .29 | ? | E | | | | | |
| 731 | | | 08 55 06.5 | -55 23 23 | 273.83 | -6.61 | 165 | 51.6 | -21.4 | 15x 7 | 19.0 | .30 | | | | | | | in st.diffr.pat. |
| 732 | | | 08 55 11.1 | -55 22 46 | 273.83 | -6.59 | 165 | 52.1 | -20.8 | 13x 13 | 18.2 | .30 | ? | E | | ? | | | neighb.of 731 |
| 733 | | | 08 55 11.1 | -59 46 40 | 277.22 | -9.43 | 125 | 18.7 | 12.5 | 20x 17 | 17.4 | .20 | | F | | | | | |
| 734 | | | 08 55 11.8 | -53 26 30 | 272.35 | -5.33 | 165 | 55.2 | 83.0 | 16x 9 | 18.5 | .48 | | | | | | | |
| 735 | | | 08 55 17.6 | -59 47 04 | 277.24 | -9.42 | 125 | 19.4 | 12.1 | 12x 9 | 18.4 | .20 | ? | E | | | | | |
| 736 | | | 08 55 18.4 | -54 43 32 | 273.34 | -6.16 | 165 | 54.1 | 14.2 | 23x 9 | 18.1 | .36 | | | | | | | |
| 737 | | | 08 55 19.0 | -57 36 09 | 275.55 | -8.02 | 125 | 21.9 | 129.0 | 36x 31 | 16.5 | .25 | | S | | | | S | dist.blg, v.obs. |
| 738 | | | 08 55 20.3 | -58 49 42 | 276.50 | -8.80 | 125 | 20.7 | 63.4 | 13x 12 | 18.3 | .28 | | S | | ? | | S | |
| 739 | | | 08 55 22.0 | -53 46 56 | 272.63 | -5.54 | 165 | 56.0 | 64.7 | 20x 8 | 18.4 | .37 | | | | | | | |
| 740 | | | 08 55 23.1 | -53 35 35 | 272.48 | -5.41 | 165 | 56.5 | 74.9 | 13x 5 | 19.4 | .41 | | | | | | | |
| 741 | | | 08 55 24.0 | -54 44 33 | 273.36 | -6.16 | 165 | 54.8 | 13.3 | 17x 8 | 18.7 | .37 | | | | | | S | |
| 742 | | | 08 55 25.9 | -55 15 04 | 273.75 | -6.48 | 165 | 54.2 | -14.0 | 13x 5 | 19.4 | .31 | | | | | | | in st.diffr.pat. |
| 743 | | | 08 55 28.9 | -57 37 25 | 275.58 | -8.01 | 125 | 23.0 | 127.9 | 13x 4 | 20.1 | .25 | | | | | | | p.cov.by * |
| 744 | | | 08 55 33.6 | -60 40 06 | 277.95 | -9.97 | 125 | 20.2 | -35.3 | 22x 13 | 17.9 | .22 | | S | | | | 1 | dist.blg |
| 745 | | | 08 55 35.8 | -54 19 47 | 273.07 | -5.87 | 165 | 57.0 | 35.3 | 20x 9 | 18.0 | .31 | ? | | | | | | |
| 746 | | | 08 55 35.9 | -58 49 16 | 276.52 | -8.77 | 125 | 22.6 | 63.7 | 12x 5 | 19.1 | .28 | | S | E | | | | w.comp. |
| 747 | | | 08 55 37.5 | -58 09 53 | 276.01 | -8.35 | 125 | 23.5 | 98.9 | 27x 4 | 19.0 | .29 | | S | | | E | | dist.blg |
| 748 | | | 08 55 39.0 | -57 38 49 | 275.61 | -8.01 | 125 | 24.2 | 126.6 | 13x 7 | 18.7 | .25 | | E | | ? | | S | |
| 749 | | | 08 55 39.0 | -58 09 10 | 276.00 | -8.34 | 125 | 23.7 | 99.5 | 12x 7 | 19.2 | .29 | | | | | | | neighb.of 747 |
| 750 | | | 08 55 49.9 | -57 35 40 | 275.59 | -7.96 | 125 | 25.6 | 129.4 | 17x 11 | 18.3 | .25 | | | | | | | |
| 751 | | | 08 55 52. | -60 18.1 : | 277.69 | -9.70 | 125 | 22.7 | -15.7 | 22x 15 | 17.5 | .17 | | S | | ? | | 1 | |
| 752 | | | 08 55 53.3 | -57 52 30 | 275.81 | -8.13 | 125 | 25.7 | 114.4 | 20x 16 | 17.4 | .29 | | S | E | | | | |
| 753 | | | 08 55 54.0 | -54 02 07 | 272.87 | -5.64 | 165 | 59.8 | 51.1 | 16x 16 | 18.0 | .40 | | S | | ? | F | | |
| 754 | | | 08 55 58.1 | -56 13 37 | 274.55 | -7.06 | 165 | 56.7 | -66.4 | 16x 4 | 19.5 | .34 | | S | | | E | | dist.nucl. |
| 755 | | | 08 56 02.6 | -58 04 21 | 275.98 | -8.25 | 125 | 26.5 | 103.8 | 13x 7 | 19.2 | .30 | ? | S | | ? | | | |
| 756 | | | 08 56 05.4 | -54 54 15 | 273.55 | -6.19 | 165 | 59.8 | 4.4 | 15x 7 | 18.8 | .39 | | | | | | | |
| 757 | | I | 08 56 05.8 | -54 00 13 | 272.86 | -5.60 | 165 | 61.4 | 52.7 | 27x 7 | 18.3 | .43 | | S | | | | | |
| 758 | | | 08 56 15.4 | -54 59 58 | 273.64 | -6.23 | 165 | 61.0 | -.7 | 16x 4 | 19.4 | .35 | | | | | | | |
| 759 | | | 08 56 18.0 | -54 56 04 | 273.59 | -6.18 | 165 | 61.4 | 2.8 | 19x 5 | 18.9 | .37 | | | | | | | |
| 760 | | | 08 56 18.5 | -57 48 11 | 275.79 | -8.05 | 125 | 28.7 | 118.1 | 16x 8 | 18.4 | .27 | | S | E | | | | or blg |
| 761 | | | 08 56 22.5 | -54 40 29 | 273.40 | -6.01 | 165 | 62.4 | 16.7 | 16x 4 | 19.7 | .34 | | S | | | E | | |
| 762 | | | 08 56 22.6 | -60 22 12 | 277.78 | -9.70 | 125 | 26.0 | -19.4 | 13x 9 | 19.0 | .17 | | S | | | | | |
| 763 | | | 08 56 22.9 | -53 38 52 | 272.62 | -5.33 | 165 | 64.3 | 71.7 | 16x 4 | 19.4 | .41 | | S | | | E | | |
| 764 | L165-008 | | 08 56 25.8 | -53 40 51 | 272.65 | -5.35 | 165 | 64.6 | 69.9 | 60x 54 | 15.1 | .41 | ? | D | | ? | | S | smooth |
| 765 | | | 08 56 25.8 | -57 53 05 | 275.87 | -8.09 | 125 | 29.5 | 113.7 | 13x 5 | 19.5 | .28 | ? | E | | | | | |
| 766 | | | 08 56 28.3 | -59 29 08 | 277.10 | -9.12 | 125 | 27.8 | 28.0 | 15x 7 | 18.7 | .21 | | S | | | | | |
| 767 | | | 08 56 33.0 | -59 04 45 | 276.80 | -8.85 | 125 | 28.8 | 49.7 | 15x 15 | 17.9 | .24 | | E | | ? | | | in st.diffr.pat. |
| 768 | | | 08 56 33.9 | -57 34 32 | 275.64 | -7.87 | 125 | 30.9 | 130.3 | 17x 4 | 19.4 | .25 | | S | | | | | |
| 769 | | | 08 56 36.3 | -58 14 52 | 276.16 | -8.30 | 125 | 30.3 | 94.3 | 17x 8 | 18.7 | .28 | | | | | | 1 | |
| 770 | | | 08 56 36.5 | -58 21 29 | 276.25 | -8.37 | 125 | 30.2 | 88.4 | 28x 13 | 17.6 | .29 | | S | E | | | S | |
| 771 | | | 08 56 37.2 | -58 19 22 | 276.22 | -8.35 | 125 | 30.3 | 90.2 | 13x 13 | 18.4 | .28 | | E | 0 | | | | |
| 772 | | | 08 56 38.2 | -58 18 11 | 276.21 | -8.34 | 125 | 30.4 | 91.3 | 16x 13 | 18.3 | .28 | | S | M | | | S | |
| 773 | | | 08 56 39.7 | -58 39 34 | 276.48 | -8.56 | 125 | 30.1 | 72.2 | 36x 5 | 18.5 | .31 | ? | S | L | | E | | cl.to br.* |
| 774 | | | 08 56 42.0 | -58 12 58 | 276.14 | -8.27 | 125 | 31.0 | 95.9 | 17x 8 | 18.7 | .28 | | S | | | | 1 | w.comp. |
| 775 | | | 08 56 43.2 | -59 05 35 | 276.82 | -8.84 | 125 | 30.0 | 49.0 | 30x 27 | 16.5 | .23 | | S | M | | F | | in st.diffr.pat. |
| 776 | | | 08 56 43.3 | -55 35 23 | 274.13 | -6.56 | 165 | 63.5 | -32.5 | 15x 8 | 18.4 | .37 | | | | | | | |
| 777 | | | 08 56 46.2 | -60 25 48 | 277.86 | -9.70 | 125 | 28.5 | -22.7 | 13x 4 | 19.7 | .17 | ? | S | | | | | |
| 778 | | | 08 56 46.6 | -60 03 25 | 277.57 | -9.46 | 125 | 29.1 | -2.7 | 12x 12 | 18.1 | .18 | | E | 0 | | | | |
| 779 | | | 08 56 46.7 | -54 32 23 | 273.34 | -5.87 | 165 | 65.8 | 23.8 | 20x 9 | 18.1 | .36 | | S | | | | S | |
| 780 | | | 08 56 46.7 | -54 40 58 | 273.45 | -5.97 | 165 | 65.6 | 16.1 | 13x 12 | 18.5 | .35 | | | | | | | |
| 781 | | | 08 56 47.9 | -57 55 37 | 275.93 | -8.08 | 125 | 32.1 | 111.4 | 17x 8 | 18.7 | .29 | | S | | | | | dist.blg |
| 782 | | | 08 57 01.3 | -61 00 10 | 278.32 | -10.04 | 125 | 29.4 | -53.4 | 20x 17 | 17.6 | .23 | | L | | ? | | | |
| 783 | CSRG0560 | | 08 57 02.1 | -58 26 37 | 276.35 | -8.39 | 125 | 33.0 | 83.7 | 35x 20 | 16.9 | .31 | | S | M | | | S | dist.blg |
| 784 | | | 08 57 02.8 | -57 47 22 | 275.85 | -7.96 | 125 | 34.0 | 118.7 | 15x 4 | 19.7 | .28 | | S | | | E | 1 | |
| 785 | | | 08 57 03.8 | -58 04 01 | 276.06 | -8.14 | 125 | 33.8 | 103.9 | 12x 3 | 20.5 | .29 | | S | | | E | | |
| 786 | | | 08 57 04.8 | -55 31 17 | 274.11 | -6.48 | 165 | 66.3 | -28.9 | 17x 5 | 19.1 | .35 | | S | | | | | |
| 787 | | | 08 57 04.8 | -59 41 43 | 277.32 | -9.19 | 125 | 31.6 | 16.6 | 54x 30 | 15.6 | .22 | | S | M | | | | dist.blg |
| 788 | | | 08 57 05.4 | -59 37 01 | 277.26 | -9.14 | 125 | 31.8 | 20.8 | 13x 8 | 19.0 | .20 | ? | S | | | | S | |
| 789 | | | 08 57 06.1 | -59 46 45 | 277.38 | -9.25 | 125 | 31.6 | 12.1 | 13x 8 | 18.4 | .20 | | S | E | | | | |
| 790 | | | 08 57 07.5 | -55 56 48 | 274.44 | -6.75 | 165 | 65.8 | -51.7 | 22x 4 | 19.0 | .36 | | S | | | | | |
| 791 | | | 08 57 10.8 | -59 11 47 | 276.94 | -8.86 | 125 | 33.0 | 43.3 | 15x 7 | 18.7 | .21 | | | | | | | blg |
| 792 | | | 08 57 16.5 | -57 35 35 | 275.72 | -7.81 | 125 | 35.9 | 129.2 | 27x 27 | 16.7 | .25 | | S | | ? | F | | v.obs. |
| 793 | | | 08 57 21.4 | -59 08 20 | 276.91 | -8.81 | 125 | 34.3 | 46.4 | 23x 9 | 17.9 | .22 | | E | | ? | | | |
| 794 | | | 08 57 22.0 | -59 01 20 | 276.82 | -8.73 | 125 | 34.5 | 52.6 | 13x 12 | 18.9 | .24 | | | | | | 1 | vLSB |
| 795 | | | 08 57 24.7 | -55 08 32 | 273.86 | -6.20 | 165 | 69.6 | -8.7 | 30x 23 | 16.9 | .34 | | S | | | F | | smooth |
| 796 | | | 08 57 29.7 | -57 44 47 | 275.85 | -7.89 | 125 | 37.3 | 121.0 | 16x 11 | 18.4 | .28 | | S | M | | | 1 | |
| 797 | | | 08 57 30.3 | -57 54 32 | 275.98 | -7.99 | 125 | 37.1 | 112.3 | 42x 16 | 16.8 | .30 | | S | M | | | S | dist.blg, v.obs. |
| 798 | | | 08 57 30.8 | -57 44 19 | 275.85 | -7.88 | 125 | 37.4 | 121.4 | 11x 7 | 18.8 | .28 | | | | | | | dist.blg |
| 799 | | I | 08 57 31.4 | -55 12 26 | 273.92 | -6.23 | 165 | 70.3 | -12.2 | 24x 19 | 17.0 | .34 | | | | | | S | |
| 800 | | | 08 57 32.2 | -57 43 59 | 275.85 | -7.88 | 125 | 37.6 | 121.7 | 11x 7 | 19.2 | .28 | | | | | | | between ** |

**Table 1.** Galaxies in the southern ZOA – I. The Hydra/Antlia Extension (continued)

| RKK Ident (1) | Other Ident (2) | IR (3) | R.A. (h m s) (4) | Dec. (° ′ ″) (5) | gal ℓ (°) (6) | gal $b$ (°) (7) | SRC (8) | X (mm) (9) | Y (mm) (10) | D x d (″) (11) | $B_J$ ($^m$) (12) | $N_{HI}$ (13) | u | Type class. (14) | o | * | Remarks (15) |
|---|---|---|---|---|---|---|---|---|---|---|---|---|---|---|---|---|---|
| 801 | | P2 | 08 57 32.5 | -53 02 58 | 272.28 | -4.81 | 165 | 74.7 | 103.5 | 9x 8 | 19.2 | .84 | | | | | pair w.802 |
| 802 | | P2 | 08 57 33.0 | -53 03 07 | 272.28 | -4.81 | 165 | 74.7 | 103.3 | 13x 7 | 19.0 | .84 | | | | | pair w.801 |
| 803 | | | 08 57 38.0 | -57 16 03 | 275.50 | -7.56 | 165 | 67.0 | -122.6 | 15x 11 | 18.3 | .25 | ? | S 5 | | | |
| 804 | | | 08 57 40.9 | -59 42 54 | 277.38 | -9.15 | 125 | 35.6 | 15.5 | 15x 5 | 19.2 | .22 | | | | | blg |
| 805 | | | 08 57 45.2 | -59 43 33 | 277.40 | -9.15 | 125 | 36.1 | 14.9 | 17x 12 | 18.2 | .21 | | S M | | | |
| 806 | | | 08 57 46.9 | -60 32 32 | 278.03 | -9.68 | 125 | 35.0 | -28.9 | 17x 12 | 18.1 | .18 | | | | | |
| 807 | | | 08 57 52.6 | -57 31 58 | 275.72 | -7.71 | 125 | 40.4 | 132.3 | 11x 8 | 19.0 | .25 | ? | | | 1 | |
| 808 | | I | 08 57 58. | -52 39.7 : | 272.03 | -4.51 | 211 | -16.9 | -142.6 | 17x 12 | 17.8 | .97 | | S 3 : | | | |
| 809 | | | 08 57 58.7 | -57 26 44 | 275.66 | -7.64 | 165 | 69.1 | -132.2 | 27x 7 | 18.2 | .25 | | S M | E | | |
| 810 | | | 08 58 00.0 | -60 35 34 | 278.09 | -9.69 | 125 | 36.4 | -31.6 | 15x 9 | 18.7 | .19 | | | | | between ** |
| 811 | | | 08 58 02.5 | -58 23 45 | 276.40 | -8.26 | 125 | 40.2 | 86.1 | 16x 4 | 19.3 | .31 | | S | E | | w.comp. |
| 812 | | | 08 58 03.3 | -61 01 59 | 278.43 | -9.97 | 125 | 36.0 | -55.2 | 13x 9 | 19.1 | .24 | | | | | |
| 813 | | | 08 58 03.5 | -59 11 14 | 277.01 | -8.77 | 125 | 39.0 | 43.7 | 17x 7 | 18.6 | .21 | | S | | | asym.spiral |
| 814 | | | 08 58 05.4 | -60 30 36 | 278.03 | -9.63 | 125 | 37.1 | -27.2 | 16x 15 | 17.7 | .18 | | E | | | |
| 815 | | | 08 58 07.2 | -53 11 38 | 272.45 | -4.84 | 165 | 79.0 | 95.5 | 20x 11 | 17.9 | .58 | | | | | |
| 816 | | | 08 58 07.4 | -60 10 02 | 277.77 | -9.40 | 125 | 37.9 | -8.8 | 11x 8 | 18.9 | .17 | | E | | | or blg |
| 817 | | | 08 58 07.5 | -57 31 45 | 275.74 | -7.68 | 165 | 70.0 | -136.8 | 30x 9 | 17.8 | .25 | | S 2 : | | | blg or s.p.* |
| 818 | | | 08 58 07.7 | -60 09 01 | 277.75 | -9.39 | 125 | 38.0 | -7.9 | 13x 4 | 20.0 | .17 | | S | | | |
| 819 | | | 08 58 09.6 | -60 30 40 | 278.03 | -9.62 | 125 | 37.6 | -27.3 | 11x 5 | 19.3 | .18 | | S | | | cl.to br.* |
| 820 | | | 08 58 10.0 | -60 15 45 | 277.84 | -9.46 | 125 | 38.0 | -14.0 | 16x 15 | 17.7 | .17 | | E | | | |
| 821 | | | 08 58 11.1 | -56 20 18 | 274.84 | -6.90 | 165 | 73.0 | -73.0 | 13x 3 | 20.2 | .36 | | S | E | | |
| 822 | | | 08 58 11.6 | -60 14 26 | 277.83 | -9.44 | 125 | 38.2 | -12.8 | 11x 9 | 18.4 | .17 | | E | | | or blg |
| 823 | | | 08 58 11.8 | -58 25 23 | 276.43 | -8.26 | 125 | 41.2 | 84.6 | 15x 8 | 18.4 | .31 | | S M | | | |
| 824 | | I | 08 58 13.7 | -60 51 24 | 278.31 | -9.84 | 125 | 37.4 | -45.8 | 16x 8 | 18.4 | .22 | | S E | | 1 | |
| 825 | | | 08 58 15.8 | -61 12 03 | 278.58 | -10.06 | 125 | 37.1 | -64.2 | 13x 8 | 19.0 | .22 | | | | | |
| 826 | | | 08 58 16.3 | -60 08 29 | 277.76 | -9.37 | 125 | 38.9 | -7.5 | 13x 4 | 19.7 | .17 | | S | | | |
| 827 | | | 08 58 16.6 | -59 18 32 | 277.12 | -8.83 | 125 | 40.3 | 37.1 | 22x 17 | 17.3 | .20 | | S | F | | |
| 828 | | | 08 58 21.6 | -57 43 37 | 275.91 | -7.79 | 125 | 43.5 | 121.8 | 13x 13 | 18.9 | .29 | | | | | |
| 829 | | | 08 58 25.6 | -57 37 10 | 275.84 | -7.71 | 125 | 44.2 | 127.6 | 12x 8 | 18.9 | .27 | ? | S | | S | |
| 830 | | | 08 58 28.5 | -58 03 00 | 276.17 | -7.99 | 125 | 43.8 | 104.5 | 15x 9 | 18.4 | .29 | | S | | | |
| 831 | | | 08 58 32.1 | -57 43 18 | 275.93 | -7.77 | 125 | 44.8 | 122.1 | 16x 11 | 18.3 | .29 | | S ? | | | |
| 832 | | | 08 58 36.1 | -53 24 46 | 272.66 | -4.93 | 165 | 82.4 | 83.7 | 13x 7 | 18.5 | .50 | | | | | dist.blg |
| 833 | | | 08 58 40.2 | -59 26 26 | 277.25 | -8.88 | 125 | 42.8 | 30.0 | 17x 13 | 18.0 | .20 | | S | | | |
| 834 | | | 08 58 41.9 | -54 13 01 | 273.28 | -5.45 | 165 | 81.4 | 40.5 | 17x 5 | 18.8 | .45 | | S ? | | | |
| 835 | | | 08 58 44.0 | -60 28 01 | 278.05 | -9.54 | 125 | 41.4 | -25.0 | 13x 11 | 18.7 | .18 | | | | | |
| 836 | | | 08 58 48.7 | -59 26 14 | 277.26 | -8.86 | 125 | 43.8 | 30.1 | 9x 9 | 18.6 | .20 | | E | | | neighb.of 833 |
| 837 | | | 08 59 00.7 | -54 11 40 | 273.29 | -5.40 | 165 | 83.9 | 41.6 | 22x 7 | 18.8 | .43 | | | | | |
| 838 | | | 08 59 05.8 | -57 18 08 | 275.66 | -7.44 | 165 | 77.5 | -124.9 | 30x 8 | 17.9 | .27 | | S M | | | |
| 839 | | | 08 59 06.8 | -58 54 20 | 276.88 | -8.48 | 125 | 46.8 | 58.5 | 13x 9 | 19.3 | .26 | | | | | vLSB |
| 840 | | | 08 59 07.5 | -54 30 29 | 273.54 | -5.60 | 165 | 84.1 | 24.8 | 15x 5 | 19.3 | .38 | | | | | |
| 841 | | | 08 59 08.6 | -60 34 42 | 278.17 | -9.57 | 125 | 43.9 | -31.1 | 13x 4 | 19.8 | .19 | | S ? | | | |
| 842 | | | 08 59 10.1 | -60 29 41 | 278.10 | -9.52 | 125 | 44.2 | -26.6 | 20x 11 | 17.8 | .18 | | S E | | | |
| 843 | | | 08 59 18.3 | -57 42 09 | 275.98 | -7.68 | 125 | 50.3 | 122.9 | 27x 8 | 18.1 | .30 | | S M | | | |
| 844 | | | 08 59 21.0 | -60 31 28 | 278.14 | -9.52 | 125 | 45.4 | -28.3 | 16x 11 | 18.6 | .19 | | S ? | | 1 | vLSB |
| 845 | | | 08 59 21.2 | -59 27 18 | 277.32 | -8.82 | 125 | 47.4 | 29.0 | 23x 9 | 18.0 | .21 | | S | E | | |
| 846 | | | 08 59 22.6 | -60 06 14 | 277.82 | -9.24 | 125 | 46.4 | -5.7 | 15x 15 | 18.6 | .17 | ? | | | | vLSB, Y-shaped |
| 847 | | | 08 59 24.5 | -59 40 37 | 277.50 | -8.96 | 125 | 47.4 | 17.1 | 13x 11 | 18.4 | .19 | | | | | |
| 848 | | | 08 59 31.7 | -59 02 08 | 277.01 | -8.53 | 125 | 49.4 | 51.5 | 13x 8 | 19.3 | .23 | | S M | | | LSB |
| 849 | | | 08 59 33.4 | -58 46 59 | 276.82 | -8.36 | 125 | 50.1 | 65.0 | 13x 7 | 19.5 | .29 | | | | | LSB, w.comp. |
| 850 | | | 08 59 34.2 | -57 21 30 | 275.74 | -7.42 | 165 | 80.8 | -128.0 | 16x 7 | 18.4 | .28 | | | | | |
| 851 | | | 08 59 37.7 | -58 53 59 | 276.92 | -8.43 | 125 | 50.4 | 58.7 | 13x 4 | 20.1 | .25 | | S | | | LSB, cl.to br.* |
| 852 | | | 08 59 42.9 | -58 48 16 | 276.85 | -8.36 | 125 | 51.1 | 63.8 | 13x 9 | 19.2 | .28 | | | | | LSB |
| 853 | | | 08 59 43.2 | -60 33 24 | 278.20 | -9.50 | 125 | 47.8 | -30.1 | 13x 7 | 19.5 | .19 | | | | | vLSB |
| 854 | | | 08 59 43.8 | -59 45 35 | 277.59 | -8.98 | 125 | 49.4 | 12.6 | 12x 8 | 18.6 | .19 | ? | E ? | | | |
| 855 | | | 08 59 50.4 | -60 33 13 | 278.20 | -9.49 | 125 | 48.6 | -29.9 | 17x 15 | 17.8 | .19 | | E | | | |
| 856 | | | 08 59 51.1 | -55 17 38 | 274.20 | -6.04 | 165 | 87.9 | -17.5 | 19x 16 | 17.4 | .36 | | E ? | | | |
| 857 | | | 08 59 52.0 | -53 31 48 | 272.87 | -4.87 | 165 | 92.2 | 76.9 | 35x 5 | 18.4 | .51 | | S | E | S | |
| 858 | | | 08 59 54.2 | -55 14 50 | 274.17 | -6.00 | 165 | 88.4 | -15.1 | 19x 8 | 18.2 | .36 | | | | | |
| 859 | | | 08 59 55.4 | -57 52 45 | 276.17 | -7.73 | 125 | 54.4 | 113.3 | 27x 4 | 18.8 | .28 | | S | E | | |
| 860 | | | 08 59 56.6 | -57 52 38 | 276.17 | -7.73 | 125 | 54.5 | 113.4 | 27x 5 | 18.3 | .28 | | S | E | | |
| 861 | | | 08 59 56.9 | -55 16 02 | 274.19 | -6.01 | 165 | 88.7 | -16.2 | 15x 13 | 18.0 | .36 | | S M | F | | |
| 862 | | | 08 59 59.7 | -57 54 25 | 276.19 | -7.74 | 125 | 54.9 | 111.8 | 13x 4 | 19.7 | .28 | | S | | | |
| 863 | | | 08 59 59.8 | -59 48 30 | 277.65 | -8.99 | 125 | 51.1 | 9.9 | 32x 30 | 16.4 | .19 | | S 5 | | S | dist.nucl, LSBdisk |
| 864 | | | 09 00 01.4 | -57 53 14 | 276.18 | -7.73 | 125 | 55.1 | 112.8 | 9x 5 | 19.8 | .28 | | S | | | |
| 865 | | | 09 00 02.5 | -57 54 25 | 276.20 | -7.74 | 125 | 55.2 | 111.8 | 11x 5 | 19.5 | .28 | | S | | | |
| 866 | | | 09 00 03.4 | -55 49 11 | 274.62 | -6.36 | 165 | 88.2 | -45.8 | 27x 15 | 17.1 | .34 | | S E: | | | |
| 867 | | | 09 00 04.1 | -60 28 35 | 278.16 | -9.42 | 125 | 50.2 | -25.8 | 16x 4 | 19.5 | .19 | | S | E | | |
| 868 | | | 09 00 04.6 | -57 53 27 | 276.19 | -7.72 | 125 | 55.5 | 112.6 | 13x 4 | 19.6 | .28 | | S | | | |
| 869 | | | 09 00 05.2 | -57 31 44 | 275.92 | -7.48 | 165 | 84.1 | -137.3 | 20x 17 | 17.6 | .31 | | S | F | S | |
| 870 | | | 09 00 05.5 | -58 05 20 | 276.34 | -7.85 | 125 | 55.2 | 102.0 | 43x 4 | 18.9 | .29 | | S | E | | vLSB |
| 871 | | | 09 00 06.1 | -59 59 49 | 277.80 | -9.10 | 125 | 51.4 | -.2 | 27x 11 | 17.5 | .17 | | S | | | |
| 872 | | | 09 00 07.5 | -60 35 35 | 278.26 | -9.49 | 125 | 50.4 | -32.1 | 12x 7 | 18.7 | .20 | | | | | blg |
| 873 | | | 09 00 08.1 | -59 17 44 | 277.27 | -8.64 | 125 | 53.1 | 37.4 | 23x 9 | 17.9 | .22 | | S | | | br.blg, vLSBdisk |
| 874 | | | 09 00 14.5 | -55 16 07 | 274.22 | -5.98 | 165 | 90.9 | -16.3 | 19x 4 | 19.3 | .34 | | S | E | S | |
| 875 | | | 09 00 15.3 | -59 46 51 | 277.65 | -8.95 | 125 | 52.9 | 11.4 | 27x 7 | 18.3 | .19 | | S 5 | | 1 | |
| 876 | | | 09 00 19.1 | -57 39 25 | 276.03 | -7.54 | 125 | 57.7 | 125.1 | 17x 3 | 19.9 | .32 | | S | E | | |
| 877 | | | 09 00 23.0 | -58 38 31 | 276.79 | -8.19 | 125 | 56.1 | 72.3 | 19x 11 | 18.6 | .31 | | | | | LSB |
| 878 | | | 09 00 23.9 | -60 30 57 | 278.22 | -9.41 | 125 | 52.3 | -28.0 | 20x 5 | 18.9 | .19 | | S M | E | | |
| 879 | | | 09 00 24.4 | -58 12 33 | 276.46 | -7.90 | 125 | 57.2 | 95.5 | 12x 5 | 19.3 | .30 | | | | | dist.blg |
| 880 | | | 09 00 33.6 | -55 42 14 | 274.58 | -6.23 | 165 | 92.2 | -39.8 | 13x 7 | 19.1 | .34 | | | | | |
| 881 | | | 09 00 34.7 | -54 19 15 | 273.54 | -5.31 | 165 | 95.9 | 34.3 | 15x 4 | 19.4 | .37 | ? | | | | |
| 882 | | | 09 00 41.9 | -60 01 50 | 277.87 | -9.07 | 125 | 55.3 | -2.1 | 52x 7 | 17.6 | .17 | | S 5 | E | | |
| 883 | | | 09 00 42.1 | -61 14 07 | 278.80 | -9.86 | 125 | 52.8 | -66.7 | 15x 7 | 19.1 | .22 | | S M | | | i.a.w.884 |
| 884 | | | 09 00 43.2 | -61 13 59 | 278.80 | -9.85 | 125 | 52.9 | -66.5 | 15x 7 | 19.1 | .22 | | S M | | | i.a.w.883 |
| 885 | | | 09 00 49.6 | -61 14 11 | 278.81 | -9.85 | 125 | 53.6 | -66.7 | 15x 5 | 19.6 | .22 | | S | | | |
| 886 | | | 09 00 51.0 | -59 29 28 | 277.47 | -8.70 | 125 | 57.5 | 26.7 | 13x 5 | 19.7 | .21 | ? | | | | dbl syst. |
| 887 | | I | 09 00 54.1 | -55 45 04 | 274.65 | -6.23 | 165 | 94.7 | -42.4 | 17x 15 | 17.8 | .33 | | | | S | |
| 888 | | I | 09 00 56.0 | -57 33 11 | 276.01 | -7.41 | 165 | 90.2 | -138.9 | 16x 7 | 18.5 | .30 | | S L: | | S | |
| 889 | | | 09 00 56.2 | -58 42 54 | 276.89 | -8.18 | 125 | 59.8 | 68.3 | 62x 28 | 15.9 | .30 | | S 1 | | S | br.blg, vLSBdisk |
| 890 | | | 09 00 59.0 | -61 11 01 | 278.78 | -9.80 | 125 | 54.7 | -64.0 | 28x 17 | 17.4 | .21 | | S M | | S | |
| 891 | | | 09 01 02.4 | -56 11 28 | 274.99 | -6.50 | 165 | 94.6 | -66.0 | 13x 3 | 19.9 | .34 | | S | E | | |
| 892 | | | 09 01 06.0 | -58 30 00 | 276.74 | -8.02 | 125 | 61.4 | 79.7 | 11x 7 | 18.7 | .31 | | | | | blg |
| 893 | | | 09 01 09.2 | -55 23 44 | 274.40 | -5.96 | 165 | 97.5 | -23.5 | 20x 20 | 17.3 | .30 | | | | S | |
| 894 | | | 09 01 09.6 | -59 56 42 | 277.85 | -8.97 | 125 | 58.6 | 2.3 | 12x 5 | 19.2 | .17 | | | | | blg |
| 895 | | | 09 01 09.7 | -53 32 23 | 273.01 | -4.73 | 165 | 102.5 | 75.9 | 13x 4 | 19.8 | .66 | ? | | | | |
| 896 | | | 09 01 14.4 | -55 33 15 | 274.53 | -6.06 | 165 | 97.8 | -32.0 | 17x 4 | 19.4 | .33 | | S | | 1 | |
| 897 | | | 09 01 20.4 | -60 23 55 | 278.21 | -9.25 | 125 | 58.8 | -22.0 | 13x 7 | 18.7 | .19 | | S | | | |
| 898 | | | 09 01 21. | -61 08.2 : | 278.77 | -9.73 | 125 | 57.2 | -61.5 | 13x 4 | 19.8 | .21 | | S ? | | | |
| 899 | | | 09 01 22.1 | -60 57 42 | 278.64 | -9.62 | 125 | 57.7 | -52.2 | 15x 4 | 19.7 | .23 | | S | E | | |
| 900 | | | 09 01 25.1 | -56 22 09 | 275.16 | -6.58 | 165 | 96.9 | -75.7 | 19x 8 | 17.9 | .31 | | | | S | |

**Table 1.** Galaxies in the southern ZOA – I. The Hydra/Antlia Extension (continued)

| RKK Ident | Other Ident | IR | R.A. (h m s) | Dec. (° ′ ″) | gal ℓ (°) | gal $b$ (°) | SRC | X (mm) | Y (mm) | D x d (″) | $B_J$ ($^m$) | $N_{HI}$ | u | Type | class. | o | * | Remarks |
|---|---|---|---|---|---|---|---|---|---|---|---|---|---|---|---|---|---|---|
| (1) | (2) | (3) | (4) | (5) | (6) | (7) | (8) | (9) | (10) | (11) | (12) | (13) | | (14) | | | | (15) |
| 901 | | | 09 01 25.1 | -59 21 45 | 277.42 | -8.56 | 125 | 61.7 | 33.4 | 15x 13 | 17.8 | .21 | | E | | | | |
| 902 | | | 09 01 26.8 | -53 17 15 | 272.85 | -4.53 | 165 | 105.5 | 89.3 | 40x 11 | 17.3 | .83 | | S | ? | | S | |
| 903 | | | 09 01 26.9 | -60 18 52 | 278.15 | -9.18 | 125 | 59.7 | -17.6 | 15x 13 | 17.9 | .19 | | E | ? | | | |
| 904 | | | 09 01 27.6 | -54 50 24 | 274.01 | -5.56 | 165 | 101.4 | 6.2 | 13x 13 | 18.1 | .36 | ? | | | | S | |
| 905 | | | 09 01 28.2 | -60 38 55 | 278.41 | -9.40 | 125 | 59.1 | -35.5 | 16x 9 | 18.1 | .21 | ? | S | ? | | | blg |
| 906 | | | 09 01 29.0 | -53 56 48 | 273.35 | -4.97 | 165 | 104.0 | 54.0 | 51x 8 | 17.1 | .50 | | S | | N | S | |
| 907 | | | 09 01 30.4 | -57 51 09 | 276.29 | -7.55 | 125 | 65.7 | 114.3 | 17x 16 | 17.9 | .31 | | | | | S | |
| 908 | | | 09 01 31.8 | -60 19 18 | 278.16 | -9.18 | 125 | 60.2 | -18.0 | 12x 5 | 19.2 | .19 | | | | | | blg |
| 909 | | | 09 01 33.9 | -60 29 10 | 278.29 | -9.29 | 125 | 60.1 | -26.8 | 17x 8 | 18.6 | .19 | | | | | 1 | |
| 910 | | | 09 01 34.3 | -54 11 53 | 273.54 | -5.12 | 165 | 104.0 | 40.5 | 17x 9 | 18.4 | .44 | | S | ? | | | |
| 911 | | | 09 01 36.3 | -58 17 12 | 276.62 | -7.83 | 125 | 65.4 | 91.0 | 19x 11 | 17.7 | .29 | | | | | | |
| 912 | | | 09 01 37.2 | -58 00 44 | 276.42 | -7.65 | 125 | 66.2 | 105.7 | 12x 7 | 19.1 | .30 | | | | | | |
| 913 | | | 09 01 37.8 | -59 57 14 | 277.89 | -8.93 | 125 | 61.7 | 1.7 | 12x 8 | 18.6 | .17 | | | | | | |
| 914 | | | 09 01 38.6 | -58 18 30 | 276.64 | -7.84 | 125 | 65.7 | 89.8 | 17x 16 | 17.9 | .29 | | S | | F | S | |
| 915 | | | 09 01 38.7 | -53 10 45 | 272.79 | -4.44 | 165 | 107.3 | 95.0 | 15x 9 | 18.4 | .99 | ? | | | | | |
| 916 | | | 09 01 39.0 | -59 56 39 | 277.89 | -8.92 | 125 | 61.9 | 2.2 | 13x 11 | 18.3 | .17 | | | | | S | |
| 917 | | | 09 01 40.0 | -57 39 09 | 276.15 | -7.41 | 125 | 67.3 | 124.9 | 13x 8 | 18.7 | .30 | | S | ? | | | |
| 918 | | | 09 01 43.8 | -59 30 12 | 277.56 | -8.62 | 125 | 63.5 | 25.8 | 17x 16 | 17.9 | .20 | | S | M | F | | |
| 919 | | | 09 01 46.3 | -58 01 48 | 276.45 | -7.65 | 125 | 67.2 | 104.7 | 13x 9 | 18.6 | .30 | | | | | 1 | |
| 920 | | | 09 01 47.7 | -59 21 07 | 277.45 | -8.52 | 125 | 64.3 | 33.9 | 11x 8 | 18.8 | .21 | | | | | | blg |
| 921 | | | 09 01 48.2 | -59 50 44 | 277.82 | -8.84 | 125 | 63.2 | 7.4 | 31x 8 | 17.5 | .18 | | L | | | | |
| 922 | | | 09 01 48.7 | -58 55 16 | 277.12 | -8.23 | 125 | 65.4 | 57.0 | 13x 11 | 18.9 | .23 | | S | M | F | | vLSBdisk |
| 923 | | | 09 01 52.0 | -58 47 58 | 277.04 | -8.15 | 125 | 66.1 | 63.4 | 15x 15 | 17.9 | .26 | ? | | | | | |
| 924 | | | 09 01 54.8 | -55 58 31 | 274.91 | -6.27 | 165 | 101.7 | -54.8 | 28x 8 | 17.8 | .32 | | | | | | |
| 925 | | | 09 01 55.0 | -53 52 36 | 273.34 | -4.87 | 165 | 107.6 | 57.6 | 16x 7 | 18.6 | .59 | ? | | | | S | |
| 926 | | | 09 01 56.0 | -58 10 18 | 276.57 | -7.72 | 125 | 68.0 | 97.0 | 22x 5 | 19.1 | .29 | | S | | | S | |
| 927 | | | 09 01 56.1 | -57 46 55 | 276.27 | -7.47 | 125 | 69.0 | 117.9 | 22x 9 | 18.0 | .32 | | E | ? | | | |
| 928 | | | 09 01 58.2 | -55 44 41 | 274.74 | -6.11 | 165 | 102.8 | -42.5 | 24x 13 | 18.0 | .33 | ? | | | | | |
| 929 | | | 09 01 58.5 | -61 14 50 | 278.91 | -9.75 | 125 | 60.9 | -67.6 | 13x 7 | 19.0 | .23 | ? | | | | | |
| 930 | | | 09 02 01.7 | -60 57 15 | 278.69 | -9.55 | 125 | 62.0 | -52.0 | 13x 5 | 19.4 | .21 | | S | | | | |
| 931 | | | 09 02 07.3 | -59 41 16 | 277.73 | -8.71 | 125 | 65.7 | 15.8 | 13x 7 | 19.2 | .19 | | S | | | | dist.nucl. |
| 932 | | | 09 02 11.9 | -57 52 43 | 276.37 | -7.50 | 125 | 70.6 | 112.7 | 17x 9 | 18.5 | .31 | | | | | 1 | |
| 933 | | | 09 02 15.6 | -60 59 31 | 278.74 | -9.56 | 125 | 63.4 | -54.1 | 13x 4 | 19.6 | .20 | | S | ? | | 1 | |
| 934 | | | 09 02 21.4 | -56 58 06 | 275.70 | -6.88 | 165 | 102.1 | -108.2 | 20x 5 | 19.0 | .32 | ? | S | ? | | S | |
| 935 | | | 09 02 26.6 | -60 19 41 | 278.25 | -9.10 | 125 | 66.3 | -18.6 | 16x 13 | 18.1 | .18 | | E | | | | |
| 936 | | | 09 02 27.0 | -56 05 49 | 275.05 | -6.29 | 165 | 105.3 | -61.5 | 24x 16 | 17.1 | .33 | | F | | | S | |
| 937 | | | 09 02 28.0 | -58 17 06 | 276.70 | -7.75 | 125 | 71.5 | 90.8 | 19x 8 | 18.7 | .28 | | | | | | |
| 938 | | | 09 02 31.2 | -55 26 00 | 274.56 | -5.84 | 165 | 107.8 | -26.0 | 15x 7 | 18.8 | .35 | ? | S | ? | | | |
| 939 | | | 09 02 36.5 | -56 24 33 | 275.30 | -6.49 | 165 | 105.6 | -78.3 | 13x 7 | 18.7 | .32 | ? | | | | S | |
| 940 | | | 09 02 37.9 | -59 52 43 | 277.92 | -8.79 | 125 | 68.6 | 5.4 | 12x 5 | 19.8 | .18 | | | | | | |
| 941 | | | 09 02 43.2 | -56 24 05 | 275.30 | -6.47 | 165 | 106.4 | -78.0 | 16x 8 | 18.4 | .32 | | | | | 1 | |
| 942 | | | 09 02 45.4 | -59 16 34 | 277.47 | -8.38 | 125 | 71.0 | 37.6 | 12x 9 | 18.9 | .22 | | | | | | |
| 943 | | I | 09 02 47.4 | -57 57 38 | 276.48 | -7.50 | 125 | 74.6 | 108.1 | 17x 9 | 18.3 | .31 | | S | M | | 1 | |
| 944 | | | 09 02 48.6 | -57 55 13 | 276.45 | -7.47 | 125 | 74.9 | 110.2 | 13x 7 | 19.5 | .31 | | S | | | | vLSB |
| 945 | | | 09 02 59.2 | -57 38 25 | 276.26 | -7.27 | 125 | 76.8 | 125.1 | 23x 9 | 17.9 | .31 | | S | M | | 1 | dist.blg |
| 946 | | | 09 03 01.3 | -60 19 21 | 278.29 | -9.04 | 125 | 70.1 | -18.5 | 19x 11 | 18.6 | .18 | | S | ? | | S | vLSB |
| 947 | | | 09 03 02.3 | -57 50 35 | 276.42 | -7.40 | 125 | 76.7 | 114.3 | 42x 8 | 17.3 | .32 | | S | M | | | dist.blg, LSBdisk |
| 948 | | | 09 03 03.5 | -59 08 21 | 277.40 | -8.26 | 125 | 73.4 | 44.9 | 20x 11 | 18.2 | .22 | | S | ? | | | |
| 949 | | | 09 03 06.2 | -60 18 13 | 278.28 | -9.02 | 125 | 70.7 | -17.5 | 20x 19 | 17.7 | .18 | | S | ? | | | LSB |
| 950 | | | 09 03 08.7 | -59 37 17 | 277.77 | -8.57 | 125 | 72.8 | 19.0 | 15x 4 | 19.4 | .21 | | S | | E | | |
| 951 | | | 09 03 08.9 | -55 47 23 | 274.89 | -6.02 | 165 | 111.5 | -45.4 | 13x 9 | 18.7 | .36 | ? | | | | S | |
| 952 | | | 09 03 10.2 | -60 38 56 | 278.55 | -9.25 | 125 | 70.2 | -36.0 | 34x 4 | 18.7 | .20 | | S | M | E | | |
| 953 | | | 09 03 12.7 | -54 49 51 | 274.18 | -5.37 | 166 | -137.3 | 4.4 | 27x 7 | 17.9 | .41 | | S | | E | | |
| 954 | | | 09 03 15.2 | -60 10 20 | 278.19 | -8.92 | 125 | 72.0 | -10.5 | 22x 11 | 18.4 | .18 | | S | M | | | dist.blg, vLSBdisk |
| 955 | | | 09 03 26.7 | -54 11 09 | 273.72 | -4.91 | 166 | -137.7 | 39.0 | 11x 11 | 18.3 | .50 | | S | ? | F | | |
| 956 | | | 09 03 27.4 | -56 17 21 | 275.29 | -6.32 | 165 | 112.3 | -72.2 | 16x 13 | 18.0 | .34 | | | | | S | |
| 957 | | | 09 03 28.4 | -58 35 14 | 277.02 | -7.85 | 125 | 77.8 | 74.3 | 17x 11 | 18.7 | .31 | | S | | V | | vLSB |
| 958 | | | 09 03 30.0 | -58 07 31 | 276.67 | -7.54 | 125 | 79.2 | 99.0 | 17x 5 | 18.9 | .31 | | | | | 2 | |
| 959 | | Q | 09 03 32.9 | -59 31 50 | 277.73 | -8.47 | 125 | 75.8 | 23.8 | 17x 7 | 18.7 | .22 | ? | | | | | dist.blg, vLSBdisk |
| 960 | | | 09 03 34.7 | -56 58 21 | 275.81 | -6.76 | 166 | -127.0 | -110.0 | 20x 3 | 19.7 | .30 | | S | | E | | |
| 961 | | | 09 03 35.4 | -57 45 16 | 276.40 | -7.28 | 125 | 80.8 | 118.8 | 13x 7 | 19.4 | .32 | | S | ? | | | |
| 962 | | | 09 03 35.9 | -58 39 40 | 277.08 | -7.89 | 125 | 78.4 | 70.3 | 15x 15 | 18.3 | .30 | | | | | 1 | |
| 963 | | | 09 03 38.6 | -60 46 51 | 278.69 | -9.29 | 125 | 73.0 | -43.2 | 13x 13 | 18.0 | .20 | | E | 0 | | | |
| 964 | | | 09 03 38.8 | -59 56 32 | 278.05 | -8.73 | 125 | 75.3 | 1.7 | 17x 12 | 17.9 | .18 | | F | | | | |
| 965 | | | 09 03 41.8 | -55 09 13 | 274.46 | -5.53 | 166 | -132.4 | -12.6 | 23x 7 | 17.9 | .37 | | S | | | | |
| 966 | | I | 09 03 43.8 | -54 14 28 | 273.79 | -4.92 | 165 | 120.8 | 37.3 | 40x 7 | 17.5 | .48 | | S | L: | | S | |
| 967 | | | 09 03 46.7 | -59 26 17 | 277.68 | -8.39 | 125 | 77.6 | 28.6 | 17x 16 | 18.0 | .24 | ? | S | | | 1 | round, br.nucl. |
| 968 | | | 09 03 47.4 | -53 24 05 | 273.17 | -4.35 | 165 | 123.9 | 82.2 | 22x 3 | 19.7 | .63 | ? | S | | E | | |
| 969 | | | 09 03 53.3 | -57 24 39 | 276.17 | -7.02 | 125 | 83.9 | 137.1 | 13x 3 | 19.5 | .34 | | S | | | | |
| 970 | | | 09 03 55.4 | -57 26 23 | 276.19 | -7.04 | 125 | 84.1 | 135.5 | 13x 11 | 18.4 | .34 | ? | E | ? | | | |
| 971 | | | 09 04 02.1 | -57 42 17 | 276.40 | -7.20 | 125 | 84.2 | 121.3 | 16x 8 | 19.2 | .31 | ? | | | | | vLSB |
| 972 | | | 09 04 02.7 | -55 08 45 | 274.49 | -5.49 | 165 | 120.3 | -11.3 | 54x 12 | 16.6 | .37 | ? | S | E: | | | |
| 973 | | | 09 04 03.4 | -57 52 25 | 276.53 | -7.32 | 125 | 83.8 | 112.3 | 16x 7 | 19.1 | .33 | | S | ? | | | |
| 974 | | | 09 04 06.6 | -59 32 20 | 277.79 | -8.42 | 125 | 79.5 | 23.1 | 15x 5 | 18.7 | .23 | ? | S | | | | or blended ** |
| 975 | | | 09 04 07.3 | -54 04 30 | 273.70 | -4.76 | 165 | 124.4 | 46.0 | 13x 8 | 19.0 | .51 | ? | | | | | |
| 976 | | | 09 04 07.7 | -60 06 22 | 278.22 | -8.80 | 125 | 78.1 | -7.2 | 19x 13 | 17.9 | .17 | | S | ? | | | dist.blg, vLSBdisk |
| 977 | | | 09 04 09.4 | -59 52 47 | 278.05 | -8.64 | 125 | 78.9 | 4.9 | 15x 5 | 18.8 | .19 | | | | | | |
| 978 | | | 09 04 11.1 | -58 07 38 | 276.73 | -7.47 | 125 | 84.0 | 98.6 | 22x 9 | 18.0 | .30 | | S | | | | |
| 979 | | | 09 04 14.2 | -58 07 13 | 276.73 | -7.46 | 125 | 84.4 | 99.0 | 20x 4 | 19.0 | .30 | | S | | | | |
| 980 | | | 09 04 15.8 | -55 36 46 | 274.86 | -5.78 | 165 | 120.5 | -36.4 | 54x 40 | 15.8 | .35 | ? | D | ? | | | LSB |
| 981 | | | 09 04 15.8 | -60 22 08 | 278.43 | -8.96 | 125 | 78.2 | -21.4 | 31x 23 | 16.9 | .19 | | S | M | | | dist.blg, vLSBdisk |
| 982 | | | 09 04 15.9 | -59 52 46 | 278.06 | -8.63 | 125 | 79.6 | 4.8 | 19x 9 | 18.3 | .19 | | S | ? | | | |
| 983 | | | 09 04 20.1 | -60 22 24 | 278.44 | -8.96 | 125 | 78.7 | -21.6 | 15x 5 | 18.8 | .19 | | S | | | | |
| 984 | | | 09 04 20.8 | -60 12 51 | 278.32 | -8.85 | 125 | 79.2 | -13.1 | 11x 8 | 18.4 | .18 | ? | E | ? | | | or * |
| 985 | | | 09 04 23.5 | -60 32 59 | 278.57 | -9.07 | 125 | 78.5 | -31.1 | 12x 8 | 18.8 | .20 | | S | | | | dist.blg, vLSBdisk |
| 986 | | | 09 04 25.2 | -55 10 51 | 274.55 | -5.47 | 166 | -126.8 | -13.7 | 40x 9 | 17.1 | .37 | | S | 1 | | | |
| 987 | | | 09 04 26.8 | -54 43 37 | 274.22 | -5.17 | 165 | 124.8 | 10.9 | 13x 7 | 19.2 | .47 | ? | S | | F | | |
| 988 | | | 09 04 28.4 | -58 52 21 | 277.32 | -7.94 | 125 | 83.9 | 58.7 | 17x 4 | 19.4 | .29 | | S | 5 | | | * on sp.arm |
| 989 | | | 09 04 29.1 | -60 10 56 | 278.30 | -8.81 | 125 | 80.2 | -11.4 | 36x 8 | 17.7 | .18 | | S | M | E | | * on sp.arm |
| 990 | | | 09 04 29.6 | -60 47 36 | 278.77 | -9.22 | 125 | 78.5 | -44.2 | 46x 9 | 17.5 | .20 | | S | | E | | w.comp. |
| 991 | | | 09 04 30.9 | -58 24 36 | 276.97 | -7.63 | 125 | 85.6 | 83.4 | 43x 9 | 17.3 | .30 | | S | M | | 1 | |
| 992 | | | 09 04 33.0 | -56 57 39 | 275.89 | -6.66 | 166 | -119.9 | -108.9 | 20x 11 | 18.0 | .31 | | S | | | | dist.nucl, LSBdisk |
| 993 | | | 09 04 33.3 | -59 29 18 | 277.79 | -8.35 | 125 | 82.7 | 25.7 | 15x 7 | 19.3 | .24 | | S | | | | vLSB, cl.to br.* |
| 994 | | | 09 04 34.5 | -59 37 09 | 277.89 | -8.43 | 125 | 82.5 | 18.6 | 15x 5 | 19.3 | .22 | | S | | | S | |
| 995 | | | 09 04 36.6 | -57 23 03 | 276.21 | -6.93 | 166 | -118.1 | -131.6 | 27x 4 | 18.8 | .34 | | S | | E | | |
| 996 | | | 09 04 40.9 | -59 56 07 | 278.14 | -8.63 | 125 | 82.2 | 1.7 | 19x 4 | 19.6 | .18 | | S | | | | vLSB |
| 997 | | | 09 04 43.5 | -59 55 39 | 278.13 | -8.62 | 125 | 82.6 | 2.1 | 15x 5 | 19.1 | .18 | | | | | 1 | |
| 998 | L166-001 | | 09 04 45. | -55 10.9 : | 274.59 | -5.44 | 166 | -124.3 | -13.6 | 34x 9 | 17.4 | .39 | | I | | | | |
| 999 | | | 09 04 45.8 | -58 28 44 | 277.05 | -7.65 | 125 | 87.1 | 79.6 | 17x 7 | 18.7 | .31 | | S | | | | |
| 1000 | | | 09 04 49.4 | -59 41 01 | 277.96 | -8.45 | 125 | 83.9 | 15.1 | 20x 4 | 19.6 | .22 | | S | | E | | vLSB |

**Table 1.** Galaxies in the southern ZOA – I. The Hydra/Antlia Extension (continued)

| RKK Ident (1) | Other Ident (2) | IR (3) | R.A. (h m s) (4) | Dec. (° ′ ″) (5) | gal ℓ (°) (6) | gal $b$ (°) (7) | SRC (8) | X (mm) (9) | Y (mm) (10) | D x d (″) (11) | $B_J$ ($^m$) (12) | $N_{HI}$ (13) | u | Type class. (14) | o | * | Remarks (15) |
|---|---|---|---|---|---|---|---|---|---|---|---|---|---|---|---|---|---|
| 1001 | | | 09 04 52.1 | -62 01 49 | 279.73 | -10.01 | 125 | 77.2 | -110.5 | 17x 13 | 18.1 | .19 | | S M | | | br.blg, vLSBdisk |
| 1002 | | | 09 04 56.1 | -58 56 19 | 277.41 | -7.94 | 125 | 86.9 | 54.9 | 24x 13 | 18.3 | .28 | ? | | | | vLSB |
| 1003 | | | 09 04 56.1 | -59 23 16 | 277.74 | -8.24 | 125 | 85.6 | 30.9 | 22x 9 | 18.0 | .26 | | S E | | | or dbl syst. |
| 1004 | | | 09 04 57.1 | -59 04 58 | 277.52 | -8.04 | 125 | 86.6 | 47.2 | 13x 9 | 18.7 | .26 | | S ? | | | pair w.1005 |
| 1005 | | | 09 04 57.9 | -59 05 11 | 277.52 | -8.04 | 125 | 86.7 | 47.0 | 17x 13 | 18.1 | .26 | | | | | pair w.1004 |
| 1006 | | | 09 04 58.1 | -59 37 48 | 277.93 | -8.40 | 125 | 85.1 | 17.9 | 15x 5 | 19.8 | .22 | | S ? | | | vLSB |
| 1007 | | | 09 05 00.5 | -58 20 51 | 276.97 | -7.54 | 125 | 89.2 | 86.5 | 13x 9 | 18.5 | .29 | | | | | |
| 1008 | | | 09 05 06.5 | -59 07 13 | 277.56 | -8.05 | 125 | 87.6 | 45.1 | 13x 9 | 18.6 | .26 | | | | | |
| 1009 | | | 09 05 10.6 | -58 22 07 | 277.00 | -7.54 | 125 | 90.3 | 85.3 | 16x 7 | 18.6 | .29 | | S E: | | | |
| 1010 | | | 09 05 11.3 | -57 45 59 | 276.55 | -7.13 | 125 | 92.2 | 117.6 | 17x 5 | 19.2 | .30 | | | | | |
| 1011 | | | 09 05 12.0 | -59 07 44 | 277.57 | -8.04 | 125 | 88.2 | 44.6 | 9x 8 | 18.9 | .26 | | | | | neighb.of 1011 |
| 1012 | | | 09 05 20.8 | -57 28 02 | 276.34 | -6.92 | 166 | -112.5 | -135.7 | 13x 9 | 18.6 | .31 | | E ? | | | |
| 1013 | | | 09 05 31.0 | -60 27 36 | 278.60 | -8.90 | 125 | 86.2 | -26.7 | 17x 3 | 19.7 | .19 | | S | E | | |
| 1014 | | | 09 05 34.7 | -59 32 23 | 277.91 | -8.28 | 125 | 89.5 | 22.5 | 19x 13 | 18.0 | .24 | | S M? | F | | vLSBdisk |
| 1015 | | | 09 05 43.5 | -58 22 33 | 277.06 | -7.49 | 125 | 94.1 | 84.7 | 16x 7 | 19.1 | .30 | | | | 1 | |
| 1016 | | | 09 05 48.1 | -60 13 55 | 278.45 | -8.73 | 125 | 88.8 | -14.6 | 15x 7 | 18.6 | .18 | | S | | | |
| 1017 | | | 09 05 55.3 | -60 14 22 | 278.47 | -8.72 | 125 | 89.6 | -15.1 | 20x 12 | 17.7 | .18 | | S ? | | | |
| 1018 | | | 09 05 56.3 | -59 20 39 | 277.80 | -8.12 | 125 | 92.6 | 32.8 | 15x 5 | 19.4 | .25 | ? | | | | dbl syst.w.1020 |
| 1019 | | | 09 05 57.2 | -55 24 44 | 274.87 | -5.47 | 166 | -114.4 | -25.4 | 16x 8 | 18.3 | .42 | | | | | cl.to br.* |
| 1020 | | | 09 05 58.5 | -60 04 31 | 278.35 | -8.61 | 125 | 90.5 | -6.3 | 20x 13 | 17.5 | .18 | | | | | dbl syst.w.1018 |
| 1021 | | | 09 06 00.2 | -55 23 30 | 274.86 | -5.45 | 166 | -114.1 | -24.3 | 9x 4 | 19.8 | .42 | | | | | |
| 1022 | | | 09 06 08.7 | -57 30 44 | 276.45 | -6.87 | 125 | 99.9 | 130.8 | 17x 9 | 18.3 | .31 | | | | | |
| 1023 | | | 09 06 15.1 | -59 10 55 | 277.70 | -7.98 | 125 | 95.2 | 41.4 | 17x 11 | 18.3 | .25 | | | | | dist.blg |
| 1024 | | | 09 06 17.3 | -59 26 26 | 277.90 | -8.15 | 125 | 94.6 | 27.5 | 15x 7 | 19.0 | .26 | | | | | |
| 1025 | | | 09 06 22.0 | -59 11 18 | 277.72 | -7.97 | 125 | 96.0 | 41.0 | 22x 8 | 18.3 | .25 | | | | 1 | LSB |
| 1026 | | | 09 06 23.8 | -60 24 50 | 278.64 | -8.79 | 125 | 92.2 | -24.6 | 16x 11 | 18.3 | .19 | | | | | w.comp. |
| 1027 | | | 09 06 29.2 | -59 07 42 | 277.68 | -7.92 | 125 | 97.0 | 44.2 | 28x 8 | 17.8 | .25 | | S E | | | |
| 1028 | | | 09 06 31.2 | -60 25 50 | 278.66 | -8.79 | 125 | 92.9 | -25.5 | 13x 8 | 19.0 | .19 | | | | | neighb.of 1026 |
| 1029 | | | 09 06 34.8 | -59 50 29 | 278.23 | -8.39 | 125 | 95.3 | 6.0 | 19x 9 | 18.1 | .19 | ? | | | S | cl.to br.* |
| 1030 | | | 09 06 39.5 | -60 10 00 | 278.48 | -8.60 | 125 | 94.7 | -11.5 | 16x 7 | 19.1 | .18 | | | | 1 | w.comp. |
| 1031 | | | 09 06 40.8 | -61 07 26 | 279.20 | -9.24 | 125 | 91.6 | -62.7 | 27x 11 | 17.7 | .24 | | S M | | | |
| 1032 | | Q | 09 06 42.4 | -53 58 58 | 273.89 | -4.42 | 166 | -112.8 | 51.4 | 24x 5 | 18.5 | .58 | | S M | | | |
| 1033 | | | 09 06 54.7 | -60 45 34 | 278.94 | -8.98 | 125 | 94.4 | -43.3 | 17x 13 | 18.1 | .19 | | L | | | |
| 1034 | | | 09 06 57.5 | -62 05 13 | 279.94 | -9.86 | 125 | 90.1 | -114.3 | 23x 20 | 17.5 | .20 | | S L | F | | dist.nucl. |
| 1035 | | | 09 07 03.2 | -59 58 51 | 278.37 | -8.44 | 125 | 98.0 | -1.7 | 27x 16 | 17.2 | .19 | | L | | | |
| 1036 | | | 09 07 04.5 | -60 03 59 | 278.44 | -8.50 | 125 | 97.8 | -6.3 | 15x 4 | 19.3 | .19 | | S | | | |
| 1037 | | | 09 07 09.9 | -61 37 23 | 279.61 | -9.54 | 125 | 93.0 | -89.6 | 67x 60 | 15.0 | .21 | | S 4 | | S | w.comp. |
| 1038 | | | 09 07 19.4 | -60 00 50 | 278.42 | -8.44 | 125 | 99.7 | -3.6 | 13x 7 | 18.9 | .19 | | S | | | |
| 1039 | | | 09 07 22.7 | -59 04 44 | 277.73 | -7.80 | 125 | 103.3 | 46.4 | 20x 11 | 18.0 | .28 | | S | | | |
| 1040 | | | 09 07 22.9 | -60 48 05 | 279.01 | -8.96 | 125 | 97.3 | -45.7 | 20x 11 | 18.0 | .19 | | E | | 1 | |
| 1041 | | | 09 07 26.1 | -54 47 04 | 274.55 | -4.89 | 166 | -104.8 | 8.8 | 27x 20 | 17.3 | .50 | | S 5 | F | | dist.nucl, vLSBdisk |
| 1042 | | | 09 07 32.3 | -60 57 16 | 279.14 | -9.05 | 125 | 97.8 | -54.0 | 30x 11 | 17.5 | .21 | | S ? | | | |
| 1043 | | | 09 07 32.9 | -61 13 14 | 279.34 | -9.23 | 125 | 100.8 | -68.6 | 23x 17 | 17.5 | .23 | | | | | |
| 1044 | | | 09 07 36.7 | -59 08 26 | 277.79 | -7.82 | 125 | 104.7 | 43.0 | 13x 5 | 19.4 | .26 | | | | | w.comp. |
| 1045 | | | 09 07 37.0 | -60 52 42 | 279.09 | -8.99 | 125 | 98.6 | -49.9 | 20x 16 | 17.5 | .20 | | E 2 | | | |
| 1046 | | | 09 07 40.4 | -59 51 16 | 278.33 | -8.30 | 125 | 102.6 | 4.8 | 16x 8 | 18.4 | .19 | | | | 1 | |
| 1047 | | | 09 07 42.1 | -60 56 16 | 279.14 | -9.03 | 125 | 98.9 | -53.1 | 11x 11 | 18.8 | .20 | | E ? | | | |
| 1048 | | | 09 07 45.1 | -61 48 29 | 279.80 | -9.61 | 125 | 96.1 | -99.7 | 17x 9 | 18.1 | .20 | | | | 1 | |
| 1049 | | | 09 07 46.5 | -60 08 26 | 278.55 | -8.48 | 125 | 102.2 | -10.6 | 30x 8 | 17.8 | .19 | | S 5 | | | |
| 1050 | | | 09 07 48.4 | -59 15 32 | 277.90 | -7.88 | 125 | 105.6 | 36.6 | 16x 9 | 18.3 | .26 | | | | | |
| 1051 | L166-003 | | 09 08 03.4 | -55 43 10 | 275.30 | -5.46 | 166 | -97.6 | -41.1 | 19x 5 | 18.7 | .32 | | S | | | w.LSB comp.(= L3?) |
| 1052 | | | 09 08 03.8 | -55 31 38 | 275.16 | -5.33 | 166 | -98.1 | -30.8 | 27x 5 | 18.4 | .37 | | S | | | |
| 1053 | | | 09 08 06.7 | -61 20 00 | 279.47 | -9.26 | 125 | 100.1 | -74.5 | 17x 16 | 18.5 | .23 | | S ? | F | | |
| 1054 | | | 09 08 10.6 | -60 44 55 | 279.04 | -8.86 | 125 | 102.7 | -43.3 | 16x 8 | 18.5 | .19 | | | | | |
| 1055 | | | 09 08 16.1 | -58 05 30 | 277.07 | -7.05 | 125 | 113.0 | 98.8 | 11x 7 | 18.9 | .28 | | | | | between ** |
| 1056 | | | 09 08 17.4 | -59 23 16 | 278.03 | -7.93 | 125 | 108.4 | 29.5 | 19x 5 | 18.7 | .28 | | S | | | dist.blg |
| 1057 | | | 09 08 18.0 | -61 02 28 | 279.27 | -9.04 | 125 | 102.4 | -59.0 | 28x 7 | 17.9 | .21 | | S M | | | |
| 1058 | | | 09 08 19.0 | -59 18 20 | 277.98 | -7.87 | 125 | 108.9 | 33.9 | 17x 8 | 18.4 | .28 | ? | | | 1 | |
| 1059 | | | 09 08 24.1 | -60 34 47 | 278.93 | -8.72 | 125 | 104.8 | -34.3 | 13x 7 | 18.9 | .19 | | S | | | w.comp. |
| 1060 | | P | 09 08 25.3 | -59 12 14 | 277.91 | -7.79 | 125 | 110.0 | 39.2 | 13x 9 | 18.8 | .29 | ? | S | | | |
| 1061 | | | 09 08 29.3 | -61 02 47 | 279.29 | -9.03 | 125 | 103.6 | -59.3 | 19x 12 | 17.8 | .22 | | E 0 | | | |
| 1062 | | | 09 08 31.0 | -59 32 03 | 278.16 | -8.00 | 125 | 109.4 | 21.5 | 16x 4 | 19.1 | .25 | | S | E | 1 | w.2 comp. |
| 1063 | | | 09 08 31.4 | -61 34 10 | 279.68 | -9.38 | 125 | 101.9 | -87.3 | 22x 17 | 18.3 | .21 | | S | | | vLSB |
| 1064 | | | 09 08 34.3 | -59 08 00 | 277.87 | -7.73 | 125 | 111.3 | 42.9 | 34x 27 | 16.5 | .28 | ? | I 0 | | S | or gas |
| 1065 | | | 09 08 36.4 | -55 45 37 | 275.38 | -5.43 | 166 | -93.4 | -43.0 | 16x 9 | 17.8 | .33 | | S | | | br.blg or * w.gas |
| 1066 | | | 09 08 37.5 | -61 33 22 | 279.68 | -9.36 | 125 | 102.6 | -86.6 | 20x 17 | 17.8 | .22 | | S | | | dist.blg, vLSBdisk |
| 1067 | | | 09 08 37.6 | -60 58 35 | 279.25 | -8.97 | 125 | 104.8 | -55.6 | 27x 8 | 17.9 | .20 | | S 5 | | | dist.blg, * on sp.arm |
| 1068 | | | 09 08 39.7 | -62 13 23 | 280.18 | -9.81 | 126 | -133.6 | -123.9 | 12x 7 | 18.7 | .23 | | S | | | |
| 1069 | | | 09 08 41.2 | -61 01 30 | 279.29 | -9.00 | 125 | 105.0 | -58.3 | 17x 7 | 18.6 | .21 | | | | | |
| 1070 | | | 09 08 44.0 | -59 42 41 | 278.31 | -8.10 | 125 | 110.2 | 12.0 | 30x 5 | 18.4 | .23 | | S M | | | vLSBdisk, cl.to br.* |
| 1071 | | | 09 08 46.0 | -58 16 32 | 277.25 | -7.13 | 125 | 115.9 | 88.7 | 15x 9 | 18.3 | .29 | | | | | dbl syst.w.1072 |
| 1072 | | | 09 08 47.0 | -58 16 34 | 277.25 | -7.13 | 125 | 116.0 | 88.7 | 8x 7 | 19.4 | .29 | | | | | dbl syst.w.1071 |
| 1073 | | | 09 08 47.1 | -61 13 13 | 279.44 | -9.12 | 125 | 108.8 | -69.0 | 13x 11 | 18.9 | .22 | ? | | | 1 | smooth |
| 1074 | | | 09 08 47.2 | -60 14 22 | 278.71 | -8.46 | 125 | 108.6 | -16.3 | 15x 4 | 19.4 | .19 | | | | | |
| 1075 | | | 09 08 47.6 | -60 15 23 | 278.72 | -8.47 | 125 | 108.6 | -17.2 | 23x 8 | 18.1 | .19 | | S M | | | |
| 1076 | | | 09 08 48.9 | -57 52 33 | 276.96 | -6.85 | 125 | 117.7 | 110.1 | 13x 5 | 19.8 | .34 | | | | | |
| 1077 | | | 09 08 50.4 | -61 24 21 | 279.58 | -9.24 | 125 | 104.5 | -78.7 | 17x 7 | 18.9 | .25 | | S ? | | | |
| 1078 | | | 09 08 52.1 | -61 02 09 | 279.31 | -8.99 | 125 | 106.1 | -58.9 | 19x 4 | 19.6 | .22 | | S | E | 1 | |
| 1079 | | | 09 08 53.7 | -62 14 37 | 280.22 | -9.80 | 126 | -132.1 | -124.9 | 13x 5 | 18.9 | .23 | | S | | | |
| 1080 | | | 09 08 59.0 | -61 38 06 | 279.77 | -9.38 | 126 | -134.0 | -92.3 | 24x 7 | 18.1 | .22 | | S L | | | |
| 1081 | | | 09 08 59.9 | -57 27 01 | 276.66 | -6.54 | 166 | -86.3 | -133.4 | 13x 5 | 19.6 | .46 | | S | | | |
| 1082 | | | 09 09 04.2 | -59 40 16 | 278.31 | -8.05 | 125 | 112.7 | 13.9 | 12x 8 | 18.7 | .24 | | | | | pair w.1084 |
| 1083 | L125-009 | | 09 09 04.5 | -58 21 35 | 277.34 | -7.15 | 126 | -146.6 | 82.8 | 70x 35 | 15.3 | .31 | | S R5 | | | tightly wound sp.arms |
| 1084 | | | 09 09 06.6 | -59 40 31 | 278.32 | -8.04 | 125 | 112.9 | 13.7 | 11x 7 | 19.1 | .24 | | | | S | pair w.1082 |
| 1085 | | | 09 09 07.9 | -61 19 13 | 279.54 | -9.16 | 125 | 106.7 | -74.3 | 11x 11 | 18.6 | .25 | | E 0 | | | w.comp. (E) |
| 1086 | | | 09 09 12.4 | -55 50 57 | 275.51 | -5.43 | 166 | -88.6 | -47.6 | 12x 5 | 19.0 | .40 | | | | | |
| 1087 | | | 09 09 12.7 | -60 40 25 | 279.07 | -8.71 | 126 | -136.4 | -40.8 | 15x 7 | 18.3 | .19 | | S E | | | |
| 1088 | | | 09 09 17.4 | -57 56 47 | 277.06 | -6.85 | 125 | 120.8 | 106.1 | 19x 5 | 18.9 | .34 | | S M | E | 1 | |
| 1089 | | | 09 09 18.4 | -62 24 10 | 280.37 | -9.87 | 126 | -128.9 | -133.2 | 19x 11 | 18.5 | .21 | | | | | smooth, vLSB |
| 1090 | | | 09 09 18.8 | -60 38 26 | 279.05 | -8.68 | 126 | -135.9 | -39.0 | 16x 7 | 18.3 | .19 | | S M? | E | 1 | |
| 1091 | | | 09 09 19.0 | -60 20 29 | 278.83 | -8.48 | 126 | -137.1 | -23.0 | 36x 4 | 18.3 | .19 | | S | E | | |
| 1092 | | | 09 09 20.1 | -60 19 18 | 278.82 | -8.46 | 126 | -137.0 | -22.0 | 13x 12 | 17.8 | .19 | | E 2 | | | |
| 1093 | | | 09 09 28.4 | -56 16 14 | 275.84 | -5.69 | 166 | -85.6 | -70.1 | 13x 8 | 18.2 | .43 | | E 4 | | | |
| 1094 | | | 09 09 30.8 | -57 38 45 | 276.86 | -6.62 | 126 | -146.2 | 121.2 | 13x 4 | 19.8 | .36 | | S | | | |
| 1095 | | | 09 09 32.0 | -61 21 07 | 279.60 | -9.14 | 126 | -131.7 | -77.0 | 15x 13 | 18.4 | .27 | | S ? | | | |
| 1096 | | | 09 09 35.2 | -62 20 21 | 280.34 | -9.81 | 125 | 105.6 | -128.9 | 13x 5 | 20.0 | .22 | | S | | 1 | LSB |
| 1097 | | I | 09 09 37.1 | -55 41 13 | 275.43 | -5.28 | 166 | -85.9 | -38.8 | 27x 7 | 17.9 | .40 | | S ? | | | br.blg or s.p.* |
| 1098 | | | 09 09 38.7 | -56 54 20 | 276.32 | -6.11 | 166 | -82.9 | -104.0 | 13x 3 | 19.7 | .31 | | S | | 1 | |
| 1099 | | | 09 09 45.8 | -60 13 32 | 278.78 | -8.36 | 126 | -134.6 | -16.7 | 16x 12 | 17.6 | .21 | | S M | | S | |
| 1100 | | | 09 09 46.0 | -60 24 00 | 278.91 | -8.48 | 126 | -133.9 | -25.9 | 15x 5 | 19.1 | .19 | | S ? | | | br.blg or s.p.* |

**Table 1.** Galaxies in the southern ZOA – I. The Hydra/Antlia Extension (continued)

| RKK Ident (1) | Other Ident (2) | IR (3) | R.A. (h m s) (4) | Dec. (° ′ ″) (5) | gal $\ell$ (°) (6) | gal $b$ (°) (7) | SRC (8) | X (mm) (9) | Y (mm) (10) | D x d (″) (11) | $B_J$ ($^m$) (12) | $N_{HI}$ (13) | u | Type (14) | class. (14) | o | * | Remarks (15) |
|---|---|---|---|---|---|---|---|---|---|---|---|---|---|---|---|---|---|---|
| 1101 | | | 09 09 46.7 | -59 36 43 | 278.33 | -7.94 | 126 | -136.8 | 16.2 | 13x 12 | 17.8 | .24 | | E | ? | | | |
| 1102 | | NG | 09 09 48.3 | -55 42 27 | 275.46 | -5.27 | 166 | -84.4 | -39.8 | 13x 5 | 18.8 | .40 | ? | S | ? | | | |
| 1103 | | | 09 09 52.3 | -61 05 17 | 279.43 | -8.93 | 126 | -130.5 | -62.7 | 16x 5 | 19.1 | .22 | | L | ? | | | |
| 1104 | | | 09 09 57.9 | -55 40 15 | 275.45 | -5.23 | 166 | -83.3 | -37.8 | 16x 16 | 17.2 | .41 | | E | 0 : | | 1 | |
| 1105 | | | 09 09 58.6 | -60 37 36 | 279.10 | -8.61 | 125 | 114.9 | -37.6 | 13x 7 | 18.7 | .19 | | S | | | | |
| 1106 | | | 09 10 00.8 | -60 10 56 | 278.77 | -8.31 | 126 | -133.1 | -14.2 | 16x 8 | 18.6 | .22 | | S | M | | | |
| 1107 | | | 09 10 01.2 | -59 03 27 | 277.94 | -7.54 | 126 | -137.3 | 46.0 | 20x 16 | 17.2 | .26 | ? | S | M | | 1 | blg, vLSBdisk |
| 1108 | | | 09 10 01.6 | -59 35 57 | 278.34 | -7.91 | 126 | -135.2 | 17.0 | 16x 4 | 18.8 | .24 | | S | | | | |
| 1109 | | | 09 10 03.1 | -57 33 00 | 276.84 | -6.51 | 126 | -142.7 | 126.6 | 12x 12 | 17.8 | .38 | | E | ? | | | |
| 1110 | | | 09 10 15.8 | -57 30 46 | 276.83 | -6.46 | 126 | -141.4 | 128.7 | 19x 5 | 18.7 | .40 | | S | | | | |
| 1111 | | | 09 10 16.5 | -57 39 24 | 276.93 | -6.56 | 126 | -140.7 | 121.0 | 12x 12 | 18.4 | .36 | | S | | | 1 | smooth |
| 1112 | | | 09 10 20.4 | -59 34 29 | 278.35 | -7.86 | 126 | -133.2 | 18.5 | 13x 11 | 17.9 | .24 | | E | ? | | | |
| 1113 | | | 09 10 30.3 | -60 21 23 | 278.94 | -8.38 | 125 | 119.5 | -23.4 | 13x 5 | 19.1 | .19 | ? | | | | 1 | |
| 1114 | | | 09 10 30.8 | -60 40 36 | 279.18 | -8.60 | 125 | 118.3 | -40.5 | 20x 5 | 19.3 | .19 | ? | S | ? | | | br.blg or s.p.* |
| 1115 | | | 09 10 38.0 | -60 23 46 | 278.98 | -8.40 | 126 | -126.8 | -25.4 | 16x 5 | 18.9 | .19 | ? | | | | 1 | cl.to br.* |
| 1116 | | | 09 10 41.6 | -62 41 26 | 280.69 | -9.95 | 91 | -132.6 | 120.1 | 11x 5 | 19.5 | .20 | | | | | | dbl syst.w.1118 |
| 1117 | | | 09 10 46.9 | -60 24 03 | 279.00 | -8.39 | 126 | -127.2 | -25.5 | 13x 13 | 18.2 | .19 | | E | 0 ? | | | |
| 1118 | | | 09 10 49.0 | -62 42 45 | 280.72 | -9.96 | 91 | -131.8 | 119.0 | 13x 7 | 19.0 | .20 | | S | ? | | | dbl syst.w.1116 |
| 1119 | | | 09 10 59.0 | -60 13 27 | 278.89 | -8.25 | 126 | -126.5 | -16.0 | 22x 9 | 17.8 | .21 | | S | M | | | |
| 1120 | | | 09 11 00.9 | -60 08 24 | 278.83 | -8.19 | 125 | 123.8 | -12.1 | 17x 12 | 18.4 | .23 | ? | | | | 1 | |
| 1121 | | | 09 11 01.3 | -60 30 15 | 279.10 | -8.43 | 126 | -125.2 | -30.9 | 16x 5 | 18.8 | .19 | | S | M? | | | br.blg, LSBdisk |
| 1122 | | | 09 11 01.6 | -56 25 23 | 276.10 | -5.64 | 166 | -73.8 | -77.8 | 13x 13 | 18.3 | .38 | | | | | | |
| 1123 | | | 09 11 11.5 | -60 15 29 | 278.93 | -8.25 | 126 | -125.0 | -17.7 | 20x 8 | 18.2 | .21 | | S | M? | | | |
| 1124 | | | 09 11 14.9 | -59 28 10 | 278.35 | -7.71 | 126 | -127.4 | 24.5 | 17x 3 | 19.9 | .24 | | S | | E | | |
| 1125 | L126-001 | I | 09 11 17.4 | -58 38 06 | 277.74 | -7.13 | 126 | -130.1 | 69.2 | 56x 34 | 15.2 | .32 | | S | 3 : | | | |
| 1126 | | | 09 11 18.7 | -59 50 35 | 278.63 | -7.96 | 125 | 127.0 | 3.6 | 13x 11 | 18.7 | .25 | | | | | | in st.diffr.pat. |
| 1127 | | | 09 11 23.2 | -60 03 33 | 278.80 | -8.10 | 126 | -124.4 | -7.0 | 22x 4 | 19.0 | .25 | | S | | E | | |
| 1128 | | | 09 11 24.8 | -61 55 41 | 280.18 | -9.37 | 125 | 118.7 | -107.8 | 13x 9 | 19.2 | .25 | | | | | 1 | |
| 1129 | L091-002 | | 09 11 27.8 | -62 59 50 | 280.98 | -10.10 | 91 | -126.6 | 104.1 | 60x 20 | 15.9 | .20 | | S | 3 | | | |
| 1130 | | | 09 11 29.1 | -59 19 10 | 278.26 | -7.58 | 125 | 130.5 | 31.5 | 22x 5 | 19.1 | .27 | | S | | | | |
| 1131 | | P | 09 11 44.5 | -59 30 08 | 278.42 | -7.68 | 125 | 131.4 | 21.6 | 19x 3 | 19.9 | .25 | | S | | | 1 | |
| 1132 | | | 09 11 47.8 | -62 43 39 | 280.81 | -9.88 | 91 | -125.7 | 118.7 | 34x 13 | 17.2 | .20 | | | | | | 2 i.a.S (2 blg, 3 arms) |
| 1133 | | | 09 11 49.7 | -62 49 56 | 280.89 | -9.95 | 91 | -125.1 | 113.1 | 23x 11 | 17.6 | .19 | | S | ? | | | |
| 1134 | | | 09 11 50.1 | -59 06 09 | 278.14 | -7.40 | 126 | -124.7 | 44.4 | 27x 4 | 18.6 | .28 | | S | | E | | |
| 1135 | | | 09 11 51.9 | -60 09 14 | 278.91 | -8.12 | 126 | -120.9 | -11.8 | 24x 4 | 18.7 | .22 | | S | M | | | w.comp. (LSB S) |
| 1136 | | | 09 11 52.7 | -58 42 46 | 277.85 | -7.13 | 126 | -125.7 | 65.3 | 12x 9 | 18.2 | .29 | | S | ? | | | |
| 1137 | | | 09 11 57.5 | -60 47 42 | 279.39 | -8.55 | 126 | -118.0 | -46.1 | 20x 5 | 18.5 | .20 | | S | | | | br.blg or s.p.* |
| 1138 | | | 09 12 00.8 | -60 57 59 | 279.52 | -8.66 | 126 | -117.1 | -55.2 | 15x 7 | 18.4 | .21 | | S | | | | |
| 1139 | | | 09 12 01.0 | -60 08 43 | 278.92 | -8.10 | 126 | -119.9 | -11.3 | 16x 7 | 18.8 | .22 | | S | | | | |
| 1140 | | | 09 12 03.2 | -62 55 34 | 280.97 | -10.00 | 91 | -123.3 | 108.2 | 43x 7 | 17.6 | .20 | | S | | E | | |
| 1141 | | | 09 12 04.7 | -59 04 47 | 278.14 | -7.36 | 126 | -123.1 | 45.7 | 19x 7 | 18.3 | .29 | | SB | 5 | | | dist.nucl, dist.sp.arms |
| 1142 | | | 09 12 05.0 | -59 26 13 | 278.40 | -7.61 | 126 | -121.8 | 26.6 | 19x 4 | 19.3 | .25 | | S | | | | |
| 1143 | | | 09 12 08.0 | -60 20 51 | 279.08 | -8.23 | 125 | 130.3 | -23.8 | 20x 5 | 18.9 | .20 | | S | M | | | br.blg |
| 1144 | | | 09 12 15.3 | -62 21 40 | 280.57 | -9.60 | 126 | -110.8 | -129.8 | 11x 3 | 19.8 | .21 | | S | | | | dbl syst.w.1145 |
| 1145 | | | 09 12 15.5 | -62 21 32 | 280.57 | -9.59 | 126 | -110.7 | -129.6 | 11x 4 | 19.5 | .21 | | S | | | | dbl syst.w.1144 |
| 1146 | L126-002 | I | 09 12 16.1 | -60 34 56 | 279.26 | -8.38 | 126 | -116.7 | -34.6 | 67x 60 | 14.2 | .20 | | SB | 2 | | S | |
| 1147 | | P | 09 12 20.3 | -61 11 40 | 279.72 | -8.79 | 126 | -114.2 | -67.3 | 27x 4 | 18.5 | .25 | | S | ? | | | |
| 1148 | | | 09 12 39.1 | -63 05 32 | 281.15 | -10.06 | 91 | -119.1 | 99.6 | 17x 11 | 18.2 | .21 | ? | | | | | |
| 1149 | | | 09 12 42.3 | -57 10 35 | 276.81 | -5.99 | 126 | -124.9 | 147.9 | 13x 3 | 19.7 | .36 | | S | | | | patchy |
| 1150 | | | 09 12 42.9 | -57 12 28 | 276.83 | -6.01 | 126 | -124.7 | 146.2 | 16x 4 | 19.3 | .35 | | S | | | | patchy |
| 1151 | | | 09 12 45.4 | -61 47 46 | 280.20 | -9.17 | 126 | -109.5 | -99.3 | 16x 5 | 19.3 | .20 | | S | | E | 1 | |
| 1152 | | | 09 12 58.1 | -59 25 46 | 278.48 | -7.52 | 126 | -115.8 | 27.4 | 13x 11 | 17.9 | .25 | | E | ? | | | |
| 1153 | | | 09 13 03.1 | -62 11 36 | 280.51 | -9.41 | 126 | -106.4 | -120.5 | 20x 13 | 17.8 | .19 | | S | M | | 1 | |
| 1154 | | | 09 13 04.9 | -60 12 49 | 279.06 | -8.05 | 126 | -112.6 | -14.5 | 20x 4 | 18.8 | .21 | | S | ? | | | |
| 1155 | | | 09 13 10.2 | -61 49 44 | 280.25 | -9.15 | 126 | -106.8 | -100.9 | 13x 7 | 18.9 | .19 | | | | | | |
| 1156 | | | 09 13 12.9 | -59 42 15 | 278.70 | -7.69 | 126 | -113.3 | 12.8 | 16x 7 | 18.1 | .30 | | S | ? | | | |
| 1157 | | | 09 13 14.9 | -63 11 40 | 281.27 | -10.08 | 91 | -115.1 | 94.4 | 22x 17 | 17.5 | .22 | | S | 4 | | | LSBdisk |
| 1158 | | | 09 13 18.0 | -60 46 00 | 279.48 | -8.41 | 126 | -109.4 | -44.0 | 20x 16 | 17.9 | .20 | | | | | | |
| 1159 | L126-003 | I | 09 13 23.0 | -60 13 30 | 279.09 | -8.03 | 126 | -111.0 | -19.5 | 114x 67 | 13.5 | .21 | | SB | 4 | | | |
| 1160 | | | 09 13 27.6 | -60 49 06 | 279.53 | -8.43 | 126 | -108.2 | -46.7 | 13x 3 | 20.1 | .20 | | S | | E | | |
| 1161 | L126-004 | | 09 13 28.6 | -60 32 12 | 279.33 | -8.24 | 126 | -108.9 | -31.7 | 67x 60 | 15.1 | .20 | | SBS | 5 | | | |
| 1162 | | | 09 13 29.7 | -62 51 57 | 281.04 | -9.84 | 91 | -114.8 | 112.0 | 15x 8 | 18.2 | .20 | | | | | | |
| 1163 | | I | 09 13 36.3 | -59 16 51 | 278.42 | -7.36 | 126 | -112.0 | 35.6 | 16x 12 | 17.1 | .25 | | | | | | dbl syst? |
| 1164 | | | 09 13 36.8 | -59 24 44 | 278.52 | -7.45 | 126 | -111.5 | 28.6 | 24x 20 | 16.8 | .25 | ? | | | | | PN? |
| 1165 | | | 09 13 39.2 | -63 12 02 | 281.30 | -10.05 | 91 | -112.6 | 94.2 | 16x 4 | 19.4 | .22 | | S | | E | | |
| 1166 | | | 09 13 41.2 | -62 56 54 | 281.12 | -9.88 | 91 | -113.3 | 107.7 | 16x 4 | 19.3 | .20 | | S | | | | |
| 1167 | | | 09 13 41.5 | -54 12 31 | 274.76 | -3.83 | 166 | -57.4 | 41.5 | 17x 13 | 18.5 | .81 | | S | 5 : | F | | dist.nucl, vLSBdisk |
| 1168 | | | 09 13 54.0 | -61 49 34 | 280.31 | -9.09 | 126 | -102.2 | -100.5 | 13x 9 | 18.1 | .19 | ? | E | ? | | | |
| 1169 | | | 09 14 09.7 | -60 21 38 | 279.26 | -8.05 | 126 | -104.9 | -22.0 | 24x 7 | 17.6 | .20 | | S | E | | | |
| 1170 | L091-004 | I | 09 14 29.1 | -62 51 37 | 281.12 | -9.75 | 91 | -108.8 | 112.7 | 121x 94 | 13.5 | .20 | | S | 1 | | S | |
| 1171 | | | 09 14 34.0 | -61 47 28 | 280.34 | -9.01 | 126 | -98.1 | -98.4 | 22x 7 | 17.8 | .21 | | S | | | | |
| 1172 | | | 09 14 37.8 | -60 06 54 | 279.12 | -7.84 | 126 | -102.6 | -8.7 | 16x 4 | 19.1 | .24 | | S | | | | |
| 1173 | | | 09 14 38.3 | -61 20 48 | 280.02 | -8.69 | 126 | -98.9 | -74.6 | 22x 5 | 18.5 | .19 | | S | | | 1 | |
| 1174 | | | 09 14 45.4 | -57 16 10 | 277.07 | -5.85 | 166 | -45.0 | -122.3 | 11x 5 | 19.6 | .48 | ? | | | | | |
| 1175 | | | 09 14 52.4 | -61 49 08 | 280.39 | -9.00 | 126 | -96.1 | -99.8 | 13x 8 | 18.8 | .21 | | S | ? | | | |
| 1176 | | | 09 14 55.7 | -62 15 57 | 280.72 | -9.30 | 126 | -91.7 | -124.9 | 13x 3 | 19.7 | .18 | | S | | E | | |
| 1177 | | | 09 14 57.9 | -56 18 08 | 276.39 | -5.16 | 166 | -44.7 | -70.4 | 20x 5 | 18.7 | .60 | | S | | | | |
| 1178 | | | 09 15 01.3 | -56 18 04 | 276.40 | -5.15 | 166 | -44.3 | -70.3 | 9x 5 | 19.7 | .61 | | | | | | neighb.of 1177 |
| 1179 | | | 09 15 06.5 | -60 36 24 | 279.52 | -8.14 | 126 | -98.0 | -34.8 | 27x 20 | 17.4 | .21 | | | | | | |
| 1180 | | | 09 15 13.3 | -63 00 16 | 281.28 | -9.79 | 91 | -103.8 | 105.3 | 40x 27 | 15.9 | .20 | ? | | | | S | |
| 1181 | | | 09 15 20.7 | -55 32 55 | 275.89 | -4.60 | 166 | -42.8 | -29.9 | 13x 4 | 19.6 | .61 | | S | ? | | | |
| 1182 | | | 09 15 28.7 | -58 07 27 | 277.76 | -6.38 | 126 | -102.2 | 98.2 | 15x 5 | 19.2 | .36 | | S | | | | dist.blg |
| 1183 | | | 09 15 41.9 | -60 04 21 | 279.18 | -7.72 | 126 | -95.5 | -6.0 | 16x 7 | 18.1 | .25 | | S | E | | | |
| 1184 | | | 09 15 43.0 | -62 47 01 | 281.16 | -9.60 | 91 | -101.5 | 117.3 | 17x 5 | 18.9 | .19 | | S | ? | | | |
| 1185 | | | 09 15 45.8 | -60 24 09 | 279.43 | -7.94 | 126 | -94.2 | -23.7 | 16x 8 | 18.0 | .21 | | E | 4 | | | cl.to br.* |
| 1186 | | | 09 15 46.7 | -62 52 07 | 281.23 | -9.65 | 91 | -100.8 | 112.8 | 34x 27 | 16.8 | .20 | | D | ? | | 1 | LSB |
| 1187 | | | 09 15 47.2 | -61 39 51 | 280.35 | -8.81 | 126 | -88.8 | -92.3 | 15x 7 | 18.4 | .19 | | S | E | | | |
| 1188 | | | 09 15 50.4 | -60 54 22 | 279.80 | -8.28 | 126 | -92.4 | -50.6 | 20x 19 | 17.7 | .21 | | S | ? | | | |
| 1189 | | | 09 15 51.9 | -56 48 48 | 276.85 | -5.43 | 166 | -37.5 | -97.6 | 20x 5 | 18.5 | .56 | | S | | E | | br.blg |
| 1190 | | | 09 15 59.5 | -60 43 52 | 279.69 | -8.15 | 126 | -91.9 | -41.2 | 20x 13 | 17.5 | .25 | ? | | | | | |
| 1191 | | I | 09 16 13.5 | -59 32 58 | 278.85 | -7.31 | 126 | -94.3 | 21.0 | 16x 9 | 18.0 | .30 | | | | | | |
| 1192 | | | 09 16 16.6 | -60 15 12 | 279.37 | -7.79 | 126 | -91.2 | -15.5 | 15x 3 | 19.6 | .20 | | S | | | | |
| 1193 | | | 09 16 17.6 | -60 35 44 | 279.62 | -8.03 | 126 | -90.2 | -33.8 | 16x 3 | 19.5 | .25 | | S | ? | | | |
| 1194 | L126-005 | | 09 16 21.3 | -59 24 19 | 278.76 | -7.19 | 126 | -92.8 | 29.9 | 60x 27 | 15.5 | .28 | | S | 4 | | S | |
| 1195 | | NG | 09 16 21.8 | -59 31 31 | 278.85 | -7.28 | 126 | -92.5 | 23.5 | 16x 4 | 19.1 | .30 | | S | ? | | | |
| 1196 | L091-007 | I | 09 16 21.9 | -62 40 19 | 281.13 | -9.47 | 126 | -85.5 | -144.0 | 121x 81 | 13.4 | .19 | | S | 5 | | S | |
| 1197 | | | 09 16 29.9 | -61 54 17 | 280.58 | -8.92 | 126 | -85.6 | -103.9 | 9x 7 | 18.7 | .19 | | E | | | | |
| 1198 | | | 09 16 31.7 | -59 58 10 | 279.18 | -7.57 | 126 | -90.3 | -.2 | 13x 3 | 19.8 | .26 | | S | | E | | |
| 1199 | | | 09 16 38.1 | -60 17 46 | 279.43 | -7.79 | 126 | -88.7 | -17.7 | 23x 13 | 17.9 | .21 | | | | | | |
| 1200 | | | 09 16 39.9 | -56 31 37 | 276.72 | -5.15 | 166 | -31.9 | -82.2 | 20x 11 | 18.2 | .57 | | S | | | | blg, vLSBdisk |

**Table 1.** Galaxies in the southern ZOA – I. The Hydra/Antlia Extension (continued)

| RKK Ident (1) | Other Ident (2) | IR (3) | R.A. (h m s) (4) | Dec. (° ′ ″) (5) | gal $\ell$ (°) (6) | gal $b$ (°) (7) | SRC (8) | X (mm) (9) | Y (mm) (10) | D x d (″) (11) | $B_J$ ($^m$) (12) | $N_{HI}$ (13) | u | Type class. (14) | o | * | Remarks (15) |
|---|---|---|---|---|---|---|---|---|---|---|---|---|---|---|---|---|---|
| 1201 | | | 09 16 40.4 | -59 24 45 | 278.79 | -7.17 | 126 | -90.7 | 29.6 | 13x 8 | 18.4 | .29 | | | | | |
| 1202 | | | 09 16 42.8 | -60 34 20 | 279.63 | -7.98 | 126 | -87.5 | -32.5 | 13x 5 | 19.4 | .25 | | | | | |
| 1203 | | | 09 16 44.1 | -57 36 33 | 277.50 | -5.90 | 126 | -94.6 | 126.2 | 16x 5 | 18.9 | .48 | | S ? | | | |
| 1204 | | | 09 16 44.3 | -56 27 21 | 276.68 | -5.09 | 166 | -31.4 | -78.3 | 27x 12 | 17.8 | .61 | ? | S 5 | | | vLSBdisk |
| 1205 | | I | 09 16 52.2 | -61 41 30 | 280.46 | -8.74 | 126 | -83.7 | -92.4 | 22x 5 | 18.1 | .18 | | | | S | |
| 1206 | | | 09 17 03.9 | -60 26 07 | 279.57 | -7.85 | 126 | -85.5 | -25.0 | 13x 4 | 19.4 | .25 | | | | | |
| 1207 | | | 09 17 07.3 | -61 47 22 | 280.55 | -8.79 | 126 | -81.9 | -97.5 | 16x 11 | 18.1 | .19 | | S | | | |
| 1208 | | | 09 17 07.8 | -63 08 40 | 281.54 | -9.73 | 91 | -91.8 | 98.5 | 20x 8 | 18.3 | .21 | | S ? | | | |
| 1209 | | | 09 17 08.4 | -61 48 20 | 280.56 | -8.80 | 126 | -81.8 | -98.4 | 20x 5 | 18.8 | .19 | | S | | | br.blg, LSBdisk |
| 1210 | | | 09 17 11.2 | -61 01 05 | 280.00 | -8.24 | 126 | -83.3 | -56.2 | 20x 8 | 18.0 | .20 | | S ? | | | i.a.w.1218 |
| 1211 | | | 09 17 14.4 | -61 49 18 | 280.58 | -8.80 | 126 | -81.1 | -99.2 | 13x 5 | 19.2 | .19 | | S 1 | | | |
| 1212 | | | 09 17 17.0 | -58 41 24 | 278.33 | -6.61 | 126 | -88.2 | 68.5 | 13x 5 | 19.4 | .30 | | S | | | w.sev.comp. |
| 1213 | | | 09 17 20.7 | -62 35 23 | 281.15 | -9.33 | 91 | -92.1 | 128.3 | 16x 7 | 18.9 | .19 | | S ? | | | |
| 1214 | | | 09 17 28.2 | -62 27 19 | 281.06 | -9.22 | 126 | -78.3 | -133.1 | 27x 13 | 17.1 | .20 | | S ? | | | |
| 1215 | | P | 09 17 31.2 | -62 12 15 | 280.89 | -9.04 | 126 | -75.8 | -120.4 | 27x 20 | 16.9 | .19 | ? | | | | |
| 1216 | | | 09 17 36.7 | -59 20 02 | 278.82 | -7.03 | 126 | -84.4 | 34.1 | 19x 5 | 18.3 | .29 | | | | S | |
| 1217 | | | 09 17 39.8 | -58 18 36 | 278.09 | -6.31 | 126 | -86.4 | 89.0 | 16x 5 | 19.1 | .40 | | S L | | | |
| 1218 | | | 09 17 45.6 | -61 06 29 | 280.11 | -8.26 | 126 | -79.4 | -60.9 | 22x 9 | 17.7 | .20 | | S 4 | | | i.a.w.1210 |
| 1219 | | | 09 17 45.7 | -62 01 07 | 280.77 | -8.89 | 126 | -77.3 | -109.6 | 13x 4 | 19.3 | .19 | | S | | | dist.nucl. |
| 1220 | | | 09 17 46.3 | -62 13 47 | 280.92 | -9.04 | 126 | -76.8 | -120.9 | 16x 3 | 19.5 | .19 | | S | | | |
| 1221 | | | 09 17 48.6 | -58 38 41 | 278.34 | -6.53 | 126 | -84.6 | 71.1 | 34x 7 | 18.2 | .30 | | S | E | | |
| 1222 | | | 09 17 54.7 | -60 34 08 | 279.74 | -7.87 | 126 | -79.7 | -32.0 | 13x 9 | 18.4 | .27 | | | | | |
| 1223 | | | 09 17 55.2 | -59 06 26 | 278.69 | -6.84 | 126 | -82.8 | 46.4 | 20x 7 | 17.9 | .28 | | S ? | | | |
| 1224 | | | 09 17 55.3 | -60 15 33 | 279.51 | -7.65 | 126 | -80.3 | -15.3 | 17x 11 | 17.6 | .23 | | E ? | | | |
| 1225 | | | 09 17 57.2 | -61 32 45 | 280.44 | -8.55 | 126 | -77.2 | -84.3 | 17x 8 | 18.0 | .17 | | | | | |
| 1226 | | | 09 17 57.6 | -61 36 44 | 280.49 | -8.59 | 126 | -77.0 | -87.8 | 16x 4 | 19.2 | .17 | | | | | box-shaped |
| 1227 | | | 09 17 58.0 | -57 04 50 | 277.24 | -5.41 | 166 | -21.8 | -111.7 | 20x 13 | 18.1 | .62 | | S 5 : | | | |
| 1228 | | | 09 18 05.5 | -61 14 11 | 280.23 | -8.32 | 126 | -77.0 | -67.7 | 11x 11 | 18.0 | .21 | | E | | | |
| 1229 | | | 09 18 06.8 | -59 21 39 | 278.89 | -7.00 | 126 | -81.0 | 32.8 | 27x 19 | 16.4 | .31 | | E 2 | | | |
| 1230 | | | 09 18 08.7 | -59 20 19 | 278.87 | -6.98 | 126 | -80.8 | 34.0 | 13x 5 | 19.2 | .31 | | S | | | |
| 1231 | | | 09 18 13.7 | -60 27 09 | 279.68 | -7.76 | 126 | -77.8 | -25.6 | 30x 20 | 16.8 | .25 | | S ? | | | |
| 1232 | | | 09 18 13.8 | -61 03 54 | 280.12 | -8.19 | 126 | -76.5 | -58.4 | 16x 7 | 18.5 | .20 | | S | | | |
| 1233 | | | 09 18 24.0 | -62 36 34 | 281.25 | -9.25 | 91 | -85.5 | 127.5 | 13x 5 | 19.0 | .22 | ? | | | | |
| 1234 | | | 09 18 27.9 | -62 44 57 | 281.36 | -9.35 | 91 | -84.7 | 120.1 | 27x 22 | 16.6 | .21 | | S 5 : | | S | |
| 1235 | | | 09 18 30.9 | -63 09 25 | 281.66 | -9.63 | 91 | -83.4 | 98.3 | 47x 27 | 16.7 | .22 | ? | S | | | vLSB |
| 1236 | | | 09 18 52.5 | -63 53 17 | 282.22 | -10.11 | 91 | -79.3 | 59.2 | 15x 4 | 19.4 | .29 | ? | S ? | | | |
| 1237 | SGC0918 | | 09 18 55.5 | -60 01 11 | 279.43 | -7.39 | 126 | -74.1 | -2.3 | 27x 13 | 17.8 | .25 | | S | | S | |
| 1238 | | | 09 18 56.4 | -58 35 40 | 278.41 | -6.39 | 126 | -76.8 | 74.1 | 13x 5 | 19.2 | .32 | | S ? | | | |
| 1239 | | | 09 18 58.6 | -62 01 22 | 280.87 | -8.80 | 126 | -69.7 | -109.5 | 16x 5 | 18.7 | .20 | | | | | p.cov.by br.* |
| 1240 | | | 09 19 06.3 | -59 20 47 | 278.96 | -6.90 | 126 | -74.2 | 33.9 | 15x 4 | 19.2 | .33 | | S | E | | |
| 1241 | | | 09 19 09.8 | -58 53 56 | 278.65 | -6.58 | 126 | -74.7 | 57.8 | 16x 9 | 18.3 | .29 | | | | | dist.nucl. |
| 1242 | | | 09 19 11.7 | -59 22 37 | 278.99 | -6.92 | 126 | -73.5 | 32.2 | 13x 7 | 18.8 | .33 | | | | | |
| 1243 | | | 09 19 12.8 | -60 22 49 | 279.71 | -7.62 | 126 | -71.5 | -21.5 | 17x 3 | 19.4 | .24 | | S | | | |
| 1244 | | I | 09 19 39.0 | -61 12 57 | 280.35 | -8.17 | 126 | -65.7 | -66.7 | 20x 4 | 18.9 | .21 | | S | E | | w.comp. |
| 1245 | | | 09 19 46.5 | -58 56 35 | 278.74 | -6.56 | 126 | -70.4 | 55.6 | 20x 5 | 18.5 | .31 | | S | E | | |
| 1246 | | | 09 19 59.2 | -60 13 09 | 279.67 | -7.44 | 126 | -66.6 | -12.7 | 13x 7 | 18.6 | .25 | | S | | | |
| 1247 | | | 09 20 10.8 | -61 07 55 | 280.34 | -8.07 | 126 | -62.7 | -62.0 | 27x 16 | 17.1 | .21 | | S 4 | | | |
| 1248 | | | 09 20 13.8 | -63 52 07 | 282.31 | -9.99 | 91 | -71.4 | 60.6 | 17x 16 | 18.0 | .28 | ? | S ? | | | |
| 1249 | | | 09 20 14.8 | -63 15 25 | 281.87 | -9.56 | 91 | -72.7 | 93.4 | 13x 4 | 19.2 | .23 | | S | | | |
| 1250 | | | 09 20 17.4 | -59 13 10 | 278.98 | -6.71 | 126 | -66.3 | 40.9 | 60x 13 | 16.0 | .33 | | S 2 | | | |
| 1251 | L126-009 | | 09 20 23.6 | -59 16 12 | 279.03 | -6.73 | 126 | -65.5 | 38.3 | 40x 30 | 15.9 | .31 | | SBR5 | | | i.a.w.1253 |
| 1252 | | | 09 20 23.6 | -61 49 05 | 280.85 | -8.53 | 126 | -61.1 | -98.3 | 17x 7 | 18.1 | .20 | | E 5 | | | |
| 1253 | L126-009 | | 09 20 26.8 | -59 16 12 | 279.03 | -6.73 | 126 | -65.2 | 38.3 | 34x 23 | 16.3 | .31 | | S 5 | | | i.a.w.1251 |
| 1254 | | | 09 20 29.6 | -59 04 31 | 278.90 | -6.58 | 126 | -65.2 | 48.7 | 13x 3 | 19.7 | .33 | | S ? | | | |
| 1255 | | | 09 20 31.9 | -63 14 35 | 281.88 | -9.53 | 91 | -71.0 | 94.2 | 16x 4 | 19.4 | .23 | | S | | | |
| 1256 | | | 09 20 35.3 | -59 14 12 | 279.02 | -6.69 | 126 | -64.3 | 40.1 | 27x 7 | 17.3 | .32 | | S 3 ? | | | |
| 1257 | | | 09 20 38.5 | -58 15 06 | 278.33 | -5.99 | 126 | -65.5 | 92.9 | 13x 11 | 18.5 | .42 | | S ? | | | blg, LSBdisk |
| 1258 | | | 09 20 43.0 | -59 43 57 | 279.39 | -7.03 | 126 | -62.6 | 13.5 | 22x 4 | 18.8 | .32 | | S | E | | |
| 1259 | | | 09 20 51.4 | -59 15 35 | 279.06 | -6.68 | 126 | -62.4 | 38.9 | 27x 17 | 16.4 | .31 | | S 0 | | | |
| 1260 | | | 09 20 51.5 | -59 18 22 | 279.10 | -6.72 | 126 | -62.3 | 36.4 | 16x 11 | 18.1 | .31 | | E 5 | | | or blg |
| 1261 | | | 09 20 54.9 | -56 26 04 | 277.07 | -4.67 | 166 | -.5 | -76.9 | 16x 11 | 17.6 | .56 | ? | L ? | | | |
| 1262 | | | 09 20 57.0 | -60 12 42 | 279.75 | -7.35 | 126 | -60.2 | -12.1 | 13x 7 | 18.8 | .27 | | | | | |
| 1263 | | | 09 20 58.1 | -62 20 27 | 281.27 | -8.85 | 126 | -56.7 | -126.2 | 13x 4 | 19.4 | .24 | | | | | |
| 1264 | | | 09 21 07.1 | -63 19 12 | 281.98 | -9.53 | 91 | -67.3 | 90.2 | 13x 9 | 18.7 | .22 | | | | | |
| 1265 | L126-010 | I | 09 21 08.3 | -60 50 05 | 280.21 | -7.77 | 126 | -58.0 | -45.4 | 121x 40 | 14.0 | .22 | | S 3 | | | |
| 1266 | | | 09 21 10.0 | -62 44 32 | 281.57 | -9.12 | 91 | -68.2 | 121.2 | 17x 8 | 18.1 | .24 | | S ? | | | * on edge of gal. |
| 1267 | | P | 09 21 10.8 | -60 22 46 | 279.89 | -7.45 | 126 | -58.4 | -21.0 | 24x 5 | 18.4 | .34 | | S | | | |
| 1268 | | | 09 21 11.4 | -61 26 44 | 280.65 | -8.20 | 126 | -56.7 | -78.2 | 17x 5 | 18.6 | .28 | | S | | | |
| 1269 | | | 09 21 12.2 | -62 28 48 | 281.39 | -8.93 | 126 | -55.0 | -133.6 | 16x 13 | 17.9 | .24 | | S M | F | | w.comp. |
| 1270 | | | 09 21 14.8 | -61 22 31 | 280.60 | -8.15 | 126 | -56.4 | -74.4 | 13x 13 | 18.6 | .27 | ? | | | | |
| 1271 | | | 09 21 22.9 | -59 19 19 | 279.15 | -6.68 | 126 | -58.7 | 35.7 | 40x 9 | 17.3 | .31 | | S 4 | | | dist.nucl, dist.sp.arms |
| 1272 | | I | 09 21 28.3 | -56 55 44 | 277.47 | -4.97 | 166 | 3.7 | -103.4 | 24x 15 | 17.1 | .67 | | S L | | S | |
| 1273 | | | 09 21 30.9 | -63 36 24 | 282.22 | -9.70 | 91 | -64.3 | 74.9 | 15x 12 | 18.0 | .22 | | E | | | |
| 1274 | | | 09 21 33.9 | -63 08 58 | 281.90 | -9.38 | 91 | -64.9 | 99.4 | 34x 13 | 17.2 | .24 | | S 4 | | | |
| 1275 | | I | 09 21 43.2 | -56 48 22 | 277.41 | -4.86 | 166 | 5.5 | -96.8 | 23x 7 | 18.4 | .80 | | S | | | blg, vLSBdisk |
| 1276 | | | 09 21 45.6 | -62 19 59 | 281.33 | -8.78 | 126 | -51.7 | -125.6 | 16x 16 | 17.2 | .25 | ? | E 0 : | | | |
| 1277 | | | 09 21 47.2 | -57 33 53 | 277.95 | -5.39 | 126 | -58.4 | 129.9 | 13x 4 | 19.8 | .59 | | S | | | blg, LSBdisk |
| 1278 | | | 09 21 48.8 | -60 02 34 | 279.70 | -7.15 | 126 | -54.7 | -2.9 | 20x 7 | 18.3 | .28 | | | | | |
| 1279 | | | 09 22 02.4 | -59 26 11 | 279.30 | -6.70 | 126 | -54.0 | 29.7 | 16x 3 | 19.7 | .31 | | S | | | |
| 1280 | | | 09 22 04.1 | -59 23 51 | 279.27 | -6.67 | 126 | -53.9 | 31.7 | 20x 9 | 18.0 | .31 | | S | | 1 | |
| 1281 | | | 09 22 04.9 | -59 28 37 | 279.33 | -6.73 | 126 | -53.7 | 27.5 | 13x 13 | 18.6 | .32 | | | | | |
| 1282 | | | 09 22 06.3 | -63 42 37 | 282.34 | -9.73 | 91 | -60.6 | 69.5 | 27x 9 | 18.2 | .24 | | | | | |
| 1283 | | P | 09 22 10.7 | -62 51 55 | 281.74 | -9.13 | 91 | -61.7 | 114.8 | 27x 19 | 17.0 | .26 | | S ? | | | br.* s.p. |
| 1284 | | | 09 22 10.7 | -63 36 56 | 282.28 | -9.66 | 91 | -60.4 | 74.6 | 40x 7 | 17.5 | .23 | | S | E | | in halo of 1286 |
| 1285 | | | 09 22 13.8 | -63 39 23 | 282.31 | -9.68 | 91 | -60.0 | 72.4 | 23x 8 | 18.3 | .23 | | S M | | S | LSB |
| 1286 | L091-009 | | 09 22 16.2 | -63 35 49 | 282.27 | -9.64 | 91 | -59.8 | 75.6 | 134x 94 | 13.6 | .22 | | E | | S | |
| 1287 | L091-010 | | 09 22 17.2 | -62 40 00 | 281.61 | -8.98 | 126 | -48.0 | -143.4 | 74x 20 | 15.5 | .26 | | S 3 | | | |
| 1288 | L091-011 | I | 09 22 18.2 | -63 27 50 | 282.18 | -9.54 | 91 | -59.9 | 82.8 | 101x 12 | 15.8 | .22 | | S 3 | E | S | |
| 1289 | | | 09 22 19.3 | -59 02 56 | 279.05 | -6.40 | 126 | -52.6 | 50.5 | 20x 12 | 17.8 | .35 | | S | | | |
| 1290 | | | 09 22 21.9 | -60 19 10 | 279.95 | -7.30 | 126 | -50.7 | -17.6 | 13x 4 | 19.6 | .29 | | S ? | | | |
| 1291 | | | 09 22 21.9 | -63 06 26 | 281.93 | -9.28 | 91 | -60.2 | 101.9 | 19x 15 | 17.4 | .24 | ? | E | | | |
| 1292 | | | 09 22 24.6 | -61 02 06 | 280.46 | -7.81 | 126 | -49.4 | -56.0 | 13x 4 | 19.4 | .22 | | | | | w.comp. |
| 1293 | | | 09 22 30.1 | -63 35 24 | 282.29 | -9.61 | 91 | -58.5 | 76.0 | 19x 13 | 17.5 | .22 | | E ? | | | in halo of 1286 |
| 1294 | | | 09 22 47.0 | -60 22 21 | 280.02 | -7.30 | 126 | -47.7 | -20.4 | 12x 9 | 18.3 | .26 | | | | | |
| 1295 | | | 09 22 48.2 | -63 24 15 | 282.18 | -9.46 | 91 | -57.0 | 86.1 | 34x 5 | 18.4 | .22 | | S | | | |
| 1296 | | | 09 22 55.1 | -60 20 44 | 280.02 | -7.27 | 126 | -46.9 | -19.0 | 13x 11 | 18.1 | .26 | | S E | | | |
| 1297 | | | 09 22 55.1 | -60 24 27 | 280.06 | -7.32 | 126 | -46.7 | -22.3 | 22x 16 | 17.3 | .26 | | SB 5 : | | | dist.sp.arms |
| 1298 | | | 09 22 58.5 | -62 37 15 | 281.63 | -8.89 | 91 | -57.3 | 128.1 | 16x 9 | 18.1 | .26 | | | | | |
| 1299 | | | 09 23 10.7 | -58 42 05 | 278.88 | -6.08 | 126 | -47.1 | 69.2 | 16x 8 | 19.1 | .39 | ? | | | | |
| 1300 | L091-012 | | 09 23 20.9 | -63 01 40 | 281.96 | -9.15 | 91 | -54.3 | 106.3 | 74x 63 | 14.7 | .25 | | S 5 | | | LSB, dist.sp.struct. |

**Table 1.** Galaxies in the southern ZOA – I. The Hydra/Antlia Extension (continued)

| RKK Ident (1) | Other Ident (2) | IR (3) | R.A. (h m s) (4) | Dec. (° ′ ″) (5) | gal ℓ (°) (6) | gal $b$ (°) (7) | SRC (8) | X (mm) (9) | Y (mm) (10) | D x d (″) (11) | $B_J$ ($^m$) (12) | $N_{HI}$ (13) | u | Type class. (14) | o * | Remarks (15) |
|---|---|---|---|---|---|---|---|---|---|---|---|---|---|---|---|---|
| 1301 | | | 09 23 23.6 | -59 33 20 | 279.50 | -6.67 | 126 | -44.7 | 23.5 | 23x 7 | 18.1 | .31 | | | | |
| 1302 | | | 09 23 25.9 | -62 40 51 | 281.72 | -8.89 | 126 | -40.9 | -144.0 | 16x 11 | 17.6 | .27 | | S ? | S | |
| 1303 | | | 09 23 29.3 | -64 00 07 | 282.66 | -9.83 | 91 | -51.9 | 54.1 | 17x 5 | 18.8 | .22 | | S M | | |
| 1304 | | | 09 23 32.7 | -63 41 52 | 282.45 | -9.61 | 91 | -52.1 | 70.5 | 19x 4 | 19.2 | .23 | | S | E | |
| 1305 | | | 09 23 33.4 | -60 27 00 | 280.15 | -7.29 | 126 | -42.4 | -24.4 | 19x 8 | 17.8 | .26 | | S | | |
| 1306 | | | 09 23 37.1 | -60 26 37 | 280.15 | -7.28 | 126 | -42.0 | -24.0 | 9x 4 | 19.7 | .26 | | | | |
| 1307 | | | 09 23 39.5 | -61 19 47 | 280.78 | -7.91 | 126 | -41.0 | -71.6 | 16x 16 | 17.6 | .23 | | | | on st.diffr.ring |
| 1308 | | | 09 23 44.5 | -61 29 22 | 280.90 | -8.02 | 126 | -40.3 | -80.1 | 16x 5 | 19.0 | .29 | | S | | |
| 1309 | | | 09 23 46.6 | -56 18 18 | 277.26 | -4.30 | 166 | 20.8 | -70.0 | 12x 5 | 19.7 | .72 | | S ? | | |
| 1310 | | | 09 23 47.4 | -64 17 09 | 282.89 | -10.01 | 91 | -49.7 | 39.0 | 13x 4 | 19.6 | .27 | ? | | | |
| 1311 | | | 09 23 53.8 | -63 48 05 | 282.55 | -9.66 | 91 | -49.9 | 65.0 | 20x 15 | 17.5 | .23 | | E | | |
| 1312 | | | 09 23 56.1 | -59 17 13 | 279.36 | -6.43 | 126 | -41.3 | 38.0 | 54x 7 | 17.1 | .36 | | S | E | |
| 1313 | | | 09 23 56.2 | -63 24 32 | 282.27 | -9.37 | 91 | -50.2 | 86.0 | 20x 12 | 18.1 | .22 | ? | | | |
| 1314 | | | 09 23 59.5 | -62 27 57 | 281.61 | -8.70 | 126 | -37.7 | -132.4 | 15x 5 | 18.9 | .28 | | | | |
| 1315 | | | 09 24 00.5 | -60 24 41 | 280.16 | -7.23 | 126 | -39.7 | -22.3 | 13x 7 | 18.7 | .26 | ? | | | |
| 1316 | | | 09 24 01.0 | -62 31 15 | 281.65 | -8.73 | 126 | -37.5 | -135.4 | 15x 8 | 18.0 | .27 | | | S | |
| 1317 | | | 09 24 01.3 | -58 38 22 | 278.92 | -5.95 | 126 | -43.5 | 72.7 | 16x 3 | 20.1 | .45 | ? | S | | |
| 1318 | | | 09 24 11.1 | -60 25 33 | 280.19 | -7.22 | 126 | -38.5 | -23.1 | 16x 8 | 18.6 | .26 | | | | |
| 1319 | | | 09 24 13.2 | -57 50 39 | 278.38 | -5.36 | 126 | -40.6 | 115.3 | 16x 13 | 17.9 | .55 | | S M | V | |
| 1320 | | | 09 24 21.1 | -62 06 30 | 281.39 | -8.41 | 126 | -35.8 | -113.2 | 13x 11 | 18.1 | .26 | | | | |
| 1321 | | | 09 24 24.4 | -64 00 21 | 282.74 | -9.76 | 91 | -46.5 | 54.1 | 20x 3 | 19.9 | .23 | | S | E | LSB |
| 1322 | | | 09 24 33.8 | -59 21 23 | 279.47 | -6.42 | 126 | -36.9 | 34.3 | 27x 7 | 18.0 | .35 | | S | | |
| 1323 | | | 09 24 37.6 | -60 32 05 | 280.30 | -7.26 | 126 | -35.5 | -28.8 | 13x 11 | 17.9 | .25 | | | | p.cov.by br.* |
| 1324 | | | 09 24 41.1 | -59 36 31 | 279.66 | -6.59 | 126 | -35.9 | 20.8 | 16x 3 | 19.7 | .34 | | S | | |
| 1325 | | | 09 24 48.1 | -59 59 39 | 279.94 | -6.86 | 126 | -34.8 | .1 | 20x 11 | 17.8 | .36 | | | | |
| 1326 | | | 09 24 53.2 | -57 54 04 | 278.49 | -5.34 | 126 | -35.9 | 112.3 | 16x 5 | 18.7 | .58 | | F ? | | |
| 1327 | | | 09 24 57.5 | -60 23 33 | 280.23 | -7.13 | 126 | -33.4 | -21.2 | 20x 13 | 17.2 | .27 | | E ? | | |
| 1328 | L126-011 | | 09 25 04.2 | -60 23 50 | 280.25 | -7.12 | 126 | -32.7 | -21.4 | 74x 27 | 15.7 | .28 | | S L | | |
| 1329 | | | 09 25 10.2 | -59 03 41 | 279.32 | -6.15 | 126 | -33.0 | 50.2 | 27x 22 | 16.7 | .40 | | S ? | | |
| 1330 | | | 09 25 10.9 | -60 27 10 | 280.29 | -7.15 | 126 | -31.9 | -24.4 | 20x 15 | 17.2 | .28 | | | | br.blg, LSBdisk |
| 1331 | | | 09 25 13.4 | -64 24 40 | 283.09 | -9.99 | 91 | -41.3 | 32.5 | 19x 7 | 18.5 | .28 | | S ? | | |
| 1332 | | | 09 25 20.9 | -64 20 38 | 283.05 | -9.93 | 91 | -40.6 | 36.1 | 13x 11 | 18.1 | .29 | ? | | 1 | |
| 1333 | | | 09 25 22.4 | -62 32 01 | 281.77 | -8.63 | 126 | -29.1 | -135.9 | 17x 15 | 17.9 | .29 | | S R | | |
| 1334 | | | 09 25 23.7 | -61 07 15 | 280.78 | -7.62 | 126 | -30.0 | -60.2 | 17x 11 | 17.6 | .23 | | E 4 : | | |
| 1335 | | I | 09 25 27.1 | -63 18 50 | 282.33 | -9.19 | 91 | -41.2 | 91.3 | 15x 8 | 19.1 | .23 | ? | | S | |
| 1336 | | | 09 25 31.5 | -60 45 17 | 280.54 | -7.34 | 126 | -29.4 | -40.6 | 16x 11 | 18.2 | .25 | | S ? | | |
| 1337 | | | 09 25 33.5 | -60 29 04 | 280.35 | -7.15 | 126 | -29.4 | -26.1 | 16x 5 | 18.7 | .28 | | S ? | | |
| 1338 | | | 09 25 36.7 | -59 05 17 | 279.38 | -6.13 | 126 | -29.9 | 48.8 | 15x 11 | 17.7 | .39 | | E ? | | |
| 1339 | | | 09 25 39.1 | -64 38 00 | 283.28 | -10.12 | 91 | -38.5 | 20.6 | 34x 20 | 16.8 | .27 | | S | V1 | pair w.1341 |
| 1340 | | | 09 25 39.7 | -64 32 31 | 283.22 | -10.05 | 91 | -38.6 | 25.5 | 22x 5 | 18.9 | .28 | | S | | |
| 1341 | | | 09 25 46.3 | -64 37 19 | 283.28 | -10.10 | 91 | -37.8 | 21.2 | 23x 7 | 18.3 | .27 | | S M | E | pair w.1339 |
| 1342 | | | 09 25 47.4 | -61 58 56 | 281.42 | -8.20 | 126 | -26.9 | -106.3 | 13x 8 | 18.6 | .29 | | | | |
| 1343 | | | 09 25 48.9 | -61 00 03 | 280.73 | -7.50 | 126 | -27.3 | -53.7 | 16x 5 | 18.7 | .23 | | | | |
| 1344 | | | 09 25 50.8 | -60 38 18 | 280.48 | -7.23 | 126 | -27.8 | -33.8 | 24x 3 | 19.1 | .27 | | S | E | |
| 1345 | L126-012 | | 09 25 53.2 | -62 20 59 | 281.69 | -8.46 | 126 | -26.0 | -126.0 | 54x 20 | 16.1 | .32 | | S E | | |
| 1346 | | | 09 25 53.7 | -60 33 45 | 280.43 | -7.17 | 126 | -26.7 | -29.8 | 16x 7 | 18.3 | .28 | | | 2 | smooth |
| 1347 | | | 09 25 54.1 | -60 33 44 | 280.43 | -7.17 | 126 | -27.0 | -30.2 | 19x 13 | 17.7 | .28 | | S 5 : | | neighb.of 1353 |
| 1348 | | | 09 25 57.4 | -60 42 58 | 280.55 | -7.28 | 126 | -26.6 | -38.5 | 23x 4 | 18.8 | .25 | | S | E | |
| 1349 | | | 09 25 57.5 | -61 46 49 | 281.29 | -8.04 | 126 | -25.9 | -95.5 | 27x 3 | 19.4 | .31 | | S | E | |
| 1350 | | | 09 25 59.4 | -59 03 06 | 279.39 | -6.07 | 126 | -27.3 | 50.8 | 24x 16 | 16.8 | .38 | | E 4 : | | |
| 1351 | | | 09 26 02.8 | -61 12 02 | 280.89 | -7.62 | 126 | -25.7 | -64.4 | 13x 13 | 18.1 | .23 | | E ? | | |
| 1352 | | | 09 26 03.1 | -60 27 56 | 280.38 | -7.09 | 126 | -26.1 | -25.0 | 36x 16 | 17.0 | .29 | | SB 3 | S | vLSBdisk |
| 1353 | L126-013 | I | 09 26 12.6 | -60 33 07 | 280.46 | -7.14 | 126 | -24.7 | -29.1 | 128x 47 | 14.2 | .28 | | S 1 | | fantastic abs. lane |
| 1354 | | | 09 26 15.0 | -60 49 03 | 280.64 | -7.33 | 126 | -24.6 | -43.9 | 17x 7 | 18.5 | .24 | | S ? | | |
| 1355 | | | 09 26 16.4 | -62 35 42 | 281.89 | -8.61 | 91 | -37.0 | 130.0 | 30x 16 | 16.6 | .30 | ? | E ? | | |
| 1356 | | I | 09 26 25.7 | -58 53 16 | 279.32 | -5.92 | 126 | -24.4 | 59.6 | 16x 9 | 17.8 | .46 | | E | | |
| 1357 | | | 09 26 29.8 | -64 41 41 | 283.39 | -10.10 | 91 | -33.6 | 17.4 | 16x 5 | 19.4 | .28 | | S | | |
| 1358 | | Q | 09 26 30.6 | -57 07 47 | 278.11 | -4.64 | 166 | 40.4 | -114.5 | 11x 11 | 18.1 | .80 | ? | | | |
| 1359 | | | 09 26 31.0 | -60 28 06 | 280.42 | -7.05 | 126 | -23.0 | -25.1 | 16x 8 | 18.4 | .30 | | S 5 | 1 | |
| 1360 | | | 09 26 35.2 | -60 29 08 | 280.44 | -7.06 | 126 | -22.8 | -25.6 | 8x 5 | 20.0 | .30 | ? | | | dist.nucl. |
| 1361 | | | 09 26 37.5 | -60 07 26 | 280.19 | -6.79 | 126 | 22.7 | 6.1 | 13x 4 | 19.8 | .40 | | S ? | | |
| 1362 | | | 09 26 37.6 | -60 30 33 | 280.46 | -7.07 | 126 | -22.7 | -26.9 | 9x 9 | 18.8 | .29 | | E ? | | |
| 1363 | | | 09 26 37.8 | -60 37 53 | 280.55 | -7.16 | 126 | -22.2 | -33.9 | 17x 9 | 18.2 | .28 | | F | | |
| 1364 | | | 09 26 44.9 | -60 45 27 | 280.65 | -7.24 | 126 | -21.4 | -40.6 | 16x 7 | 18.3 | .25 | | E ? | | |
| 1365 | | | 09 26 52.1 | -61 51 50 | 281.43 | -8.03 | 126 | -20.1 | -99.9 | 20x 15 | 17.5 | .31 | ? | S | | sharp edge |
| 1366 | | | 09 26 52.8 | -58 53 46 | 279.37 | -5.88 | 126 | -21.3 | 59.2 | 16x 9 | 18.1 | .46 | | F | | |
| 1367 | | | 09 26 54.9 | -58 54 40 | 279.38 | -5.89 | 126 | -21.0 | 58.4 | 20x 16 | 17.4 | .45 | | S M | | |
| 1368 | | | 09 26 58.6 | -60 12 25 | 280.28 | -6.82 | 126 | -20.2 | -10.5 | 13x 5 | 18.9 | .36 | ? | | | |
| 1369 | | | 09 27 00.2 | -61 40 22 | 281.31 | -7.88 | 126 | -19.4 | -89.7 | 54x 34 | 16.2 | .31 | | S 5 | | |
| 1370 | | | 09 27 01.9 | -60 57 54 | 280.82 | -7.37 | 126 | -19.5 | -51.7 | 13x 7 | 18.6 | .23 | | | S | |
| 1371 | L126-014 | | 09 27 06.5 | -60 34 56 | 280.56 | -7.08 | 126 | -19.1 | -31.2 | 87x 47 | 14.3 | .29 | | E 5 | | |
| 1372 | | | 09 27 07.5 | -60 28 06 | 280.48 | -7.00 | 126 | -19.0 | -25.1 | 15x 4 | 18.8 | .31 | ? | S | | |
| 1373 | | | 09 27 17.1 | -61 03 33 | 280.90 | -7.41 | 126 | -17.8 | -56.8 | 20x 11 | 17.6 | .24 | | | S | |
| 1374 | | | 09 27 18.1 | -58 03 40 | 278.83 | -5.24 | 126 | -18.6 | 103.9 | 16x 5 | 19.4 | .55 | | | | |
| 1375 | | | 09 27 20.0 | -64 49 03 | 283.54 | -10.12 | 91 | -26.4 | 8.5 | 17x 9 | 19.0 | .26 | | S ? | | LSB |
| 1376 | L126-015 | | 09 27 24.2 | -60 38 08 | 280.62 | -7.10 | 126 | -17.1 | -34.1 | 81x 16 | 15.7 | .28 | | S E | | |
| 1377 | | | 09 27 27.5 | -60 56 24 | 280.84 | -7.31 | 126 | -16.7 | -50.4 | 16x 7 | 18.3 | .24 | | | | |
| 1378 | | | 09 27 36.6 | -61 03 33 | 280.93 | -7.39 | 126 | -15.7 | -56.8 | 23x 19 | 16.9 | .25 | ? | SB 5 | | dbl syst? |
| 1379 | | | 09 27 41.1 | -64 57 02 | 283.66 | -10.19 | 91 | -26.6 | 3.8 | 27x 9 | 17.7 | .25 | | S | | |
| 1380 | | | 09 27 51.0 | -65 04 43 | 283.77 | -10.27 | 91 | -23.5 | -5.5 | 15x 7 | 19.2 | .26 | | | | dist.nucl. |
| 1381 | | | 09 27 51.9 | -64 52 31 | 283.62 | -10.12 | 91 | -25.7 | 7.9 | 27x 16 | 17.1 | .25 | | | S | |
| 1382 | | | 09 27 54.8 | -59 46 56 | 280.08 | -6.44 | 126 | -14.0 | 11.7 | 17x 13 | 17.4 | .38 | | E 3 | | |
| 1383 | | | 09 27 58.5 | -60 27 15 | 280.55 | -6.92 | 126 | -13.4 | -24.3 | 17x 11 | 17.9 | .31 | | | | |
| 1384 | L126-016 | | 09 28 00.5 | -60 28 25 | 280.56 | -6.93 | 126 | -13.2 | -25.4 | 81x 38 | 14.9 | .31 | | S | | w.comp. |
| 1385 | | | 09 28 04.9 | -57 55 05 | 278.81 | -5.06 | 126 | -13.1 | 111.6 | 16x 5 | 18.9 | .61 | | S | | |
| 1386 | | | 09 28 06.4 | -64 55 20 | 283.68 | -10.14 | 91 | -24.2 | 5.3 | 13x 8 | 18.7 | .25 | | | | neighb.of 1387 |
| 1387 | | | 09 28 09.9 | -64 56 14 | 283.69 | -10.15 | 91 | -23.9 | 4.6 | 27x 16 | 16.9 | .25 | | F | | w.2 comp. |
| 1388 | | | 09 28 16.9 | -64 28 20 | 283.37 | -9.80 | 91 | -23.5 | 29.5 | 20x 5 | 19.1 | .28 | | S M | | br.blg, LSBdisk |
| 1389 | | | 09 28 19.2 | -62 41 47 | 282.13 | -8.52 | 126 | -10.9 | -144.5 | 13x 13 | 18.3 | .28 | | | | |
| 1390 | | | 09 28 19.7 | -64 48 00 | 283.61 | -10.04 | 91 | -23.1 | 11.9 | 27x 17 | 16.6 | .25 | | S E | | |
| 1391 | | | 09 28 21.4 | -58 30 37 | 279.24 | -5.47 | 126 | -11.1 | 79.9 | 20x 5 | 19.1 | .56 | | | | |
| 1392 | | | 09 28 28.7 | -64 44 11 | 283.57 | -9.98 | 91 | -19.8 | 12.8 | 9x 4 | 20.1 | .25 | | | 1 | |
| 1393 | | | 09 28 33.0 | -61 53 14 | 281.59 | -7.91 | 126 | -9.5 | -101.1 | 20x 16 | 17.1 | .29 | | E ? | | w.comp. |
| 1394 | | | 09 28 33.7 | -62 50 44 | 282.26 | -8.60 | 91 | -22.7 | 116.7 | 13x 11 | 18.0 | .25 | ? | | | or * |
| 1395 | | | 09 28 40.4 | -60 52 44 | 280.90 | -7.17 | 126 | -8.8 | -47.1 | 16x 13 | 17.4 | .26 | | E 2 | | |
| 1396 | | | 09 28 41.8 | -59 35 30 | 280.02 | -6.23 | 126 | -8.7 | 21.9 | 16x 11 | 18.1 | .43 | ? | | S | |
| 1397 | L126-017 | I | 09 28 42.0 | -61 57 44 | 281.66 | -7.95 | 126 | -8.6 | -105.1 | 108x 60 | 13.8 | .31 | | S 0 | | |
| 1398 | | | 09 28 42.9 | -64 42 42 | 283.58 | -9.94 | 91 | -20.9 | 16.7 | 16x 4 | 19.7 | .25 | | | | vLSB |
| 1399 | | NG | 09 28 45.8 | -61 59 23 | 281.68 | -7.97 | 126 | -8.2 | -106.6 | 20x 7 | 18.1 | .31 | | S | | |
| 1400 | | | 09 28 47.1 | -58 53 08 | 279.54 | -5.71 | 126 | -8.1 | 59.8 | 24x 20 | 17.1 | .47 | | | S | |

**Table 1.** Galaxies in the southern ZOA – I. The Hydra/Antlia Extension (continued)

| RKK Ident (1) | Other Ident (2) | IR (3) | R.A. (h m s) (4) | Dec. (° ′ ″) (5) | gal ℓ (°) (6) | gal b (°) (7) | SRC (8) | X (mm) (9) | Y (mm) (10) | D x d (″) (11) | $B_J$ ($^m$) (12) | $N_{HI}$ (13) | u | Type class. (14) | o | * | Remarks (15) |
|---|---|---|---|---|---|---|---|---|---|---|---|---|---|---|---|---|---|
| 1401 | | | 09 28 55.8 | -65 09 07 | 283.90 | -10.25 | 91 | -17.5 | -9.5 | 13x 8 | 18.8 | .29 | | E | | | |
| 1402 | L091-014 | | 09 28 59.1 | -62 51 59 | 282.31 | -8.59 | 91 | -20.1 | 115.6 | 74x 27 | 15.6 | .26 | | S 7 | | | |
| 1403 | | | 09 29 00.1 | -62 24 13 | 281.99 | -8.25 | 126 | -6.7 | -128.8 | 12x 7 | 18.7 | .28 | | S | | | trpl syst.w.1404,1407 |
| 1404 | | | 09 29 02.0 | -62 24 15 | 281.99 | -8.25 | 126 | -6.5 | -128.8 | 11x 11 | 18.0 | .28 | | E | | | trpl syst.w.1403,1407 |
| 1405 | | | 09 29 04.4 | -64 54 22 | 283.74 | -10.06 | 91 | -16.5 | 3.7 | 11x 8 | 18.7 | .25 | | E | | | |
| 1406 | | | 09 29 04.7 | -65 11 44 | 283.94 | -10.27 | 91 | -16.7 | -11.8 | 19x 5 | 18.9 | .30 | | S | | | |
| 1407 | | | 09 29 06.0 | -62 24 34 | 282.00 | -8.25 | 126 | -6.0 | -129.1 | 16x 4 | 18.7 | .28 | | S | | | trpl syst.w.1403,1404 |
| 1408 | | I | 09 29 18.8 | -64 48 49 | 283.69 | -9.97 | 91 | -17.4 | 11.2 | 38x 11 | 16.9 | .25 | | S E | | | |
| 1409 | | I | 09 29 20.6 | -57 50 15 | 278.87 | -4.89 | 126 | -4.1 | 116.0 | 34x 5 | 18.0 | .67 | | S M | E | | dist.dust lane |
| 1410 | | I3 | 09 29 25.3 | -63 04 32 | 282.49 | -8.70 | 91 | -17.4 | 104.4 | 8x 4 | 20.0 | .26 | | S | | | trpl syst.w.1412,1413 |
| 1411 | | I | 09 29 25.4 | -57 23 36 | 278.58 | -4.56 | 126 | -3.5 | 139.8 | 34x 27 | 16.4 | .91 | ? | S | F | | LSB, PN? |
| 1412 | | I3 | 09 29 25.4 | -63 04 20 | 282.49 | -8.70 | 91 | -17.4 | 104.6 | 24x 7 | 18.0 | .26 | | S | | | trpl syst.w.1410,1413 |
| 1413 | | I3 | 09 29 27.3 | -63 04 08 | 282.49 | -8.70 | 91 | -17.2 | 104.8 | 11x 5 | 19.1 | .26 | | S | | | trpl syst.w.1410,1412 |
| 1414 | | | 09 29 31.1 | -61 12 46 | 281.21 | -7.34 | 126 | -2.1 | -63.9 | 34x 13 | 17.0 | .25 | | S L | | S | |
| 1415 | | | 09 29 31.5 | -64 53 15 | 283.76 | -10.01 | 91 | -14.0 | 4.7 | 11x 9 | 19.4 | .26 | | | | | vLSB |
| 1416 | | | 09 29 34.0 | -60 44 25 | 280.89 | -6.99 | 126 | -2.9 | -39.6 | 13x 4 | 19.2 | .29 | | S | | | |
| 1417 | | | 09 29 34.8 | -65 13 39 | 284.00 | -10.25 | 91 | -13.9 | -13.6 | 9x 4 | 19.8 | .30 | | | | | blg |
| 1418 | | | 09 29 36.3 | -65 06 28 | 283.92 | -10.17 | 91 | -13.6 | -7.1 | 19x 5 | 19.1 | .28 | | S | | 1 | |
| 1419 | L126-018 | | 09 29 45.3 | -59 34 37 | 280.10 | -6.13 | 126 | -1.5 | 22.7 | 40x 16 | 16.8 | .44 | | S L | | | |
| 1420 | | | 09 29 48.0 | -64 17 05 | 283.36 | -9.55 | 91 | -14.8 | 39.6 | 24x 20 | 17.1 | .28 | | S L | | S | |
| 1421 | | | 09 29 51.6 | -64 53 45 | 283.79 | -9.99 | 91 | -12.0 | 4.2 | 9x 8 | 19.4 | .26 | | | | | p.cov.by * |
| 1422 | | | 09 29 52.4 | -63 05 46 | 282.54 | -8.68 | 91 | -14.6 | 103.4 | 34x 7 | 17.7 | .25 | | S | | | dist.blg, cl.to br.* |
| 1423 | | | 09 29 56.1 | -65 01 57 | 283.90 | -10.09 | 91 | -11.7 | -3.1 | 12x 5 | 19.3 | .27 | | S M: | | | |
| 1424 | | | 09 29 56.9 | -59 34 02 | 280.12 | -6.10 | 126 | -.2 | 23.2 | 11x 8 | 18.8 | .43 | | | | | neighb.of 1419 |
| 1425 | | | 09 30 10.2 | -59 46 04 | 280.27 | -6.23 | 126 | 1.3 | 12.5 | 16x 13 | 17.4 | .38 | | S 0 | | | |
| 1426 | | I | 09 30 13.3 | -47 24 18 | 271.88 | 2.86 | 212 | 2.7 | 139.2 | 16x 13 | 17.8 | .91 | | S | | | dist.nucl. |
| 1427 | | | 09 30 16.5 | -65 03 15 | 283.94 | -10.08 | 91 | -9.8 | -4.3 | 8x 8 | 19.1 | .27 | | E ? | | | w.comp, cl.to br.* |
| 1428 | | | 09 30 17.4 | -64 45 26 | 283.73 | -9.86 | 91 | -11.9 | 14.3 | 13x 11 | 18.4 | .25 | ? | | | | |
| 1429 | | | 09 30 22.7 | -61 49 43 | 281.71 | -7.72 | 126 | 2.1 | -98.0 | 20x 4 | 19.1 | .30 | | S | E | | |
| 1430 | | | 09 30 22.9 | -61 22 19 | 281.39 | -7.39 | 126 | 2.2 | -73.5 | 13x 8 | 18.6 | .30 | ? | S 3 | | | dist.dust lane |
| 1431 | | | 09 30 24.8 | -64 34 16 | 283.61 | -9.72 | 91 | -11.2 | 24.3 | 32x 8 | 17.6 | .26 | | S E: | | | |
| 1432 | | | 09 30 25.2 | -64 55 20 | 283.86 | -9.97 | 91 | -11.1 | 5.4 | 30x 8 | 17.7 | .26 | | | | 1 | |
| 1433 | | | 09 30 26.7 | -60 09 01 | 280.56 | -6.49 | 126 | 3.0 | -8.1 | 15x 5 | 18.6 | .35 | | S E | | | |
| 1434 | | | 09 30 28.0 | -58 35 38 | 279.50 | -5.34 | 126 | 3.7 | 75.4 | 16x 5 | 19.1 | .67 | ? | | | | |
| 1435 | | | 09 30 28.0 | -63 37 53 | 282.96 | -9.03 | 91 | -11.0 | 74.7 | 13x 11 | 18.0 | .24 | ? | | | | or * w.halo |
| 1436 | | | 09 30 29.3 | -62 59 53 | 282.52 | -8.57 | 91 | -10.9 | 108.6 | 27x 20 | 16.8 | .27 | | S M | | S | dist.blg |
| 1437 | | | 09 30 29.5 | -64 14 40 | 283.39 | -9.47 | 91 | -10.8 | 41.8 | 15x 4 | 19.4 | .27 | | S | E | | |
| 1438 | | | 09 30 30.5 | -61 01 17 | 281.16 | -7.12 | 126 | 3.1 | -54.8 | 16x 3 | 19.5 | .30 | | | | S | smooth |
| 1439 | | | 09 30 30.5 | -61 25 12 | 281.44 | -7.41 | 126 | 3.0 | -76.1 | 27x 13 | 17.1 | .31 | | S 1 | E | | dist.dust lane |
| 1440 | | | 09 30 40.7 | -60 32 54 | 280.86 | -6.76 | 126 | 4.4 | -29.4 | 13x 5 | 18.9 | .35 | | S E | | | |
| 1441 | | | 09 30 41.3 | -64 06 00 | 283.31 | -9.35 | 91 | -9.6 | 49.5 | 15x 5 | 18.8 | .24 | | | | S | |
| 1442 | | | 09 31 04.7 | -64 49 34 | 283.84 | -9.85 | 91 | -5.1 | 7.8 | 8x 3 | 20.3 | .24 | | S M | E | | |
| 1443 | | | 09 31 09.7 | -65 00 01 | 283.97 | -9.97 | 91 | -6.9 | 1.3 | 23x 23 | 17.4 | .26 | | S | F | | dist.nucl, vLSBdisk? |
| 1444 | | | 09 31 10.6 | -65 02 17 | 284.00 | -10.00 | 91 | -6.8 | -.8 | 22x 3 | 19.3 | .26 | | S | E | | |
| 1445 | FGCE0751 | | 09 31 11.1 | -59 40 14 | 280.30 | -6.07 | 126 | 8.2 | 17.6 | 87x 4 | 17.1 | .39 | | S | E | | |
| 1446 | | | 09 31 17.7 | -64 57 10 | 283.95 | -9.93 | 91 | -6.1 | 3.8 | 27x 23 | 17.3 | .25 | ? | S | F | 2 | |
| 1447 | | | 09 31 26.4 | -64 46 48 | 283.84 | -9.79 | 91 | -5.3 | 13.1 | 13x 7 | 19.1 | .24 | | | | | |
| 1448 | | | 09 31 26.6 | -64 44 49 | 283.81 | -9.77 | 91 | -2.9 | 12.0 | 7x 4 | 20.4 | .24 | | | | | |
| 1449 | | | 09 31 30.6 | -58 44 05 | 279.69 | -5.36 | 126 | 10.9 | 67.8 | 11x 11 | 18.0 | .58 | | E 0 : | | | |
| 1450 | | | 09 31 39.9 | -65 00 59 | 284.02 | -9.95 | 91 | -4.0 | .4 | 16x 4 | 19.5 | .25 | | S | | | |
| 1451 | | | 09 31 42.8 | -64 52 25 | 283.92 | -9.84 | 91 | -3.8 | 8.1 | 17x 5 | 19.1 | .24 | | S | | | |
| 1452 | | | 09 31 54.2 | -64 10 50 | 283.46 | -9.32 | 91 | -2.5 | 45.2 | 16x 8 | 18.7 | .25 | | | | 1 | |
| 1453 | | | 09 32 02.4 | -64 17 51 | 283.55 | -9.40 | 91 | -1.8 | 38.9 | 16x 13 | 18.6 | .26 | | | | | vLSB |
| 1454 | | | 09 32 02.5 | -62 53 54 | 282.59 | -8.37 | 91 | -1.4 | 114.0 | 20x 5 | 18.8 | .28 | | | | | box-shaped |
| 1455 | | | 09 32 21.1 | -47 50 26 | 272.44 | 2.78 | 212 | 21.8 | 115.7 | 13x 4 | 19.4 | .86 | | S 5 | E | | |
| 1456 | | | 09 32 25.5 | -61 12 41 | 281.47 | -7.10 | 126 | 16.7 | -63.5 | 19x 9 | 17.7 | .31 | | E 3 | | | w.LSB comp. |
| 1457 | | | 09 32 26.4 | -62 27 33 | 282.32 | -8.02 | 91 | 1.1 | 137.5 | 47x 40 | 15.9 | .30 | | S | | | dist.nucl, vLSBdisk? |
| 1458 | | | 09 32 27.7 | -60 57 10 | 281.29 | -6.91 | 126 | 16.7 | -49.7 | 13x 4 | 19.6 | .36 | | S | E | | |
| 1459 | | | 09 32 28.2 | -64 11 49 | 283.52 | -9.29 | 91 | .8 | 44.3 | 13x 7 | 18.9 | .26 | | | | | |
| 1460 | | | 09 32 28.4 | -64 32 39 | 283.76 | -9.55 | 91 | .7 | 25.7 | 22x 11 | 17.9 | .26 | ? | S ? | | | dist.nucl, or * w. gas |
| 1461 | | | 09 32 31.8 | -61 38 13 | 281.77 | -7.41 | 126 | 15.8 | -87.9 | 13x 11 | 18.1 | .36 | | S | | | |
| 1462 | | | 09 32 45.0 | -63 22 24 | 282.97 | -8.67 | 91 | 2.7 | 88.5 | 16x 16 | 17.7 | .22 | | S | F | S | |
| 1463 | | | 09 32 47.2 | -60 36 42 | 281.09 | -6.63 | 126 | 18.3 | -33.0 | 13x 11 | 17.9 | .33 | | E ? | | | |
| 1464 | L126-019 | I | 09 32 52.2 | -61 03 34 | 281.40 | -6.95 | 126 | 18.5 | -56.9 | 108x 81 | 14.2 | .36 | | S 6 | F | | LSB |
| 1465 | | | 09 33 09.0 | -61 02 38 | 281.42 | -6.92 | 126 | 20.3 | -56.2 | 17x 5 | 18.6 | .36 | | S L | E | | neighb.of 1464 |
| 1466 | | | 09 33 13.0 | -64 29 13 | 283.78 | -9.45 | 91 | 5.0 | 28.7 | 13x 4 | 19.8 | .28 | ? | | | | |
| 1467 | | | 09 33 19.3 | -64 42 30 | 283.94 | -9.60 | 91 | 5.5 | 16.9 | 22x 7 | 18.4 | .27 | | S | | | |
| 1468 | | | 09 33 23.1 | -60 23 46 | 281.00 | -6.42 | 126 | 22.5 | -19.8 | 20x 11 | 17.6 | .35 | | | | | blg, LSBdisk |
| 1469 | | | 09 33 43.2 | -64 30 12 | 283.83 | -9.42 | 91 | 7.9 | 27.8 | 27x 11 | 17.3 | .29 | | S | | | obs.disk |
| 1470 | | | 09 33 47.9 | -62 06 51 | 282.20 | -7.66 | 126 | 23.4 | -113.6 | 30x 11 | 17.4 | .34 | | S E | | | |
| 1471 | | | 09 33 52.1 | -64 42 41 | 283.98 | -9.57 | 91 | 8.6 | 16.7 | 17x 9 | 18.1 | .28 | | E ? | | | |
| 1472 | | | 09 33 53.4 | -63 12 59 | 282.96 | -8.46 | 91 | 9.7 | 96.8 | 13x 4 | 19.6 | .22 | ? | S | | 2 | |
| 1473 | | | 09 34 02.7 | -50 09 03 | 274.20 | 1.25 | 212 | 34.9 | -8.4 | 13x 4 | 19.4 | .38 | | S ? | | 1 | vLSB |
| 1474 | | | 09 34 23.2 | -63 02 25 | 282.88 | -8.30 | 91 | 12.8 | 106.2 | 16x 9 | 18.5 | .27 | | | | | p.cov.by * |
| 1475 | | | 09 34 24.9 | -62 40 24 | 282.64 | -8.02 | 126 | 26.7 | -143.6 | 16x 11 | 18.1 | .27 | | SB 5 | | 1 | or gas |
| 1476 | | | 09 34 25.2 | -64 50 11 | 284.11 | -9.62 | 91 | 11.7 | 9.9 | 20x 4 | 19.5 | .26 | | S M | E | | |
| 1477 | | | 09 34 25.6 | -61 50 11 | 282.07 | -7.40 | 126 | 27.7 | -98.8 | 19x 8 | 18.4 | .39 | | S | | | neighb.of 1479 |
| 1478 | | | 09 34 30.2 | -62 21 56 | 282.43 | -7.79 | 126 | 27.6 | -127.1 | 13x 13 | 17.7 | .30 | ? | E 0 ? | | | or * w.halo |
| 1479 | L126-020 | | 09 34 30.4 | -61 46 52 | 282.04 | -7.35 | 126 | 28.2 | -95.8 | 54x 47 | 15.3 | .38 | | S 5 | | | |
| 1480 | | | 09 34 44.9 | -59 56 51 | 280.82 | -5.97 | 126 | 32.0 | 2.4 | 13x 5 | 19.4 | .42 | | | | | |
| 1481 | | | 09 34 46.4 | -64 48 13 | 284.12 | -9.57 | 91 | 13.7 | 11.6 | 13x 5 | 19.0 | .28 | | S | | | |
| 1482 | | | 09 34 52.0 | -62 10 30 | 282.34 | -7.62 | 126 | 30.0 | -117.0 | 60x 54 | 15.2 | .31 | | S 3 | | | |
| 1483 | | | 09 34 57.5 | -56 16 17 | 278.38 | -3.22 | 167 | -141.4 | -73.0 | 13x 7 | 18.3 | .27 | ? | S | | | |
| 1484 | L126-021 | | 09 35 10.9 | -61 52 25 | 282.16 | -7.37 | 126 | 32.4 | -100.9 | 51x 47 | 14.8 | .36 | | L | | | |
| 1485 | | | 09 35 16.5 | -58 45 37 | 280.08 | -5.05 | 126 | 37.0 | 66.0 | 16x 7 | 19.0 | .69 | | | | | |
| 1486 | L091-015 | | 09 35 18.6 | -63 43 13 | 283.42 | -8.73 | 91 | 17.8 | 69.7 | 81x 40 | 15.1 | .25 | | S 7 | | S | LSB |
| 1487 | | | 09 35 20.1 | -61 33 08 | 281.96 | -7.12 | 126 | 33.8 | -83.7 | 20x 5 | 18.7 | .40 | | S | | | |
| 1488 | | | 09 35 26.4 | -60 33 17 | 281.29 | -6.37 | 126 | 35.8 | -30.2 | 19x 13 | 17.7 | .36 | | S | | | dist.nucl. |
| 1489 | | | 09 35 29.4 | -63 53 45 | 283.56 | -8.85 | 91 | 18.7 | 60.2 | 20x 7 | 18.4 | .27 | | | | | |
| 1490 | | | 09 35 34.3 | -64 04 02 | 283.68 | -8.97 | 91 | 19.0 | 51.1 | 16x 7 | 18.5 | .28 | | S | | | |
| 1491 | | | 09 35 37.1 | -62 24 10 | 282.56 | -7.73 | 126 | 34.4 | -129.3 | 13x 5 | 19.2 | .31 | | | | | |
| 1492 | L126-022 | | 09 35 37.4 | -60 42 23 | 281.41 | -6.47 | 126 | 36.8 | -38.4 | 81x 34 | 14.9 | .38 | | SB 4 | | | dist.nucl, smooth |
| 1493 | | | 09 35 37.9 | -50 11 52 | 274.42 | 1.39 | 212 | 48.5 | -11.2 | 13x 5 | 19.7 | .33 | | S ? | | | dist.blg, vLSBdisk, w.comp |
| 1494 | | | 09 35 42.4 | -64 42 55 | 284.13 | -9.44 | 91 | 19.1 | 16.3 | 15x 5 | 19.6 | .31 | ? | | | | LSB |
| 1495 | | | 09 35 51.1 | -61 22 58 | 281.89 | -6.95 | 126 | 37.3 | -74.7 | 27x 9 | 17.9 | .40 | | S M | | | |
| 1496 | | | 09 35 52.6 | -65 26 30 | 284.64 | -9.96 | 91 | 19.4 | -22.7 | 30x 11 | 17.6 | .25 | ? | | | | gas? |
| 1497 | | | 09 35 54.7 | -64 49 00 | 284.22 | -9.50 | 91 | 20.2 | 10.8 | 13x 8 | 19.2 | .30 | ? | S ? | | | |
| 1498 | | | 09 35 57.8 | -60 27 10 | 281.27 | -6.25 | 126 | 39.4 | -24.9 | 23x 9 | 18.3 | .36 | | | | S | |
| 1499 | | | 09 36 05.2 | -60 05 16 | 281.04 | -5.97 | 126 | 40.7 | -5.3 | 54x 5 | 17.6 | .34 | | S L | E | | |
| 1500 | | | 09 36 06.1 | -65 33 16 | 284.74 | -10.03 | 91 | 20.5 | -28.7 | 13x 5 | 19.2 | .22 | | | | | |

**Table 1.** Galaxies in the southern ZOA – I. The Hydra/Antlia Extension (continued)

| RKK Ident (1) | Other Ident (2) | IR (3) | R.A. (h m s) (4) | Dec. (° ′ ″) (5) | gal $\ell$ (°) (6) | gal $b$ (°) (7) | SRC (8) | X (mm) (9) | Y (mm) (10) | D x d (″) (11) | $B_J$ ($^m$) (12) | $N_{HI}$ (13) | u | Type class. (14) | | o | * | Remarks (15) |
|---|---|---|---|---|---|---|---|---|---|---|---|---|---|---|---|---|---|---|
| 1501 | | | 09 36 14.3 | -59 25 32 | 280.61 | -5.46 | 126 | 42.7 | 30.1 | 36x 5 | 18.5 | .51 | | S | | E | | br.* s.p. |
| 1502 | | | 09 36 17.7 | -60 40 08 | 281.45 | -6.38 | 126 | 41.3 | -36.5 | 13x 8 | 18.6 | .38 | | S | | | | |
| 1503 | | | 09 36 20.4 | -65 02 38 | 284.41 | -9.64 | 91 | 22.4 | -1.4 | 20x 7 | 18.6 | .23 | | S | ? | | | |
| 1504 | | | 09 36 20.5 | -63 11 56 | 283.16 | -8.27 | 91 | 24.5 | 97.5 | 22x 11 | 18.3 | .23 | ? | | | | | |
| 1505 | | | 09 36 27.3 | -65 42 34 | 284.87 | -10.12 | 91 | 22.3 | -37.1 | 17x 5 | 18.9 | .22 | | S | | | | |
| 1506 | L126-023 | I | 09 36 31.6 | -61 55 31 | 282.31 | -7.30 | 126 | 40.8 | -103.9 | 114x 47 | 14.1 | .38 | | S | 3 | | | |
| 1507 | | | 09 36 35.1 | -60 18 42 | 281.24 | -6.09 | 126 | 43.7 | -17.4 | 22x 11 | 17.3 | .37 | | S | E | | | |
| 1508 | | | 09 36 35.4 | -48 08 01 | 273.17 | 3.04 | 212 | 59.5 | 99.3 | 13x 8 | 18.5 | .71 | | S | | | 1 | |
| 1509 | | | 09 36 37.4 | -65 33 54 | 284.78 | -10.00 | 91 | 23.4 | -29.4 | 19x 8 | 18.8 | .22 | | | | | | smooth |
| 1510 | | | 09 36 40.4 | -60 18 23 | 281.24 | -6.08 | 126 | 44.3 | -17.1 | 11x 5 | 18.9 | .36 | | | | | | w.comp. |
| 1511 | | | 09 36 45.4 | -61 19 06 | 281.93 | -6.83 | 126 | 43.2 | -71.4 | 34x 5 | 18.2 | .41 | | S | | E | | dist.nucl. |
| 1512 | | | 09 36 49.7 | -64 18 34 | 283.95 | -9.06 | 91 | 26.1 | 37.9 | 17x 4 | 19.8 | .30 | | S | | | | |
| 1513 | | | 09 36 56.3 | -63 07 59 | 283.16 | -8.17 | 91 | 28.2 | 101.0 | 35x 8 | 17.5 | .23 | | S | M | | | |
| 1514 | L091-016 | I | 09 36 57.6 | -63 15 38 | 283.25 | -8.26 | 91 | 28.1 | 94.1 | 108x 27 | 14.9 | .23 | | S | 0 | | S | |
| 1515 | | I2 | 09 36 59.1 | -64 54 33 | 284.37 | -9.49 | 91 | 26.2 | 5.7 | 23x 5 | 18.7 | .29 | | S | | E | | |
| 1516 | | I2 | 09 36 59.9 | -64 55 53 | 284.38 | -9.51 | 91 | 26.3 | 4.6 | 23x 16 | 17.4 | .28 | | L | | | | |
| 1517 | L126-024 | I | 09 37 08.0 | -61 36 09 | 282.15 | -7.01 | 126 | 45.2 | -86.7 | 108x 74 | 13.7 | .34 | | S | 4 | | | |
| 1518 | | | 09 37 10.4 | -63 48 44 | 283.64 | -8.66 | 91 | 28.7 | 64.5 | 13x 5 | 19.5 | .26 | | S | ? | | | |
| 1519 | L126-025 | I | 09 37 12.5 | -61 30 38 | 282.10 | -6.94 | 126 | 45.8 | -81.8 | 54x 40 | 14.9 | .34 | | L | | | | |
| 1520 | | NG | 09 37 13.4 | -63 15 30 | 283.27 | -8.24 | 91 | 29.7 | 94.2 | 13x 11 | 18.0 | .22 | | | | | | |
| 1521 | | NG | 09 37 14.8 | -61 35 13 | 282.15 | -6.99 | 126 | 45.9 | -85.9 | 20x 4 | 19.1 | .34 | | S | ? | | | |
| 1522 | L091-017 | | 09 37 17.1 | -63 42 27 | 283.58 | -8.57 | 91 | 29.5 | 70.1 | 87x 60 | 14.6 | .26 | | S | 2 | | S | |
| 1523 | | NG | 09 37 17.5 | -61 37 25 | 282.18 | -7.02 | 126 | 46.2 | -87.9 | 16x 8 | 18.0 | .34 | | E | 4 | | | neighb.of 1517 |
| 1524 | | I | 09 37 19.5 | -63 44 46 | 283.61 | -8.60 | 91 | 29.7 | 68.0 | 54x 40 | 15.4 | .26 | | SB | 1 | | S | |
| 1525 | | | 09 37 20.1 | -61 56 47 | 282.40 | -7.26 | 126 | 45.9 | -105.2 | 17x 11 | 17.6 | .40 | | E | | | | |
| 1526 | | | 09 37 28.2 | -63 49 23 | 283.67 | -8.65 | 91 | 30.5 | 63.9 | 22x 13 | 17.1 | .27 | | E | | | | |
| 1527 | | | 09 37 35.8 | -61 01 24 | 281.81 | -6.54 | 126 | 49.2 | -55.8 | 27x 5 | 18.0 | .38 | | S | | | | in st.diffr.pat. |
| 1528 | | | 09 37 42.3 | -65 01 58 | 284.51 | -9.53 | 91 | 30.1 | -1.0 | 17x 5 | 19.1 | .26 | | | | | | p.cov.by *, w.comp. |
| 1529 | | | 09 37 43.5 | -65 47 39 | 285.02 | -10.10 | 91 | 29.2 | -41.8 | 19x 5 | 19.4 | .23 | | S | | E | | |
| 1530 | | | 09 37 48.2 | -60 07 49 | 281.23 | -5.86 | 126 | 52.1 | -8.0 | 11x 4 | 19.4 | .38 | | S | E | | | |
| 1531 | | | 09 37 51.8 | -65 37 51 | 284.92 | -9.97 | 91 | 30.2 | -33.1 | 19x 11 | 18.1 | .22 | | L | ? | | | |
| 1532 | | | 09 37 54.5 | -64 01 32 | 283.84 | -8.77 | 91 | 32.8 | 53.0 | 17x 5 | 19.2 | .28 | ? | | | | | p.cov.by * |
| 1533 | | | 09 38 02.0 | -62 50 08 | 283.06 | -7.87 | 91 | 35.3 | 116.7 | 19x 5 | 18.8 | .23 | | S | | | | |
| 1534 | | I | 09 38 08.9 | -61 58 06 | 282.49 | -7.21 | 126 | 51.0 | -106.5 | 34x 4 | 18.6 | .46 | | S | | | | |
| 1535 | | | 09 38 13.1 | -61 42 10 | 282.32 | -7.00 | 126 | 51.9 | -92.3 | 16x 8 | 18.3 | .35 | | S | E | | | |
| 1536 | | | 09 38 14.1 | -60 04 35 | 281.24 | -5.78 | 126 | 55.1 | -5.2 | 15x 4 | 19.5 | .43 | | S | | E | | dist.nucl. or s.p.* |
| 1537 | | | 09 38 21.2 | -54 38 32 | 277.66 | -1.68 | 167 | -121.2 | 15.8 | 16x 5 | 18.9 | .00 | | S | 3 | | | |
| 1538 | | | 09 38 25.8 | -49 07 30 | 274.05 | 2.49 | 212 | 74.3 | 45.7 | 16x 16 | 17.8 | .88 | | S | E | F | | |
| 1539 | | | 09 38 27.3 | -65 25 48 | 284.83 | -9.77 | 91 | 33.7 | -22.4 | 16x 4 | 19.4 | .25 | | S | | | | w.comp. |
| 1540 | PMNJ0939 | | 09 38 42.1 | -63 56 05 | 283.85 | -8.64 | 91 | 37.6 | 57.7 | 16x 7 | 18.4 | .28 | | | | | | |
| 1541 | | | 09 38 49.2 | -63 59 52 | 283.90 | -8.68 | 91 | 38.2 | 54.3 | 54x 8 | 17.2 | .28 | | S | M | E | | |
| 1542 | | | 09 38 53.7 | -49 49 23 | 274.57 | 2.01 | 212 | 77.1 | 8.1 | 17x 4 | 19.1 | .78 | | S | | E | 1 | vLSB |
| 1543 | | | 09 38 55.9 | -49 17 56 | 274.23 | 2.41 | 212 | 78.4 | 36.2 | 31x 7 | 17.8 | .67 | | S | ? | | | |
| 1544 | | | 09 38 59.0 | -60 16 52 | 281.44 | -5.87 | 126 | 59.7 | -16.3 | 13x 5 | 18.9 | .43 | | S | ? | | | |
| 1545 | | | 09 39 05.0 | -63 18 47 | 283.46 | -8.15 | 91 | 40.9 | 91.0 | 27x 11 | 17.3 | .24 | | S | | | S | |
| 1546 | | | 09 39 07.5 | -49 27 09 | 274.35 | 2.32 | 212 | 79.8 | 27.9 | 20x 7 | 18.1 | .65 | | S | | | S | |
| 1547 | | | 09 39 25.8 | -48 46 09 | 273.94 | 2.87 | 212 | 83.7 | 64.4 | 20x 20 | 17.0 | .79 | | S | E | | | br.blg, LSBdisk |
| 1548 | | | 09 39 29.4 | -64 07 29 | 284.04 | -8.73 | 91 | 41.9 | 47.4 | 28x 4 | 19.0 | .26 | | S | | E | | |
| 1549 | | | 09 39 32.6 | -60 16 33 | 281.49 | -5.82 | 126 | 63.4 | -16.2 | 13x 4 | 19.8 | .45 | | I | ? | | | larger vLSBdisk? |
| 1550 | | | 09 39 33.3 | -63 12 46 | 283.44 | -8.04 | 91 | 43.9 | 96.2 | 13x 5 | 19.4 | .24 | ? | | | | | |
| 1551 | | | 09 39 36.1 | -65 04 33 | 284.69 | -9.43 | 91 | 40.8 | -3.6 | 40x 19 | 16.6 | .26 | | S | 1 | | | |
| 1552 | | | 09 39 40.7 | -62 12 43 | 282.78 | -7.27 | 126 | 60.1 | -119.9 | 13x 13 | 18.3 | .35 | | | | | | |
| 1553 | | | 09 39 45.4 | -63 58 24 | 283.96 | -8.59 | 91 | 43.7 | 55.4 | 13x 4 | 19.6 | .28 | ? | | | | | |
| 1554 | | | 09 39 52.0 | -49 19 24 | 274.36 | 2.50 | 212 | 86.5 | 34.6 | 34x 27 | 16.5 | .75 | | SB | 5 | | | |
| 1555 | | | 09 39 58.6 | -64 54 00 | 284.60 | -9.27 | 91 | 43.2 | 5.7 | 22x 8 | 18.3 | .29 | | S | | | | |
| 1556 | | | 09 40 01.6 | -48 01 38 | 273.53 | 3.49 | 212 | 90.5 | 104.0 | 27x 7 | 18.0 | .82 | | S | | | 1 | |
| 1557 | | | 09 40 01.8 | -61 11 55 | 282.14 | -6.48 | 126 | 64.6 | -65.8 | 34x 16 | 16.7 | .40 | | SB | 5 | | | |
| 1558 | | | 09 40 14.0 | -65 28 53 | 285.01 | -9.69 | 91 | 43.6 | -25.5 | 13x 7 | 18.7 | .22 | | | | | | br.blg, vLSBdisk |
| 1559 | | | 09 40 14.8 | -61 18 36 | 282.24 | -6.55 | 126 | 65.7 | -71.8 | 20x 4 | 19.1 | .37 | | S | | | | |
| 1560 | | | 09 40 15.2 | -61 12 44 | 282.17 | -6.47 | 126 | 66.0 | -66.5 | 27x 5 | 18.2 | .40 | | S | | | | |
| 1561 | | I | 09 40 18.1 | -60 34 10 | 281.75 | -5.98 | 126 | 67.8 | -32.1 | 34x 20 | 16.2 | .46 | | E | 4 | | | |
| 1562 | | | 09 40 20.2 | -61 17 33 | 282.23 | -6.53 | 126 | 66.3 | -70.9 | 13x 8 | 18.4 | .37 | | | | | | dbl E? |
| 1563 | | | 09 40 20.7 | -65 18 41 | 284.90 | -9.56 | 91 | 44.5 | -16.4 | 32x 7 | 18.1 | .26 | | S | | E | | |
| 1564 | | | 09 40 33.0 | -61 20 39 | 282.29 | -6.55 | 126 | 67.6 | -73.7 | 20x 7 | 18.1 | .37 | | S | ? | | | |
| 1565 | | | 09 40 34.6 | -65 08 34 | 284.81 | -9.41 | 91 | 46.1 | -7.4 | 16x 7 | 18.4 | .28 | | S | E | | | |
| 1566 | | | 09 40 37.7 | -64 28 01 | 284.36 | -8.90 | 91 | 47.8 | 28.8 | 17x 16 | 17.8 | .27 | ? | | | | 2 | |
| 1567 | | | 09 40 39.2 | -50 18 51 | 275.10 | 1.83 | 212 | 91.3 | -18.8 | 16x 5 | 18.9 | .05 | | S | | | 1 | LSB |
| 1568 | | | 09 40 45.3 | -60 22 10 | 281.67 | -5.80 | 126 | 71.3 | -21.5 | 30x 7 | 18.3 | .48 | | S | L | E | | |
| 1569 | | Q | 09 40 48.5 | -48 40 20 | 274.05 | 3.09 | 212 | 96.1 | 69.2 | 20x 7 | 18.1 | .78 | | S | M | | | cl.to br.* |
| 1570 | | | 09 40 51.0 | -65 29 32 | 285.07 | -9.66 | 91 | 47.0 | -26.2 | 16x 5 | 19.5 | .22 | | S | L ? | | | LSB |
| 1571 | | | 09 40 58.1 | -60 52 53 | 282.02 | -6.17 | 126 | 71.4 | -49.0 | 15x 8 | 18.2 | .45 | | S | E | | | |
| 1572 | | | 09 41 05.3 | -48 46 17 | 274.15 | 3.04 | 212 | 98.4 | 63.7 | 17x 7 | 18.3 | .72 | | S | | | | |
| 1573 | | | 09 41 06.0 | -63 45 54 | 283.94 | -8.34 | 91 | 52.0 | 66.3 | 23x 11 | 17.6 | .28 | | | | | | |
| 1574 | | | 09 41 12.9 | -49 21 10 | 274.54 | 2.62 | 212 | 98.2 | 32.5 | 13x 4 | 19.2 | .80 | | S | | E | | |
| 1575 | | | 09 41 13.7 | -48 37 46 | 274.08 | 3.17 | 212 | 99.9 | 71.3 | 13x 9 | 18.4 | .76 | ? | S | 3 : | | | br.blg |
| 1576 | | | 09 41 14.9 | -62 23 09 | 283.04 | -7.28 | 126 | 72.2 | -126.8 | 20x 11 | 17.6 | .30 | | S | ? | | S | |
| 1577 | | | 09 41 15.8 | -63 47 32 | 283.97 | -8.35 | 91 | 52.9 | 64.8 | 31x 27 | 16.3 | .28 | | S | 0 | | | |
| 1578 | | | 09 41 19.0 | -65 54 46 | 285.38 | -9.94 | 91 | 48.6 | -48.8 | 13x 7 | 19.0 | .22 | ? | | | | | diff. |
| 1579 | | | 09 41 22.3 | -49 29 39 | 274.66 | 2.53 | 212 | 99.3 | 24.9 | 27x 7 | 17.9 | .76 | | S | E | E | | br.blg, LSBdisk |
| 1580 | | | 09 41 35.5 | -64 50 37 | 284.69 | -9.12 | 91 | 52.5 | 8.4 | 16x 13 | 17.5 | .30 | | E | | | | or dbl* w.halo |
| 1581 | | | 09 41 40.4 | -65 15 37 | 284.98 | -9.43 | 91 | 52.1 | -13.9 | 24x 16 | 17.2 | .27 | ? | E | | | 2 | v.diff. |
| 1582 | | | 09 41 45.3 | -62 31 32 | 283.17 | -7.35 | 126 | 72.2 | -137.3 | 13x 5 | 19.3 | .29 | | | | | | |
| 1583 | | I | 09 42 09.6 | -66 06 27 | 285.58 | -10.03 | 91 | 52.8 | -59.4 | 16x 7 | 18.8 | .24 | | S | M | | S | |
| 1584 | | | 09 42 11.2 | -66 03 29 | 285.55 | -9.99 | 91 | 53.0 | -56.8 | 13x 7 | 18.5 | .24 | | S | | | | |
| 1585 | | | 09 42 13.3 | -65 47 24 | 285.37 | -9.79 | 91 | 53.9 | -42.5 | 27x 13 | 17.6 | .23 | | S | | | | dist.nucl. |
| 1586 | | | 09 42 14.9 | -64 47 31 | 284.71 | -9.03 | 91 | 56.4 | 11.0 | 12x 8 | 18.6 | .29 | ? | | | | | |
| 1587 | | | 09 42 17.1 | -66 04 23 | 285.57 | -10.00 | 91 | 53.5 | -57.6 | 15x 4 | 19.7 | .24 | | S | | E | | |
| 1588 | | | 09 42 20.3 | -60 37 17 | 281.98 | -5.86 | 126 | 81.0 | -35.5 | 11x 8 | 18.4 | .47 | | F | | | | |
| 1589 | | | 09 42 20.3 | -64 46 40 | 284.71 | -9.02 | 91 | 56.9 | 11.7 | 13x 8 | 19.1 | .29 | ? | S | ? | | 1 | v.fuzzy |
| 1590 | | | 09 42 26.6 | -55 20 46 | 278.57 | -1.83 | 167 | -87.9 | -20.3 | 19x 3 | 19.5 | .90 | ? | S | | E | | vLSB |
| 1591 | | | 09 42 26.7 | -48 18 13 | 274.02 | 3.55 | 212 | 111.5 | 88.3 | 17x 17 | 17.3 | .66 | | S | E | F | | |
| 1592 | | | 09 42 31.4 | -66 07 06 | 285.61 | -10.02 | 91 | 54.7 | -60.1 | 13x 7 | 18.7 | .24 | | S | ? | | | |
| 1593 | | | 09 42 38.0 | -49 05 49 | 274.55 | 2.96 | 212 | 111.2 | 45.7 | 27x 19 | 17.0 | .68 | | S | 5 | | S | |
| 1594 | | | 09 42 40.8 | -64 49 38 | 284.77 | -9.03 | 91 | 58.8 | 9.0 | 17x 4 | 19.3 | .29 | | S | | | | |
| 1595 | | | 09 42 41.3 | -62 22 51 | 283.16 | -7.17 | 126 | 78.4 | -129.8 | 34x 9 | 17.1 | .28 | | S | 1 ? | | S | |
| 1596 | | | 09 42 52.5 | -65 51 21 | 285.47 | -9.79 | 91 | 57.3 | -46.1 | 16x 13 | 18.1 | .23 | ? | | | | | cl.to br.* |
| 1597 | | | 09 42 53.3 | -48 05 15 | 273.93 | 3.76 | 212 | 116.0 | 99.7 | 17x 13 | 17.8 | .62 | | S | M | F | | |
| 1598 | | | 09 42 53.8 | -62 15 25 | 283.10 | -7.06 | 126 | 80.0 | -123.3 | 16x 8 | 18.2 | .31 | | | | | | in st.diffr.pat. |
| 1599 | | | 09 42 56.7 | -63 48 08 | 284.12 | -8.23 | 91 | 62.9 | 63.8 | 17x 12 | 18.1 | .28 | | | | | | diff. |
| 1600 | | | 09 42 57.5 | -64 43 57 | 284.73 | -8.94 | 91 | 60.6 | 14.0 | 22x 16 | 17.2 | .28 | | L | | | | |

**Table 1.** Galaxies in the southern ZOA – I. The Hydra/Antlia Extension (continued)

| RKK Ident (1) | Other Ident (2) | IR (3) | R.A. (h m s) (4) | Dec. (° ′ ″) (5) | gal $\ell$ (°) (6) | gal $b$ (°) (7) | SRC (8) | X (mm) (9) | Y (mm) (10) | D x d (″) (11) | $B_J$ ($^m$) (12) | $N_{HI}$ (13) | u | Type class. (14) | | o | * | Remarks (15) |
|---|---|---|---|---|---|---|---|---|---|---|---|---|---|---|---|---|---|---|
| 1601 | | | 09 42 59.7 | -63 33 21 | 283.96 | -8.04 | 91 | 63.8 | 77.0 | 16x 5 | 19.3 | .27 | | S | ? | | | |
| 1602 | | | 09 43 09.5 | -62 14 26 | 283.11 | -7.03 | 126 | 81.7 | -122.5 | 17x 16 | 17.4 | .32 | | | | | | in st.diffr.pat. |
| 1603 | | | 09 43 09.8 | -48 10 22 | 274.02 | 3.72 | 212 | 118.2 | 95.0 | 16x 4 | 19.1 | .63 | | S | | | | |
| 1604 | | | 09 43 12.2 | -60 46 51 | 282.16 | -5.91 | 126 | 86.3 | -44.4 | 19x 8 | 18.4 | .46 | | S | | | | blg or s.p.* |
| 1605 | | | 09 43 12.8 | -63 32 55 | 283.97 | -8.02 | 91 | 65.1 | 77.3 | 26x 8 | 18.2 | .27 | | | | | | |
| 1606 | | | 09 43 16.3 | -49 29 22 | 274.89 | 2.73 | 212 | 115.8 | 24.4 | 16x 16 | 17.8 | .66 | | S | E | | | br.blg, LSBdisk |
| 1607 | | | 09 43 19.8 | -62 52 58 | 283.55 | -7.51 | 91 | 67.5 | 113.0 | 15x 9 | 18.2 | .26 | ? | | | | 1 | |
| 1608 | | | 09 43 23.0 | -63 32 20 | 283.98 | -8.00 | 91 | 66.2 | 77.8 | 12x 8 | 19.1 | .27 | ? | | | | | |
| 1609 | | | 09 43 30.6 | -64 25 10 | 284.57 | -8.66 | 91 | 64.6 | 30.6 | 13x 5 | 19.4 | .27 | ? | | | | S | cl.to br.* |
| 1610 | | | 09 43 32.5 | -47 54 37 | 273.90 | 3.96 | 212 | 122.3 | 108.9 | 54x 13 | 16.9 | .48 | | | | | | |
| 1611 | | | 09 43 38.0 | -61 05 51 | 282.41 | -6.12 | 126 | 88.1 | -61.5 | 16x 7 | 18.1 | .40 | | | | | | br.blg or blended ** |
| 1612 | | | 09 43 38.6 | -65 35 35 | 285.35 | -9.54 | 91 | 62.2 | -32.3 | 16x 4 | 19.3 | .24 | ? | S | M | | | |
| 1613 | | | 09 43 39.1 | -65 20 46 | 285.19 | -9.36 | 91 | 62.9 | -19.1 | 16x 16 | 18.0 | .27 | | E | ? | | | |
| 1614 | | | 09 43 41.9 | -64 29 23 | 284.63 | -8.70 | 91 | 65.5 | 26.8 | 31x 13 | 17.1 | .27 | | S | E | | | |
| 1615 | | | 09 43 58.1 | -63 06 30 | 283.75 | -7.63 | 91 | 70.8 | 100.7 | 15x 5 | 19.6 | .26 | ? | | | | | |
| 1616 | | | 09 43 59.9 | -64 36 15 | 284.73 | -8.77 | 91 | 66.9 | 20.6 | 22x 9 | 17.8 | .26 | | S | ? | | | |
| 1617 | | | 09 44 02.9 | -48 55 21 | 274.62 | 3.24 | 213 | -140.2 | 54.3 | 13x 9 | 18.5 | .54 | | | | | | |
| 1618 | | | 09 44 03.1 | -64 35 38 | 284.73 | -8.76 | 91 | 67.2 | 21.1 | 20x 8 | 17.9 | .26 | | E | ? | | 1 | |
| 1619 | | | 09 44 06.3 | -48 16 42 | 274.21 | 3.74 | 212 | 126.4 | 88.9 | 13x 9 | 18.5 | .56 | | | | | | |
| 1620 | | | 09 44 09.6 | -47 31 29 | 273.73 | 4.33 | 212 | 128.9 | 129.3 | 20x 3 | 19.7 | .45 | | S | | E | | |
| 1621 | | | 09 44 14.8 | -49 07 23 | 274.77 | 3.11 | 212 | 125.3 | 43.6 | 24x 20 | 17.2 | .53 | | SB | 5 | | | LSBdisk |
| 1622 | | | 09 44 18.6 | -49 13 05 | 274.84 | 3.04 | 212 | 125.6 | 38.5 | 16x 16 | 17.4 | .59 | | S | ? | | S | |
| 1623 | | | 09 44 43.2 | -64 35 11 | 284.78 | -8.70 | 91 | 71.1 | 21.3 | 13x 12 | 18.8 | .27 | ? | | | | | |
| 1624 | | | 09 44 46.1 | -47 53 55 | 274.05 | 4.11 | 212 | 133.3 | 109.0 | 17x 17 | 17.1 | .51 | | E | ? | | | |
| 1625 | | | 09 44 46.3 | -66 02 11 | 285.74 | -9.81 | 91 | 67.2 | -56.3 | 16x 16 | 18.1 | .23 | ? | | | | 1 | |
| 1626 | | | 09 44 46.4 | -47 39 11 | 273.90 | 4.29 | 212 | 134.1 | 122.1 | 13x 7 | 18.3 | .41 | | S | E | | 2 | |
| 1627 | | | 09 44 47.5 | -57 20 27 | 280.10 | -3.15 | 167 | -66.3 | -126.5 | 13x 7 | 19.2 | .24 | | S | | E | S | |
| 1628 | L091-018 | I | 09 44 48.0 | -63 02 24 | 283.78 | -7.52 | 91 | 76.1 | 104.1 | 141x 16 | 15.1 | .25 | | S | 2 | E | | |
| 1629 | | | 09 44 51. | -62 34.8 : | 283.48 | -7.16 | 127 | -142.4 | -147.2 | 13x 9 | 18.6 | .28 | | S | E | | | |
| 1630 | | | 09 44 54.2 | -65 23 09 | 285.32 | -9.30 | 91 | 69.8 | -21.6 | 15x 9 | 18.4 | .25 | | | | | | |
| 1631 | | | 09 45 01.5 | -60 42 07 | 282.28 | -5.71 | 126 | 98.4 | -40.8 | 13x 11 | 17.9 | .47 | | E | ? | | | |
| 1632 | | | 09 45 05.3 | -47 32 34 | 273.87 | 4.41 | 212 | 137.2 | 127.9 | 20x 7 | 17.7 | .40 | | S | 3 : | E | | |
| 1633 | | | 09 45 10. | -61 01.3 : | 282.50 | -5.94 | 127 | -147.6 | -60.0 | 13x 7 | 18.6 | .35 | ? | S | M | | | |
| 1634 | | | 09 45 10.2 | -48 09 34 | 274.27 | 3.95 | 212 | 136.2 | 94.8 | 13x 9 | 18.5 | .52 | | S | | | | br.blg, LSBdisk |
| 1635 | | | 09 45 13.5 | -48 54 52 | 274.76 | 3.37 | 213 | -129.8 | 55.2 | 13x 8 | 18.8 | .50 | ? | | | | | |
| 1636 | | | 09 45 19.5 | -60 51 16 | 282.41 | -5.80 | 126 | 99.9 | -49.1 | 15x 8 | 18.5 | .41 | | | | | S | |
| 1637 | | | 09 45 19.7 | -60 20 21 | 282.08 | -5.41 | 126 | 101.6 | -21.5 | 22x 17 | 17.0 | .47 | | S | M | | | |
| 1638 | | | 09 45 21.6 | -48 07 15 | 274.27 | 4.00 | 212 | 138.0 | 96.8 | 27x 4 | 18.8 | .50 | | S | | E | | |
| 1639 | | | 09 45 27.5 | -48 30 35 | 274.53 | 3.71 | 212 | 137.7 | 75.9 | 16x 16 | 17.7 | .60 | | S | 5 : | F | | |
| 1640 | | | 09 45 48.2 | -63 57 52 | 284.47 | -8.15 | 91 | 79.3 | 54.2 | 20x 4 | 19.5 | .26 | | S | | E | | LSB |
| 1641 | | | 09 45 48.8 | -60 36 54 | 282.30 | -5.58 | 126 | 103.9 | -36.5 | 23x 8 | 18.5 | .48 | | | | | | |
| 1642 | | | 09 45 49.5 | -51 44 29 | 276.64 | 1.26 | 213 | -117.7 | -96.0 | 23x 3 | 19.1 | .00 | ? | S | | E | | sharp edge |
| 1643 | | | 09 45 52.4 | -60 04 21 | 281.96 | -5.16 | 126 | 106.1 | -7.5 | 13x 8 | 18.9 | .53 | ? | | | | | w.3 comp. |
| 1644 | | | 09 45 52.9 | -51 35 42 | 276.56 | 1.38 | 212 | 132.0 | -89.5 | 15x 11 | 17.9 | .92 | ? | S | | | | |
| 1645 | | | 09 45 58.0 | -61 51 23 | 283.12 | -6.52 | 127 | -138.9 | -104.4 | 13x 7 | 18.7 | .37 | | S | | | | |
| 1646 | | | 09 45 58.3 | -47 11 25 | 273.76 | 4.78 | 212 | 146.3 | 146.3 | 13x 5 | 19.6 | .48 | | S | 5 : | | | LSBdisk |
| 1647 | | | 09 46 15.5 | -64 16 13 | 284.70 | -8.36 | 91 | 80.9 | 37.7 | 15x 5 | 19.2 | .29 | | | | | | |
| 1648 | | | 09 46 15.8 | -49 48 34 | 275.47 | 2.79 | 213 | -118.6 | 7.7 | 17x 5 | 18.6 | .51 | | S | E | | | |
| 1649 | | | 09 46 16.5 | -47 17 27 | 273.86 | 4.74 | 213 | -124.3 | 142.6 | 27x 20 | 17.1 | .41 | | S | 5 | | | |
| 1650 | | | 09 46 25.5 | -48 24 01 | 274.59 | 3.90 | 213 | -120.5 | 83.2 | 13x 8 | 18.6 | .46 | ? | S | | | | nucl.or s.p.* |
| 1651 | | | 09 46 28.9 | -65 14 22 | 285.35 | -9.08 | 91 | 79.1 | -14.2 | 15x 5 | 19.3 | .25 | ? | | | | S | diff. |
| 1652 | | | 09 46 31.8 | -66 15 31 | 286.02 | -9.86 | 91 | 76.0 | -68.8 | 24x 8 | 18.2 | .23 | | S | ? | | S | |
| 1653 | | | 09 46 41.1 | -64 11 00 | 284.68 | -8.26 | 91 | 83.7 | 42.2 | 32x 7 | 17.8 | .28 | | | | | | dbl syst? arc |
| 1654 | | | 09 46 42.3 | -48 02 24 | 274.39 | 4.20 | 213 | -118.8 | 102.6 | 16x 8 | 18.4 | .46 | | S | | | | br.blg, LSBdisk |
| 1655 | | | 09 46 42.8 | -49 12 27 | 275.14 | 3.30 | 213 | -116.1 | 40.1 | 24x 5 | 18.7 | .46 | | S | | | | |
| 1656 | | | 09 46 46.9 | -47 29 17 | 274.05 | 4.64 | 213 | -119.3 | 132.3 | 16x 4 | 18.9 | .35 | | S | | E | | |
| 1657 | | | 09 47 00.1 | -66 13 44 | 286.04 | -9.81 | 91 | 78.6 | -67.4 | 19x 5 | 19.3 | .23 | | S | | | | |
| 1658 | | | 09 47 01.3 | -61 26 22 | 282.95 | -6.12 | 127 | -134.0 | -81.6 | 17x 7 | 18.4 | .34 | | S | | | | dist.nucl. |
| 1659 | | | 09 47 02.2 | -48 29 52 | 274.73 | 3.89 | 213 | -114.8 | 78.2 | 13x 13 | 18.1 | .48 | | S | | F | | w.comp. |
| 1660 | | | 09 47 03.5 | -47 31 48 | 274.12 | 4.64 | 213 | -116.7 | 130.1 | 16x 16 | 17.7 | .34 | | S | R1 | F | | |
| 1661 | | | 09 47 03.9 | -65 13 12 | 285.39 | -9.03 | 91 | 82.4 | -13.4 | 20x 11 | 18.2 | .25 | | | | | 2 | diff. |
| 1662 | | | 09 47 12.2 | -49 49 13 | 275.59 | 2.88 | 213 | -110.5 | 7.4 | 20x 8 | 18.2 | .53 | | I | | | | |
| 1663 | | | 09 47 14.3 | -66 10 02 | 286.01 | -9.74 | 91 | 80.1 | -64.1 | 22x 11 | 18.2 | .23 | ? | S | ? | | | larger vLSBdisk? |
| 1664 | L213-002 | I | 09 47 22.9 | -47 41 09 | 274.26 | 4.55 | 213 | -113.4 | 121.9 | 101x 94 | 13.2 | .35 | | L | | | | dbl syst: S0 + small S... |
| 1665 | | | 09 47 24.1 | -47 18 24 | 274.02 | 4.84 | 213 | -114.0 | 142.2 | 13x 8 | 18.5 | .47 | | S | | | | |
| 1666 | | P | 09 47 31.4 | -47 47 33 | 274.34 | 4.48 | 213 | -111.9 | 116.2 | 43x 16 | 16.4 | .35 | | E | 7 : | | | |
| 1667 | | | 09 47 32.6 | -65 09 23 | 285.38 | -8.95 | 91 | 85.3 | -10.2 | 16x 7 | 18.6 | .25 | | S | ? | | | |
| 1668 | | | 09 47 34. | -47 41.4 : | 274.28 | 4.57 | 213 | -111.3 | 114.7 | 20x 4 | 18.9 | .36 | | S | | E | | |
| 1669 | | | 09 47 35.4 | -48 54 24 | 275.06 | 3.63 | 213 | -109.1 | 56.5 | 34x 34 | 16.0 | .54 | | S | 4 | F | | |
| 1670 | | | 09 47 38.2 | -66 30 43 | 286.27 | -9.98 | 91 | 81.0 | -82.7 | 15x 8 | 18.4 | .21 | ? | S | ? | | | |
| 1671 | | | 09 47 39.8 | -61 27 40 | 283.02 | -6.09 | 127 | -129.8 | -82.5 | 27x 20 | 17.1 | .36 | | SBR4 | | | | |
| 1672 | | | 09 47 56.0 | -48 17 39 | 274.71 | 4.14 | 213 | -107.3 | 89.4 | 16x 11 | 17.8 | .43 | | F | | | | |
| 1673 | | | 09 48 04.0 | -65 06 27 | 285.39 | -8.88 | 91 | 88.4 | -7.7 | 30x 30 | 16.9 | .24 | ? | S | 5 | | S | LSB |
| 1674 | | | 09 48 05.4 | -48 02 55 | 274.58 | 4.34 | 213 | -106.3 | 102.7 | 36x 7 | 17.2 | .35 | ? | S | | E | | smooth |
| 1675 | | | 09 48 07.6 | -60 58 28 | 282.75 | -5.68 | 127 | -128.7 | -56.3 | 23x 8 | 17.8 | .39 | | S | | | | |
| 1676 | | | 09 48 08.7 | -63 32 57 | 284.40 | -7.67 | 91 | 94.5 | 75.6 | 16x 9 | 18.7 | .30 | ? | | | | | |
| 1677 | | | 09 48 15.5 | -52 30 17 | 277.41 | .90 | 213 | -96.0 | -136.1 | 16x 5 | 19.1 | .44 | | S | | | | smooth |
| 1678 | | | 09 48 16.1 | -51 30 14 | 276.79 | 1.68 | 213 | -97.9 | -82.5 | 16x 13 | 17.9 | .91 | ? | | | | S | |
| 1679 | | | 09 48 18.0 | -48 31 06 | 274.90 | 4.00 | 213 | -103.5 | 77.6 | 15x 15 | 17.9 | .49 | | SB | 5 | | | |
| 1680 | | | 09 48 18.8 | -48 39 05 | 274.99 | 3.90 | 213 | -103.2 | 70.4 | 13x 13 | 17.9 | .47 | | E | ? | | | |
| 1681 | | | 09 48 30.3 | -62 53 49 | 284.01 | -7.14 | 91 | 99.1 | 110.3 | 20x 13 | 17.9 | .25 | ? | | | | | p.cov.by * |
| 1682 | | I | 09 48 32.9 | -48 31 40 | 274.94 | 4.02 | 213 | -101.3 | 77.1 | 17x 13 | 18.0 | .49 | | | | | | |
| 1683 | | | 09 48 37.6 | -49 51 38 | 275.79 | 2.99 | 213 | -98.1 | 5.7 | 27x 5 | 18.1 | .49 | | S | | E | | |
| 1684 | | | 09 48 40.7 | -50 11 56 | 276.01 | 2.74 | 213 | -97.0 | -12.4 | 13x 13 | 18.0 | .53 | | S | M | | | |
| 1685 | L213-003 | I | 09 48 43.1 | -49 28 40 | 275.56 | 3.30 | 213 | -98.0 | 26.3 | 54x 40 | 15.2 | .48 | | S | 5 : | | | |
| 1686 | | | 09 48 45.2 | -65 21 21 | 285.61 | -9.02 | 91 | 91.4 | -21.3 | 13x 7 | 18.4 | .26 | ? | | | | | |
| 1687 | | | 09 49 00.7 | -48 30 25 | 274.99 | 4.09 | 213 | -97.3 | 78.4 | 13x 5 | 19.2 | .49 | | S | | | | LSBdisk |
| 1688 | | | 09 49 03.9 | -48 28 39 | 274.99 | 4.13 | 213 | -96.0 | 80.1 | 47x 35 | 15.6 | .48 | | S | 0 | | | br.blg, LSBdisk |
| 1689 | | | 09 49 04.8 | -48 29 12 | 274.98 | 4.11 | 213 | -96.7 | 79.5 | 13x 13 | 17.5 | .49 | | L | ? | | | |
| 1690 | | | 09 49 05.1 | -64 08 35 | 284.86 | -8.06 | 91 | 97.9 | 43.4 | 36x 27 | 17.1 | .28 | ? | | | | | blg, vLSBdisk, or * w.halo |
| 1691 | | | 09 49 09.0 | -48 06 58 | 274.76 | 4.40 | 213 | -96.7 | 99.4 | 20x 7 | 18.5 | .37 | | | | | | dist.nucl. |
| 1692 | | | 09 49 09.8 | -48 20 17 | 274.90 | 4.23 | 213 | -96.2 | 87.5 | 27x 9 | 17.3 | .45 | | S | 1 | | | |
| 1693 | | | 09 49 14.9 | -48 24 49 | 274.96 | 4.18 | 213 | -98.5 | 81.8 | 16x 11 | 18.1 | .48 | | | | | | |
| 1694 | | | 09 49 15.3 | -48 24 45 | 274.96 | 4.18 | 213 | -95.3 | 83.5 | 20x 13 | 17.6 | .48 | | S | | | | |
| 1695 | | | 09 49 16.0 | -48 29 03 | 275.01 | 4.13 | 213 | -95.0 | 79.7 | 13x 13 | 17.7 | .48 | | E | ? | | | |
| 1696 | | | 09 49 16.2 | -53 50 27 | 278.37 | -.04 | 167 | -37.4 | 61.9 | 13x 7 | 18.2 | .14 | ? | S | 0 | | | |
| 1697 | | | 09 49 20.7 | -64 06 22 | 284.86 | -8.01 | 91 | 99.5 | 45.3 | 20x 9 | 18.6 | .29 | | | | | | LSB |
| 1698 | | | 09 49 21.2 | -48 43 57 | 275.17 | 3.95 | 213 | -93.8 | 66.4 | 16x 5 | 18.8 | .47 | | S | | | | br.blg |
| 1699 | | | 09 49 21.4 | -48 23 37 | 274.96 | 4.21 | 213 | -94.4 | 84.6 | 13x 7 | 18.8 | .47 | | S | | | | dist.nucl, LSBdisk |
| 1700 | | | 09 49 21.7 | -48 09 34 | 274.81 | 4.39 | 213 | -94.8 | 97.1 | 16x 5 | 18.5 | .38 | | E | 7 | | | |

**Table 1.** Galaxies in the southern ZOA – I. The Hydra/Antlia Extension (continued)

| RKK Ident (1) | Other Ident (2) | IR (3) | R.A. (h m s) (4) | Dec. (° ′ ″) (5) | gal ℓ (°) (6) | gal $b$ (°) (7) | SRC (8) | X (mm) (9) | Y (mm) (10) | D x d (″) (11) | $B_J$ ($^m$) (12) | $N_{HI}$ (13) | u | Type class. (14) | o * | Remarks (15) |
|---|---|---|---|---|---|---|---|---|---|---|---|---|---|---|---|---|
| 1701 | | | 09 49 22.5 | -48 40 23 | 275.14 | 3.99 | 213 | -94.3 | 71.2 | 9x 9 | 18.8 | .47 | | S | | dist.blg |
| 1702 | | | 09 49 23.7 | -63 08 21 | 284.25 | -7.26 | 91 | 103.5 | 97.0 | 16x 7 | 18.5 | .27 | | S E | | |
| 1703 | | | 09 49 28.6 | -48 42 53 | 275.18 | 3.97 | 213 | -92.8 | 67.4 | 27x 16 | 16.7 | .47 | | S 1 | | |
| 1704 | | | 09 49 32.5 | -48 33 18 | 275.09 | 4.10 | 213 | -92.5 | 76.0 | 20x 16 | 17.0 | .49 | | E 4 | | |
| 1705 | | | 09 49 34.6 | -48 04 49 | 274.79 | 4.48 | 213 | -93.0 | 101.4 | 13x 4 | 19.2 | .40 | | S | | dist.blg |
| 1706 | | | 09 49 37.9 | -62 45 06 | 284.02 | -6.94 | 91 | 106.5 | 117.6 | 13x 3 | 20.3 | .22 | ? | S | | |
| 1707 | | | 09 49 47.9 | -47 07 52 | 274.22 | 5.24 | 213 | -92.6 | 152.4 | 81x 13 | 16.1 | .41 | | I | | |
| 1708 | | | 09 49 49.8 | -62 07 11 | 283.64 | -6.44 | 127 | -113.7 | -116.7 | 16x 7 | 18.8 | .31 | | S | | dist.nucl. |
| 1709 | | | 09 49 54.7 | -49 26 19 | 275.69 | 3.45 | 213 | -87.7 | 28.7 | 20x 13 | 17.8 | .46 | | S | | |
| 1710 | | | 09 49 54.8 | -47 26 52 | 274.44 | 5.00 | 213 | -91.0 | 135.4 | 13x 13 | 17.7 | .46 | | E 0 : | | |
| 1711 | | | 09 50 05.1 | -63 46 36 | 284.71 | -7.71 | 91 | 105.2 | 62.6 | 38x 16 | 16.9 | .31 | | | S | |
| 1712 | | | 09 50 09.5 | -49 23 27 | 275.69 | 3.52 | 213 | -85.7 | 31.3 | 13x 5 | 18.9 | .44 | | | | |
| 1713 | | | 09 50 14.6 | -48 45 26 | 275.30 | 4.02 | 213 | -86.5 | 66.9 | 11x 7 | 19.1 | .50 | | | | dbl syst? 2 nucl? |
| 1714 | | | 09 50 15.9 | -49 32 03 | 275.79 | 3.41 | 213 | -86.4 | 22.2 | 13x 9 | 18.4 | .44 | | S | | dist.blg, LSBdisk |
| 1715 | | | 09 50 17. | -62 39.8 : | 284.02 | -6.83 | 127 | -109.7 | -145.7 | 20x 12 | 17.5 | .24 | | S E | | |
| 1716 | | | 09 50 19.0 | -48 45 27 | 275.31 | 4.03 | 213 | -85.8 | 66.9 | 9x 5 | 19.8 | .50 | | S ? | | |
| 1717 | | | 09 50 19.8 | -63 00 28 | 284.24 | -7.09 | 91 | 109.7 | 103.6 | 34x 30 | 16.4 | .26 | ? | | S | |
| 1718 | | | 09 50 25.4 | -48 29 59 | 275.17 | 4.24 | 213 | -84.7 | 79.2 | 13x 7 | 18.8 | .46 | | S 0 : | | |
| 1719 | | | 09 50 40.9 | -48 28 04 | 275.18 | 4.29 | 213 | -82.5 | 80.9 | 17x 8 | 18.2 | .46 | | S 3 : | | |
| 1720 | | I | 09 50 43.3 | -48 46 16 | 275.38 | 4.06 | 213 | -81.7 | 64.7 | 19x 8 | 18.0 | .48 | | | | |
| 1721 | | | 09 50 45.6 | -48 36 26 | 275.28 | 4.19 | 213 | -81.6 | 73.5 | 20x 8 | 18.0 | .46 | | S ? | | cl.to br.* |
| 1722 | | | 09 50 49.5 | -47 33 12 | 274.63 | 5.02 | 213 | -82.6 | 130.0 | 16x 8 | 18.6 | .49 | | | | smooth |
| 1723 | | | 09 51 01.9 | -63 01 12 | 284.31 | -7.05 | 91 | 113.9 | 102.6 | 34x 20 | 17.1 | .27 | | S | S | |
| 1724 | | | 09 51 02.7 | -63 04 11 | 284.35 | -7.09 | 91 | 113.8 | 100.0 | 16x 5 | 18.8 | .28 | | S E | | |
| 1725 | | | 09 51 02.8 | -63 13 12 | 284.44 | -7.21 | 91 | 113.2 | 91.9 | 15x 5 | 19.8 | .26 | ? | | | |
| 1726 | | | 09 51 04.5 | -48 34 40 | 275.30 | 4.24 | 213 | -78.8 | 75.1 | 20x 13 | 17.6 | .45 | | S 3 | | smooth |
| 1727 | | | 09 51 06.3 | -66 48 26 | 286.73 | -9.99 | 91 | 98.3 | -99.9 | 20x 5 | 18.9 | .24 | | S | | dist.nucl. |
| 1728 | | Q | 09 51 08.3 | -48 42 09 | 275.39 | 4.15 | 213 | -78.1 | 68.5 | 34x 16 | 16.5 | .45 | | SX 5 : | | |
| 1729 | | | 09 51 12.2 | -48 38 50 | 275.36 | 4.20 | 213 | -77.6 | 71.5 | 16x 13 | 17.7 | .44 | | S | | |
| 1730 | | | 09 51 15.6 | -65 31 58 | 285.93 | -8.99 | 91 | 104.6 | -31.8 | 13x 5 | 19.8 | .22 | | | | LSB |
| 1731 | | | 09 51 15.9 | -63 12 32 | 284.45 | -7.18 | 91 | 114.6 | 92.4 | 13x 5 | 19.5 | .27 | ? | | | |
| 1732 | | | 09 51 18.3 | -49 05 36 | 275.65 | 3.86 | 213 | -76.1 | 47.6 | 27x 13 | 17.3 | .46 | | S 5 | | dist.sp.arms |
| 1733 | | P | 09 51 28. | -61 18.1 : | 283.27 | -5.68 | 127 | -106.0 | -72.4 | 20x 13 | 17.9 | .46 | | E 3 : | | or S w.LSBdisk |
| 1734 | | | 09 51 28.3 | -66 45 38 | 286.73 | -9.93 | 91 | 100.4 | -97.5 | 17x 7 | 18.8 | .24 | | S | | |
| 1735 | | | 09 51 33.2 | -50 47 33 | 276.74 | 2.56 | 213 | -71.5 | -43.5 | 16x 16 | 17.2 | .59 | | E 0 | | |
| 1736 | | | 09 51 34.9 | -48 27 28 | 275.29 | 4.39 | 213 | -74.5 | 81.7 | 20x 5 | 18.6 | .45 | | S | | |
| 1737 | | | 09 51 37.2 | -66 56 47 | 286.86 | -10.07 | 92 | -129.5 | -108.2 | 13x 11 | 18.4 | .25 | | S ? | | |
| 1738 | | | 09 51 43.5 | -66 58 01 | 286.88 | -10.08 | 92 | -128.9 | -109.2 | 15x 5 | 19.3 | .25 | | S ? | | |
| 1739 | | | 09 51 48.8 | -65 31 06 | 285.96 | -8.95 | 91 | 107.7 | -31.3 | 17x 5 | 18.8 | .22 | | | | |
| 1740 | | | 09 51 58.9 | -49 46 42 | 276.16 | 3.40 | 213 | -69.2 | 11.0 | 13x 7 | 18.6 | .40 | | | | |
| 1741 | | | 09 52 10.0 | -67 30 41 | 287.26 | -10.47 | 92 | -123.7 | -138.1 | 36x 11 | 17.4 | .20 | | S | | |
| 1742 | | | 09 52 24.0 | -48 49 50 | 275.63 | 4.18 | 213 | -66.8 | 61.9 | 15x 7 | 18.4 | .43 | | | | |
| 1743 | | | 09 52 25.7 | -66 14 27 | 286.47 | -9.47 | 91 | 107.8 | -70.2 | 13x 5 | 19.7 | .23 | ? | S | | dist.nucl. |
| 1744 | | | 09 52 27.5 | -49 40 02 | 276.16 | 3.53 | 213 | -65.2 | 17.0 | 20x 16 | 17.2 | .38 | | I ? | | |
| 1745 | | | 09 52 28. | -61 46.3 : | 283.66 | -5.97 | 127 | -98.2 | -97.2 | 17x 9 | 18.0 | .33 | | S 3 : | | |
| 1746 | | | 09 52 29.7 | -48 17 05 | 275.30 | 4.62 | 213 | -66.6 | 91.2 | 20x 8 | 17.7 | .50 | | S 0 | | |
| 1747 | | | 09 52 30.3 | -50 04 50 | 276.42 | 3.21 | 213 | -64.3 | -5.1 | 16x 13 | 17.6 | .52 | | | | |
| 1748 | | | 09 52 46.0 | -48 20 14 | 275.37 | 4.61 | 213 | -64.1 | 88.4 | 13x 13 | 17.9 | .51 | | E 0 | | |
| 1749 | | | 09 52 54.3 | -49 31 39 | 276.13 | 3.69 | 213 | -61.5 | 24.6 | 47x 5 | 17.6 | .38 | | S L | E | |
| 1750 | | | 09 53 04.0 | -48 37 57 | 275.59 | 4.41 | 213 | -61.1 | 72.6 | 24x 7 | 18.2 | .49 | | | | |
| 1751 | | | 09 53 04.5 | -62 41 32 | 284.29 | -6.65 | 91 | 127.8 | 119.1 | 13x 13 | 18.3 | .24 | | S ? | S | |
| 1752 | | | 09 53 06.9 | -65 29 04 | 286.05 | -8.83 | 92 | -129.1 | -29.4 | 12x 7 | 18.8 | .22 | | | | |
| 1753 | | | 09 53 06.9 | -66 24 24 | 286.63 | -9.56 | 92 | -124.4 | -78.6 | 15x 4 | 19.9 | .23 | | | | |
| 1754 | | | 09 53 09.2 | -48 20 21 | 275.42 | 4.64 | 213 | -60.7 | 88.4 | 13x 11 | 17.9 | .50 | | E 3 | | p.cov.by * |
| 1755 | | | 09 53 09.7 | -48 06 46 | 275.28 | 4.82 | 213 | -60.8 | 100.5 | 13x 3 | 19.7 | .45 | | S | | smooth |
| 1756 | | | 09 53 12.7 | -62 24 39 | 284.13 | -6.42 | 91 | 130.0 | 134.1 | 15x 8 | 18.7 | .26 | | S ? | 1 | |
| 1757 | | I | 09 53 16.6 | -49 06 51 | 275.92 | 4.05 | 213 | -58.7 | 46.8 | 19x 27 | 16.8 | .44 | ? | SB | 1 | |
| 1758 | | I | 09 53 16.8 | -49 11 57 | 275.97 | 3.98 | 213 | -58.6 | 42.3 | 24x 13 | 17.0 | .44 | | S 2 | | |
| 1759 | | | 09 53 18.8 | -48 57 41 | 275.83 | 4.17 | 213 | -58.6 | 55.0 | 27x 13 | 17.4 | .49 | ? | S | | faint sp.arms |
| 1760 | | | 09 53 20.3 | -64 24 28 | 285.39 | -7.98 | 92 | -133.1 | 28.2 | 24x 11 | 17.6 | .27 | | I | S | or dbl syst. |
| 1761 | | | 09 53 26.4 | -67 29 48 | 287.35 | -10.39 | 92 | -117.3 | -136.6 | 24x 23 | 17.2 | .21 | | S | F 1 | dwarf? |
| 1762 | | | 09 53 29.3 | -48 13 47 | 275.40 | 4.77 | 213 | -57.8 | 94.3 | 15x 4 | 19.2 | .46 | | S | | dist.nucl. |
| 1763 | | | 09 53 34.4 | -67 06 48 | 287.12 | -10.08 | 92 | -118.5 | -116.1 | 35x 31 | 16.3 | .23 | | SB 4 | S | or i.a.dbl syst. |
| 1764 | | | 09 53 35. | -62 28.4 : | 284.20 | -6.44 | 127 | -89.5 | -134.4 | 16x 5 | 18.8 | .25 | ? | S 1 | | |
| 1765 | | | 09 53 35.4 | -66 39 15 | 286.83 | -9.72 | 92 | -120.7 | -91.6 | 16x 7 | 18.2 | .22 | | S ? | | |
| 1766 | | | 09 53 37. | -60 26.0 : | 282.94 | -4.84 | 127 | -94.6 | -25.2 | 40x 27 | 16.1 | .53 | | S 4 : | | |
| 1767 | | | 09 53 47. | -61 42.9 : | 283.75 | -5.83 | 127 | -90.2 | -93.7 | 27x 20 | 16.4 | .35 | | S 4 | V | |
| 1768 | | | 09 53 50.1 | -62 09 59 | 284.03 | -6.18 | 127 | -89.0 | -117.7 | 27x 13 | 17.3 | .26 | | S ? | | or E..5 |
| 1769 | | | 09 53 52.4 | -65 16 22 | 285.98 | -8.62 | 92 | -125.9 | -17.7 | 30x 7 | 18.0 | .23 | | S | S | |
| 1770 | | | 09 53 54.1 | -66 12 55 | 286.57 | -9.36 | 91 | 115.9 | -69.5 | 13x 9 | 18.6 | .25 | ? | | | |
| 1771 | | | 09 53 54.7 | -61 32 53 | 283.66 | -5.69 | 127 | -89.8 | -84.8 | 20x 9 | 17.8 | .38 | | S 3 | | |
| 1772 | | | 09 54 02.2 | -48 48 24 | 275.83 | 4.37 | 213 | -52.4 | 63.5 | 13x 9 | 18.5 | .49 | | S ? | | |
| 1773 | | | 09 54 05.5 | -62 58 03 | 284.56 | -6.80 | 92 | -135.6 | 105.5 | 13x 4 | 19.8 | .25 | | S | E | |
| 1774 | | | 09 54 11.5 | -65 11 58 | 285.96 | -8.54 | 92 | -124.5 | -13.7 | 20x 8 | 18.0 | .23 | | S E | | |
| 1775 | | | 09 54 12.4 | -48 07 22 | 275.43 | 4.92 | 213 | -51.5 | 100.2 | 13x 4 | 19.3 | .45 | | S | | |
| 1776 | | | 09 54 12.7 | -49 59 17 | 276.58 | 3.45 | 213 | -49.7 | .1 | 16x 8 | 18.3 | .45 | | S | | br.blg, LSBdisk |
| 1777 | | | 09 54 13.4 | -65 32 59 | 286.18 | -8.81 | 92 | -122.6 | -32.3 | 24x 22 | 16.8 | .22 | | S M | | |
| 1778 | | | 09 54 14.1 | -50 00 31 | 276.59 | 3.44 | 213 | -49.5 | -1.0 | 13x 11 | 18.2 | .45 | | S | | |
| 1779 | | | 09 54 17.1 | -66 19 37 | 286.68 | -9.42 | 92 | -118.6 | -73.8 | 15x 4 | 19.2 | .25 | | S ? | | |
| 1780 | FGCE0780 | | 09 54 19.0 | -66 53 37 | 287.04 | -9.86 | 92 | -115.7 | -104.0 | 71x 8 | 16.8 | .23 | | S | E | |
| 1781 | | | 09 54 21. | -61 12.7 : | 283.49 | -5.40 | 127 | -87.9 | -66.6 | 15x 3 | 19.6 | .40 | | S | E | |
| 1782 | | | 09 54 21.9 | -67 04 28 | 287.15 | -10.00 | 92 | -114.6 | -113.7 | 20x 8 | 18.0 | .23 | | S ? | | |
| 1783 | | | 09 54 22.7 | -62 47 14 | 284.47 | -6.63 | 92 | -134.7 | 115.3 | 13x 7 | 18.7 | .24 | | S | | |
| 1784 | | | 09 54 23.6 | -49 59 31 | 276.60 | 3.47 | 213 | -48.2 | .0 | 13x 7 | 18.1 | .45 | | E 4 | | pair w.1786 |
| 1785 | | | 09 54 24.8 | -50 01 46 | 276.63 | 3.44 | 213 | -48.0 | -2.0 | 16x 11 | 17.8 | .46 | | S | | br.blg, 1 dist.sp.arm |
| 1786 | | | 09 54 25.2 | -49 59 41 | 276.61 | 3.47 | 213 | -48.0 | -.2 | 34x 32 | 15.4 | .46 | | E | | pair w.1784 |
| 1787 | | | 09 54 28. | -61 47.5 : | 283.86 | -5.84 | 127 | -85.6 | -97.6 | 15x 15 | 17.8 | .32 | | S 4 | F | |
| 1788 | | | 09 54 28.6 | -51 52 41 | 277.77 | 1.99 | 213 | -45.8 | -101.1 | 13x 7 | 18.8 | .81 | | S ? | | |
| 1789 | | | 09 54 31.8 | -66 55 11 | 287.07 | -9.87 | 92 | -114.4 | -105.4 | 22x 5 | 18.8 | .23 | | S | E S | |
| 1790 | | | 09 54 36.5 | -65 39 54 | 286.29 | -8.88 | 92 | -119.9 | -38.3 | 15x 12 | 18.0 | .20 | | | F | |
| 1791 | | | 09 54 36.7 | -67 14 52 | 287.28 | -10.12 | 92 | -112.5 | -122.8 | 13x 11 | 19.0 | .21 | ? | | | LSB |
| 1792 | | | 09 54 39.9 | -49 36 28 | 276.40 | 3.80 | 213 | -46.2 | 20.6 | 27x 7 | 17.8 | .44 | | S | | |
| 1793 | | | 09 54 42.5 | -49 47 31 | 276.52 | 3.66 | 213 | -45.6 | 10.7 | 16x 13 | 17.9 | .46 | | | | dist.nucl. |
| 1794 | | | 09 54 50.7 | -65 42 42 | 286.33 | -8.90 | 92 | -118.4 | -40.7 | 31x 4 | 19.3 | .20 | | S | E | vLSB |
| 1795 | | | 09 54 58.8 | -61 30 22 | 283.73 | -5.58 | 127 | -83.1 | -82.2 | 27x 4 | 18.6 | .38 | | S M | | |
| 1796 | | | 09 55 00.4 | -66 43 09 | 286.98 | -9.68 | 91 | 119.2 | -97.0 | 16x 5 | 19.4 | .21 | ? | S ? | | |
| 1797 | | | 09 55 02.5 | -49 03 12 | 276.11 | 4.27 | 213 | -43.3 | 50.4 | 13x 5 | 18.8 | .43 | | S E | | |
| 1798 | | | 09 55 06. | -62 26.7 : | 284.32 | -6.31 | 127 | -80.1 | -132.5 | 13x 8 | 18.4 | .24 | | | | |
| 1799 | | | 09 55 06.2 | -65 46 04 | 286.39 | -8.93 | 92 | -116.7 | -43.6 | 17x 8 | 18.6 | .19 | | S M | 1 | p.cov.by * |
| 1800 | | | 09 55 09.1 | -49 41 41 | 276.52 | 3.78 | 213 | -41.9 | 16.0 | 16x 7 | 18.3 | .44 | | S | | br.blg, LSBdisk |

**Table 1.** Galaxies in the southern ZOA – I. The Hydra/Antlia Extension (continued)

| RKK Ident (1) | Other Ident (2) | IR (3) | R.A. (h m s) (4) | Dec. (° ′ ″) (5) | gal ℓ (°) (6) | gal $b$ (°) (7) | SRC (8) | X (mm) (9) | Y (mm) (10) | D x d (″) (11) | $B_J$ ($^m$) (12) | $N_{HI}$ (13) | u | Type class. (14) | | o | * | Remarks (15) |
|---|---|---|---|---|---|---|---|---|---|---|---|---|---|---|---|---|---|---|
| 1801 | | | 09 55 11.2 | -50 16 27 | 276.88 | 3.32 | 213 | -41.1 | -15.1 | 16x 7 | 18.8 | .55 | | | | | | cl.to br.* |
| 1802 | | | 09 55 12.2 | -67 17 56 | 287.36 | -10.13 | 92 | -109.2 | -125.3 | 20x 11 | 17.7 | .21 | | S | E | | | |
| 1803 | | | 09 55 14.1 | -63 26 59 | 284.96 | -7.10 | 92 | -126.5 | 80.3 | 13x 7 | 19.0 | .28 | | S | | | | |
| 1804 | | | 09 55 14.4 | -65 42 18 | 286.36 | -8.87 | 92 | -116.3 | -40.1 | 16x 16 | 18.1 | .20 | | | | | | |
| 1805 | | | 09 55 15.6 | -50 04 03 | 276.76 | 3.49 | 213 | -40.7 | -4.0 | 13x 8 | 18.4 | .51 | | | | | | br.blg, LSBdisk |
| 1806 | | | 09 55 28.6 | -49 52 50 | 276.67 | 3.66 | 213 | -38.9 | 6.1 | 16x 9 | 18.3 | .51 | | S | E | | | |
| 1807 | | | 09 55 28.8 | -63 16 51 | 284.87 | -6.95 | 91 | 139.6 | 86.3 | 20x 11 | 18.4 | .29 | ? | | | | S | |
| 1808 | L092-001 | | 09 55 29.1 | -65 23 17 | 286.18 | -8.61 | 92 | -116.4 | -23.1 | 74x 54 | 14.6 | .21 | | E | 3 | | | |
| 1809 | | | 09 55 34.9 | -50 26 03 | 277.02 | 3.23 | 213 | -37.6 | -23.6 | 13x 9 | 18.5 | .57 | | | | | | |
| 1810 | | | 09 55 37.8 | -67 02 25 | 287.23 | -9.90 | 92 | -108.1 | -111.3 | 19x 7 | 18.5 | .21 | | | | | | |
| 1811 | | | 09 55 38.2 | -64 33 56 | 285.68 | -7.95 | 92 | -119.2 | 20.9 | 22x 4 | 19.6 | .26 | | S | | E | | LSB |
| 1812 | | | 09 55 47.9 | -65 42 01 | 286.40 | -8.83 | 92 | -113.2 | -39.6 | 19x 3 | 19.9 | .20 | | S | | E | | |
| 1813 | | | 09 55 48.9 | -66 33 00 | 286.94 | -9.50 | 92 | -109.4 | -85.0 | 23x 4 | 19.3 | .22 | | S | | E | 1 | |
| 1814 | | | 09 55 51.9 | -62 42 33 | 284.56 | -6.47 | 92 | -126.0 | 120.2 | 13x 7 | 19.0 | .23 | ? | | | | | |
| 1815 | | | 09 55 56.6 | -66 20 36 | 286.82 | -9.33 | 92 | -109.6 | -73.9 | 19x 7 | 18.9 | .23 | | S | | | | LSB |
| 1816 | | | 09 56 07.8 | -63 18 32 | 284.95 | -6.92 | 92 | -121.8 | 88.3 | 31x 5 | 18.3 | .28 | | S | | E | | |
| 1817 | | | 09 56 09.6 | -65 56 06 | 286.58 | -8.99 | 92 | -110.2 | -52.0 | 22x 8 | 18.0 | .20 | | | | | | |
| 1818 | | | 09 56 13.2 | -62 53 43 | 284.70 | -6.59 | 92 | -123.0 | 110.5 | 13x 4 | 19.8 | .26 | | S | | E | | |
| 1819 | | | 09 56 14.1 | -62 52 11 | 284.69 | -6.57 | 92 | -123.1 | 111.8 | 19x 5 | 18.7 | .26 | | S | | | 1 | |
| 1820 | | | 09 56 17. | -62 23.5 : | 284.40 | -6.19 | 127 | -72.8 | -129.2 | 20x 7 | 18.6 | .27 | | | | | | nucl.or s.p.*, LSBdisk |
| 1821 | | | 09 56 18.3 | -50 25 12 | 277.11 | 3.32 | 213 | -31.5 | -22.8 | 23x 23 | 17.3 | .57 | ? | | | | | smooth |
| 1822 | | | 09 56 18.6 | -50 53 39 | 277.40 | 2.94 | 213 | -31.2 | -48.2 | 13x 7 | 18.7 | .62 | | S | | | | |
| 1823 | | | 09 56 20. | -61 23.2 : | 283.79 | -5.39 | 127 | -74.7 | -75.5 | 20x 4 | 18.9 | .38 | | S | M | E | | |
| 1824 | | | 09 56 32.4 | -47 18 11 | 275.23 | 5.81 | 213 | -31.0 | 144.4 | 13x 7 | 18.8 | .32 | | S | ? | | | |
| 1825 | | | 09 56 33.7 | -47 20 13 | 275.26 | 5.79 | 213 | -30.8 | 142.6 | 13x 7 | 18.6 | .32 | | S | 5 | | | |
| 1826 | | | 09 56 41.4 | -67 05 13 | 287.34 | -9.87 | 92 | -102.4 | -113.3 | 16x 13 | 18.2 | .22 | | E | ? | | | |
| 1827 | | | 09 56 42. | -61 14.1 : | 283.73 | -5.24 | 127 | -72.7 | -67.3 | 20x 8 | 18.6 | .42 | | S | 5 | | S | |
| 1828 | | | 09 56 51.4 | -64 00 54 | 285.45 | -7.43 | 92 | -114.5 | 50.9 | 16x 11 | 17.9 | .25 | | E | | | | |
| 1829 | | | 09 56 51.4 | -64 27 22 | 285.72 | -7.78 | 92 | -112.7 | 27.3 | 22x 7 | 18.8 | .24 | | S | L | | | LSB |
| 1830 | | | 09 56 55.0 | -63 24 50 | 285.09 | -6.95 | 92 | -116.6 | 83.1 | 13x 9 | 18.2 | .25 | ? | | | | | |
| 1831 | | | 09 56 55.5 | -66 27 43 | 286.97 | -9.36 | 92 | -103.8 | -79.8 | 13x 11 | 18.1 | .23 | | | | | | p.cov.by * |
| 1832 | | | 09 56 57.8 | -64 02 08 | 285.47 | -7.44 | 92 | -113.8 | 49.9 | 20x 8 | 18.1 | .25 | | | | | | |
| 1833 | | NS | 09 56 58.3 | -63 41 21 | 285.26 | -7.17 | 92 | -115.2 | 68.4 | 20x 20 | 17.5 | .25 | ? | | | | | smooth, sharp edge, PN? |
| 1834 | | | 09 56 59.1 | -49 13 34 | 276.47 | 4.33 | 213 | -26.2 | 41.3 | 17x 11 | 17.8 | .50 | | S | E | | | |
| 1835 | | | 09 57 06.3 | -49 57 27 | 276.93 | 3.76 | 213 | -24.8 | 2.1 | 13x 7 | 18.3 | .49 | | S | | | | |
| 1836 | | | 09 57 07.3 | -50 00 19 | 276.96 | 3.72 | 213 | -24.7 | -.5 | 13x 7 | 18.3 | .50 | | S | | | | |
| 1837 | | | 09 57 10.0 | -64 16 38 | 285.64 | -7.62 | 92 | -111.6 | 37.0 | 13x 13 | 18.6 | .25 | | | | | | |
| 1838 | | | 09 57 11.1 | -49 22 21 | 276.58 | 4.23 | 213 | -24.3 | 33.5 | 13x 7 | 18.5 | .49 | | SB | 0 | | | |
| 1839 | | | 09 57 13. | -61 15.7 : | 283.80 | -5.22 | 127 | -69.3 | -68.6 | 16x 5 | 18.8 | .40 | | | | | | |
| 1840 | | | 09 57 13.0 | -65 11 39 | 286.21 | -8.34 | 92 | -107.5 | -12.0 | 13x 9 | 18.7 | .24 | ? | | | | | cl.to br.* |
| 1841 | | | 09 57 29.0 | -50 18 12 | 277.19 | 3.52 | 213 | -21.4 | -16.4 | 13x 13 | 18.3 | .56 | | | | | | |
| 1842 | | | 09 57 31.0 | -64 27 23 | 285.78 | -7.74 | 92 | -108.8 | 27.6 | 20x 11 | 17.5 | .23 | | E | ? | | | |
| 1843 | | | 09 57 31.3 | -64 19 43 | 285.70 | -7.64 | 92 | -109.3 | 34.4 | 19x 8 | 18.7 | .25 | | S | | | | |
| 1844 | | | 09 57 35.5 | -64 27 44 | 285.79 | -7.74 | 92 | -108.4 | 27.3 | 13x 8 | 18.4 | .23 | | S | E | | | |
| 1845 | | | 09 57 37.1 | -50 33 46 | 277.36 | 3.33 | 213 | -20.2 | -30.3 | 27x 13 | 17.1 | .63 | | S | M | | | |
| 1846 | | | 09 57 38.7 | -48 12 39 | 275.94 | 5.20 | 213 | -20.7 | 95.8 | 20x 7 | 18.2 | .63 | | S | | | | |
| 1847 | | | 09 57 42.3 | -66 59 56 | 287.37 | -9.74 | 92 | -97.5 | -108.2 | 13x 5 | 19.7 | .20 | | I | ? | | | |
| 1848 | | | 09 57 42.9 | -66 29 21 | 287.05 | -9.34 | 92 | -99.5 | -81.0 | 17x 11 | 17.8 | .23 | | L | | | | |
| 1849 | | | 09 57 43.4 | -49 56 36 | 277.00 | 3.83 | 213 | -19.5 | 2.9 | 20x 9 | 17.7 | .56 | | S | | | | |
| 1850 | | | 09 57 46.3 | -47 29 46 | 275.52 | 5.79 | 213 | -19.7 | 134.1 | 20x 9 | 17.9 | .34 | | S | | | 2 | smooth |
| 1851 | | | 09 57 47.4 | -65 24 51 | 286.39 | -8.48 | 92 | -103.4 | -23.5 | 16x 8 | 18.6 | .21 | | S | ? | | | w.comp. |
| 1852 | | | 09 58 01.2 | -65 15 45 | 286.32 | -8.35 | 92 | -102.8 | -15.3 | 23x 12 | 17.7 | .22 | | | | | 1 | |
| 1853 | | | 09 58 04.7 | -64 29 26 | 285.85 | -7.73 | 92 | -105.5 | 26.0 | 47x 34 | 15.4 | .23 | | E | | | 1 | |
| 1854 | | | 09 58 05.9 | -49 29 16 | 276.77 | 4.23 | 213 | -16.3 | 27.3 | 27x 13 | 17.0 | .47 | | S | M | | | |
| 1855 | | | 09 58 12.3 | -67 18 46 | 287.60 | -9.96 | 92 | -93.6 | -124.8 | 15x 15 | 18.4 | .24 | | | | | | LSB |
| 1856 | L092-003 | | 09 58 14.7 | -67 03 53 | 287.45 | -9.76 | 92 | -94.4 | -111.5 | 101x 34 | 14.9 | .21 | | S | 1 | | S | |
| 1857 | | | 09 58 17.2 | -47 48 52 | 275.78 | 5.58 | 213 | -15.0 | 117.0 | 20x 8 | 18.2 | .46 | | S | | | | smooth |
| 1858 | | | 09 58 20.6 | -65 13 13 | 286.32 | -8.29 | 92 | -101.1 | -12.9 | 34x 7 | 17.5 | .23 | | S | | | | |
| 1859 | | | 09 58 21.1 | -65 56 11 | 286.76 | -8.86 | 92 | -98.3 | -51.2 | 17x 3 | 19.9 | .21 | | S | | E | | |
| 1860 | | | 09 58 25.2 | -48 16 28 | 276.08 | 5.23 | 213 | -13.7 | 92.4 | 13x 8 | 18.6 | .54 | | S | L : | | | |
| 1861 | | | 09 58 26. | -61 45.0 : | 284.21 | -5.52 | 127 | -60.6 | -94.4 | 20x 7 | 18.3 | .31 | | | | | | p.cov.by br.* |
| 1862 | L092-004 | | 09 58 32.6 | -67 24 07 | 287.68 | -10.01 | 92 | -91.5 | -129.4 | 114x 47 | 14.6 | .24 | | S | 1 | | S | |
| 1863 | | | 09 58 37.3 | -49 08 58 | 276.63 | 4.55 | 213 | -11.8 | 45.5 | 16x 5 | 18.5 | .51 | | S | E | | | |
| 1864 | | | 09 58 39.2 | -64 11 59 | 285.72 | -7.46 | 92 | -103.3 | 41.8 | 81x 8 | 17.0 | .24 | | S | M | E | S | |
| 1865 | | | 09 58 40.5 | -65 05 55 | 286.27 | -8.17 | 92 | -99.7 | -6.2 | 15x 4 | 19.7 | .25 | | | | | | |
| 1866 | | | 09 58 42.1 | -49 03 35 | 276.59 | 4.63 | 213 | -11.2 | 50.3 | 16x 11 | 17.7 | .55 | | E | 2 | | | |
| 1867 | | | 09 58 53. | -61 00.7 : | 283.80 | -4.90 | 127 | -59.0 | -54.9 | 17x 9 | 18.4 | .43 | | | | | | smooth |
| 1868 | | | 09 58 56.8 | -48 14 06 | 276.12 | 5.32 | 213 | -9.0 | 94.5 | 20x 5 | 18.5 | .54 | | S | E | | | |
| 1869 | | | 09 58 58. | -60 40.0 : | 283.60 | -4.62 | 127 | -59.0 | -36.4 | 20x 13 | 17.4 | .43 | | E | ? | | | |
| 1870 | | | 09 59 00.1 | -48 58 27 | 276.58 | 4.73 | 213 | -8.5 | 54.9 | 13x 8 | 18.2 | .56 | | S | 0 | | | |
| 1871 | | | 09 59 05.5 | -64 49 56 | 286.14 | -7.93 | 92 | -98.4 | 8.2 | 24x 19 | 17.1 | .27 | | F | | | 1 | |
| 1872 | | | 09 59 08.7 | -65 40 51 | 286.67 | -8.61 | 92 | -94.9 | -37.2 | 16x 3 | 19.8 | .24 | ? | S | ? | | 1 | |
| 1873 | | | 09 59 09.9 | -65 43 58 | 286.70 | -8.65 | 92 | -94.6 | -40.0 | 16x 13 | 17.9 | .23 | | | | | | |
| 1874 | | | 09 59 18.3 | -67 17 46 | 287.67 | -9.88 | 92 | -88.0 | -123.5 | 15x 12 | 18.6 | .22 | ? | S | | | | or trpl syst. |
| 1875 | | | 09 59 24.3 | -63 49 11 | 285.55 | -7.11 | 92 | -100.3 | 62.4 | 51x 22 | 15.9 | .26 | | S | 1 : | | S | |
| 1876 | | I | 09 59 28.0 | -49 16 22 | 276.82 | 4.54 | 213 | -4.4 | 38.9 | 27x 7 | 17.8 | .52 | | S | 0 | | | |
| 1877 | | | 09 59 28.7 | -64 51 40 | 286.20 | -7.93 | 92 | -96.1 | 6.8 | 38x 30 | 16.4 | .27 | | | | | S | |
| 1878 | | NS | 09 59 36. | -60 45.0 : | 283.72 | -4.64 | 127 | -54.7 | -40.7 | 16x 5 | 18.7 | .42 | | S | 1 : | | | |
| 1879 | | | 09 59 40.7 | -64 50 28 | 286.20 | -7.90 | 92 | -95.0 | 7.9 | 16x 4 | 20.0 | .27 | | | | | | LSB, w.2 comp. |
| 1880 | | | 09 59 41. | -60 47.4 : | 283.75 | -4.67 | 127 | -54.1 | -42.9 | 15x 11 | 18.1 | .41 | | S | 5 | F | | |
| 1881 | | | 09 59 51.5 | -64 49 55 | 286.21 | -7.89 | 92 | -94.0 | 8.5 | 13x 13 | 18.9 | .27 | | | | | | neighb.of 1879 |
| 1882 | | | 09 59 54.1 | -65 58 05 | 286.91 | -8.79 | 92 | -89.7 | -52.3 | 15x 11 | 18.1 | .20 | | | | | | |
| 1883 | | | 09 59 59.9 | -47 24 16 | 275.76 | 6.09 | 213 | .5 | 139.1 | 16x 16 | 17.9 | .28 | | S | 5 | | | |
| 1884 | | | 10 00 00.8 | -65 59 37 | 286.93 | -8.80 | 92 | -89.0 | -53.6 | 27x 15 | 17.1 | .20 | | S | ? | | | br.blg |
| 1885 | | | 10 00 17.4 | -65 38 36 | 286.74 | -8.51 | 92 | -88.7 | -34.8 | 13x 8 | 18.8 | .26 | ? | | | | | |
| 1886 | | | 10 00 33.4 | -65 38 29 | 286.76 | -8.49 | 92 | -87.3 | -34.6 | 20x 17 | 17.9 | .26 | | S | | F | S | LSB |
| 1887 | | | 10 00 34.0 | -49 37 14 | 277.17 | 4.37 | 213 | 5.1 | 20.2 | 13x 5 | 19.2 | .54 | | S | | | | |
| 1888 | | | 10 00 39.5 | -67 21 00 | 287.81 | -9.84 | 92 | -80.8 | -125.9 | 20x 17 | 17.4 | .22 | | | | | | br.nucl. |
| 1889 | | | 10 00 55.9 | -63 10 22 | 285.30 | -6.49 | 92 | -93.4 | 97.7 | 13x 5 | 19.6 | .29 | ? | S | ? | | S | |
| 1890 | | | 10 00 57.9 | -47 19 53 | 275.85 | 6.24 | 213 | 9.2 | 142.9 | 17x 7 | 18.2 | .26 | | S | E | | | |
| 1891 | | | 10 00 59.3 | -64 51 34 | 286.32 | -7.83 | 92 | -87.5 | 7.4 | 22x 7 | 18.2 | .24 | | S | | | | |
| 1892 | | | 10 01 07.7 | -65 27 07 | 286.69 | -8.30 | 92 | -84.7 | -24.2 | 27x 17 | 17.1 | .26 | | S | M | | | 1 dist.sp.arm |
| 1893 | | | 10 01 11.7 | -63 08 41 | 285.31 | -6.45 | 92 | -91.9 | 99.2 | 20x 7 | 18.6 | .29 | | S | ? | | 1 | |
| 1894 | | | 10 01 16.6 | -53 01 01 | 279.29 | 1.70 | 167 | 58.6 | 105.8 | 24x 20 | 17.0 | .90 | ? | I | ? | | S | |
| 1895 | | | 10 01 18.6 | -63 47 20 | 285.71 | -6.96 | 92 | -89.1 | 64.8 | 15x 11 | 18.0 | .25 | | | | | | |
| 1896 | | | 10 01 22.9 | -47 16 15 | 275.87 | 6.33 | 213 | 13.1 | 146.1 | 23x 11 | 17.6 | .26 | | | | | | dist.nucl, smooth |
| 1897 | | | 10 01 23.0 | -47 33 10 | 276.04 | 6.11 | 213 | 13.0 | 131.0 | 20x 9 | 17.7 | .29 | | E | 3 | | | |
| 1898 | | | 10 01 23.9 | -65 34 44 | 286.79 | -8.38 | 92 | -82.8 | -30.9 | 19x 4 | 19.2 | .26 | | S | | E | S | |
| 1899 | | | 10 01 31.6 | -65 13 54 | 286.59 | -8.10 | 92 | -83.2 | -12.3 | 30x 7 | 18.0 | .26 | | S | ? | | | |
| 1900 | | | 10 01 31.7 | -49 50 43 | 277.43 | 4.28 | 213 | 13.4 | 8.1 | 38x 4 | 17.8 | .47 | | S | 9 | E | | |

**Table 1.** Galaxies in the southern ZOA – I. The Hydra/Antlia Extension (continued)

| RKK Ident (1) | Other Ident (2) | IR (3) | R.A. (h m s) (4) | Dec. (° ′ ″) (5) | gal ℓ (°) (6) | gal b (°) (7) | SRC (8) | X (mm) (9) | Y (mm) (10) | D x d (″) (11) | $B_J$ ($^m$) (12) | $N_{HI}$ (13) | u | Type class. (14) | o | * | Remarks (15) |
|---|---|---|---|---|---|---|---|---|---|---|---|---|---|---|---|---|---|
| 1901 | | | 10 01 34.3 | -67 34 28 | 288.02 | -9.97 | 92 | -75.4 | -137.6 | 15x 11 | 18.1 | .25 | | S ? | | | |
| 1902 | | I | 10 01 41.8 | -65 08 29 | 286.55 | -8.02 | 92 | -82.6 | -7.4 | 16x 4 | 19.3 | .25 | | S ? | | | cl.to br.* |
| 1903 | | | 10 01 42.0 | -66 59 12 | 287.67 | -9.49 | 92 | -76.6 | -106.1 | 20x 5 | 18.6 | .21 | | S | E | | |
| 1904 | | | 10 01 47. | -61 43.9 : | 284.52 | -5.27 | 127 | -39.5 | -92.9 | 20x 16 | 17.3 | .34 | | S ? | | S | |
| 1905 | | | 10 01 47.4 | -47 19 00 | 275.96 | 6.34 | 213 | 16.7 | 143.7 | 13x 11 | 17.9 | .25 | | S ? | | | |
| 1906 | | | 10 01 47.6 | -65 39 13 | 286.87 | -8.42 | 92 | -80.4 | -34.8 | 16x 5 | 18.8 | .25 | | | | | |
| 1907 | L092-006 | I | 10 01 54.6 | -64 43 26 | 286.32 | -7.67 | 92 | -82.7 | 15.0 | 148x 87 | 13.5 | .23 | | SXR3 | | S | |
| 1908 | | | 10 02 16.2 | -65 30 06 | 286.82 | -8.27 | 92 | -78.2 | -26.5 | 13x 5 | 19.5 | .24 | ? | | | 1 | |
| 1909 | L092-007 | | 10 02 17.1 | -67 12 20 | 287.85 | -9.63 | 92 | -72.9 | -117.7 | 134x 34 | 15.0 | .21 | | I | | S | LSB |
| 1910 | | | 10 02 23.4 | -65 57 29 | 287.11 | -8.63 | 92 | -76.2 | -50.9 | 40x 12 | 16.8 | .22 | | S M | | S | w.comp. |
| 1911 | | | 10 02 29.8 | -48 48 41 | 276.94 | 5.20 | 213 | 22.3 | 63.4 | 16x 4 | 19.2 | .40 | | S | E | | |
| 1912 | | | 10 02 35.6 | -64 16 14 | 286.11 | -7.26 | 92 | -80.1 | 39.5 | 27x 5 | 18.2 | .21 | | S E | | | |
| 1913 | | | 10 02 36.6 | -47 49 15 | 276.37 | 6.01 | 213 | 23.9 | 116.5 | 20x 4 | 18.7 | .32 | | S 5 : | E | | dist.nucl. |
| 1914 | | | 10 02 37.8 | -67 02 31 | 287.78 | -9.48 | 92 | -71.6 | -108.8 | 16x 8 | 18.4 | .21 | | | | | |
| 1915 | | | 10 02 38.4 | -65 54 44 | 287.10 | -8.58 | 92 | -75.0 | -48.4 | 13x 4 | 19.5 | .23 | | S | | | |
| 1916 | | | 10 02 40.3 | -48 49 00 | 276.97 | 5.22 | 213 | 23.9 | 63.2 | 40x 4 | 18.0 | .38 | | S | E | | |
| 1917 | | | 10 02 47.1 | -67 19 24 | 287.96 | -9.70 | 92 | -70.0 | -123.8 | 36x 4 | 18.6 | .21 | | S L | | 1 | |
| 1918 | | | 10 02 49.9 | -48 54 48 | 277.05 | 5.16 | 213 | 25.2 | 58.0 | 13x 7 | 18.3 | .47 | | | | | |
| 1919 | AM1002 | | 10 02 50.4 | -48 29 50 | 276.80 | 5.49 | 213 | 25.6 | 80.3 | 40x 40 | 15.3 | .47 | | S 6 | | | |
| 1920 | | | 10 02 55.2 | -66 34 07 | 287.52 | -9.09 | 92 | -71.5 | -83.4 | 56x 56 | 15.0 | .23 | | L | | S | |
| 1921 | | | 10 03 01.1 | -64 25 20 | 286.24 | -7.35 | 92 | -77.2 | 31.5 | 23x 15 | 17.2 | .18 | | | | 1 | |
| 1922 | | | 10 03 04.2 | -64 28 00 | 286.27 | -7.39 | 92 | -76.8 | 29.1 | 16x 9 | 18.3 | .19 | | | | | cl.to br.* |
| 1923 | | | 10 03 05.3 | -49 32 51 | 277.46 | 4.67 | 213 | 27.1 | 23.9 | 20x 8 | 18.0 | .51 | | | | | br.blg, LSBdisk |
| 1924 | | | 10 03 10.4 | -52 14 02 | 279.06 | 2.51 | 213 | 25.9 | -120.1 | 16x 5 | 18.7 | .69 | | | | 2 | |
| 1925 | | | 10 03 18.5 | -67 31 14 | 288.12 | -9.83 | 92 | -66.7 | -134.2 | 22x 9 | 17.8 | .22 | | | | | |
| 1926 | | | 10 03 25.4 | -51 06 10 | 278.42 | 3.44 | 213 | 28.8 | -59.5 | 13x 8 | 18.4 | .53 | ? | S ? | | | br.blg, LSBdisk |
| 1927 | | | 10 03 29.4 | -65 32 45 | 286.95 | -8.23 | 92 | -71.3 | -28.5 | 16x 5 | 19.1 | .23 | | S ? | | | |
| 1928 | | | 10 03 36.7 | -49 34 46 | 277.55 | 4.69 | 213 | 31.6 | 22.1 | 13x 4 | 19.4 | .44 | | S ? | | | dist.nucl. |
| 1929 | | | 10 03 37.3 | -64 54 51 | 286.58 | -7.71 | 92 | -72.4 | 5.4 | 16x 3 | 19.9 | .25 | | S | E | | |
| 1930 | | | 10 03 37.6 | -49 09 06 | 277.30 | 5.04 | 213 | 32.1 | 45.1 | 13x 13 | 17.7 | .45 | | F | | | |
| 1931 | | | 10 03 40.3 | -62 30 04 | 285.15 | -5.76 | 127 | .0 | .0 | 23x 15 | 17.6 | .25 | | | | | LSB |
| 1932 | | | 10 03 40.5 | -48 43 23 | 277.05 | 5.39 | 213 | 32.8 | 68.1 | 16x 13 | 17.4 | .41 | | | | | p.cov.by br.* |
| 1933 | | | 10 03 41.3 | -62 40 20 | 285.26 | -5.90 | 92 | -78.1 | 125.4 | 20x 7 | 18.6 | .27 | | S | E | | |
| 1934 | | | 10 03 43.8 | -47 31 07 | 276.34 | 6.37 | 213 | 34.2 | 132.6 | 13x 7 | 18.6 | .22 | | | | | smooth |
| 1935 | | | 10 03 45.1 | -47 48 05 | 276.51 | 6.14 | 213 | 34.2 | 117.5 | 20x 7 | 17.9 | .30 | | F | | | |
| 1936 | | | 10 03 48.7 | -66 13 19 | 287.38 | -8.75 | 92 | -67.7 | -64.6 | 30x 12 | 17.1 | .19 | | E ? | | S | |
| 1937 | | | 10 03 51.1 | -67 37 51 | 288.23 | -9.89 | 92 | -63.6 | -140.0 | 12x 9 | 18.9 | .21 | | | | | LSB |
| 1938 | L062-002 | I | 10 03 59.3 | -67 30 14 | 288.17 | -9.78 | 92 | -63.3 | -133.2 | 60x 54 | 15.1 | .21 | | SX 3 : | | S | LSBdisk |
| 1939 | L213-006 | | 10 04 11.8 | -50 21 59 | 278.09 | 4.11 | 213 | 36.0 | -20.1 | 54x 16 | 16.0 | .44 | | I | | | |
| 1940 | | | 10 04 13.7 | -63 53 03 | 286.02 | -6.84 | 92 | -71.6 | 60.7 | 16x 8 | 18.6 | .23 | | I | | | |
| 1941 | | | 10 04 21.6 | -49 07 10 | 277.37 | 5.14 | 213 | 38.5 | 46.7 | 20x 7 | 18.0 | .38 | | S | E | | br.blg |
| 1942 | | I | 10 04 22.4 | -47 48 28 | 276.60 | 6.20 | 213 | 39.8 | 117.0 | 13x 8 | 18.6 | .30 | | I | | | |
| 1943 | L092-008 | | 10 04 30.8 | -67 08 02 | 287.99 | -9.45 | 92 | -61.5 | -113.2 | 215x161 | 12.3 | .24 | | E 3 | | S | |
| 1944 | | | 10 04 31.7 | -48 56 39 | 277.29 | 5.29 | 213 | 40.2 | 56.1 | 13x 7 | 18.3 | .32 | | S | E | | |
| 1945 | | | 10 04 34.2 | -49 59 53 | 277.92 | 4.44 | 213 | 39.5 | -.4 | 47x 40 | 15.1 | .44 | | SBR4 | | | |
| 1946 | | Q | 10 04 41.6 | -65 03 33 | 286.76 | -7.76 | 92 | -65.9 | -2.1 | 13x 4 | 19.1 | .27 | | S ? | | | w.comp. |
| 1947 | L092-009 | | 10 04 43.6 | -66 49 08 | 287.81 | -9.18 | 92 | -61.2 | -96.3 | 81x 13 | 16.3 | .24 | | I | | S | |
| 1948 | | | 10 04 46. | -62 08.7 : | 285.05 | -5.40 | 127 | -21.8 | -115.7 | 27x 7 | 18.2 | .29 | | S | | 1 | |
| 1949 | L263-004 | Q2 | 10 04 47.4 | -47 27 05 | 276.45 | 6.53 | 213 | 43.9 | 136.1 | 74x 47 | 14.5 | .20 | | SB 3 | | | |
| 1950 | | | 10 04 48.0 | -49 29 56 | 277.66 | 4.87 | 213 | 42.0 | 26.3 | 17x 16 | 17.3 | .37 | | SB 5 | F | | |
| 1951 | | | 10 04 48.9 | -49 53 43 | 277.89 | 4.55 | 213 | 40.2 | 6.4 | 16x 16 | 17.1 | .42 | | F | | | |
| 1952 | | | 10 04 48.9 | -62 01 12 | 284.98 | -5.29 | 127 | -20.2 | -108.2 | 17x 13 | 17.8 | .32 | | S P | | | |
| 1953 | | | 10 04 52.6 | -50 49 16 | 278.44 | 3.80 | 213 | 41.3 | -44.6 | 16x 7 | 17.9 | .40 | | E | | | cl.to br.* |
| 1954 | | Q2 | 10 04 59.4 | -47 28 58 | 276.49 | 6.52 | 213 | 45.7 | 134.4 | 20x 13 | 17.0 | .19 | | S | | | |
| 1955 | | Q | 10 05 01.5 | -49 42 05 | 277.80 | 4.73 | 213 | 43.7 | 15.4 | 13x 5 | 19.2 | .40 | | | | | |
| 1956 | | | 10 05 05.3 | -47 17 25 | 276.39 | 6.69 | 213 | 46.8 | 144.7 | 27x 9 | 17.5 | .26 | | S L : | | | |
| 1957 | | | 10 05 12.9 | -66 21 41 | 287.58 | -8.78 | 92 | -59.8 | -71.7 | 11x 7 | 19.3 | .21 | | | | | |
| 1958 | | | 10 05 14.3 | -48 52 18 | 277.35 | 5.42 | 213 | 46.5 | 59.9 | 16x 7 | 17.9 | .29 | | S 1 | | | br.blg |
| 1959 | | | 10 05 18.7 | -66 52 08 | 287.89 | -9.19 | 92 | -58.0 | -98.8 | 20x 20 | 17.6 | .23 | ? | | | S | LSB |
| 1960 | | | 10 05 22.8 | -62 02 07 | 285.04 | -5.27 | 127 | -16.6 | -109.0 | 16x 9 | 17.8 | .30 | | E 3 : | | | sharp edge |
| 1961 | | | 10 05 31.0 | -48 58 32 | 277.44 | 5.36 | 213 | 48.8 | 54.2 | 13x 11 | 18.2 | .30 | | | F | | |
| 1962 | | | 10 05 31.8 | -51 06 48 | 278.70 | 3.63 | 213 | 46.5 | -60.4 | 17x 15 | 17.4 | .55 | ? | | | | |
| 1963 | L092-010 | I | 10 05 32.4 | -64 07 06 | 286.28 | -6.95 | 92 | -63.4 | 48.5 | 74x 74 | 14.3 | .22 | | S 3 | | S | 92-EN010 |
| 1964 | | | 10 05 33.7 | -65 48 19 | 287.28 | -8.31 | 92 | -59.2 | -41.8 | 16x 13 | 18.1 | .23 | | S | | 1 | |
| 1965 | | | 10 05 38.5 | -48 50 47 | 277.38 | 5.48 | 213 | 50.0 | 61.1 | 13x 7 | 18.7 | .29 | | | | | |
| 1966 | | | 10 05 47.9 | -64 46 18 | 286.69 | -7.46 | 92 | -60.3 | 13.6 | 20x 7 | 18.2 | .23 | | S | | | |
| 1967 | | | 10 05 48.2 | -62 19 02 | 285.25 | -5.47 | 127 | -14.0 | -124.0 | 24x 4 | 18.6 | .25 | | S | E | | |
| 1968 | | | 10 05 50.3 | -47 40 29 | 276.72 | 6.45 | 213 | 53.1 | 123.9 | 13x 13 | 18.1 | .19 | | S | | | |
| 1969 | | | 10 05 53.2 | -49 25 12 | 277.75 | 5.04 | 213 | 51.5 | 30.3 | 16x 4 | 19.2 | .33 | | S 3 | | | |
| 1970 | | I | 10 05 54.0 | -62 49 34 | 285.55 | -5.88 | 92 | -64.2 | 117.8 | 20x 5 | 18.5 | .27 | ? | S | | S | or blended ** |
| 1971 | | | 10 05 55.9 | -47 35 58 | 276.69 | 6.52 | 213 | 54.1 | 127.9 | 13x 7 | 18.6 | .19 | ? | | | | |
| 1972 | | | 10 05 58.5 | -50 30 06 | 278.40 | 4.17 | 213 | 51.0 | -27.7 | 20x 9 | 17.6 | .43 | | I ? | | | dbl syst? |
| 1973 | | | 10 06 00.0 | -64 48 05 | 286.72 | -7.47 | 92 | -59.1 | 12.0 | 17x 7 | 18.6 | .23 | | | | | |
| 1974 | | | 10 06 03.0 | -47 24 12 | 276.59 | 6.69 | 213 | 55.4 | 138.4 | 27x 7 | 17.7 | .17 | | S ? | | | |
| 1975 | | | 10 06 07.6 | -47 56 14 | 276.92 | 6.27 | 213 | 55.4 | 109.8 | 13x 13 | 18.0 | .24 | | S 5 : | | | dist.nucl. |
| 1976 | | | 10 06 11.8 | -65 51 52 | 287.37 | -8.32 | 92 | -55.6 | -44.9 | 30x 16 | 17.2 | .22 | ? | I | | | or HII-reg. |
| 1977 | | I | 10 06 19.2 | -48 58 24 | 277.55 | 5.44 | 213 | 55.9 | 54.2 | 20x 13 | 17.2 | .29 | ? | | | | dbl syst, w.comp. |
| 1978 | | | 10 06 23.4 | -48 23 17 | 277.22 | 5.93 | 213 | 57.2 | 85.5 | 20x 13 | 17.5 | .27 | ? | | | | |
| 1979 | | | 10 06 30.7 | -64 44 18 | 286.73 | -7.39 | 92 | -56.3 | 15.5 | 27x 7 | 18.2 | .23 | | S | E | 1 | pair w.1981 |
| 1980 | | | 10 06 32.4 | -66 09 36 | 287.57 | -8.54 | 92 | -53.1 | -60.6 | 13x 4 | 19.6 | .19 | | S | N | | |
| 1981 | | | 10 06 32.5 | -64 45 17 | 286.74 | -7.40 | 92 | -56.1 | 14.6 | 34x 27 | 16.6 | .23 | | S | F | S | pair w.1979 |
| 1982 | | | 10 06 34.9 | -47 46 08 | 276.88 | 6.45 | 213 | 59.7 | 118.7 | 27x 3 | 18.9 | .21 | | S | E | | |
| 1983 | | | 10 06 39.9 | -64 58 26 | 286.88 | -7.57 | 92 | -55.0 | 2.9 | 17x 9 | 18.8 | .27 | | | | | |
| 1984 | | | 10 06 42.6 | -49 44 32 | 278.05 | 4.85 | 213 | 58.3 | 12.9 | 11x 5 | 19.1 | .36 | | | | | trpl syst.w.1990,1991 |
| 1985 | | | 10 06 44.2 | -48 35 02 | 277.38 | 5.80 | 213 | 60.1 | 75.0 | 40x 7 | 17.3 | .28 | | S 3 : | | S | |
| 1986 | | | 10 06 45.5 | -65 36 43 | 287.26 | -8.08 | 92 | -53.1 | -31.2 | 19x 3 | 19.9 | .24 | | S | E | | |
| 1987 | | | 10 06 46.4 | -47 35 06 | 276.80 | 6.62 | 213 | 61.7 | 128.5 | 16x 3 | 19.3 | .19 | | S | E | | blg or s.p.* |
| 1988 | | I | 10 06 49.2 | -64 43 03 | 286.74 | -7.35 | 92 | -54.6 | 16.7 | 15x 13 | 17.8 | .22 | | S E | | | |
| 1989 | | | 10 06 49.3 | -63 17 36 | 285.91 | -6.19 | 92 | -57.6 | 93.0 | 20x 7 | 18.8 | .36 | | S ? | | | |
| 1990 | | | 10 06 49.7 | -49 44 28 | 278.07 | 4.86 | 213 | 59.3 | 12.9 | 15x 8 | 18.3 | .35 | | S 3 : | | | trpl syst.w.1984,1991 |
| 1991 | | | 10 06 50.4 | -49 45 13 | 278.07 | 4.86 | 213 | 59.4 | 12.3 | 13x 5 | 18.9 | .35 | | S 1 | | | trpl syst.w.1984,1990 |
| 1992 | | | 10 06 57.3 | -62 46 17 | 285.62 | -5.76 | 92 | -57.8 | 121.0 | 13x 7 | 18.9 | .29 | | | | | |
| 1993 | | | 10 06 57.4 | -50 45 24 | 278.67 | 4.05 | 213 | 59.0 | -41.5 | 17x 7 | 18.1 | .37 | | S | | | |
| 1994 | | | 10 07 05.9 | -64 49 27 | 286.83 | -7.42 | 92 | -52.8 | 11.1 | 16x 13 | 17.7 | .23 | | E | | | dbl syst.w.1996 |
| 1995 | | | 10 07 06.5 | -61 11 34 | 284.72 | -4.46 | 127 | -5.7 | -63.8 | 34x 20 | 16.3 | .37 | ? | S ? | | | or blended ** |
| 1996 | | | 10 07 07.8 | -64 49 16 | 286.83 | -7.42 | 92 | -52.6 | 11.2 | 13x 9 | 18.4 | .23 | | E | | | dbl syst.w.1994 |
| 1997 | | | 10 07 21.3 | -66 27 20 | 287.81 | -8.73 | 92 | -48.1 | -76.3 | 23x 20 | 17.1 | .19 | | S M | | S | |
| 1998 | | | 10 07 24.5 | -49 48 27 | 278.18 | 4.86 | 213 | 64.3 | 9.2 | 13x 13 | 18.3 | .35 | | S ? | | | |
| 1999 | | | 10 07 27.0 | -48 21 18 | 277.34 | 6.05 | 213 | 66.7 | 87.1 | 20x 7 | 17.9 | .29 | | S 0 | | | |
| 2000 | L092-012 | I | 10 07 30.2 | -66 47 02 | 288.02 | -8.99 | 92 | -46.7 | -93.8 | 195x 74 | 13.1 | .20 | | SB 3 : | | S | or Im? or i.a.dbl syst? |

**Table 1.** Galaxies in the southern ZOA – I. The Hydra/Antlia Extension (continued)

| RKK Ident (1) | Other Ident (2) | IR (3) | R.A. (h m s) (4) | Dec. (° ′ ″) (5) | gal ℓ (°) (6) | gal $b$ (°) (7) | SRC (8) | X (mm) (9) | Y (mm) (10) | D x d (″) (11) | $B_J$ ($^m$) (12) | $N_{HI}$ (13) | u | Type class. (14) | o * | Remarks (15) |
|---|---|---|---|---|---|---|---|---|---|---|---|---|---|---|---|---|
| 2001 | | | 10 07 32.9 | -48 24 08 | 277.38 | 6.03 | 213 | 67.5 | 84.5 | 13x 3 | 19.7 | .28 | | S | E | |
| 2002 | | | 10 07 41.7 | -51 55 09 | 279.44 | 3.17 | 213 | 63.5 | -104.0 | 23x 5 | 18.3 | .84 | | S | E | |
| 2003 | | | 10 07 42.8 | -49 28 22 | 278.03 | 5.17 | 213 | 67.4 | 27.1 | 17x 3 | 19.0 | .27 | | S | E | |
| 2004 | | | 10 07 43.3 | -48 46 10 | 277.62 | 5.74 | 213 | 68.5 | 64.8 | 13x 5 | 18.9 | .26 | | S | | dist.nucl. |
| 2005 | | | 10 07 43.5 | -64 45 12 | 286.84 | -7.33 | 92 | -49.4 | 15.0 | 15x 5 | 18.8 | .22 | | S ? | | |
| 2006 | | | 10 07 43.6 | -48 44 31 | 277.60 | 5.76 | 213 | 68.6 | 66.3 | 47x 13 | 16.7 | .26 | | S 3 : | | |
| 2007 | | | 10 07 44.0 | -50 41 28 | 278.73 | 4.17 | 213 | 65.7 | -38.2 | 27x 11 | 17.2 | .37 | | I | | |
| 2008 | | | 10 07 54.8 | -48 59 27 | 277.77 | 5.58 | 213 | 69.9 | 52.9 | 16x 11 | 17.8 | .25 | | SB 5 | | |
| 2009 | | | 10 07 58.0 | -48 46 26 | 277.66 | 5.76 | 213 | 70.7 | 64.5 | 17x 9 | 18.1 | .24 | | S 0 | | |
| 2010 | | | 10 08 02.8 | -64 54 42 | 286.97 | -7.44 | 92 | -47.2 | 6.6 | 20x 7 | 18.3 | .26 | | S | | |
| 2011 | | | 10 08 04.0 | -47 14 09 | 276.78 | 7.03 | 213 | 73.9 | 146.9 | 19x 19 | 16.8 | .15 | ? | S 3 : | | dbl syst? |
| 2012 | | | 10 08 09.0 | -63 51 33 | 286.36 | -6.57 | 92 | -48.6 | 63.0 | 13x 13 | 18.7 | .25 | ? | | | |
| 2013 | | | 10 08 05. | -49 44.0 : | 278.23 | 4.99 | 213 | 70.1 | 13.0 | 16x 7 | 18.1 | .30 | | S 5 : | | dist.nucl. |
| 2014 | | | 10 08 11.8 | -52 27 21 | 279.81 | 2.77 | 213 | 66.8 | -132.9 | 20x 13 | 17.9 | .74 | | | | |
| 2015 | | | 10 08 15.0 | -62 23 25 | 285.52 | -5.36 | 127 | 1.3 | -128.0 | 20x 13 | 17.4 | .27 | | S | | |
| 2016 | | | 10 08 15.2 | -49 45 15 | 278.26 | 4.99 | 213 | 71.6 | 11.9 | 13x 8 | 18.4 | .31 | | | | |
| 2017 | HEN373 | I | 10 08 16.4 | -56 47 07 | 282.30 | -.77 | 167 | 105.1 | -98.1 | 67x 27 | 15.8 | .30 | ? | S ? | | vLSB, diff, gas or neb? |
| 2018 | | | 10 08 25.1 | -49 56 06 | 278.39 | 4.85 | 213 | 72.8 | 2.2 | 16x 11 | 18.1 | .38 | | S 5 : | S | dist.nucl, dist.sp.arms |
| 2019 | | | 10 08 31.9 | -63 13 31 | 286.03 | -6.03 | 92 | -47.4 | 97.0 | 24x 4 | 19.3 | .39 | | S | E | |
| 2020 | | | 10 08 36.7 | -64 40 36 | 286.88 | -7.21 | 92 | -44.4 | 19.3 | 13x 8 | 18.7 | .23 | ? | | | br.blg or s.p.* |
| 2021 | | | 10 08 36.7 | -67 29 16 | 288.52 | -9.50 | 92 | -39.6 | -131.3 | 22x 7 | 18.4 | .21 | | S ? | S | |
| 2022 | | | 10 08 42.0 | -50 25 13 | 278.70 | 4.48 | 213 | 74.4 | -23.9 | 13x 7 | 18.6 | .36 | | S 1 : | | br.blg |
| 2023 | | | 10 08 42.1 | -53 41 20 | 280.58 | 1.80 | 167 | 116.6 | 67.6 | 20x 20 | 17.6 | .21 | ? | S | | vLSB |
| 2024 | | | 10 08 43.5 | -50 24 41 | 278.70 | 4.49 | 213 | 74.6 | -23.5 | 20x 16 | 17.3 | .36 | | E ? | | |
| 2025 | | | 10 08 51.8 | -64 52 31 | 287.02 | -7.36 | 92 | -42.7 | 8.7 | 13x 11 | 18.1 | .25 | ? | E | | |
| 2026 | L092-013 | | 10 08 51.9 | -66 45 30 | 288.11 | -8.89 | 92 | -39.5 | -92.2 | 98x 81 | 13.8 | .20 | | E 4 | S | |
| 2027 | | | 10 08 53.9 | -47 18 46 | 276.94 | 7.05 | 213 | 81.4 | 142.6 | 19x 11 | 17.6 | .14 | | S 2 | | w.comp. |
| 2028 | | | 10 08 54.1 | -64 43 40 | 286.93 | -7.23 | 92 | -42.7 | 16.6 | 13x 13 | 18.1 | .24 | ? | E | 1 | |
| 2029 | | | 10 08 57.0 | -65 05 22 | 287.15 | -7.53 | 92 | -41.8 | -2.8 | 19x 9 | 17.8 | .30 | | | | |
| 2030 | | | 10 08 58.9 | -48 53 25 | 277.86 | 5.76 | 213 | 79.4 | 58.0 | 8x 8 | 18.9 | .23 | | F | | |
| 2031 | | | 10 09 01.6 | -48 14 08 | 277.49 | 6.30 | 213 | 81.0 | 93.1 | 22x 13 | 17.3 | .26 | | SX 3 | | |
| 2032 | | | 10 09 07.4 | -48 53 26 | 277.88 | 5.78 | 213 | 80.7 | 57.9 | 34x 13 | 16.8 | .23 | | S ? | | |
| 2033 | | | 10 09 08.6 | -48 52 51 | 277.88 | 5.79 | 213 | 80.9 | 58.4 | 20x 17 | 17.1 | .23 | | L | | |
| 2034 | | | 10 09 11.2 | -48 11 16 | 277.48 | 6.36 | 213 | 82.5 | 95.6 | 20x 8 | 18.0 | .24 | | S E: | | |
| 2035 | | | 10 09 12.9 | -66 57 44 | 288.26 | -9.04 | 92 | -37.3 | -103.1 | 19x 8 | 18.3 | .20 | ? | S ? | S | |
| 2036 | | | 10 09 15.4 | -65 22 05 | 287.34 | -7.74 | 92 | -39.7 | -17.7 | 31x 27 | 16.7 | .28 | | E | | dwarf? |
| 2037 | | Q | 10 09 17.5 | -48 14 11 | 277.53 | 6.33 | 213 | 83.3 | 92.9 | 30x 11 | 17.3 | .25 | | E ? | | smooth |
| 2038 | | | 10 09 26.8 | -67 32 35 | 288.62 | -9.50 | 92 | -35.2 | -134.2 | 12x 12 | 18.4 | .19 | ? | S ? | F | |
| 2039 | | | 10 09 27.0 | -64 56 15 | 287.10 | -7.37 | 92 | -39.2 | 5.4 | 40x 5 | 18.0 | .26 | | S | ES | |
| 2040 | L092-014 | | 10 09 29.4 | -66 24 02 | 287.96 | -8.57 | 92 | -36.8 | -73.0 | 87x 51 | 14.2 | .21 | | F | S | |
| 2041 | | | 10 09 35.9 | -49 12 56 | 278.13 | 5.55 | 213 | 84.3 | 40.4 | 27x 13 | 17.2 | .27 | | I | | smooth |
| 2042 | | | 10 09 37.7 | -48 56 38 | 277.98 | 5.78 | 213 | 85.0 | 54.9 | 13x 13 | 17.9 | .23 | | | | smooth |
| 2043 | L092-015 | I | 10 09 39.4 | -66 53 50 | 288.26 | -8.96 | 92 | -35.1 | -99.5 | 60x 47 | 15.0 | .19 | | SX 3 | S | |
| 2044 | | | 10 09 42.4 | -47 21 09 | 277.07 | 7.09 | 213 | 88.6 | 140.2 | 16x 4 | 19.4 | .16 | | S | E | smooth |
| 2045 | | | 10 09 48.0 | -64 25 40 | 286.84 | -6.93 | 92 | -38.0 | 32.8 | 20x 3 | 19.3 | .23 | | S | E | |
| 2046 | | | 10 09 51.1 | -67 02 00 | 288.35 | -9.06 | 92 | -33.9 | -106.8 | 17x 3 | 19.6 | .20 | | S | E1 | |
| 2047 | | | 10 09 56.3 | -48 26 01 | 277.73 | 6.23 | 213 | 88.7 | 82.2 | 13x 13 | 17.9 | .22 | | S M | | |
| 2048 | | | 10 09 58.4 | -64 06 04 | 286.67 | -6.65 | 92 | -37.5 | 50.3 | 17x 4 | 19.4 | .26 | | S | ES | |
| 2049 | | | 10 10 09.3 | -48 46 38 | 277.95 | 5.97 | 213 | 90.0 | 63.7 | 13x 7 | 18.6 | .21 | ? | I ? | | |
| 2050 | | | 10 10 10.7 | -50 30 28 | 278.95 | 4.55 | 213 | 86.8 | -29.0 | 13x 13 | 18.0 | .36 | | | | br.blg |
| 2051 | | | 10 10 16.8 | -51 04 59 | 279.29 | 4.08 | 213 | 86.6 | -59.9 | 16x 5 | 18.9 | .44 | | | | smooth |
| 2052 | | | 10 10 19.5 | -64 55 06 | 287.17 | -7.30 | 92 | -34.3 | 6.6 | 16x 11 | 18.2 | .25 | | S ? | S | dbl syst. |
| 2053 | | | 10 10 21.5 | -50 30 14 | 278.97 | 4.56 | 213 | 88.4 | -28.9 | 13x 7 | 18.6 | .35 | | I | | |
| 2054 | | | 10 10 21.7 | -47 40 34 | 277.35 | 6.89 | 213 | 94.0 | 122.7 | 13x 7 | 18.1 | .18 | | | | |
| 2055 | | | 10 10 23.7 | -67 10 52 | 288.48 | -9.15 | 92 | -30.9 | -114.6 | 19x 16 | 17.7 | .20 | | S | N | |
| 2056 | | | 10 10 26.8 | -65 12 53 | 287.35 | -7.54 | 92 | -33.2 | -9.3 | 13x 9 | 18.8 | .29 | | S ? | 1 | |
| 2057 | | | 10 10 27.2 | -63 52 36 | 286.58 | -6.44 | 92 | -34.9 | 62.4 | 40x 12 | 17.1 | .27 | | S M | S | dbl syst? |
| 2058 | | | 10 10 30. | -53 00.8 : | 280.41 | 2.51 | 167 | 131.0 | 103.1 | 20x 5 | 18.2 | .63 | | S M | | |
| 2059 | | | 10 10 31.6 | -65 32 12 | 287.54 | -7.80 | 92 | -32.3 | -26.5 | 16x 9 | 18.1 | .25 | | E ? | | |
| 2060 | | | 10 10 35.3 | -49 35 58 | 278.48 | 5.33 | 213 | 92.2 | 19.5 | 13x 4 | 19.4 | .26 | ? | S ? | E | |
| 2061 | | | 10 10 35.4 | -63 58 24 | 286.65 | -6.51 | 92 | -34.0 | 57.3 | 20x 5 | 18.9 | .25 | | S ? | | |
| 2062 | | | 10 10 36.2 | -52 45 24 | 280.28 | 2.73 | 167 | 134.5 | 116.7 | 13x 11 | 17.9 | .69 | | F | | |
| 2063 | L127-011 | I | 10 10 36.5 | -62 17 11 | 285.69 | -5.12 | 127 | 15.9 | -122.5 | 67x 27 | 15.1 | .29 | ? | S 3 | | plate flaw? |
| 2064 | | | 10 10 36.6 | -51 22 24 | 279.50 | 3.87 | 213 | 88.7 | -75.6 | 34x 16 | 16.7 | .47 | ? | S ? | S | |
| 2065 | | | 10 10 41.3 | -50 29 46 | 279.01 | 4.60 | 213 | 91.2 | -28.6 | 27x 5 | 18.1 | .34 | | S 4 | | |
| 2066 | | | 10 10 41.9 | -52 28 50 | 280.14 | 2.97 | 167 | 136.1 | 131.4 | 13x 7 | 19.2 | .82 | | | | |
| 2067 | | | 10 10 44.7 | -49 07 24 | 278.23 | 5.74 | 213 | 94.5 | 45.0 | 13x 9 | 18.3 | .26 | | F ? | | |
| 2068 | | | 10 10 45.2 | -49 37 14 | 278.52 | 5.33 | 213 | 93.6 | 18.3 | 27x 17 | 16.6 | .26 | ? | S L | | |
| 2069 | | | 10 10 45.6 | -48 39 38 | 277.97 | 6.12 | 213 | 95.6 | 69.7 | 13x 7 | 18.6 | .22 | | S L ? | | |
| 2070 | | | 10 10 45.9 | -50 32 19 | 279.04 | 4.57 | 213 | 90.6 | -30.4 | 11x 9 | 18.8 | .31 | | | 1 | |
| 2071 | | | 10 10 46.5 | -50 31 01 | 279.03 | 4.59 | 213 | 91.9 | -29.7 | 20x 7 | 17.9 | .32 | | S 1 | 1 | |
| 2072 | | | 10 10 47.6 | -65 00 38 | 287.26 | -7.35 | 92 | -31.5 | 1.7 | 27x 23 | 16.7 | .27 | | S M | F | |
| 2073 | | | 10 10 49. | -62 08.8 : | 285.63 | -4.99 | 127 | 17.4 | -115.1 | 16x 8 | 18.2 | .29 | | S M | | |
| 2074 | | | 10 10 49.3 | -63 10 23 | 286.21 | -5.84 | 92 | -33.6 | 100.2 | 13x 8 | 18.7 | .38 | | | | |
| 2075 | | | 10 10 53.1 | -47 31 56 | 277.34 | 7.06 | 213 | 99.0 | 130.2 | 13x 5 | 19.2 | .17 | | S | | dbl syst. of S..L |
| 2076 | | | 10 10 55.4 | -50 24 56 | 278.99 | 4.69 | 213 | 93.4 | -24.3 | 20x 5 | 18.5 | .33 | | S L | | |
| 2077 | | | 10 10 55.9 | -49 46 22 | 278.63 | 5.22 | 213 | 94.8 | 10.1 | 27x 8 | 17.8 | .28 | ? | I | S | |
| 2078 | | | 10 10 58.2 | -50 16 31 | 278.92 | 4.81 | 213 | 94.1 | -16.9 | 13x 4 | 19.4 | .31 | ? | I | | |
| 2079 | | | 10 10 58.2 | -50 30 57 | 279.06 | 4.61 | 213 | 93.6 | -29.7 | 27x 13 | 17.1 | .31 | | F | | |
| 2080 | | | 10 11 00.4 | -48 08 52 | 277.71 | 6.56 | 213 | 98.8 | 97.1 | 20x 13 | 17.4 | .21 | | F | | |
| 2081 | | I | 10 11 08.4 | -63 46 59 | 286.59 | -6.32 | 92 | -31.0 | 67.5 | 34x 11 | 16.9 | .29 | | S 0 | | |
| 2082 | | | 10 11 13.4 | -48 25 55 | 277.90 | 6.35 | 213 | 100.1 | 81.8 | 13x 5 | 19.0 | .21 | | I ? | | br.knots, LSBdisk |
| 2083 | | | 10 11 16.2 | -47 40 41 | 277.48 | 6.98 | 213 | 102.2 | 122.2 | 27x 13 | 16.7 | .18 | | S | | dbl syst.w.2084 |
| 2084 | | | 10 11 16.3 | -47 40 23 | 277.48 | 6.98 | 213 | 102.2 | 122.5 | 26x 3 | 18.7 | .18 | | S | E | dist.nucl, pair w.2083 |
| 2085 | | | 10 11 18.2 | -67 14 55 | 288.60 | -9.16 | 92 | -26.0 | -118.1 | 17x 7 | 18.3 | .22 | | S ? | S | |
| 2086 | | | 10 11 21.3 | -49 41 14 | 278.64 | 5.33 | 213 | 98.6 | 14.5 | 20x 7 | 18.1 | .27 | | S | E | |
| 2087 | | | 10 11 23.8 | -47 57 31 | 277.66 | 6.76 | 213 | 102.7 | 107.1 | 13x 9 | 18.1 | .19 | | S ? | | |
| 2088 | | | 10 11 24.4 | -50 18 47 | 279.00 | 4.82 | 213 | 97.7 | -19.0 | 17x 17 | 17.3 | .29 | | F ? | | |
| 2089 | | | 10 11 24.4 | -64 15 59 | 286.89 | -6.70 | 92 | -28.9 | 41.7 | 17x 9 | 18.2 | .25 | | S E | | |
| 2090 | | | 10 11 27.6 | -49 24 15 | 278.49 | 5.57 | 213 | 100.2 | 29.7 | 13x 5 | 18.7 | .24 | | F | | |
| 2091 | | | 10 11 28.3 | -52 12 08 | 280.08 | 3.26 | 168 | -127.9 | 146.3 | 13x 11 | 18.3 | .54 | ? | S 5 | | |
| 2092 | | | 10 11 30.3 | -50 27 01 | 279.09 | 4.71 | 213 | 98.3 | -26.4 | 20x 4 | 18.7 | .29 | | S | S | pair w.2094 |
| 2093 | | | 10 11 30.4 | -63 35 24 | 286.51 | -6.14 | 92 | -29.0 | 77.9 | 13x 8 | 18.4 | .33 | ? | | | |
| 2094 | | | 10 11 30.6 | -50 27 51 | 279.10 | 4.70 | 213 | 98.3 | -27.2 | 17x 7 | 18.3 | .29 | | | | pair w.2092,br.blg or s.p. |
| 2095 | | I | 10 11 31.0 | -52 16 27 | 280.12 | 3.21 | 168 | -127.3 | 142.5 | 40x 20 | 16.0 | .54 | ? | S 3 | | or gas |
| 2096 | | | 10 11 33.5 | -48 33 59 | 278.03 | 6.27 | 213 | 102.8 | 74.5 | 15x 4 | 19.0 | .19 | | S ? | E | |
| 2097 | | | 10 11 33.6 | -62 01 58 | 285.63 | -4.85 | 127 | 22.2 | -109.1 | 40x 24 | 15.6 | .35 | ? | | | cl.to br.* |
| 2098 | | | 10 11 34.0 | -50 01 21 | 278.86 | 5.07 | 213 | 99.7 | -3.5 | 16x 11 | 18.1 | .32 | | I | | |
| 2099 | | | 10 11 35.6 | -50 39 14 | 279.22 | 4.55 | 213 | 97.4 | -36.9 | 15x 13 | 17.8 | .27 | | | 1 | |
| 2100 | | | 10 11 38.1 | -51 06 34 | 279.48 | 4.18 | 213 | 97.9 | -61.8 | 16x 5 | 18.7 | .36 | | S | | |

**Table 1.** Galaxies in the southern ZOA – I. The Hydra/Antlia Extension (continued)

| RKK Ident (1) | Other Ident (2) | IR (3) | R.A. (h m s) (4) | Dec. (° ′ ″) (5) | gal ℓ (°) (6) | gal b (°) (7) | SRC (8) | X (mm) (9) | Y (mm) (10) | D x d (″) (11) | $B_J$ ($^m$) (12) | $N_{HI}$ (13) | u | Type class. (14) | o * | Remarks (15) |
|---|---|---|---|---|---|---|---|---|---|---|---|---|---|---|---|---|
| 2101 | | | 10 11 42.1 | -50 34 14 | 279.18 | 4.63 | 213 | 98.5 | -32.5 | 15x 9 | 18.1 | .28 | | S E? | S | |
| 2102 | | | 10 11 43.5 | -50 37 21 | 279.22 | 4.59 | 213 | 99.1 | -33.0 | 20x 7 | 18.0 | .27 | | S E | E | |
| 2103 | L213-008 | | 10 11 44.8 | -49 48 17 | 278.76 | 5.27 | 213 | 101.8 | 8.1 | 34x 27 | 15.8 | .29 | | SB 3 | | |
| 2104 | | Q | 10 11 45.4 | -50 41 33 | 279.26 | 4.54 | 213 | 99.9 | -39.5 | 20x 16 | 17.2 | .26 | | L | | |
| 2105 | | | 10 11 49.6 | -49 24 36 | 278.54 | 5.60 | 213 | 103.4 | 29.2 | 20x 20 | 17.1 | .23 | | E ? | | |
| 2106 | | | 10 11 49.6 | -50 31 24 | 279.17 | 4.68 | 213 | 100.8 | -30.4 | 38x 15 | 16.1 | .28 | | S 3 : | | |
| 2107 | | | 10 11 50.4 | -50 15 46 | 279.03 | 4.90 | 213 | 100.5 | -16.0 | 12x 8 | 18.7 | .30 | | | | |
| 2108 | | | 10 11 50.8 | -48 41 01 | 278.13 | 6.20 | 213 | 105.1 | 68.1 | 40x 16 | 16.0 | .19 | | SX | | |
| 2109 | L213-009 | I | 10 11 50.9 | -52 02 16 | 280.03 | 3.43 | 213 | 97.6 | -111.6 | 47x 40 | 15.3 | .49 | | SB 3 | | |
| 2110 | | | 10 11 53.3 | -61 55 56 | 285.61 | -4.74 | 127 | 24.4 | -103.8 | 27x 16 | 17.1 | .38 | | S | | larger LSBdisk? |
| 2111 | | | 10 11 53.6 | -49 25 27 | 278.56 | 5.59 | 213 | 103.9 | 28.4 | 40x 8 | 16.9 | .23 | | S M | | |
| 2112 | | | 10 11 54.4 | -67 41 33 | 288.90 | -9.49 | 92 | -22.5 | -141.9 | 23x 8 | 18.1 | .21 | | S E | | |
| 2113 | | | 10 11 57.0 | -50 00 40 | 278.90 | 5.12 | 213 | 103.1 | -3.0 | 16x 7 | 18.1 | .32 | | F | | |
| 2114 | | | 10 12 04.7 | -51 30 41 | 279.76 | 3.89 | 213 | 100.7 | -83.5 | 34x 9 | 17.0 | .45 | | S 3 : | | |
| 2115 | | | 10 12 05.1 | -50 30 15 | 279.20 | 4.72 | 213 | 103.1 | -29.5 | 16x 4 | 18.9 | .29 | | S | E | |
| 2116 | | | 10 12 05.4 | -67 00 45 | 288.52 | -8.92 | 92 | -22.2 | -105.4 | 16x 15 | 18.1 | .20 | | | | |
| 2117 | | | 10 12 06.6 | -50 32 55 | 279.22 | 4.69 | 213 | 103.2 | -31.9 | 13x 5 | 18.9 | .28 | | | | dbl syst. |
| 2118 | | | 10 12 06.7 | -62 06 10 | 285.73 | -4.87 | 127 | 25.6 | -112.9 | 54x 5 | 17.6 | .35 | | S 5 | ES | |
| 2119 | | | 10 12 09.0 | -49 25 35 | 278.60 | 5.62 | 213 | 106.1 | 28.2 | 20x 16 | 17.0 | .24 | | S M | | |
| 2120 | | | 10 12 11.1 | -50 26 01 | 279.17 | 4.79 | 213 | 104.1 | -25.7 | 16x 16 | 17.7 | .31 | | S | | dist.nucl, LSBdisk, dwarf? |
| 2121 | | | 10 12 19.9 | -48 22 32 | 278.02 | 6.50 | 213 | 110.1 | 84.4 | 17x 7 | 18.5 | .21 | | S L | | dist.nucl. |
| 2122 | | | 10 12 21.2 | -50 08 11 | 279.02 | 5.05 | 213 | 106.2 | -9.9 | 13x 7 | 18.8 | .33 | | S ? | S | |
| 2123 | | | 10 12 23.1 | -62 47 14 | 286.14 | -5.42 | 127 | 26.5 | -149.6 | 20x 7 | 18.4 | .40 | | S M | | dist.blg or s.p.* |
| 2124 | | | 10 12 26.6 | -48 24 24 | 278.06 | 6.49 | 213 | 111.1 | 82.7 | 13x 7 | 18.2 | .21 | | E ? | | |
| 2125 | | | 10 12 26.8 | -50 41 26 | 279.35 | 4.60 | 213 | 104.5 | -39.2 | 11x 5 | 19.3 | .26 | ? | | | or blended ** |
| 2126 | | | 10 12 28.9 | -50 14 43 | 279.10 | 4.97 | 213 | 107.1 | -15.8 | 13x 9 | 18.1 | .31 | | E ? | | |
| 2127 | | | 10 12 31.0 | -50 49 17 | 279.43 | 4.50 | 213 | 106.0 | -46.7 | 16x 16 | 17.4 | .29 | | F ? | | |
| 2128 | | | 10 12 31.5 | -63 17 21 | 286.44 | -5.82 | 92 | -23.2 | 94.2 | 15x 4 | 19.2 | .36 | | S | N | |
| 2129 | | | 10 12 31.9 | -50 06 54 | 279.04 | 5.08 | 213 | 107.8 | -8.8 | 16x 13 | 17.7 | .34 | | SB 3 : | | |
| 2130 | | | 10 12 33.4 | -49 16 05 | 278.56 | 5.78 | 213 | 110.0 | 36.5 | 17x 13 | 17.4 | .20 | | S L | | |
| 2131 | | | 10 12 34.7 | -64 56 05 | 287.38 | -7.18 | 92 | -21.5 | 6.0 | 13x 11 | 18.3 | .28 | ? | E ? | | w.comp. |
| 2132 | | | 10 12 35.6 | -49 22 09 | 278.62 | 5.70 | 213 | 110.1 | 31.1 | 24x 8 | 17.5 | .22 | | S ? | | |
| 2133 | | | 10 12 36.6 | -67 28 29 | 288.83 | -9.27 | 92 | -19.1 | -130.1 | 13x 5 | 19.4 | .21 | | S ? | | |
| 2134 | | | 10 12 37.7 | -50 03 36 | 279.02 | 5.14 | 213 | 108.8 | -5.9 | 13x 5 | 18.5 | .35 | | S 1 | | |
| 2135 | | | 10 12 40.0 | -49 12 31 | 278.54 | 5.84 | 213 | 111.1 | 39.7 | 13x 7 | 18.6 | .21 | | | | |
| 2136 | | | 10 12 40.5 | -65 15 17 | 287.57 | -7.44 | 92 | -20.7 | -11.2 | 74x 22 | 16.0 | .28 | | S M | S | LSB |
| 2137 | | | 10 12 41.1 | -49 57 15 | 278.97 | 5.23 | 213 | 109.5 | -.3 | 27x 8 | 17.4 | .35 | | S L? | E | |
| 2138 | | Q | 10 12 45.8 | -52 21 42 | 280.33 | 3.24 | 213 | 104.3 | -129.3 | 13x 5 | 19.2 | .50 | | S | | |
| 2139 | | | 10 12 48.3 | -61 13 09 | 285.30 | -4.09 | 127 | 31.0 | -65.7 | 20x 15 | 17.5 | .43 | | S ? | | |
| 2140 | L092-017 | | 10 12 48.4 | -64 39 25 | 287.24 | -6.94 | 92 | -20.4 | 20.9 | 74x 40 | 15.3 | .26 | | S 1 | S | |
| 2141 | | | 10 12 49.1 | -49 17 28 | 278.61 | 5.79 | 213 | 112.3 | 35.2 | 16x 7 | 18.5 | .20 | ? | S | | br.blg |
| 2142 | | | 10 12 52.4 | -47 28 55 | 277.59 | 7.29 | 213 | 117.1 | 132.1 | 40x 20 | 16.2 | .19 | | SB 4 | | LSBdisk |
| 2143 | | | 10 12 55.0 | -49 14 30 | 278.59 | 5.84 | 213 | 113.2 | 37.8 | 27x 20 | 16.6 | .20 | ? | F ? | | or blended ** |
| 2144 | | | 10 12 55.7 | -67 12 37 | 288.70 | -9.04 | 92 | -17.7 | -115.9 | 17x 9 | 18.8 | .21 | | S | S | LSB |
| 2145 | | | 10 12 57.0 | -65 53 10 | 287.95 | -7.94 | 92 | -18.6 | -45.0 | 13x 12 | 17.9 | .25 | | E 1? | | |
| 2146 | | | 10 12 58.5 | -65 52 17 | 287.94 | -7.93 | 92 | -18.5 | -44.2 | 13x 4 | 19.2 | .25 | | S | | |
| 2147 | | | 10 13 04.2 | -49 20 14 | 278.67 | 5.77 | 213 | 114.4 | 32.6 | 13x 5 | 18.7 | .22 | | F | | |
| 2148 | | | 10 13 06.1 | -50 35 01 | 279.38 | 4.75 | 213 | 111.6 | -34.1 | 34x 8 | 17.4 | .30 | | S 3 | E | dist.dust lane |
| 2149 | | | 10 13 06.8 | -48 38 58 | 278.29 | 6.35 | 213 | 116.4 | 69.5 | 19x 7 | 18.2 | .19 | | S M | | |
| 2150 | | | 10 13 08.1 | -48 57 42 | 278.47 | 6.09 | 213 | 115.9 | 52.7 | 17x 9 | 17.8 | .21 | | E ? | | w.comp. |
| 2151 | | | 10 13 11.7 | -48 47 57 | 278.38 | 6.23 | 213 | 116.8 | 61.4 | 20x 13 | 17.2 | .21 | | S L | | or i.a.dbl syst. |
| 2152 | | | 10 13 12.9 | -51 57 37 | 280.16 | 3.61 | 213 | 109.0 | -107.9 | 13x 7 | 18.7 | .47 | | | | smooth |
| 2153 | | | 10 13 16.4 | -48 58 54 | 278.50 | 6.09 | 213 | 117.0 | 51.6 | 20x 20 | 16.7 | .21 | ? | S | | p.cov.by br.* |
| 2154 | | | 10 13 21.2 | -62 07 08 | 285.86 | -4.80 | 127 | 33.3 | -114.0 | 40x 34 | 16.2 | .37 | | S | FS | |
| 2155 | | | 10 13 23.4 | -49 00 39 | 278.53 | 6.07 | 213 | 117.3 | 50.4 | 16x 11 | 18.4 | .21 | | S L? | | LSB |
| 2156 | | P | 10 13 23.6 | -49 53 22 | 279.02 | 5.35 | 213 | 115.8 | 2.9 | 20x 5 | 18.6 | .36 | | S L | | |
| 2157 | | | 10 13 31.5 | -51 04 25 | 279.71 | 4.38 | 213 | 113.9 | -60.6 | 34x 23 | 16.1 | .47 | | S M | | |
| 2158 | | | 10 13 32.5 | -47 23 09 | 277.63 | 7.43 | 213 | 123.4 | 137.0 | 13x 13 | 17.9 | .21 | | E 0? | | |
| 2159 | | | 10 13 33.2 | -47 30 40 | 277.70 | 7.33 | 213 | 123.2 | 130.3 | 16x 16 | 17.3 | .18 | | S L | | |
| 2160 | | | 10 13 33.8 | -49 00 37 | 278.55 | 6.09 | 213 | 118.8 | 50.3 | 13x 8 | 18.4 | .20 | | S E? | | |
| 2161 | | | 10 13 35.2 | -50 09 48 | 279.20 | 5.14 | 213 | 116.8 | -11.8 | 13x 4 | 18.7 | .32 | | S E | | |
| 2162 | | | 10 13 38.5 | -48 41 47 | 278.39 | 6.36 | 213 | 120.4 | 67.1 | 16x 4 | 19.1 | .20 | | S M | | |
| 2163 | | | 10 13 43.3 | -49 28 52 | 278.84 | 5.71 | 213 | 119.7 | 24.7 | 24x 4 | 18.5 | .29 | | S L | E | no struct. |
| 2164 | | | 10 13 44.9 | -50 23 47 | 279.36 | 4.96 | 213 | 117.5 | -24.4 | 13x 7 | 18.6 | .28 | | S | F | dist.nucl. |
| 2165 | | | 10 13 46.2 | -51 08 09 | 279.77 | 4.35 | 213 | 115.8 | -64.0 | 13x 7 | 18.7 | .40 | ? | S ? | | |
| 2166 | | | 10 13 47.3 | -50 09 44 | 279.23 | 5.16 | 213 | 118.5 | -11.8 | 20x 11 | 17.8 | .32 | | S | | |
| 2167 | | | 10 13 47.8 | -62 32 16 | 286.13 | -5.12 | 127 | 35.5 | -136.5 | 24x 7 | 18.2 | .46 | | S 3 | S | w.2 comp. |
| 2168 | | | 10 13 52.6 | -61 32 34 | 285.58 | -4.29 | 127 | 37.5 | -83.2 | 54x 11 | 16.5 | .43 | | S L | E | |
| 2169 | | | 10 13 54.4 | -48 46 58 | 278.47 | 6.31 | 213 | 122.5 | 62.3 | 9x 7 | 19.3 | .21 | ? | | | |
| 2170 | | | 10 13 58.7 | -48 41 29 | 278.43 | 6.39 | 213 | 124.0 | 66.9 | 13x 5 | 18.9 | .21 | | S 4 | | |
| 2171 | | | 10 13 58.9 | -48 41 51 | 278.43 | 6.39 | 213 | 124.0 | 66.5 | 12x 5 | 19.1 | .21 | | | | |
| 2172 | | | 10 14 06.8 | -50 49 26 | 279.64 | 4.64 | 213 | 119.5 | -47.4 | 16x 8 | 17.9 | .25 | | F | | |
| 2173 | | | 10 14 07.6 | -51 55 12 | 280.26 | 3.73 | 213 | 116.7 | -106.1 | 40x 5 | 18.2 | .46 | ? | S | E | |
| 2174 | | | 10 14 09.5 | -64 52 25 | 287.48 | -7.04 | 92 | -12.5 | 9.4 | 13x 9 | 18.5 | .30 | | | | |
| 2175 | | | 10 14 12.6 | -49 02 42 | 278.66 | 6.12 | 213 | 125.1 | 47.8 | 16x 4 | 18.7 | .20 | | S | E | i.a.w.2176 |
| 2176 | | | 10 14 13.1 | -49 02 46 | 278.66 | 6.12 | 213 | 125.2 | 47.8 | 13x 8 | 18.4 | .20 | | S | F | i.a.w.2175 |
| 2177 | | | 10 14 14.2 | -48 48 53 | 278.53 | 6.31 | 213 | 125.9 | 60.1 | 13x 13 | 17.9 | .21 | | S L? | | |
| 2178 | | | 10 14 14.2 | -48 55 10 | 278.59 | 6.23 | 213 | 125.7 | 54.5 | 27x 20 | 16.4 | .20 | | S L | | |
| 2179 | | | 10 14 14.5 | -50 39 18 | 279.57 | 4.79 | 213 | 121.1 | -38.4 | 20x 13 | 17.5 | .26 | | SX 4 : | | |
| 2180 | | | 10 14 16.6 | -49 14 24 | 278.78 | 5.96 | 213 | 125.2 | 37.3 | 13x 5 | 18.8 | .20 | | S 0 | | blg or s.p.* |
| 2181 | | | 10 14 16.8 | -49 13 59 | 278.77 | 5.97 | 213 | 125.2 | 37.7 | 13x 3 | 19.4 | .20 | | S | E | |
| 2182 | | | 10 14 22.0 | -67 40 28 | 289.08 | -9.34 | 92 | -10.0 | -140.7 | 40x 15 | 16.5 | .21 | | S 1 | | |
| 2183 | | | 10 14 22.7 | -48 47 31 | 278.54 | 6.35 | 213 | 127.2 | 61.3 | 19x 16 | 17.4 | .22 | | SB 0 | | |
| 2184 | | | 10 14 24.5 | -48 50 33 | 278.57 | 6.31 | 213 | 127.4 | 58.6 | 17x 9 | 18.0 | .22 | | F | | |
| 2185 | | | 10 14 33.3 | -65 27 36 | 287.85 | -7.50 | 92 | -10.0 | -22.0 | 13x 3 | 20.3 | .28 | | | | |
| 2186 | | | 10 14 36.0 | -51 27 03 | 280.06 | 4.16 | 213 | 121.9 | -81.2 | 15x 11 | 18.1 | .36 | | | | |
| 2187 | | | 10 14 36.1 | -51 07 19 | 279.87 | 4.43 | 213 | 122.8 | -63.6 | 16x 11 | 17.6 | .31 | | | | |
| 2188 | | | 10 14 39.6 | -51 19 49 | 280.00 | 4.26 | 213 | 122.7 | -74.8 | 13x 11 | 17.9 | .34 | | | | |
| 2189 | | | 10 14 41.8 | -62 09 15 | 286.01 | -4.74 | 127 | 41.7 | -116.1 | 20x 13 | 17.4 | .38 | | | S | |
| 2190 | | | 10 14 43.9 | -49 40 33 | 279.08 | 5.64 | 213 | 127.9 | 13.8 | 13x 5 | 18.5 | .27 | | F | | |
| 2191 | | | 10 14 44.0 | -66 53 28 | 288.67 | -8.67 | 92 | -8.4 | -98.7 | 19x 5 | 18.6 | .22 | | S E | | |
| 2192 | L092-019 | I | 10 14 44.9 | -64 37 00 | 287.39 | -6.79 | 92 | -9.3 | 23.2 | 67x 8 | 16.9 | .29 | | S | ES | cl.to br.* |
| 2193 | | | 10 14 47.6 | -50 11 57 | 279.38 | 5.21 | 213 | 127.0 | -14.3 | 16x 5 | 18.7 | .30 | ? | S | 1 | |
| 2194 | | | 10 14 47.7 | -52 29 28 | 280.66 | 3.31 | 213 | 120.6 | -137.0 | 16x 13 | 17.7 | .56 | | S | | 1 dist.sp.arm |
| 2195 | | | 10 14 48.6 | -50 19 59 | 279.46 | 5.10 | 213 | 126.8 | -21.4 | 23x 5 | 18.3 | .27 | | S | ES | trpl syst? |
| 2196 | | NS | 10 14 49.2 | -50 56 02 | 279.80 | 4.61 | 213 | 125.2 | -53.6 | 34x 27 | 15.9 | .28 | | F | | cl.to br.* |
| 2197 | | | 10 14 50.1 | -48 49 46 | 278.62 | 6.36 | 213 | 131.2 | 59.1 | 34x 20 | 16.3 | .22 | | SB 4 | | |
| 2198 | | | 10 14 51.6 | -53 16 02 | 281.10 | 2.67 | 168 | -97.5 | 90.7 | 27x 9 | 17.9 | .64 | | S | | dist.nucl. |
| 2199 | | | 10 14 53.4 | -48 18 33 | 278.34 | 6.79 | 213 | 133.1 | 86.9 | 20x 8 | 18.0 | .16 | | S 1 | | |
| 2200 | L213-011 | I | 10 14 55.2 | -48 37 47 | 278.52 | 6.53 | 213 | 132.5 | 69.8 | 269x175 | 11.9 | .21 | | SB 4 | | |

**Table 1.** Galaxies in the southern ZOA – I. The Hydra/Antlia Extension (continued)

| RKK Ident (1) | Other Ident (2) | IR (3) | R.A. (h m s) (4) | Dec. (° ′ ″) (5) | gal $\ell$ (°) (6) | gal $b$ (°) (7) | SRC (8) | X (mm) (9) | Y (mm) (10) | D x d (″) (11) | $B_J$ ($^m$) (12) | $N_{HI}$ (13) | u | Type class. (14) | o * | Remarks (15) |
|---|---|---|---|---|---|---|---|---|---|---|---|---|---|---|---|---|
| 2201 | | | 10 14 58.0 | -48 54 46 | 278.69 | 6.30 | 213 | 131.4 | 54.8 | 11x 8 | 18.9 | .22 | | | | dbl syst? |
| 2202 | | | 10 14 59.1 | -50 07 45 | 279.37 | 5.29 | 213 | 128.9 | -10.6 | 13x 13 | 17.9 | .29 | | L ? | F S | |
| 2203 | | | 10 15 00.0 | -49 02 49 | 278.77 | 6.19 | 213 | 131.3 | 47.6 | 34x 19 | 16.3 | .23 | | S ? | | p.cov.by ** |
| 2204 | | | 10 15 01.4 | -49 15 50 | 278.89 | 6.01 | 213 | 131.6 | 35.7 | 13x 13 | 18.1 | .24 | | E 0 | | faint halo |
| 2205 | | I | 10 15 01.7 | -50 31 55 | 279.60 | 4.96 | 213 | 128.1 | -32.2 | 13x 8 | 18.5 | .23 | | S L | | |
| 2206 | | | 10 15 02.4 | -64 59 18 | 287.62 | -7.08 | 92 | -7.5 | 3.3 | 13x 7 | 18.9 | .30 | ? | | | |
| 2207 | | | 10 15 04.6 | -49 09 45 | 278.84 | 6.10 | 213 | 131.6 | 41.4 | 19x 16 | 17.7 | .23 | | F | 1 | |
| 2208 | | | 10 15 06.3 | -65 00 43 | 287.64 | -7.09 | 92 | -7.1 | 2.0 | 13x 4 | 19.7 | .30 | | S ? | | |
| 2209 | | | 10 15 07.0 | -50 09 44 | 279.41 | 5.27 | 213 | 129.9 | -12.4 | 20x 13 | 17.2 | .29 | | L | | |
| 2210 | | | 10 15 08.5 | -47 55 58 | 278.16 | 7.13 | 213 | 136.4 | 107.0 | 19x 7 | 17.7 | .17 | | S E | | |
| 2211 | | | 10 15 08.6 | -50 50 51 | 279.79 | 4.71 | 213 | 128.2 | -49.1 | 34x 7 | 17.3 | .25 | | S 5 : | E | |
| 2212 | | | 10 15 10.9 | -49 45 18 | 279.19 | 5.62 | 213 | 131.6 | 9.3 | 13x 13 | 17.9 | .30 | | S 4 : | | |
| 2213 | | I | 10 15 13.8 | -51 36 10 | 280.22 | 4.08 | 213 | 126.7 | -89.6 | 17x 8 | 17.9 | .37 | | S ? | | |
| 2214 | | | 10 15 15.6 | -50 54 44 | 279.84 | 4.66 | 213 | 129.0 | -52.7 | 13x 13 | 18.1 | .27 | | E 0 ? | | |
| 2215 | | | 10 15 16.7 | -50 40 33 | 279.71 | 4.86 | 213 | 129.8 | -40.0 | 27x 7 | 18.1 | .25 | | S 5 : | E | dist.nucl. |
| 2216 | | | 10 15 18.0 | -49 11 35 | 278.89 | 6.10 | 214 | -129.9 | 38.2 | 17x 5 | 18.8 | .26 | | S | 1 | |
| 2217 | | | 10 15 19.1 | -47 27 15 | 277.92 | 7.54 | 213 | 139.3 | 132.5 | 13x 5 | 18.5 | .18 | | S | E | |
| 2218 | | | 10 15 23.0 | -48 31 07 | 278.53 | 6.67 | 213 | 136.9 | 75.5 | 16x 5 | 18.9 | .19 | ? | S ? | | |
| 2219 | | | 10 15 23.1 | -50 49 56 | 279.82 | 4.74 | 213 | 130.3 | -48.4 | 22x 11 | 17.9 | .24 | | S | E | w.comp. |
| 2220 | | | 10 15 24.8 | -51 04 28 | 279.95 | 4.54 | 213 | 129.8 | -61.4 | 20x 13 | 17.9 | .29 | | S | | |
| 2221 | | | 10 15 26.2 | -51 17 28 | 280.08 | 4.36 | 213 | 129.4 | -73.0 | 13x 9 | 17.9 | .35 | ? | E 3 : | | |
| 2222 | | | 10 15 26.6 | -48 50 00 | 278.71 | 6.41 | 213 | 135.9 | 58.9 | 20x 8 | 18.6 | .20 | | S ? | | LSB, poss.larger |
| 2223 | | | 10 15 26.7 | -65 03 46 | 287.70 | -7.12 | 92 | -5.2 | -.7 | 11x 5 | 19.4 | .30 | | S ? | S | neighb.of 2228 |
| 2224 | | | 10 15 28.4 | -48 56 40 | 278.78 | 6.32 | 213 | 135.8 | 52.9 | 12x 7 | 19.1 | .21 | | S | | cl.to br.* |
| 2225 | | | 10 15 29.1 | -64 33 18 | 287.42 | -6.69 | 92 | -5.1 | 26.5 | 17x 7 | 18.8 | .32 | | S ? | | |
| 2226 | | | 10 15 29.4 | -48 42 55 | 278.65 | 6.51 | 213 | 137.3 | 64.9 | 20x 9 | 17.6 | .21 | | S | | |
| 2227 | | | 10 15 29.5 | -48 15 13 | 278.39 | 6.90 | 213 | 138.6 | 89.6 | 19x 8 | 17.6 | .15 | | S E: | | |
| 2228 | | | 10 15 33.5 | -65 03 34 | 287.71 | -7.11 | 92 | -4.5 | -.5 | 13x 13 | 18.1 | .30 | | S ? | | |
| 2229 | | | 10 15 37.2 | -50 29 36 | 279.66 | 5.04 | 213 | 133.2 | -30.4 | 30x 11 | 17.2 | .25 | | S S5 | | |
| 2230 | | | 10 15 38.1 | -49 49 46 | 279.29 | 5.60 | 214 | -124.6 | 4.4 | 17x 8 | 18.4 | .30 | | E ? | | |
| 2231 | | | 10 15 38.1 | -65 02 39 | 287.71 | -7.09 | 92 | -4.1 | .3 | 11x 3 | 20.2 | .30 | | S | E | neighb.of 2228 |
| 2232 | | | 10 15 40.0 | -51 39 38 | 280.31 | 4.07 | 213 | 130.2 | -92.9 | 20x 13 | 17.2 | .37 | | S E | S | |
| 2233 | | | 10 15 42.6 | -48 50 42 | 278.75 | 6.42 | 213 | 138.2 | 58.1 | 7x 7 | 19.9 | .20 | ? | | | |
| 2234 | | | 10 15 42.8 | -51 37 29 | 280.30 | 4.11 | 213 | 130.7 | -91.0 | 13x 13 | 17.7 | .37 | ? | S L | | |
| 2235 | | | 10 15 46.4 | -66 52 53 | 288.75 | -8.61 | 92 | -3.0 | -98.2 | 27x 13 | 18.1 | .22 | ? | | | vLSB |
| 2236 | | | 10 15 52.8 | -49 26 56 | 279.11 | 5.94 | 214 | -123.9 | 24.9 | 13x 11 | 18.4 | .22 | ? | | | |
| 2237 | | | 10 15 54.2 | -47 46 39 | 278.18 | 7.33 | 213 | 143.7 | 114.9 | 13x 5 | 18.9 | .16 | | S ? | | |
| 2238 | | | 10 15 57.1 | -52 00 04 | 280.54 | 3.81 | 213 | 131.5 | -111.3 | 27x 7 | 18.2 | .53 | | | | |
| 2239 | | | 10 15 58.1 | -49 40 09 | 279.25 | 5.76 | 213 | 138.7 | 13.6 | 16x 8 | 17.9 | .26 | | L | | or S |
| 2240 | | | 10 16 00.9 | -47 44 47 | 278.18 | 7.37 | 213 | 144.8 | 116.5 | 19x 13 | 17.0 | .16 | | S L | | |
| 2241 | | | 10 16 05.8 | -48 16 49 | 278.49 | 6.93 | 214 | -126.2 | 87.5 | 13x 11 | 18.5 | .16 | ? | | | |
| 2242 | | | 10 16 06.4 | -63 55 16 | 287.12 | -6.13 | 92 | -1.6 | 60.5 | 13x 8 | 18.4 | .33 | ? | | | |
| 2243 | | | 10 16 08.4 | -48 33 54 | 278.66 | 6.70 | 213 | 143.5 | 72.6 | 16x 8 | 17.5 | .22 | | S 0 | | |
| 2244 | | | 10 16 12.1 | -50 51 34 | 279.94 | 4.79 | 213 | 137.1 | -50.3 | 13x 5 | 18.5 | .25 | | S E: | | |
| 2245 | | | 10 16 16.2 | -61 57 43 | 286.05 | -4.48 | 127 | 51.9 | -106.1 | 13x 8 | 18.4 | .43 | | S | | |
| 2246 | | | 10 16 16.8 | -47 41 48 | 278.19 | 7.43 | 214 | -126.7 | 118.8 | 13x 11 | 18.7 | .16 | | | | vLSB |
| 2247 | | | 10 16 17.7 | -65 37 00 | 288.08 | -7.53 | 92 | -.3 | -30.4 | 60x 40 | 16.0 | .29 | | SX M | S | vLSB |
| 2248 | | NS | 10 16 19.7 | -52 29 33 | 280.86 | 3.44 | 214 | -108.9 | -137.7 | 16x 4 | 19.9 | .53 | | S | E | |
| 2249 | | | 10 16 21.3 | -62 30 04 | 286.36 | -4.93 | 127 | 51.4 | -135.0 | 60x 7 | 17.3 | .39 | | S 5 | E | |
| 2250 | | | 10 16 25.0 | -51 40 04 | 280.41 | 4.13 | 214 | -111.2 | -93.5 | 50x 12 | 16.9 | .38 | | S ? | S | |
| 2251 | | | 10 16 26.1 | -49 49 20 | 279.40 | 5.67 | 214 | -117.7 | 5.2 | 15x 11 | 17.9 | .30 | | E | | HSB |
| 2252 | | | 10 16 26.6 | -49 53 08 | 279.43 | 5.62 | 214 | -117.4 | 1.8 | 24x 11 | 18.0 | .30 | | | | |
| 2253 | | | 10 16 27.7 | -65 10 22 | 287.85 | -7.15 | 92 | .6 | -6.6 | 22x 13 | 17.6 | .30 | | SB M | 1 | |
| 2254 | | | 10 16 30.3 | -51 45 19 | 280.47 | 4.07 | 214 | -110.2 | -98.1 | 12x 9 | 18.4 | .42 | ? | E ? | | |
| 2255 | | | 10 16 31.1 | -50 40 46 | 279.88 | 4.97 | 214 | -113.9 | -40.6 | 17x 8 | 18.9 | .24 | | S | | dist.nucl, vLSBdisk |
| 2256 | | | 10 16 36.7 | -47 46 35 | 278.28 | 7.40 | 214 | -123.4 | 114.8 | 16x 8 | 18.4 | .16 | | S ? | | p.cov.by br.* |
| 2257 | | | 10 16 37.9 | -47 43 28 | 278.26 | 7.44 | 214 | -123.4 | 117.6 | 16x 8 | 18.1 | .16 | | S E | S | |
| 2258 | | | 10 16 38.8 | -48 08 55 | 278.50 | 7.09 | 214 | -121.8 | 94.9 | 24x 13 | 17.5 | .15 | | S M | V | dist.blg, obs.disk |
| 2259 | | | 10 16 40.1 | -50 50 17 | 279.99 | 4.85 | 214 | -112.1 | -49.0 | 13x 13 | 18.1 | .24 | ? | | | diff.halo |
| 2260 | | | 10 16 40.2 | -50 28 36 | 279.79 | 5.15 | 214 | -113.4 | -29.7 | 13x 13 | 17.8 | .25 | ? | E | 1 | |
| 2261 | | | 10 16 40.3 | -50 14 51 | 279.66 | 5.34 | 214 | -114.2 | -17.4 | 19x 9 | 18.1 | .30 | | S | 1 | or dbl syst. |
| 2262 | | | 10 16 40.4 | -47 46 21 | 278.29 | 7.41 | 214 | -122.9 | 115.0 | 22x 13 | 17.6 | .16 | | E | | |
| 2263 | | | 10 16 40.8 | -50 59 53 | 280.08 | 4.71 | 214 | -111.4 | -57.5 | 16x 7 | 18.5 | .26 | ? | | | or blended ** |
| 2264 | | | 10 16 41.4 | -50 19 29 | 279.71 | 5.28 | 214 | -113.8 | -21.5 | 30x 4 | 18.6 | .28 | | S M | E | blg, LSBdisk |
| 2265 | | | 10 16 41.6 | -47 49 38 | 278.32 | 7.36 | 214 | -122.5 | 112.1 | 17x 9 | 18.3 | .16 | ? | | | p.cov.by br.* |
| 2266 | | I | 10 16 44.5 | -66 01 09 | 288.35 | -7.84 | 92 | 2.1 | -52.0 | 13x 11 | 18.0 | .27 | ? | E ? | S | |
| 2267 | | | 10 16 45.6 | -49 46 47 | 279.42 | 5.74 | 214 | -115.1 | 7.7 | 19x 9 | 18.4 | .28 | | | | |
| 2268 | | | 10 16 45.9 | -50 25 42 | 279.78 | 5.20 | 214 | -112.7 | -27.0 | 27x 17 | 17.2 | .25 | | S ? | S | v.diff. |
| 2269 | | | 10 16 47.9 | -49 25 07 | 279.22 | 6.04 | 214 | -116.0 | 27.0 | 23x 13 | 17.7 | .22 | | | | p.cov.by **, v.obs. |
| 2270 | | | 10 16 49.8 | -49 23 53 | 279.21 | 6.06 | 214 | -115.8 | 28.1 | 24x 7 | 18.2 | .22 | | | | p.cov.by **, v.obs. |
| 2271 | | | 10 16 52.8 | -49 28 43 | 279.27 | 6.00 | 214 | -115.1 | 23.9 | 16x 5 | 18.9 | .22 | ? | | | p.cov.by ** |
| 2272 | | | 10 16 52.8 | -67 21 31 | 289.11 | -8.95 | 92 | 2.9 | -123.7 | 13x 13 | 18.9 | .23 | ? | | | vLSB |
| 2273 | | | 10 16 53.4 | -48 35 19 | 278.77 | 6.74 | 214 | -118.1 | 71.5 | 35x 9 | 17.4 | .22 | | S | | |
| 2274 | | | 10 16 57.1 | -47 53 02 | 278.39 | 7.34 | 214 | -120.0 | 109.2 | 20x 9 | 18.1 | .15 | | | | br.blg, w.comp. |
| 2275 | | | 10 16 57.2 | -49 56 54 | 279.54 | 5.62 | 214 | -112.8 | -1.2 | 13x 9 | 19.0 | .31 | | | | vvLSB |
| 2276 | | | 10 16 57.6 | -63 22 21 | 286.90 | -5.62 | 92 | 3.5 | 89.9 | 60x 8 | 17.3 | .36 | | S L | E S | |
| 2277 | | | 10 16 57.7 | -49 38 57 | 279.37 | 5.87 | 214 | -113.8 | 14.8 | 24x 15 | 17.5 | .25 | | | | v.obs. |
| 2278 | | | 10 16 58.8 | -48 28 42 | 278.73 | 6.85 | 214 | -117.7 | 77.4 | 27x 16 | 17.1 | .18 | | S M | | |
| 2279 | | | 10 17 01.9 | -49 21 15 | 279.22 | 6.12 | 214 | -114.2 | 30.6 | 16x 9 | 18.1 | .23 | | F | | |
| 2280 | | | 10 17 02.7 | -47 51 57 | 278.40 | 7.36 | 214 | -119.2 | 110.2 | 44x 40 | 15.5 | .16 | | S 1 | S | dist.abs.lane, like CenA |
| 2281 | | | 10 17 03.4 | -48 44 10 | 278.88 | 6.64 | 214 | -116.1 | 63.7 | 20x 9 | 17.9 | .24 | | | | or blended ** |
| 2282 | | | 10 17 08.0 | -50 40 49 | 279.96 | 5.02 | 214 | -108.7 | -40.3 | 20x 8 | 18.3 | .26 | | S M | | poss.larger? |
| 2283 | | | 10 17 09.4 | -50 37 52 | 279.94 | 5.06 | 214 | -108.7 | -37.7 | 30x 3 | 19.0 | .27 | | S | E S | |
| 2284 | | | 10 17 11.4 | -49 26 01 | 279.28 | 6.07 | 214 | -112.6 | 26.4 | 54x 44 | 15.4 | .22 | | | S | br.blg, fuzzy disk |
| 2285 | | | 10 17 12.2 | -49 16 19 | 279.20 | 6.20 | 214 | -113.0 | 35.1 | 13x 3 | 20.0 | .23 | | S | | |
| 2286 | | | 10 17 12.7 | -52 16 21 | 280.85 | 3.69 | 214 | -102.5 | -125.4 | 16x 3 | 20.1 | .46 | | S | | |
| 2287 | | I3 | 10 17 13.4 | -49 31 55 | 279.34 | 5.99 | 214 | -111.9 | 21.2 | 12x 3 | 19.8 | .23 | | S | | neighb.of 2291 |
| 2288 | | | 10 17 14.6 | -65 12 51 | 287.94 | -7.14 | 92 | 5.0 | -8.8 | 16x 7 | 18.5 | .29 | ? | | | |
| 2289 | | | 10 17 17.0 | -49 25 04 | 279.29 | 6.09 | 214 | -111.8 | 27.3 | 19x 5 | 18.9 | .22 | | S ? | 1 | |
| 2290 | | | 10 17 17.6 | -48 00 59 | 278.51 | 7.26 | 214 | -116.5 | 102.3 | 31x 24 | 16.7 | .16 | | S M | V S | |
| 2291 | | I3 | 10 17 19.8 | -49 31 26 | 279.35 | 6.00 | 214 | -111.0 | 21.7 | 23x 9 | 17.7 | .23 | | S M? | | |
| 2292 | | | 10 17 20.8 | -48 26 48 | 278.76 | 6.91 | 214 | -114.6 | 79.3 | 19x 13 | 17.8 | .17 | | | | |
| 2293 | | | 10 17 21.2 | -49 43 31 | 279.47 | 5.84 | 214 | -110.1 | 10.9 | 19x 5 | 18.9 | .27 | | S | | |
| 2294 | | I3 | 10 17 21.8 | -49 32 05 | 279.36 | 6.00 | 214 | -110.7 | 21.1 | 12x 7 | 18.6 | .23 | | | | neighb.of 2291 |
| 2295 | | | 10 17 22.6 | -50 47 46 | 280.06 | 4.94 | 214 | -106.3 | -46.4 | 16x 16 | 18.1 | .28 | | S M | F | v.obs, cl.to br.* |
| 2296 | | | 10 17 24.1 | -49 26 26 | 279.32 | 6.08 | 214 | -110.7 | 26.2 | 20x 5 | 18.6 | .23 | | | | p.cov.by ** |
| 2297 | | | 10 17 26.6 | -49 43 53 | 279.48 | 5.84 | 214 | -109.3 | 10.6 | 13x 7 | 19.1 | .28 | | | | |
| 2298 | | | 10 17 28.0 | -49 34 37 | 279.40 | 5.97 | 214 | -109.7 | 18.9 | 31x 7 | 17.9 | .24 | | S | E | |
| 2299 | | | 10 17 28.3 | -48 58 28 | 279.07 | 6.48 | 214 | -111.7 | 51.2 | 15x 11 | 18.1 | .22 | | E 2 | | |
| 2300 | | | 10 17 28.4 | -62 04 29 | 286.23 | -4.50 | 127 | 59.2 | -112.4 | 16x 11 | 17.8 | .42 | | | S | |

**Table 1.** Galaxies in the southern ZOA – I. The Hydra/Antlia Extension (continued)

| RKK Ident (1) | Other Ident (2) | IR (3) | R.A. (h m s) (4) | Dec. (° ′ ″) (5) | gal ℓ (°) (6) | gal b (°) (7) | SRC (8) | X (mm) (9) | Y (mm) (10) | D x d (″) (11) | $B_J$ ($^m$) (12) | $N_{HI}$ (13) | u | Type class. (14) | | o | * | Remarks (15) |
|---|---|---|---|---|---|---|---|---|---|---|---|---|---|---|---|---|---|---|
| 2301 | | | 10 17 31. | -62 14.9 : | 286.33 | -4.64 | 127 | 59.2 | -121.7 | 16x 13 | 17.7 | .36 | | S | | | | |
| 2302 | | | 10 17 36.9 | -49 38 26 | 279.46 | 5.93 | 214 | -108.2 | 15.6 | 34x 12 | 17.2 | .26 | | S | | | | p.cov.by * |
| 2303 | | | 10 17 37.3 | -64 01 31 | 287.32 | -6.12 | 92 | 7.3 | 54.9 | 17x 16 | 17.7 | .35 | | S | ? | | S | |
| 2304 | | | 10 17 37.8 | -51 40 32 | 280.57 | 4.23 | 214 | -101.1 | -93.3 | 24x 13 | 17.4 | .38 | | | | | | br.blg, diff. halo |
| 2305 | | | 10 17 39.3 | -47 52 52 | 278.49 | 7.41 | 214 | -113.7 | 109.8 | 17x 13 | 18.0 | .16 | | L | | | | |
| 2306 | | | 10 17 40.1 | -48 32 02 | 278.85 | 6.86 | 214 | -111.4 | 74.9 | 13x 9 | 18.7 | .19 | | | | | | |
| 2307 | | | 10 17 42.0 | -51 30 51 | 280.50 | 4.37 | 214 | -101.1 | -84.6 | 13x 11 | 18.4 | .35 | | | | | | w.comp. |
| 2308 | | | 10 17 42.1 | -51 18 36 | 280.38 | 4.54 | 214 | -101.8 | -73.7 | 30x 5 | 18.4 | .35 | | S | | E | | |
| 2309 | | | 10 17 42.9 | -64 27 20 | 287.57 | -6.48 | 92 | 7.8 | 31.8 | 16x 5 | 19.0 | .33 | | | | | | in st.diffr.pat. |
| 2310 | | I2 | 10 17 43.4 | -60 44 12 | 285.53 | -3.36 | 127 | 63.8 | -40.9 | 54x 24 | 15.2 | .79 | ? | E | | | | i.a.w.2314,galactic? |
| 2311 | | | 10 17 44.8 | -49 40 48 | 279.50 | 5.91 | 214 | -106.9 | 13.6 | 23x 8 | 18.0 | .28 | | | | | | |
| 2312 | | | 10 17 45.3 | -50 28 17 | 279.93 | 5.25 | 214 | -104.2 | -28.8 | 22x 7 | 18.3 | .29 | | S | | E | S | |
| 2313 | | | 10 17 45.7 | -49 47 39 | 279.56 | 5.82 | 214 | -106.4 | 7.5 | 47x 23 | 16.1 | .30 | | S | E: | | | |
| 2314 | | I2 | 10 17 47.1 | -60 44 08 | 285.53 | -3.36 | 127 | 64.2 | -40.9 | 34x 24 | 15.7 | .79 | ? | E | | | | i.a.w.2310,galactic? |
| 2315 | | | 10 17 47.6 | -49 34 57 | 279.45 | 6.00 | 214 | -106.8 | 18.8 | 20x 4 | 19.0 | .25 | | S | | | | |
| 2316 | | | 10 17 49.1 | -49 40 50 | 279.51 | 5.92 | 214 | -106.3 | 13.6 | 27x 13 | 17.6 | .28 | | | | | | dist.nucl, vLSB |
| 2317 | | I | 10 17 50.9 | -47 32 22 | 278.33 | 7.71 | 214 | -113.1 | 128.2 | 40x 22 | 16.4 | .15 | | S | M | | S | br.blg, vvLSBdisk |
| 2318 | | | 10 17 52.6 | -50 06 05 | 279.74 | 5.57 | 214 | -104.4 | -8.9 | 13x 3 | 19.9 | .30 | | S | | | | |
| 2319 | | | 10 17 52.9 | -64 25 48 | 287.57 | -6.45 | 92 | 8.8 | 33.2 | 20x 8 | 18.4 | .34 | | | | | | in st.diffr.pat. |
| 2320 | | | 10 17 57.1 | -49 58 14 | 279.68 | 5.68 | 214 | -104.2 | -1.9 | 24x 9 | 17.9 | .30 | | S | E? | | | |
| 2321 | | | 10 17 58.6 | -49 03 59 | 279.19 | 6.44 | 214 | -106.9 | 46.5 | 23x 13 | 17.5 | .25 | | F | | | | |
| 2322 | | NG | 10 17 58.9 | -47 32 28 | 278.35 | 7.72 | 214 | -111.9 | 128.2 | 19x 16 | 17.6 | .15 | | | | | S | neighb.of 2317 |
| 2323 | | | 10 18 00.1 | -62 40 11 | 286.61 | -4.96 | 127 | 61.2 | -144.4 | 20x 9 | 18.0 | .37 | | S | M | | S | blg or s.p.* |
| 2324 | | | 10 18 02.0 | -49 58 21 | 279.69 | 5.69 | 214 | -103.4 | -1.9 | 20x 8 | 18.4 | .30 | | S | | | | neighb.of 2320, p.cov.by * |
| 2325 | | | 10 18 03.2 | -48 07 51 | 278.68 | 7.23 | 214 | -109.3 | 96.6 | 16x 11 | 17.7 | .16 | | F | ? | | | |
| 2326 | | | 10 18 09.5 | -48 26 41 | 278.87 | 6.98 | 214 | -107.4 | 79.9 | 34x 7 | 17.5 | .18 | | S | 1 | | | |
| 2327 | | | 10 18 09.9 | -51 28 58 | 280.54 | 4.44 | 214 | -97.3 | -82.7 | 30x 5 | 18.1 | .35 | | S | | E | | HSB |
| 2328 | | | 10 18 11.4 | -50 12 48 | 279.85 | 5.50 | 214 | -101.3 | -14.7 | 16x 11 | 18.5 | .28 | | | | | | vLSB, v.obs |
| 2329 | | | 10 18 11.6 | -49 27 13 | 279.43 | 6.14 | 214 | -103.8 | 25.9 | 13x 7 | 18.9 | .23 | | S | ? | | | |
| 2330 | | | 10 18 12.9 | -50 38 38 | 280.09 | 5.14 | 214 | -99.7 | -37.8 | 24x 7 | 18.4 | .35 | | S | ? | | | v.obs. |
| 2331 | | | 10 18 15.9 | -65 59 40 | 288.46 | -7.73 | 92 | 10.5 | -50.7 | 13x 7 | 19.4 | .26 | ? | | | | | |
| 2332 | | | 10 18 16.7 | -51 25 05 | 280.52 | 4.50 | 214 | -96.6 | -79.2 | 12x 5 | 19.2 | .36 | | S | | | | |
| 2333 | | | 10 18 18.6 | -48 07 40 | 278.72 | 7.26 | 214 | -107.0 | 96.9 | 16x 13 | 17.7 | .16 | | | | | | p.cov.by ** |
| 2334 | | | 10 18 20.3 | -49 42 47 | 279.59 | 5.93 | 214 | -101.7 | 12.1 | 26x 3 | 19.2 | .31 | | S | | E | | |
| 2335 | | | 10 18 21.0 | -65 19 20 | 288.10 | -7.17 | 92 | 11.2 | -14.6 | 16x 16 | 17.4 | .30 | | E | | | | |
| 2336 | | | 10 18 22.4 | -51 34 20 | 280.62 | 4.38 | 214 | -95.3 | -87.4 | 17x 3 | 19.7 | .37 | | S | | E | | |
| 2337 | | | 10 18 26.9 | -49 07 47 | 279.29 | 6.43 | 214 | -102.6 | 43.4 | 59x 40 | 15.6 | .26 | | I | M | | S | dwarf |
| 2338 | | | 10 18 29.2 | -51 19 00 | 280.49 | 4.60 | 214 | -95.2 | -73.6 | 24x 5 | 18.6 | .37 | | S | | | | |
| 2339 | | NS | 10 18 34.2 | -51 12 39 | 280.44 | 4.70 | 214 | -94.8 | -67.9 | 16x 3 | 20.2 | .38 | | S | | | | |
| 2340 | | | 10 18 35.5 | -48 36 17 | 279.02 | 6.89 | 214 | -103.0 | 71.6 | 13x 8 | 18.7 | .19 | | S | | | | |
| 2341 | | I | 10 18 35.5 | -63 45 52 | 287.27 | -5.85 | 92 | 13.1 | 68.9 | 47x 31 | 15.7 | .35 | | L | | | S | |
| 2342 | | | 10 18 36.1 | -51 29 02 | 280.60 | 4.47 | 214 | -93.7 | -82.5 | 17x 5 | 19.1 | .36 | | S | | | S | v.obs. |
| 2343 | | | 10 18 37.4 | -49 06 00 | 279.30 | 6.47 | 214 | -101.2 | 45.1 | 17x 3 | 19.7 | .27 | | S | | E | | |
| 2344 | | | 10 18 39.0 | -47 59 35 | 278.69 | 7.40 | 214 | -104.4 | 104.3 | 15x 5 | 19.1 | .16 | | S | ? | | | |
| 2345 | | | 10 18 39.1 | -50 01 28 | 279.81 | 5.70 | 214 | -98.0 | -4.4 | 20x 15 | 17.8 | .28 | | S | M? | | | v.obs. |
| 2346 | | | 10 18 40.3 | -49 29 54 | 279.52 | 6.14 | 214 | -99.5 | 23.8 | 20x 12 | 17.6 | .23 | | L | | | | or br.blg |
| 2347 | | | 10 18 42.2 | -47 32 00 | 278.45 | 7.79 | 214 | -105.4 | 129.0 | 13x 8 | 18.8 | .14 | | S | ? | | | neighb.of 2350 |
| 2348 | | | 10 18 45.9 | -64 32 50 | 287.71 | -6.49 | 92 | 13.8 | 26.9 | 34x 4 | 18.8 | .33 | | S | | E | | |
| 2349 | | | 10 18 46.1 | -47 59 46 | 278.71 | 7.41 | 214 | -103.4 | 104.2 | 27x 4 | 18.8 | .16 | | S | | E | | |
| 2350 | | | 10 18 46.7 | -47 32 28 | 278.46 | 7.79 | 214 | -104.7 | 128.6 | 20x 20 | 16.9 | .14 | | | | | | |
| 2351 | | | 10 18 48.0 | -48 36 17 | 279.05 | 6.90 | 214 | -101.2 | 71.7 | 17x 5 | 18.7 | .19 | | S | | | | |
| 2352 | | | 10 18 48.1 | -48 06 42 | 278.78 | 7.32 | 214 | -102.7 | 98.1 | 47x 7 | 17.4 | .16 | | S | | | | |
| 2353 | | | 10 18 53.7 | -47 23 15 | 278.39 | 7.93 | 214 | -104.1 | 136.9 | 16x 16 | 17.6 | .16 | ? | E | 0 | | S | |
| 2354 | | | 10 18 55.1 | -49 56 13 | 279.80 | 5.80 | 214 | -95.9 | .4 | 16x 12 | 18.4 | .28 | | | | | | v.diff. |
| 2355 | | | 10 18 58.4 | -50 20 18 | 280.02 | 5.47 | 214 | -94.2 | -21.0 | 23x 11 | 17.6 | .31 | | | | | | p.cov.by ** |
| 2356 | | | 10 18 58.5 | -66 22 19 | 288.73 | -8.01 | 92 | 14.1 | -70.9 | 13x 5 | 19.4 | .24 | ? | | | | | |
| 2357 | | | 10 18 59.2 | -49 36 55 | 279.63 | 6.07 | 214 | -96.4 | 17.7 | 15x 5 | 19.2 | .27 | | | | | | |
| 2358 | | | 10 19 01.0 | -48 29 58 | 279.02 | 7.01 | 214 | -99.6 | 77.4 | 15x 4 | 19.4 | .18 | | S | ? | | | |
| 2359 | | | 10 19 05.7 | -47 24 36 | 278.43 | 7.93 | 214 | -102.2 | 135.8 | 32x 22 | 16.5 | .16 | | S | M | | | cl.to br.* |
| 2360 | | | 10 19 11.0 | -48 30 08 | 279.05 | 7.02 | 214 | -98.1 | 77.4 | 31x 19 | 16.6 | .18 | | S | M | | | |
| 2361 | | | 10 19 12.1 | -50 30 21 | 280.14 | 5.34 | 214 | -91.7 | -29.9 | 34x 24 | 16.4 | .34 | | S | M: | | S | |
| 2362 | | | 10 19 14.0 | -64 21 26 | 287.65 | -6.31 | 92 | 16.6 | 37.1 | 20x 4 | 19.3 | .38 | | S | | E | | |
| 2363 | | | 10 19 17.7 | -48 46 26 | 279.21 | 6.81 | 214 | -96.3 | 62.9 | 38x 19 | 16.6 | .25 | | L | | | | |
| 2364 | | | 10 19 18.5 | -49 37 23 | 279.68 | 6.10 | 214 | -93.6 | 17.4 | 20x 17 | 17.4 | .26 | ? | L | | | | br.blg, fuzzy envelope |
| 2365 | | | 10 19 21.5 | -50 04 30 | 279.93 | 5.72 | 214 | -91.7 | -6.7 | 24x 15 | 17.1 | .27 | | E | | | S | or S..E |
| 2366 | | | 10 19 23.7 | -50 33 28 | 280.20 | 5.32 | 214 | -89.9 | -32.6 | 20x 8 | 18.3 | .35 | | S | | | | |
| 2367 | | | 10 19 24.0 | -48 57 18 | 279.33 | 6.66 | 214 | -94.8 | 53.2 | 34x 24 | 16.4 | .26 | | L | | | | w.comp. |
| 2368 | | | 10 19 25.4 | -48 21 13 | 279.00 | 7.17 | 214 | -96.4 | 85.4 | 16x 7 | 18.2 | .17 | | S | E | | | |
| 2369 | | | 10 19 25.8 | -50 19 33 | 280.08 | 5.52 | 214 | -90.4 | -20.1 | 27x 4 | 18.8 | .31 | | S | | E | | |
| 2370 | | | 10 19 28.0 | -49 49 42 | 279.81 | 5.94 | 214 | -91.6 | 6.5 | 23x 19 | 17.0 | .28 | | | | | | v.obs. |
| 2371 | | | 10 19 28.7 | -48 21 09 | 279.01 | 7.18 | 214 | -95.9 | 85.5 | 12x 4 | 19.3 | .16 | | S | | | | neighb.of 2368 |
| 2372 | | | 10 19 29.6 | -51 33 40 | 280.76 | 4.48 | 214 | -86.0 | -86.2 | 16x 3 | 20.2 | .39 | | S | | E | | vvLSB |
| 2373 | | | 10 19 30.3 | -49 23 22 | 279.58 | 6.31 | 214 | -92.6 | 30.0 | 13x 3 | 19.5 | .23 | | S | | | | |
| 2374 | | | 10 19 31.1 | -50 02 02 | 279.93 | 5.77 | 214 | -90.5 | -4.5 | 22x 15 | 17.6 | .28 | | | | | | blg |
| 2375 | | | 10 19 31.6 | -62 32 41 | 286.69 | -4.76 | 92 | 19.4 | 134.2 | 17x 8 | 18.8 | .39 | ? | | | | S | |
| 2376 | | | 10 19 33.1 | -48 53 03 | 279.31 | 6.74 | 214 | -93.7 | 57.1 | 43x 9 | 17.2 | .28 | | S | 5 | | S | |
| 2377 | | | 10 19 34.0 | -48 19 41 | 279.01 | 7.21 | 214 | -95.2 | 86.9 | 19x 8 | 18.3 | .16 | | | | | | p.cov.by * |
| 2378 | L092-021 | | 10 19 36.2 | -66 14 22 | 288.71 | -7.86 | 92 | 17.6 | -63.9 | 74x 60 | 14.5 | .24 | | L | | | S | |
| 2379 | | | 10 19 37.9 | -47 26 07 | 278.53 | 7.96 | 214 | -97.3 | 134.7 | 20x 5 | 18.5 | .15 | | S | E | | | |
| 2380 | | I | 10 19 39.9 | -49 21 43 | 279.58 | 6.35 | 214 | -91.2 | 31.6 | 40x 30 | 16.3 | .23 | | | | | S | br.blg, v.obs. |
| 2381 | | | 10 19 40.2 | -50 48 21 | 280.37 | 5.13 | 214 | -86.8 | -45.7 | 17x 7 | 18.6 | .37 | | S | | | | |
| 2382 | | | 10 19 40.5 | -48 30 52 | 279.12 | 7.06 | 214 | -93.7 | 76.9 | 16x 7 | 18.8 | .20 | | | | | | w.comp. |
| 2383 | | | 10 19 41.5 | -50 15 55 | 280.08 | 5.59 | 214 | -88.3 | -16.8 | 13x 7 | 19.5 | .31 | ? | | | | | vvLSB, smooth |
| 2384 | | | 10 19 42.2 | -50 45 55 | 280.35 | 5.17 | 214 | -86.7 | -43.5 | 19x 13 | 17.9 | .36 | | S | M: | | S | |
| 2385 | | | 10 19 43.7 | -49 05 05 | 279.44 | 6.58 | 214 | -91.5 | 46.4 | 13x 8 | 18.6 | .25 | | | | | | |
| 2386 | | | 10 19 43.9 | -48 44 43 | 279.26 | 6.87 | 214 | -92.5 | 64.6 | 38x 3 | 18.9 | .25 | | S | | E | | |
| 2387 | | | 10 19 46.5 | -66 01 03 | 288.61 | -7.67 | 92 | 18.7 | -52.0 | 13x 8 | 18.6 | .25 | | | | | | |
| 2388 | | | 10 19 49.2 | -49 38 44 | 279.76 | 6.12 | 214 | -89.1 | 16.5 | 13x 11 | 18.4 | .26 | | | | | | |
| 2389 | | | 10 19 51.5 | -49 16 00 | 279.56 | 6.44 | 214 | -89.8 | 36.8 | 22x 13 | 18.2 | .24 | ? | | | | | vLSB |
| 2390 | | | 10 19 55.8 | -49 36 36 | 279.76 | 6.16 | 214 | -88.2 | 18.4 | 24x 4 | 18.9 | .26 | | S | | E | | |
| 2391 | | | 10 20 00.0 | -49 20 43 | 279.62 | 6.39 | 214 | -88.4 | 32.6 | 44x 16 | 17.4 | .24 | | | | | S | v.diff. & obs. |
| 2392 | | | 10 20 01.6 | -67 19 48 | 289.35 | -8.76 | 92 | 19.1 | -122.3 | 20x 5 | 18.9 | .23 | ? | S | | | S | |
| 2393 | | | 10 20 04.0 | -49 13 05 | 279.56 | 6.50 | 214 | -88.2 | 39.5 | 43x 9 | 17.1 | .25 | | S | | E | | |
| 2394 | | | 10 20 09.5 | -49 21 45 | 279.65 | 6.39 | 214 | -87.0 | 31.8 | 13x 5 | 19.0 | .24 | | S | | | | |
| 2395 | | | 10 20 14.1 | -62 42 09 | 286.85 | -4.85 | 127 | 74.8 | -146.8 | 20x 9 | 18.0 | .36 | | S | | | 1 | |
| 2396 | | | 10 20 14.7 | -47 28 20 | 278.63 | 7.99 | 214 | -91.6 | 133.0 | 20x 20 | 17.9 | .15 | | S | | F | | dist.nucl, vvLSBdisk |
| 2397 | | | 10 20 16.1 | -48 54 13 | 279.42 | 6.78 | 214 | -87.3 | 56.4 | 15x 5 | 19.2 | .28 | | | | | | |
| 2398 | | | 10 20 17.0 | -50 29 09 | 280.28 | 5.45 | 214 | -82.6 | -28.3 | 13x 12 | 18.4 | .34 | ? | | | | | v.obs, fuzzy |
| 2399 | | | 10 20 19.0 | -63 57 38 | 287.53 | -5.91 | 92 | 23.2 | 58.2 | 17x 15 | 18.0 | .37 | | S | | | | |
| 2400 | | | 10 20 19.1 | -49 26 34 | 279.72 | 6.34 | 214 | -85.3 | 27.6 | 23x 5 | 18.6 | .23 | | S | | | | |

**Table 1.** Galaxies in the southern ZOA – I. The Hydra/Antlia Extension (continued)

| RKK Ident (1) | Other Ident (2) | IR (3) | R.A. (h m s) (4) | Dec. (° ′ ″) (5) | gal $\ell$ (°) (6) | gal $b$ (°) (7) | SRC (8) | X (mm) (9) | Y (mm) (10) | D x d (″) (11) | $B_J$ ($^m$) (12) | $N_{HI}$ (13) | u | Type class. (14) | o * | Remarks (15) |
|---|---|---|---|---|---|---|---|---|---|---|---|---|---|---|---|---|
| 2401 | | | 10 20 27.4 | -67 04 31 | 289.24 | -8.52 | 92 | 21.5 | -108.7 | 13x 9 | 18.6 | .25 | | | | |
| 2402 | | | 10 20 27.5 | -49 07 08 | 279.56 | 6.62 | 214 | -85.0 | 45.0 | 40x 5 | 17.8 | .26 | | S | E | |
| 2403 | | | 10 20 28.4 | -52 01 18 | 281.13 | 4.18 | 214 | -76.5 | -110.4 | 15x 3 | 20.4 | .64 | | S | E | vvLSB |
| 2404 | | | 10 20 28.7 | -50 44 52 | 280.45 | 5.25 | 214 | -80.2 | -42.2 | 17x 4 | 19.8 | .35 | | S | E | vvLSB |
| 2405 | | | 10 20 29.3 | -51 36 32 | 280.91 | 4.52 | 214 | -77.6 | -88.3 | 20x 9 | 18.5 | .44 | | S M: | | dist.nucl, vLSBdisk |
| 2406 | | | 10 20 30.0 | -50 03 05 | 280.07 | 5.84 | 214 | -82.0 | -4.9 | 20x 19 | 17.4 | .29 | | | | diff. |
| 2407 | | Q | 10 20 30.1 | -50 08 40 | 280.12 | 5.76 | 214 | -81.7 | -9.9 | 16x 4 | 19.4 | .34 | ? | S ? | E1 | |
| 2408 | | | 10 20 32.8 | -50 38 00 | 280.39 | 5.35 | 214 | -79.9 | -36.1 | 32x 13 | 17.5 | .35 | | S M | | v.obs. |
| 2409 | | | 10 20 35.4 | -47 33 47 | 278.73 | 7.94 | 214 | -88.3 | 128.3 | 16x 4 | 19.3 | .15 | | S | E1 | |
| 2410 | | | 10 20 39.3 | -51 11 05 | 280.71 | 4.90 | 214 | -77.4 | -65.5 | 12x 4 | 19.4 | .35 | | S | | |
| 2411 | | | 10 20 39.8 | -47 37 53 | 278.78 | 7.89 | 214 | -87.4 | 124.7 | 17x 12 | 17.8 | .14 | | E 3 | | |
| 2412 | | | 10 20 42.1 | -66 28 09 | 288.93 | -8.00 | 92 | 23.3 | -76.2 | 20x 7 | 18.5 | .24 | | S ? | | br.blg or s.p.* |
| 2413 | | | 10 20 47.0 | -47 30 31 | 278.73 | 8.00 | 214 | -86.7 | 131.3 | 15x 4 | 19.6 | .15 | | S | E | |
| 2414 | | | 10 20 47.6 | -66 29 33 | 288.95 | -8.01 | 92 | 23.8 | -77.5 | 12x 3 | 20.0 | .25 | | I ? | | |
| 2415 | | | 10 20 51.1 | -49 28 08 | 279.81 | 6.36 | 214 | -80.6 | 26.4 | 20x 15 | 17.3 | .26 | | E 3 | | v.obs. |
| 2416 | | | 10 20 55.6 | -66 30 22 | 288.97 | -8.02 | 92 | 24.5 | -78.2 | 13x 5 | 19.1 | .25 | | | | |
| 2417 | | | 10 20 57.0 | -47 31 52 | 278.77 | 8.00 | 214 | -85.1 | 130.2 | 26x 13 | 17.6 | .15 | | S | VS | |
| 2418 | | | 10 20 57.8 | -49 46 13 | 279.98 | 6.12 | 214 | -78.8 | 10.3 | 16x 7 | 18.5 | .29 | | | | |
| 2419 | | | 10 21 05.2 | -65 03 56 | 288.20 | -6.79 | 92 | 26.7 | -1.1 | 20x 9 | 18.4 | .39 | | S ? | 1 | asym.spiral |
| 2420 | | | 10 21 05.3 | -47 27 34 | 278.75 | 8.07 | 214 | -84.1 | 134.1 | 17x 9 | 18.2 | .15 | | | | |
| 2421 | | | 10 21 05.3 | -52 09 34 | 281.29 | 4.11 | 214 | -71.1 | -117.5 | 19x 5 | 18.7 | .63 | | S ? | | p.cov.by * |
| 2422 | | | 10 21 05.8 | -49 35 26 | 279.90 | 6.28 | 214 | -78.1 | 20.0 | 19x 7 | 18.8 | .30 | | | | v.obs. |
| 2423 | | | 10 21 08.4 | -51 40 00 | 281.03 | 4.53 | 214 | -72.0 | -91.1 | 17x 8 | 18.3 | .42 | | | S | |
| 2424 | | | 10 21 08.8 | -49 22 07 | 279.79 | 6.47 | 214 | -78.3 | 31.9 | 24x 5 | 19.2 | .29 | | S | E1 | v.obs. |
| 2425 | | | 10 21 09.7 | -49 08 09 | 279.67 | 6.67 | 214 | -78.8 | 44.4 | 51x 13 | 17.1 | .31 | | S M | | v.obs. |
| 2426 | L214-002 | I | 10 21 10.1 | -49 13 02 | 279.71 | 6.60 | 214 | -78.5 | 40.0 | 77x 44 | 14.8 | .30 | | S 2 | S | |
| 2427 | | | 10 21 10.5 | -47 52 43 | 278.99 | 7.73 | 214 | -82.1 | 111.7 | 19x 13 | 17.8 | .15 | | | | p.cov.by br.* |
| 2428 | | | 10 21 11.4 | -50 04 24 | 280.18 | 5.88 | 214 | -76.0 | -5.8 | 13x 13 | 18.2 | .32 | ? | S M | F | or * w.halo |
| 2429 | | | 10 21 12.2 | -51 12 55 | 280.80 | 4.92 | 214 | -72.7 | -66.9 | 13x 4 | 19.5 | .36 | | S | | w.comp. |
| 2430 | | | 10 21 12.8 | -50 19 05 | 280.31 | 5.68 | 214 | -75.1 | -18.9 | 20x 3 | 19.4 | .35 | | S | E | |
| 2431 | | | 10 21 15.8 | -53 15 37 | 281.90 | 3.19 | 168 | -46.2 | 92.9 | 17x 8 | 18.1 | .57 | | S ? | | |
| 2432 | | | 10 21 16.6 | -65 31 46 | 288.47 | -7.17 | 92 | 27.3 | -25.9 | 26x 19 | 17.0 | .33 | ? | S | FS | or PN |
| 2433 | | | 10 21 20.2 | -49 44 21 | 280.02 | 6.17 | 214 | -75.6 | 12.2 | 38x 4 | 18.4 | .31 | | S | E | |
| 2434 | | | 10 21 20.8 | -53 35 44 | 282.09 | 2.92 | 168 | -45.2 | 74.9 | 20x 13 | 17.0 | .57 | | E 3 | | |
| 2435 | | | 10 21 23.6 | -49 56 07 | 280.13 | 6.01 | 214 | -74.6 | 1.7 | 34x 16 | 17.7 | .31 | ? | | | dwarf Im? |
| 2436 | | | 10 21 28.1 | -47 52 18 | 279.03 | 7.76 | 214 | -79.5 | 112.2 | 23x 20 | 17.0 | .16 | | S E? | | |
| 2437 | | | 10 21 28.1 | -51 36 05 | 281.04 | 4.61 | 214 | -69.5 | -87.5 | 24x 9 | 18.4 | .40 | | S ? | S | vLSB |
| 2438 | | | 10 21 28.8 | -51 11 52 | 280.82 | 4.95 | 214 | -70.5 | -65.9 | 13x 5 | 19.2 | .36 | | S ? | | |
| 2439 | | | 10 21 33.7 | -48 20 26 | 279.29 | 7.37 | 214 | -77.4 | 87.1 | 16x 4 | 19.3 | .19 | | S | E | w.comp. |
| 2440 | | | 10 21 34.8 | -50 04 35 | 280.23 | 5.91 | 214 | -72.7 | -5.8 | 13x 3 | 19.5 | .34 | | S | | HSB |
| 2441 | | | 10 21 43.5 | -50 19 22 | 280.39 | 5.71 | 214 | -70.7 | -18.9 | 15x 4 | 19.4 | .34 | | S | E | p.cov.by ** |
| 2442 | | | 10 21 43.6 | -50 19 21 | 280.39 | 5.72 | 214 | -70.7 | -18.9 | 13x 8 | 18.6 | .35 | ? | | | p.cov.by ** |
| 2443 | L214-003 | | 10 21 44.8 | -51 16 42 | 280.90 | 4.91 | 214 | -68.0 | -70.1 | 74x 7 | 17.1 | .39 | | S | E | |
| 2444 | | | 10 21 45.8 | -65 05 37 | 288.28 | -6.78 | 92 | 30.5 | -2.6 | 17x 15 | 17.9 | .39 | | S ? | S | |
| 2445 | | | 10 21 47.4 | -47 57 12 | 279.12 | 7.72 | 214 | -76.4 | 108.0 | 34x 34 | 16.2 | .16 | | S R2 | S | |
| 2446 | | | 10 21 47.5 | -49 42 04 | 280.06 | 6.25 | 214 | -71.8 | 14.4 | 20x 7 | 18.4 | .32 | | S ? | | |
| 2447 | | | 10 21 47.6 | -48 07 57 | 279.21 | 7.57 | 214 | -75.9 | 98.4 | 15x 4 | 19.3 | .16 | | S | E | |
| 2448 | | | 10 21 48.2 | -50 21 22 | 280.41 | 5.69 | 214 | -70.0 | -20.7 | 15x 11 | 18.3 | .34 | | S ? | F | |
| 2449 | | | 10 21 48.8 | -47 55 49 | 279.11 | 7.74 | 214 | -76.3 | 109.2 | 27x 13 | 17.2 | .16 | | | | neighb.of 2445 |
| 2450 | | | 10 21 51.4 | -49 31 07 | 279.97 | 6.41 | 214 | -71.7 | 24.2 | 34x 28 | 16.4 | .32 | ? | S RM | S | sharp edge, PN? |
| 2451 | | | 10 22 01.9 | -50 02 59 | 280.28 | 5.97 | 214 | -68.8 | -4.2 | 27x 7 | 18.1 | .35 | | S | E | |
| 2452 | | | 10 22 03.0 | -64 48 41 | 288.15 | -6.52 | 92 | 32.4 | 12.5 | 19x 15 | 17.9 | .40 | | SB 5 | | |
| 2453 | | | 10 22 03.6 | -49 09 16 | 279.80 | 6.73 | 214 | -70.9 | 43.8 | 15x 4 | 19.1 | .37 | | S ? | | |
| 2454 | | | 10 22 05.1 | -50 02 51 | 280.29 | 5.98 | 214 | -68.4 | -4.0 | 23x 7 | 18.4 | .34 | | S | E | |
| 2455 | | | 10 22 06.3 | -49 41 08 | 280.09 | 6.29 | 214 | -69.1 | 15.3 | 17x 4 | 19.4 | .32 | | | 1 | v.obs. |
| 2456 | | I2 | 10 22 09.8 | -51 43 50 | 281.20 | 4.56 | 214 | -63.4 | -94.1 | 24x 5 | 18.7 | .42 | | | | pair w.2457 |
| 2457 | | I2 | 10 22 14.2 | -51 44 02 | 281.21 | 4.56 | 214 | -62.7 | -94.3 | 22x 4 | 19.1 | .42 | | SY M | | pair w.2456 |
| 2458 | | | 10 22 14.3 | -64 47 51 | 288.16 | -6.50 | 92 | 33.5 | 13.2 | 16x 5 | 19.3 | .41 | | S | | |
| 2459 | | | 10 22 15.2 | -49 39 13 | 280.10 | 6.33 | 214 | -67.9 | 17.1 | 42x 7 | 17.8 | .32 | | S | E | |
| 2460 | | | 10 22 26.6 | -50 39 58 | 280.67 | 5.49 | 214 | -63.7 | -37.0 | 13x 11 | 18.5 | .33 | ? | | | |
| 2461 | | | 10 22 28.5 | -48 58 46 | 279.77 | 6.92 | 214 | -67.7 | 53.3 | 11x 4 | 19.9 | .33 | | | | neighb.of 2462 |
| 2462 | | | 10 22 29.5 | -48 58 18 | 279.76 | 6.92 | 214 | -67.6 | 53.7 | 16x 5 | 19.3 | .34 | | S | | |
| 2463 | | | 10 22 34.3 | -49 52 41 | 280.26 | 6.16 | 214 | -64.6 | 5.2 | 15x 12 | 18.3 | .32 | | | | |
| 2464 | | | 10 22 41.8 | -67 39 51 | 289.75 | -8.90 | 92 | 32.5 | -140.5 | 15x 4 | 19.1 | .24 | | | | |
| 2465 | | | 10 22 42.5 | -50 49 35 | 280.79 | 5.37 | 214 | -61.1 | -45.5 | 27x 3 | 19.7 | .33 | | S | E | vvLSB |
| 2466 | | | 10 22 43.5 | -51 13 47 | 281.01 | 5.03 | 214 | -59.9 | -67.1 | 20x 4 | 19.1 | .40 | | S | | |
| 2467 | | | 10 22 45.6 | -51 03 54 | 280.92 | 5.17 | 214 | -60.1 | -58.2 | 19x 4 | 19.3 | .33 | | S | E | |
| 2468 | | | 10 22 48.8 | -49 44 41 | 280.22 | 6.30 | 214 | -62.9 | 12.5 | 36x 5 | 18.3 | .33 | | S M | E | |
| 2469 | L168-002 | | 10 22 49.9 | -54 32 40 | 282.78 | 2.23 | 168 | -32.4 | 24.3 | 101x 13 | 15.7 | .63 | | S 4 | E | |
| 2470 | | | 10 22 51.1 | -49 48 29 | 280.26 | 6.25 | 214 | -62.4 | 9.1 | 24x 15 | 17.6 | .33 | | S | | dist.blg, LSBdisk |
| 2471 | | | 10 22 51.7 | -50 02 11 | 280.39 | 6.05 | 214 | -61.7 | -3.1 | 16x 9 | 18.0 | .32 | | L | | or S..E |
| 2472 | | | 10 23 02.5 | -65 22 08 | 288.54 | -6.94 | 92 | 37.3 | -17.5 | 17x 4 | 19.6 | .38 | | S | E | |
| 2473 | | | 10 23 03.8 | -65 20 55 | 288.53 | -6.92 | 92 | 37.5 | -16.5 | 13x 8 | 18.6 | .38 | | S M: | | |
| 2474 | | | 10 23 03.9 | -51 05 58 | 280.98 | 5.17 | 214 | -57.4 | -60.0 | 15x 5 | 19.3 | .36 | | | | |
| 2475 | | | 10 23 05.8 | -65 47 30 | 288.77 | -7.30 | 92 | 37.1 | -40.2 | 20x 7 | 18.6 | .29 | | S ? | N | vLSBdisk |
| 2476 | | | 10 23 10.3 | -67 13 07 | 289.55 | -8.50 | 92 | 35.5 | -116.7 | 19x 8 | 18.7 | .25 | | | | |
| 2477 | | | 10 23 11.6 | -49 58 17 | 280.40 | 6.14 | 214 | -59.0 | .5 | 13x 7 | 18.9 | .32 | | S | | |
| 2478 | | | 10 23 12.3 | -50 36 43 | 280.74 | 5.60 | 214 | -57.4 | -33.8 | 23x 8 | 18.2 | .32 | | S M | | v.obs. |
| 2479 | | | 10 23 12.6 | -50 53 07 | 280.89 | 5.36 | 214 | -56.7 | -48.5 | 13x 11 | 18.9 | .32 | ? | | | between 2 ** |
| 2480 | | | 10 23 15.2 | -49 33 04 | 280.18 | 6.50 | 214 | -59.5 | 23.0 | 23x 11 | 17.6 | .33 | | L | | |
| 2481 | | | 10 23 17.5 | -48 16 48 | 279.51 | 7.58 | 214 | -62.2 | 91.1 | 15x 11 | 18.2 | .28 | ? | | | p.cov.by * |
| 2482 | | | 10 23 17.6 | -50 13 31 | 280.55 | 5.93 | 214 | -57.6 | -13.1 | 23x 5 | 18.7 | .30 | | S | E | |
| 2483 | | | 10 23 22.7 | -66 28 25 | 289.16 | -7.86 | 92 | 37.7 | -76.8 | 15x 7 | 18.7 | .26 | | S | 1 | faint sp.arms? |
| 2484 | | | 10 23 24.6 | -51 42 48 | 281.35 | 4.68 | 214 | -53.1 | -92.7 | 26x 11 | 17.6 | .40 | | S M: | | in dense quartett |
| 2485 | L214-004 | I2 | 10 23 24.8 | -49 57 59 | 280.42 | 6.16 | 214 | -57.2 | .8 | 17x 8 | 18.3 | .32 | | | | i.a.w.2486 |
| 2486 | L214-004 | I2 | 10 23 25.9 | -49 57 24 | 280.42 | 6.17 | 214 | -57.0 | 1.4 | 34x 16 | 16.8 | .32 | | S 5 | | dist.sp.arms, i.a.w.2485 |
| 2487 | | | 10 23 27.6 | -51 43 08 | 281.36 | 4.68 | 214 | -52.6 | -93.0 | 13x 8 | 18.6 | .40 | | E | | in dense quartett |
| 2488 | | | 10 23 27.8 | -49 35 24 | 280.23 | 6.48 | 214 | -57.6 | 21.0 | 13x 11 | 18.4 | .34 | | | | p.cov.by ** |
| 2489 | | | 10 23 28.7 | -49 37 05 | 280.25 | 6.46 | 214 | -57.4 | 19.5 | 28x 8 | 17.9 | .35 | | S E | | |
| 2490 | | | 10 23 29.0 | -51 42 47 | 281.36 | 4.68 | 214 | -52.5 | -92.7 | 27x 9 | 17.7 | .40 | | S E | | in dense quartett |
| 2491 | | | 10 23 30.6 | -49 36 24 | 280.25 | 6.47 | 214 | -57.2 | 20.1 | 27x 20 | 16.9 | .35 | | F | | |
| 2492 | | | 10 23 32.0 | -51 43 15 | 281.37 | 4.68 | 214 | -52.0 | -93.1 | 12x 9 | 18.6 | .41 | | E | | in dense quartett |
| 2493 | | | 10 23 32.2 | -49 37 05 | 280.26 | 6.47 | 214 | -56.9 | 19.5 | 13x 5 | 19.4 | .35 | | S ? | | |
| 2494 | | | 10 23 32.7 | -64 22 01 | 288.05 | -6.06 | 92 | 41.6 | 36.1 | 20x 13 | 18.1 | .54 | | | | dist.nucl. or s.p.* |
| 2495 | | | 10 23 33.4 | -51 32 30 | 281.28 | 4.84 | 214 | -52.3 | -83.5 | 20x 8 | 18.4 | .38 | | | | p.cov.by ** |
| 2496 | | | 10 23 33.8 | -47 59 15 | 279.39 | 7.85 | 214 | -60.4 | 106.9 | 20x 7 | 18.5 | .23 | | S M | E | |
| 2497 | | | 10 23 34.0 | -49 34 55 | 280.24 | 6.50 | 214 | -56.7 | 21.5 | 23x 5 | 18.5 | .35 | | S | | |
| 2498 | | | 10 23 38.4 | -49 35 11 | 280.25 | 6.50 | 214 | -56.1 | 21.3 | 26x 5 | 18.6 | .35 | | I | | p.cov.by ** |
| 2499 | | | 10 23 42.8 | -49 35 55 | 280.27 | 6.50 | 214 | -55.4 | 20.6 | 16x 7 | 18.6 | .35 | ? | | | |
| 2500 | | | 10 23 51.6 | -49 29 27 | 280.23 | 6.60 | 214 | -54.4 | 26.5 | 13x 9 | 18.5 | .33 | ? | | | |

**Table 1.** Galaxies in the southern ZOA – I. The Hydra/Antlia Extension (continued)

| RKK Ident (1) | Other Ident (2) | IR (3) | R.A. (h m s) (4) | Dec. (° ′ ″) (5) | gal $\ell$ (°) (6) | gal $b$ (°) (7) | SRC (8) | X (mm) (9) | Y (mm) (10) | D x d (″) (11) | $B_J$ ($^m$) (12) | $N_{HI}$ (13) | u | Type class. (14) | o * | Remarks (15) |
|---|---|---|---|---|---|---|---|---|---|---|---|---|---|---|---|---|
| 2501 | | | 10 23 52.1 | -48 09 55 | 279.53 | 7.73 | 214 | -57.3 | 97.5 | 15x 11 | 18.6 | .28 | ? | | | between ** |
| 2502 | | | 10 23 52.3 | -51 17 22 | 281.19 | 5.08 | 214 | -50.2 | -69.9 | 27x 27 | 16.7 | .35 | | S M | F S | |
| 2503 | | | 10 23 55.8 | -53 32 09 | 282.38 | 3.17 | 168 | -24.6 | 78.5 | 17x 5 | 18.4 | .51 | | S L | E | |
| 2504 | | | 10 23 56.9 | -48 25 48 | 279.68 | 7.51 | 214 | -56.0 | 83.3 | 13x 11 | 18.3 | .33 | | | | |
| 2505 | | | 10 23 57.5 | -67 23 55 | 289.71 | -8.61 | 92 | 39.3 | -126.4 | 30x 16 | 16.8 | .25 | | S E? | | |
| 2506 | | | 10 24 01.7 | -53 22 24 | 282.31 | 3.32 | 168 | -24.0 | 87.2 | 20x 13 | 17.9 | .50 | | S E | | |
| 2507 | | I | 10 24 01.8 | -53 37 04 | 282.44 | 3.11 | 168 | -23.8 | 74.1 | 51x 34 | 15.4 | .55 | | SB 5 | | |
| 2508 | L062-010 | I | 10 24 04.9 | -67 29 26 | 289.77 | -8.68 | 92 | 39.8 | -131.4 | 60x 60 | 14.9 | .25 | | SX 7 | S | |
| 2509 | | | 10 24 09.8 | -51 40 09 | 281.43 | 4.78 | 214 | -46.9 | -90.1 | 16x 12 | 18.7 | .43 | ? | | | vvLSB |
| 2510 | | Q | 10 24 10.9 | -51 22 47 | 281.28 | 5.02 | 214 | -47.4 | -74.6 | 30x 5 | 18.1 | .35 | | S ? | | |
| 2511 | | | 10 24 12.5 | -49 44 28 | 280.41 | 6.42 | 214 | -50.8 | 13.2 | 13x 12 | 18.2 | .34 | | E ? | | |
| 2512 | | Q | 10 24 13.1 | -53 31 37 | 282.42 | 3.20 | 168 | -22.4 | 79.0 | 27x 27 | 16.1 | .50 | ? | F | 1 | |
| 2513 | | | 10 24 17.6 | -52 12 33 | 281.73 | 4.33 | 214 | -44.6 | -119.0 | 15x 8 | 18.6 | .53 | | S | | |
| 2514 | | | 10 24 28.0 | -49 35 11 | 280.37 | 6.57 | 214 | -48.9 | 21.6 | 13x 12 | 18.5 | .36 | | S M? | F | v.obs, dist.nucl. |
| 2515 | | | 10 24 28.9 | -52 11 43 | 281.75 | 4.36 | 214 | -43.1 | -118.2 | 11x 8 | 18.9 | .56 | ? | | | neighb.of 2513,larger disk |
| 2516 | | | 10 24 29.9 | -48 08 43 | 279.61 | 7.80 | 214 | -51.7 | 98.8 | 11x 4 | 19.5 | .31 | | S | | neighb.of 2518 |
| 2517 | | NG | 10 24 34.4 | -48 09 55 | 279.63 | 7.79 | 214 | -51.0 | 97.7 | 8x 8 | 19.1 | .31 | | E ? | | neighb.of 2518 |
| 2518 | | I | 10 24 34.6 | -48 10 55 | 279.64 | 7.78 | 214 | -50.9 | 96.8 | 15x 7 | 18.6 | .32 | | S ? | | |
| 2519 | | | 10 24 34.9 | -67 04 51 | 289.59 | -8.31 | 92 | 43.1 | -109.5 | 15x 5 | 19.3 | .25 | | | 1 | |
| 2520 | | | 10 24 37.5 | -49 38 32 | 280.42 | 6.54 | 214 | -47.4 | 18.6 | 13x 5 | 19.5 | .36 | | S | | |
| 2521 | | | 10 24 39.2 | -50 48 39 | 281.04 | 5.55 | 214 | -44.7 | -43.9 | 24x 11 | 18.1 | .39 | ? | S | S | vvLSB |
| 2522 | L214-005 | NG | 10 24 44.0 | -48 54 27 | 280.04 | 7.17 | 214 | -48.0 | 58.0 | 22x 19 | 17.0 | .28 | | E 0 | | blg or s.p.*, |
| 2523 | | | 10 24 44.9 | -50 10 45 | 280.72 | 6.09 | 214 | -45.2 | -10.1 | 13x 5 | 19.1 | .27 | | | 1 | |
| 2524 | | | 10 24 48.2 | -51 57 15 | 281.66 | 4.59 | 214 | -41.0 | -105.1 | 20x 15 | 17.5 | .52 | | F | | |
| 2525 | L214-005 | I | 10 24 51.7 | -48 53 34 | 280.06 | 7.20 | 214 | -46.9 | 58.8 | 47x 20 | 16.2 | .29 | | S 3 | | |
| 2526 | | | 10 24 51.8 | -51 56 33 | 281.67 | 4.60 | 214 | -40.5 | -104.5 | 13x 7 | 19.6 | .52 | | | | neighb.of 2524, vvLSB |
| 2527 | | | 10 24 52.3 | -51 18 29 | 281.33 | 5.14 | 214 | -41.8 | -70.5 | 23x 4 | 18.9 | .36 | | S | S | |
| 2528 | | | 10 24 52.5 | -49 31 44 | 280.39 | 6.66 | 214 | -45.5 | 24.8 | 27x 8 | 18.1 | .34 | | S M | | v.obs. |
| 2529 | | | 10 24 52.8 | -50 00 05 | 280.64 | 6.26 | 214 | -44.4 | -.5 | 15x 9 | 18.6 | .30 | | | | |
| 2530 | | | 10 24 53.7 | -51 10 54 | 281.27 | 5.25 | 214 | -41.9 | -63.7 | 13x 11 | 18.7 | .35 | ? | | S | |
| 2531 | | NG | 10 24 54.0 | -48 53 07 | 280.06 | 7.21 | 214 | -46.6 | 59.3 | 17x 7 | 18.3 | .29 | | S | | |
| 2532 | | | 10 24 55.7 | -53 33 47 | 282.52 | 3.23 | 168 | -16.7 | 77.1 | 13x 9 | 17.6 | .51 | ? | | | |
| 2533 | | NS | 10 24 57.5 | -49 59 56 | 280.65 | 6.26 | 214 | -43.8 | -.4 | 16x 13 | 18.1 | .30 | | | | |
| 2534 | | | 10 24 58.3 | -48 47 47 | 280.02 | 7.29 | 214 | -46.2 | 64.1 | 13x 11 | 18.5 | .32 | | E 0 ? | | between 2 ** |
| 2535 | | NG | 10 24 58.5 | -48 52 32 | 280.06 | 7.22 | 214 | -46.0 | 59.8 | 15x 4 | 19.3 | .29 | | S | E | |
| 2536 | | | 10 24 59.3 | -51 16 49 | 281.33 | 5.18 | 214 | -40.9 | -69.0 | 13x 3 | 20.0 | .36 | | | | |
| 2537 | | | 10 25 00.3 | -51 25 50 | 281.42 | 5.05 | 214 | -40.4 | -77.0 | 16x 8 | 18.4 | .39 | | | S | |
| 2538 | | | 10 25 01.9 | -48 18 51 | 279.77 | 7.70 | 214 | -46.6 | 89.9 | 22x 8 | 18.3 | .34 | | S E | | |
| 2539 | | | 10 25 03.8 | -67 25 29 | 289.81 | -8.58 | 92 | 45.0 | -128.0 | 23x 16 | 17.0 | .24 | | F ? | | |
| 2540 | | | 10 25 05.6 | -49 20 42 | 280.33 | 6.83 | 214 | -44.0 | 34.7 | 13x 5 | 19.0 | .31 | | S ? | | |
| 2541 | | | 10 25 06.1 | -49 00 31 | 280.15 | 7.12 | 214 | -44.6 | 52.7 | 13x 5 | 19.0 | .29 | | | | |
| 2542 | | | 10 25 06.6 | -48 19 47 | 279.79 | 7.70 | 214 | -45.9 | 89.1 | 11x 11 | 18.7 | .35 | | | | neighb.of 2538 |
| 2543 | | | 10 25 11.7 | -48 02 30 | 279.65 | 7.95 | 214 | -45.7 | 104.6 | 17x 7 | 18.2 | .31 | | | | |
| 2544 | L214-006 | | 10 25 14.4 | -48 52 20 | 280.10 | 7.25 | 214 | -43.6 | 60.1 | 67x 16 | 16.1 | .31 | | S 2 | NS | pair w.2546 |
| 2545 | | I | 10 25 15.5 | -67 11 34 | 289.71 | -8.37 | 92 | 46.4 | -115.6 | 13x 4 | 19.7 | .25 | | S | | blg, 1 sp.arm |
| 2546 | | | 10 25 16.6 | -48 51 37 | 280.10 | 7.26 | 214 | -43.3 | 60.7 | 51x 43 | 15.3 | .31 | | S 1 | VS | pair w.2544 |
| 2547 | | | 10 25 16.8 | -50 47 34 | 281.12 | 5.61 | 214 | -39.4 | -42.8 | 13x 11 | 18.7 | .38 | ? | S ? | F | vvLSB |
| 2548 | | NG | 10 25 21.8 | -67 13 19 | 289.73 | -8.39 | 92 | 46.9 | -117.2 | 17x 4 | 19.4 | .24 | | | | |
| 2549 | | | 10 25 22.5 | -50 13 14 | 280.83 | 6.11 | 214 | -39.8 | -12.1 | 38x 22 | 16.7 | .26 | | S 5 : | VS | v.obs. |
| 2550 | | | 10 25 25.2 | -51 42 29 | 281.62 | 4.85 | 214 | -36.4 | -91.8 | 15x 4 | 19.6 | .46 | | S | | |
| 2551 | | | 10 25 27.0 | -49 33 38 | 280.49 | 6.68 | 214 | -40.4 | 23.3 | 17x 11 | 18.0 | .36 | | S ? | | |
| 2552 | | | 10 25 30.0 | -48 45 37 | 280.08 | 7.36 | 214 | -41.6 | 66.2 | 24x 5 | 18.9 | .36 | | S | E | v.obs. |
| 2553 | | | 10 25 31.1 | -50 05 30 | 280.78 | 6.23 | 214 | -38.8 | -5.2 | 16x 4 | 19.3 | .29 | | S | | |
| 2554 | | | 10 25 33.1 | -51 51 40 | 281.71 | 4.73 | 214 | -35.0 | -99.9 | 16x 16 | 18.1 | .52 | | S | F S | |
| 2555 | | | 10 25 33.8 | -51 04 08 | 281.30 | 5.40 | 214 | -36.5 | -57.5 | 22x 13 | 17.7 | .38 | | S M | | dist.sp.arms |
| 2556 | | | 10 25 35.2 | -52 09 54 | 281.88 | 4.47 | 214 | -34.1 | -116.2 | 13x 5 | 19.5 | .53 | ? | | S | p.cov.by * |
| 2557 | | | 10 25 35.6 | -47 58 49 | 279.68 | 8.04 | 214 | -42.3 | 108.0 | 16x 12 | 17.9 | .33 | | | | |
| 2558 | | | 10 25 37.5 | -48 13 10 | 279.81 | 7.84 | 214 | -41.5 | 95.2 | 15x 8 | 18.3 | .35 | | | | |
| 2559 | | | 10 25 37.6 | -48 59 15 | 280.21 | 7.18 | 214 | -40.0 | 54.0 | 31x 8 | 17.7 | .32 | | S | | p.cov.by ** |
| 2560 | | | 10 25 38.7 | -48 13 18 | 279.81 | 7.84 | 214 | -41.3 | 95.1 | 15x 3 | 20.4 | .35 | | S | E | |
| 2561 | | | 10 25 39.3 | -51 10 29 | 281.37 | 5.32 | 214 | -35.5 | -63.1 | 16x 13 | 18.5 | .37 | ? | | S | |
| 2562 | | | 10 25 42.5 | -50 11 06 | 280.86 | 6.17 | 214 | -37.0 | -10.1 | 17x 15 | 18.2 | .27 | ? | | 1 | v.diff. |
| 2563 | | | 10 25 45.8 | -49 37 28 | 280.57 | 6.65 | 214 | -37.6 | 20.0 | 19x 13 | 17.8 | .35 | | S ? | | |
| 2564 | | | 10 25 47.0 | -49 17 08 | 280.39 | 6.94 | 214 | -38.1 | 38.1 | 22x 11 | 17.9 | .33 | | | | v.obs, p.cov.by ** |
| 2565 | | | 10 25 53.2 | -51 19 39 | 281.48 | 5.21 | 214 | -33.3 | -71.2 | 19x 17 | 17.6 | .38 | | S | F S | |
| 2566 | | | 10 25 55.9 | -48 38 38 | 280.07 | 7.50 | 214 | -38.0 | 72.5 | 27x 9 | 17.7 | .36 | | S E | S | |
| 2567 | | | 10 25 57.9 | -49 45 20 | 280.67 | 6.56 | 214 | -35.6 | 13.0 | 32x 7 | 18.5 | .32 | | S | 1 | vvLSBdisk |
| 2568 | | | 10 26 00.0 | -50 25 14 | 281.02 | 5.99 | 214 | -34.0 | -22.6 | 15x 7 | 19.0 | .29 | ? | | | p.cov.by ** |
| 2569 | | | 10 26 05.0 | -49 27 05 | 280.52 | 6.83 | 214 | -35.1 | 29.3 | 31x 5 | 18.4 | .33 | | S | E | LSB |
| 2570 | | | 10 26 05.3 | -50 09 10 | 280.89 | 6.23 | 214 | -33.8 | -8.2 | 24x 5 | 18.6 | .27 | | S | E | |
| 2571 | | | 10 26 06.0 | -49 14 48 | 280.42 | 7.00 | 214 | -35.4 | 40.3 | 17x 5 | 19.1 | .33 | | S ? | | pair w.2574 |
| 2572 | | | 10 26 07.0 | -48 28 07 | 280.01 | 7.67 | 214 | -36.6 | 82.0 | 19x 5 | 18.9 | .32 | | S | E | |
| 2573 | | | 10 26 07.2 | -48 59 21 | 280.28 | 7.22 | 214 | -35.7 | 54.1 | 27x 13 | 17.6 | .35 | | S M | V | v.obs. |
| 2574 | | | 10 26 09.1 | -49 15 28 | 280.43 | 7.00 | 214 | -34.9 | 39.7 | 23x 8 | 18.1 | .33 | | | | v.obs, pair w.2571 |
| 2575 | | | 10 26 10.7 | -48 07 31 | 279.84 | 7.96 | 214 | -36.7 | 100.4 | 15x 5 | 19.3 | .35 | | S ? | | |
| 2576 | | | 10 26 10.8 | -48 53 40 | 280.24 | 7.31 | 214 | -35.3 | 59.2 | 31x 4 | 19.1 | .36 | | S | E S | v.obs. |
| 2577 | | | 10 26 11.5 | -49 55 30 | 280.79 | 6.43 | 214 | -33.3 | 4.0 | 13x 5 | 19.5 | .29 | ? | | | between ** |
| 2578 | | | 10 26 12.7 | -50 11 11 | 280.93 | 6.21 | 214 | -32.6 | -10.0 | 20x 4 | 19.4 | .27 | | S | | |
| 2579 | L214-007 | | 10 26 13.9 | -50 15 36 | 280.97 | 6.15 | 214 | -32.3 | -13.9 | 54x 32 | 15.7 | .28 | | SB | S | polar ring, poss.larger? |
| 2580 | | Q | 10 26 15.8 | -51 05 15 | 281.40 | 5.44 | 214 | -30.5 | -58.3 | 23x 12 | 17.7 | .38 | | S ? | S | dbl syst. |
| 2581 | | | 10 26 17.8 | -51 20 07 | 281.54 | 5.24 | 214 | -29.8 | -71.5 | 13x 7 | 19.0 | .38 | ? | | | |
| 2582 | | | 10 26 19.5 | -53 17 12 | 282.56 | 3.57 | 168 | -5.6 | 92.0 | 17x 17 | 18.0 | .53 | | S | F | dist.nucl. |
| 2583 | | | 10 26 19.6 | -49 24 25 | 280.53 | 6.88 | 214 | -33.1 | 31.8 | 20x 16 | 17.5 | .33 | | | S | v.obs. |
| 2584 | | | 10 26 19.7 | -51 40 57 | 281.72 | 4.94 | 214 | -28.9 | -90.1 | 32x 15 | 16.7 | .40 | | | S | |
| 2585 | L214-008 | I | 10 26 23.3 | -50 26 07 | 281.08 | 6.01 | 214 | -30.7 | -23.3 | 65x 47 | 14.9 | .29 | | S M | | HSB |
| 2586 | L214-009 | | 10 26 23.7 | -50 32 32 | 281.14 | 5.92 | 214 | -30.4 | -29.0 | 19x 11 | 18.5 | .30 | | S | | large LSBdisk |
| 2587 | L214-009 | | 10 26 24.7 | -50 33 01 | 281.14 | 5.92 | 214 | -30.3 | -29.4 | 20x 7 | 18.9 | .30 | | S | | poss.larger? |
| 2588 | | I | 10 26 26.4 | -64 13 54 | 288.25 | -5.78 | 92 | 58.7 | 42.8 | 39x 20 | 16.9 | .66 | | S | | |
| 2589 | | | 10 26 26.6 | -51 27 16 | 281.62 | 5.15 | 214 | -28.4 | -77.9 | 19x 5 | 18.8 | .36 | | S | | |
| 2590 | | | 10 26 27.4 | -51 07 16 | 281.45 | 5.43 | 214 | -28.9 | -60.0 | 20x 5 | 18.9 | .39 | | S | | |
| 2591 | | | 10 26 31.8 | -50 08 08 | 280.94 | 6.28 | 214 | -30.0 | -7.2 | 27x 20 | 16.8 | .26 | | E 3 | | |
| 2592 | | | 10 26 37.6 | -49 13 26 | 280.48 | 7.07 | 214 | -30.8 | 41.7 | 47x 34 | 15.6 | .33 | | | S | dbl. E syst? |
| 2593 | | | 10 26 37.7 | -49 13 26 | 280.48 | 7.07 | 214 | -30.8 | 41.7 | 27x 4 | 18.8 | .33 | | S | E | |
| 2594 | | | 10 26 41.6 | -51 07 19 | 281.48 | 5.45 | 214 | -26.9 | -60.0 | 22x 8 | 18.2 | .40 | | S | | |
| 2595 | L214-010 | I | 10 26 42.5 | -50 42 53 | 281.27 | 5.80 | 214 | -27.5 | -38.2 | 81x 22 | 15.6 | .34 | | S 1 | S | |
| 2596 | | | 10 26 43.1 | -49 03 37 | 280.41 | 7.21 | 214 | -30.3 | 50.5 | 20x 3 | 19.8 | .38 | | S | E | v.obs, pair w.2597 |
| 2597 | | | 10 26 46.2 | -49 03 52 | 280.41 | 7.21 | 214 | -29.8 | 50.3 | 32x 5 | 18.7 | .38 | | S | E | v.obs, pair w.2596 |
| 2598 | L214-011 | | 10 26 46.2 | -50 26 35 | 281.14 | 6.04 | 214 | -27.4 | -23.6 | 67x 32 | 15.4 | .29 | | S 1 | | dist.dust lane |
| 2599 | | | 10 26 50.2 | -67 04 46 | 289.78 | -8.19 | 92 | 54.8 | -109.8 | 13x 5 | 19.7 | .26 | | | | LSB |
| 2600 | | | 10 26 55.1 | -64 45 30 | 288.57 | -6.21 | 92 | 60.3 | 14.5 | 17x 8 | 18.7 | .55 | | S ? | | |

**Table 1.** Galaxies in the southern ZOA – I. The Hydra/Antlia Extension (continued)

| RKK Ident (1) | Other Ident (2) | IR (3) | R.A. (h m s) (4) | Dec. (° ′ ″) (5) | gal ℓ (°) (6) | gal b (°) (7) | SRC (8) | X (mm) (9) | Y (mm) (10) | D x d (″) (11) | $B_J$ ($^m$) (12) | $N_{HI}$ (13) | u | Type class. (14) | o | * | Remarks (15) |
|---|---|---|---|---|---|---|---|---|---|---|---|---|---|---|---|---|---|
| 2601 | | | 10 26 55.3 | -67 37 02 | 290.07 | -8.65 | 92 | 54.1 | -138.7 | 20x 13 | 17.9 | .23 | | | | | |
| 2602 | | | 10 26 57.3 | -50 10 09 | 281.02 | 6.29 | 214 | -26.3 | -8.9 | 47x 20 | 16.3 | .25 | | S M | | S | br.blg, diff. LSBdisk |
| 2603 | | | 10 26 59.9 | -49 59 14 | 280.93 | 6.45 | 214 | -26.2 | .9 | 38x 27 | 16.4 | .25 | | S M: | | S | br.blg, LSBdisk |
| 2604 | | I | 10 27 02.5 | -51 25 58 | 281.69 | 5.21 | 214 | -23.4 | -76.6 | 17x 12 | 17.6 | .36 | | | | | w.sev.comp. |
| 2605 | | | 10 27 05.3 | -49 13 17 | 280.54 | 7.11 | 214 | -26.8 | 41.9 | 24x 4 | 18.7 | .34 | | S | E | | |
| 2606 | | | 10 27 08.7 | -48 50 16 | 280.35 | 7.44 | 214 | -26.9 | 62.5 | 27x 13 | 17.3 | .36 | | F | | | |
| 2607 | | | 10 27 17.0 | -51 08 43 | 281.57 | 5.48 | 214 | -21.9 | -61.1 | 20x 5 | 18.8 | .44 | | S | | 1 | |
| 2608 | | I | 10 27 27.1 | -67 15 28 | 289.92 | -8.31 | 92 | 57.6 | -119.5 | 40x 13 | 16.6 | .25 | | S 0 | | | |
| 2609 | | | 10 27 31.1 | -51 38 46 | 281.86 | 5.07 | 214 | -19.1 | -87.9 | 20x 11 | 17.7 | .36 | | L ? | | | |
| 2610 | | | 10 27 33.4 | -51 14 23 | 281.66 | 5.42 | 214 | -19.4 | -66.1 | 15x 7 | 19.2 | .41 | | S | | | v.obs, pair w.2611 |
| 2611 | | | 10 27 34.5 | -51 14 14 | 281.66 | 5.42 | 214 | -19.3 | -65.9 | 15x 5 | 19.7 | .41 | | S | | | v.obs, pair w.2610 |
| 2612 | | | 10 27 35.1 | -50 20 56 | 281.20 | 6.18 | 214 | -20.6 | -18.3 | 20x 9 | 17.8 | .27 | | S E | | | |
| 2613 | | | 10 27 39.3 | -50 38 03 | 281.36 | 5.95 | 214 | -19.6 | -33.6 | 13x 9 | 18.7 | .30 | | S ? | | S | |
| 2614 | | | 10 27 43.2 | -49 19 54 | 280.69 | 7.07 | 214 | -21.1 | 36.2 | 44x 27 | 16.0 | .34 | | L | | | |
| 2615 | | | 10 27 46.8 | -49 48 33 | 280.95 | 6.66 | 214 | -19.8 | 10.6 | 34x 5 | 18.2 | .29 | | S M | N | | |
| 2616 | | | 10 27 50.8 | -50 28 54 | 281.30 | 6.09 | 214 | -18.2 | -25.4 | 19x 3 | 19.6 | .27 | | S | E | | |
| 2617 | | | 10 27 52.1 | -48 06 55 | 280.07 | 8.12 | 214 | -21.6 | 101.4 | 15x 4 | 19.3 | .33 | | S | | | |
| 2618 | | | 10 27 53.2 | -50 03 38 | 281.09 | 6.46 | 214 | -18.5 | -2.8 | 19x 5 | 18.9 | .27 | | S | | 1 | |
| 2619 | | | 10 27 54.6 | -49 48 42 | 280.96 | 6.67 | 214 | -18.7 | 10.5 | 30x 24 | 16.9 | .29 | | SBR4 | | S | |
| 2620 | L214-012 | | 10 27 54.8 | -49 59 09 | 281.06 | 6.52 | 214 | -18.4 | 1.2 | 60x 35 | 15.6 | .27 | | S ? | | | w.sev.comp. |
| 2621 | | | 10 27 56.9 | -48 26 50 | 280.26 | 7.84 | 214 | -20.4 | 83.7 | 13x 3 | 19.9 | .34 | | S | | | |
| 2622 | | | 10 27 58.4 | -50 23 40 | 281.28 | 6.18 | 214 | -17.2 | -20.7 | 16x 5 | 19.3 | .26 | ? | | | S | LSB, smooth, arc-shaped |
| 2623 | | | 10 28 01.5 | -66 29 38 | 289.57 | -7.63 | 92 | 62.5 | -78.7 | 16x 15 | 18.2 | .31 | ? | | | S | smooth |
| 2624 | | | 10 28 04.9 | -49 17 42 | 280.72 | 7.13 | 214 | -18.0 | 38.3 | 19x 12 | 18.1 | .35 | | | | | br.blg, vLSBdisk |
| 2625 | | Q | 10 28 05.0 | -49 48 58 | 280.99 | 6.68 | 214 | -17.1 | 10.3 | 12x 9 | 18.9 | .29 | | | | | neighb.of 2619 |
| 2626 | | | 10 28 07.0 | -50 10 51 | 281.18 | 6.37 | 214 | -16.3 | -9.2 | 15x 15 | 17.6 | .25 | | E 0 | | S | |
| 2627 | | | 10 28 10.3 | -51 54 10 | 282.08 | 4.90 | 214 | -13.3 | -101.5 | 17x 8 | 18.4 | .41 | | S | | | |
| 2628 | | | 10 28 11.4 | -51 55 48 | 282.10 | 4.88 | 214 | -13.1 | -102.9 | 16x 7 | 18.8 | .42 | | S ? | | S | |
| 2629 | | | 10 28 12.1 | -53 35 44 | 282.96 | 3.45 | 168 | 9.4 | 75.5 | 20x 7 | 18.3 | .50 | | S L | E | | |
| 2630 | | | 10 28 12.6 | -49 52 23 | 281.04 | 6.64 | 214 | -16.0 | 7.3 | 30x 16 | 17.1 | .29 | | L ? | | S | |
| 2631 | | I | 10 28 13.4 | -51 41 43 | 281.98 | 5.08 | 214 | -13.1 | -90.3 | 30x 9 | 17.6 | .35 | | S E? | | | |
| 2632 | | | 10 28 13.9 | -49 11 05 | 280.68 | 7.24 | 214 | -16.8 | 44.2 | 12x 11 | 18.6 | .38 | ? | | | | |
| 2633 | | | 10 28 15.6 | -51 13 42 | 281.75 | 5.49 | 214 | -13.5 | -65.3 | 30x 22 | 16.7 | .39 | | S M | V | S | |
| 2634 | | | 10 28 18.3 | -50 11 33 | 281.22 | 6.38 | 214 | -14.7 | -9.8 | 17x 7 | 18.6 | .25 | | | | | |
| 2635 | | | 10 28 21.2 | -50 57 02 | 281.62 | 5.73 | 214 | -13.2 | -50.4 | 24x 13 | 17.4 | .41 | ? | | | 1 | v.diff. |
| 2636 | | | 10 28 26.1 | -50 00 23 | 281.14 | 6.55 | 214 | -13.8 | .2 | 17x 15 | 17.5 | .28 | | | | | p.cov.by ** |
| 2637 | | | 10 28 28.2 | -67 28 59 | 290.13 | -8.46 | 92 | 62.3 | -131.8 | 15x 13 | 18.3 | .24 | | S M? | | | |
| 2638 | | | 10 28 29.2 | -49 25 02 | 280.84 | 7.06 | 214 | -14.2 | 31.8 | 16x 7 | 18.4 | .34 | | S ? | | | p.cov.by ** |
| 2639 | | | 10 28 33.7 | -49 40 20 | 280.98 | 6.85 | 214 | -13.2 | 18.1 | 27x 9 | 17.4 | .32 | | E 7 | | | |
| 2640 | | | 10 28 34.9 | -49 38 31 | 280.97 | 6.87 | 214 | -13.1 | 19.8 | 23x 7 | 17.8 | .32 | | L ? | | | HSB, p.cov.by ** |
| 2641 | | | 10 28 42.3 | -52 26 32 | 282.43 | 4.48 | 168 | 13.6 | 137.3 | 27x 8 | 17.8 | .44 | | S | | S | smooth |
| 2642 | | | 10 28 44.6 | -50 30 54 | 281.44 | 6.14 | 214 | -10.5 | -27.0 | 24x 8 | 17.9 | .27 | | | | | br.blg |
| 2643 | | | 10 28 45.8 | -49 59 12 | 281.17 | 6.59 | 214 | -11.0 | 1.4 | 13x 4 | 19.6 | .29 | | | | | pair w.2644 |
| 2644 | | | 10 28 48.0 | -49 59 19 | 281.18 | 6.59 | 214 | -10.7 | 1.3 | 16x 9 | 18.5 | .29 | | | | | pair w.2643 |
| 2645 | | | 10 28 48.5 | -64 14 10 | 288.47 | -5.65 | 92 | 72.4 | 41.9 | 17x 13 | 18.1 | .70 | | S ? | | | dist.nucl. or s.p.* |
| 2646 | | | 10 28 48.7 | -49 48 45 | 281.09 | 6.75 | 214 | -10.9 | 10.7 | 54x 23 | 16.2 | .30 | | S 4 | | | |
| 2647 | | | 10 28 52.0 | -49 38 28 | 281.01 | 6.90 | 214 | -10.6 | 19.9 | 19x 4 | 19.1 | .32 | | S M | E | | |
| 2648 | | | 10 28 52.2 | -49 53 57 | 281.14 | 6.68 | 214 | -10.2 | 6.1 | 13x 8 | 19.2 | .29 | | S ? | | | |
| 2649 | | P | 10 28 52.6 | -51 06 54 | 281.77 | 5.63 | 214 | -8.5 | -59.1 | 36x 27 | 16.4 | .40 | | S M | | | |
| 2650 | | | 10 28 55.5 | -49 06 17 | 280.74 | 7.36 | 214 | -10.8 | 48.6 | 28x 16 | 17.0 | .38 | | S E? | | | |
| 2651 | | | 10 28 56.1 | -49 02 51 | 280.71 | 7.41 | 214 | -10.8 | 51.7 | 24x 8 | 17.9 | .38 | | | | | |
| 2652 | L214-013 | I | 10 28 57.0 | -48 47 25 | 280.58 | 7.63 | 214 | -11.0 | 65.5 | 87x 48 | 14.5 | .33 | | S 3 | | | |
| 2653 | | | 10 29 04.6 | -51 32 53 | 282.02 | 5.28 | 214 | -6.2 | -82.3 | 13x 9 | 18.7 | .34 | ? | | | 1 | |
| 2654 | | | 10 29 13.7 | -51 33 41 | 282.05 | 5.28 | 214 | -5.0 | -82.9 | 13x 5 | 19.4 | .34 | | S ? | | | |
| 2655 | | | 10 29 14.6 | -50 23 19 | 281.45 | 6.29 | 214 | -6.4 | -20.1 | 15x 11 | 18.5 | .26 | | | | | |
| 2656 | | | 10 29 16.1 | -52 15 42 | 282.41 | 4.68 | 168 | 18.2 | 147.0 | 15x 9 | 18.4 | .45 | | S | | | |
| 2657 | | | 10 29 17.2 | -48 39 30 | 280.56 | 7.78 | 214 | -8.3 | 72.6 | 24x 16 | 17.8 | .30 | | S M | | | vLSB |
| 2658 | | | 10 29 19.2 | -49 34 26 | 281.04 | 6.99 | 214 | -6.8 | 23.6 | 27x 16 | 17.3 | .34 | | S ? | | | v.obs. |
| 2659 | | | 10 29 22.1 | -51 07 14 | 281.84 | 5.67 | 214 | -4.4 | -59.3 | 40x 19 | 16.3 | .39 | | E | | S | |
| 2660 | | | 10 29 29.7 | -50 49 21 | 281.71 | 5.93 | 214 | -3.7 | -43.3 | 24x 7 | 18.1 | .37 | | S E | | | |
| 2661 | | | 10 29 32.2 | -49 37 11 | 281.09 | 6.97 | 214 | -4.8 | 21.2 | 16x 9 | 18.7 | .33 | ? | | | S | vLSB |
| 2662 | | | 10 29 35.0 | -50 43 14 | 281.66 | 6.03 | 214 | -3.1 | -37.8 | 11x 9 | 18.4 | .34 | | E | | | cl.to br.* |
| 2663 | | | 10 29 46.4 | -63 28 40 | 288.18 | -4.95 | 92 | 80.1 | 82.3 | 50x 8 | 17.4 | .47 | ? | S L | | S | |
| 2664 | | | 10 29 46.4 | -66 08 44 | 289.54 | -7.24 | 92 | 72.8 | -60.6 | 34x 16 | 17.1 | .30 | | S E | | | |
| 2665 | | | 10 29 50.6 | -51 54 52 | 282.31 | 5.02 | 214 | .6 | -101.7 | 27x 16 | 17.9 | .40 | | | | | v.diff. |
| 2666 | | | 10 29 52.0 | -49 55 10 | 281.29 | 6.74 | 214 | -1.6 | 5.2 | 17x 12 | 18.1 | .27 | | S ? | F | | br.blg |
| 2667 | | | 10 29 52.2 | -49 47 49 | 281.23 | 6.85 | 214 | -1.7 | 11.7 | 13x 5 | 19.5 | .29 | | S E | | | in group around 2672 |
| 2668 | | | 10 29 53.2 | -49 48 50 | 281.24 | 6.83 | 214 | -1.6 | 10.8 | 11x 8 | 18.9 | .29 | | | | | in group around 2672 |
| 2669 | | | 10 29 53.2 | -49 49 43 | 281.25 | 6.82 | 214 | -1.5 | 10.1 | 11x 11 | 18.6 | .28 | | | | | in group around 2672 |
| 2670 | | | 10 29 54.6 | -49 48 35 | 281.24 | 6.84 | 214 | -1.4 | 11.1 | 16x 7 | 18.6 | .29 | | | | 1 | in group around 2672 |
| 2671 | | | 10 29 54.9 | -66 27 41 | 289.72 | -7.51 | 92 | 72.6 | -77.5 | 17x 4 | 19.3 | .33 | | S | | | |
| 2672 | | | 10 29 57.1 | -49 48 15 | 281.25 | 6.85 | 214 | -1.0 | 11.4 | 27x 13 | 17.3 | .29 | | S E | | | large blg, in small group |
| 2673 | | | 10 29 57.2 | -51 36 26 | 282.17 | 5.29 | 214 | 1.1 | -85.3 | 20x 20 | 17.6 | .36 | | S L? | F | | v.obs. |
| 2674 | | | 10 29 59.7 | -50 01 12 | 281.36 | 6.66 | 214 | -.4 | -.2 | 16x 8 | 18.5 | .27 | | | | S | dist.blg, vLSBdisk |
| 2675 | | | 10 30 00.0 | -52 14 32 | 282.50 | 4.75 | 214 | 2.3 | -119.3 | 16x 4 | 19.7 | .43 | | S | | | |
| 2676 | | | 10 30 01.7 | -50 47 42 | 281.76 | 6.00 | 214 | .8 | -41.7 | 17x 12 | 17.8 | .38 | | E | | | |
| 2677 | | | 10 30 01.8 | -49 33 08 | 281.13 | 7.07 | 214 | -.6 | 24.9 | 15x 4 | 19.3 | .36 | ? | S | E | | |
| 2678 | | | 10 30 02.7 | -51 21 43 | 282.06 | 5.51 | 214 | 1.6 | -72.1 | 16x 7 | 18.4 | .41 | | | | 1 | |
| 2679 | | | 10 30 04.4 | -67 10 10 | 290.10 | -8.11 | 92 | 71.5 | -115.4 | 13x 5 | 19.4 | .34 | | | | | |
| 2680 | | | 10 30 04.7 | -53 16 10 | 283.03 | 3.87 | 168 | 24.5 | 92.9 | 24x 7 | 18.6 | .41 | | S | E | | dist.nucl. |
| 2681 | | | 10 30 06.3 | -51 35 03 | 282.18 | 5.33 | 214 | 2.4 | -84.0 | 13x 5 | 19.5 | .36 | ? | | | | |
| 2682 | | | 10 30 07.0 | -49 41 40 | 281.21 | 6.95 | 214 | .3 | 17.3 | 19x 9 | 18.3 | .32 | | S | | | |
| 2683 | | | 10 30 13.4 | -48 35 32 | 280.66 | 7.91 | 214 | .0 | 76.4 | 19x 8 | 18.3 | .28 | | S E | | | dist.blg, cl.to br.* |
| 2684 | L092-022 | I | 10 30 14.5 | -63 27 02 | 288.21 | -4.90 | 92 | 83.0 | 83.6 | 128x 47 | 14.6 | .48 | | S 7 | | S | |
| 2685 | | | 10 30 15.9 | -49 15 05 | 281.01 | 7.35 | 214 | 1.1 | 41.1 | 23x 8 | 18.1 | .35 | | S | E | | pair w.2686 |
| 2686 | | | 10 30 16.1 | -49 15 20 | 281.01 | 7.34 | 214 | 1.1 | 40.8 | 13x 11 | 18.4 | .35 | | S ? | F | | pair w.2685 |
| 2687 | | | 10 30 17.0 | -50 27 02 | 281.62 | 6.32 | 214 | 2.6 | -23.2 | 51x 38 | 15.8 | .30 | | S M | | S | dist.blg, large LSBdisk |
| 2688 | | | 10 30 17.0 | -51 35 46 | 282.21 | 5.33 | 214 | 3.9 | -84.6 | 13x 3 | 19.9 | .35 | | S | | | |
| 2689 | | | 10 30 23.4 | -50 32 38 | 281.69 | 6.25 | 214 | 3.6 | -28.2 | 12x 9 | 18.4 | .31 | | E ? | | | |
| 2690 | | | 10 30 24.0 | -53 15 11 | 283.06 | 3.91 | 168 | 27.0 | 93.8 | 20x 20 | 17.0 | .43 | | S 5 | F | | |
| 2691 | | | 10 30 25.6 | -49 52 23 | 281.35 | 6.83 | 214 | 3.2 | 7.8 | 22x 5 | 18.7 | .28 | | | | | |
| 2692 | | | 10 30 38.8 | -52 43 45 | 282.83 | 4.38 | 168 | 29.3 | 121.8 | 13x 11 | 18.7 | .37 | ? | | | | |
| 2693 | | | 10 30 40.4 | -50 22 20 | 281.64 | 6.42 | 214 | 5.8 | -18.9 | 17x 12 | 17.8 | .33 | | E | | | |
| 2694 | | | 10 30 47.5 | -50 19 34 | 281.63 | 6.47 | 214 | 6.8 | -16.5 | 23x 3 | 19.4 | .34 | | S M | E | S | v.obs. |
| 2695 | | | 10 30 49.0 | -49 57 58 | 281.45 | 6.78 | 214 | 6.6 | 2.8 | 16x 13 | 18.1 | .28 | | S | V | S | |
| 2696 | | | 10 30 59.2 | -50 17 13 | 281.64 | 6.52 | 214 | 8.4 | -14.3 | 16x 7 | 18.8 | .34 | | S ? | | | |
| 2697 | | | 10 31 01.2 | -48 30 09 | 280.73 | 8.06 | 214 | 6.9 | 81.3 | 17x 5 | 19.2 | .27 | | S | E | | |
| 2698 | | | 10 31 01.4 | -48 28 58 | 280.72 | 8.07 | 214 | 6.9 | 82.4 | 20x 5 | 18.9 | .27 | | S | N | | |
| 2699 | | | 10 31 11.5 | -52 46 23 | 282.93 | 4.39 | 168 | 33.7 | 119.4 | 17x 8 | 18.8 | .37 | | S ? | | | |
| 2700 | | | 10 31 13.2 | -49 09 11 | 281.09 | 7.51 | 214 | 9.3 | 46.5 | 34x 13 | 17.3 | .37 | | | | S | br.* s.p. |

**Table 1.** Galaxies in the southern ZOA – I. The Hydra/Antlia Extension (continued)

| RKK Ident (1) | Other Ident (2) | IR (3) | R.A. (h m s) (4) | Dec. (° ′ ″) (5) | gal ℓ (°) (6) | gal $b$ (°) (7) | SRC (8) | X (mm) (9) | Y (mm) (10) | D x d (″) (11) | $B_J$ ($^m$) (12) | $N_{HI}$ (13) | u | Type class. (14) | o | * | Remarks (15) |
|---|---|---|---|---|---|---|---|---|---|---|---|---|---|---|---|---|---|
| 2701 | | | 10 31 16.8 | -52 43 39 | 282.91 | 4.43 | 168 | 34.4 | 121.8 | 16x 13 | 17.9 | .37 | ? | S ? | F | | |
| 2702 | | | 10 31 18.8 | -49 42 07 | 281.38 | 7.05 | 214 | 10.7 | 17.1 | 13x 8 | 18.7 | .32 | | S ? | | 1 | |
| 2703 | | | 10 31 25.1 | -49 58 31 | 281.54 | 6.82 | 214 | 11.8 | 2.5 | 13x 3 | 19.9 | .31 | | S | | | br.blg |
| 2704 | | | 10 31 27.9 | -50 10 25 | 281.65 | 6.65 | 214 | 12.4 | -8.2 | 13x 11 | 18.6 | .33 | | | | 1 | |
| 2705 | | | 10 31 31.3 | -49 38 41 | 281.38 | 7.11 | 214 | 12.4 | 20.2 | 13x 8 | 18.2 | .32 | | S ? | | | HSB, cl.to br.* |
| 2706 | | | 10 31 33.8 | -50 54 29 | 282.03 | 6.03 | 214 | 13.9 | -47.5 | 20x 11 | 17.7 | .39 | | | | 1 | |
| 2707 | | | 10 31 34.3 | -48 53 03 | 281.00 | 7.77 | 214 | 12.2 | 60.9 | 16x 12 | 18.1 | .34 | | | | | cl.to br.* |
| 2708 | | | 10 31 45.3 | -49 31 45 | 281.36 | 7.23 | 214 | 14.3 | 26.4 | 17x 9 | 18.4 | .35 | | | | 1 | smooth |
| 2709 | | | 10 31 53.9 | -50 05 53 | 281.67 | 6.75 | 214 | 16.1 | -4.1 | 19x 7 | 18.7 | .32 | | S ? | | 1 | |
| 2710 | | | 10 31 56.2 | -50 21 54 | 281.81 | 6.52 | 214 | 16.6 | -18.4 | 22x 11 | 17.7 | .34 | | S | | | v.obs. |
| 2711 | | | 10 32 04.5 | -51 44 59 | 282.53 | 5.34 | 214 | 18.9 | -92.6 | 15x 3 | 20.1 | .33 | | S | E | | vvLSB |
| 2712 | | | 10 32 06.7 | -48 42 38 | 280.99 | 7.97 | 214 | 16.8 | 70.3 | 23x 7 | 17.9 | .31 | | S | | | |
| 2713 | | | 10 32 16.1 | -49 18 37 | 281.32 | 7.46 | 214 | 18.6 | 38.2 | 12x 8 | 19.3 | .43 | | | | | vLSB, trpl syst.w.2714,271 |
| 2714 | | | 10 32 17.7 | -49 18 41 | 281.32 | 7.46 | 214 | 18.9 | 38.2 | 11x 4 | 20.1 | .43 | ? | S | E | | vLSB, trpl syst.w.2713,271 |
| 2715 | | | 10 32 18.4 | -49 51 35 | 281.60 | 6.99 | 214 | 19.4 | 8.8 | 19x 8 | 18.6 | .33 | | S | | | v.obs. |
| 2716 | | | 10 32 20.8 | -49 18 33 | 281.33 | 7.47 | 214 | 19.3 | 38.3 | 23x 12 | 17.7 | .44 | | S E? | | | trpl syst.w.2713,2714 |
| 2717 | | | 10 32 34.8 | -50 19 55 | 281.88 | 6.60 | 214 | 22.1 | -16.5 | 15x 4 | 19.7 | .37 | | S | E | | |
| 2718 | | | 10 32 41.9 | -49 34 17 | 281.51 | 7.27 | 214 | 22.6 | 24.3 | 13x 7 | 19.0 | .34 | | S ? | | | |
| 2719 | | | 10 32 46.4 | -50 00 29 | 281.74 | 6.90 | 214 | 23.5 | .9 | 13x 9 | 18.6 | .33 | | | | | |
| 2720 | | | 10 32 48.1 | -50 07 27 | 281.81 | 6.80 | 214 | 23.8 | -5.3 | 17x 5 | 18.7 | .37 | | S ? | | | |
| 2721 | | | 10 32 51.4 | -50 43 33 | 282.12 | 6.29 | 214 | 24.7 | -37.6 | 15x 3 | 19.8 | .32 | | S | E | 1 | |
| 2722 | | | 10 32 53.7 | -49 29 25 | 281.50 | 7.36 | 214 | 24.2 | 28.6 | 19x 9 | 17.9 | .39 | | E | | | w.large LSBhalo? |
| 2723 | | | 10 32 53.8 | -49 41 04 | 281.60 | 7.19 | 214 | 24.4 | 18.2 | 19x 9 | 17.8 | .35 | | F | | | |
| 2724 | | | 10 32 57.3 | -52 04 40 | 282.81 | 5.12 | 214 | 26.4 | -110.0 | 17x 7 | 18.3 | .34 | | S ? | | | |
| 2725 | | | 10 33 00.1 | -51 03 11 | 282.30 | 6.01 | 214 | 26.1 | -55.1 | 34x 24 | 16.5 | .37 | | S M: | | S | |
| 2726 | | | 10 33 01.2 | -49 30 40 | 281.53 | 7.35 | 214 | 25.3 | 27.5 | 17x 7 | 18.5 | .38 | | | | | |
| 2727 | | | 10 33 03.7 | -50 18 31 | 281.94 | 6.66 | 214 | 26.2 | -15.2 | 17x 9 | 18.2 | .41 | | | | | |
| 2728 | | | 10 33 09.2 | -50 23 04 | 281.99 | 6.60 | 214 | 27.0 | -19.3 | 30x 20 | 17.1 | .44 | | S M | V | S | dist.nucl, LSBdisk |
| 2729 | | | 10 33 11.5 | -48 44 47 | 281.17 | 8.03 | 214 | 26.4 | 68.5 | 16x 7 | 18.6 | .32 | | | | S | |
| 2730 | | | 10 33 13.8 | -54 25 00 | 284.01 | 3.11 | 168 | 48.5 | 31.1 | 19x 11 | 17.4 | .78 | | S | | | |
| 2731 | | | 10 33 14.8 | -49 02 03 | 281.32 | 7.78 | 214 | 27.0 | 53.1 | 15x 9 | 18.9 | .42 | ? | | | | v.obs. |
| 2732 | | | 10 33 21.4 | -49 42 54 | 281.68 | 7.20 | 214 | 28.4 | 16.6 | 19x 4 | 19.3 | .38 | | S | E | | |
| 2733 | | I | 10 33 22.1 | -67 27 48 | 290.52 | -8.20 | 92 | 87.5 | -132.2 | 40x 35 | 15.7 | .28 | | S 1 | | S | |
| 2734 | | | 10 33 25.6 | -52 13 38 | 282.95 | 5.03 | 214 | 30.4 | -118.0 | 13x 3 | 20.0 | .38 | | S | E | | |
| 2735 | | | 10 33 27.2 | -50 14 12 | 281.95 | 6.76 | 214 | 29.5 | -11.3 | 20x 7 | 18.4 | .39 | | | | | v.obs, cl.to br.* |
| 2736 | | | 10 33 32.7 | -52 30 55 | 283.11 | 4.79 | 168 | 53.1 | 132.9 | 16x 13 | 17.9 | .41 | ? | S | | | |
| 2737 | | | 10 33 44.5 | -48 47 20 | 281.27 | 8.03 | 214 | 31.3 | 66.3 | 34x 8 | 17.7 | .37 | | S E | N | S | |
| 2738 | | NG | 10 33 44.6 | -53 49 32 | 283.78 | 3.66 | 168 | 53.2 | 62.7 | 19x 7 | 17.9 | .52 | | S 5 | E | | |
| 2739 | | | 10 33 47.2 | -50 36 25 | 282.19 | 6.46 | 214 | 32.5 | -31.1 | 26x 7 | 18.7 | .42 | | | | | vvLSB |
| 2740 | | I | 10 33 52.1 | -53 50 54 | 283.81 | 3.65 | 168 | 54.2 | 61.4 | 54x 5 | 17.6 | .52 | | S 5 | E | S | |
| 2741 | | | 10 33 54.0 | -49 53 39 | 281.85 | 7.09 | 214 | 33.2 | 7.1 | 13x 9 | 18.8 | .43 | | | | | |
| 2742 | | | 10 33 54.7 | -63 40 28 | 288.68 | -4.89 | 92 | 104.1 | 70.3 | 16x 15 | 17.6 | .64 | | E ? | | | |
| 2743 | | | 10 34 00.4 | -49 58 42 | 281.90 | 7.02 | 214 | 34.1 | 2.6 | 31x 28 | 16.4 | .41 | | S M: | | S | dist.nucl, p.cov.by * |
| 2744 | | | 10 34 08.4 | -53 48 19 | 283.82 | 3.71 | 168 | 56.4 | 63.7 | 16x 8 | 18.4 | .50 | | S | | | |
| 2745 | | | 10 34 27.9 | -49 04 55 | 281.52 | 7.84 | 214 | 37.7 | 50.6 | 16x 7 | 19.3 | .40 | ? | | | | vvLSB |
| 2746 | | | 10 34 31.6 | -53 16 29 | 283.61 | 4.20 | 168 | 60.1 | 92.0 | 19x 17 | 17.6 | .38 | | S | | | dist.nucl. |
| 2747 | | | 10 34 38.0 | -49 05 05 | 281.54 | 7.85 | 214 | 39.2 | 50.5 | 20x 4 | 19.5 | .40 | | S | E | S | vLSB |
| 2748 | | | 10 34 48.4 | -66 34 11 | 290.20 | -7.36 | 92 | 98.4 | -84.9 | 23x 23 | 17.1 | .37 | | S | F | | |
| 2749 | | | 10 34 58.3 | -52 28 08 | 283.27 | 4.93 | 214 | 43.1 | -130.9 | 13x 3 | 20.2 | .39 | ? | S | E | | |
| 2750 | | | 10 35 01.9 | -48 52 31 | 281.50 | 8.07 | 214 | 42.7 | 61.8 | 11x 4 | 19.7 | .38 | | | | | neighb.of 2753 |
| 2751 | | | 10 35 02.7 | -49 45 00 | 281.93 | 7.31 | 214 | 43.0 | 14.9 | 34x 15 | 17.1 | .58 | | S E | | | poss.larger disk? |
| 2752 | | | 10 35 04.4 | -52 27 15 | 283.28 | 4.96 | 214 | 44.0 | -130.1 | 20x 3 | 19.6 | .39 | | S | E | | |
| 2753 | | | 10 35 04.8 | -48 52 38 | 281.50 | 8.07 | 214 | 43.1 | 61.7 | 16x 7 | 18.4 | .38 | ? | | | | dbl syst? |
| 2754 | L168-009 | I | 10 35 18.1 | -54 40 31 | 284.40 | 3.04 | 168 | 64.3 | 16.8 | 54x 24 | 16.1 | .73 | | S 5 | | | |
| 2755 | | | 10 35 20.7 | -52 41 30 | 283.43 | 4.77 | 168 | 67.5 | 123.1 | 17x 13 | 17.8 | .34 | | S 5 | | | |
| 2756 | | I | 10 35 22.5 | -48 50 22 | 281.53 | 8.12 | 214 | 45.7 | 63.7 | 13x 5 | 19.2 | .38 | ? | | | | |
| 2757 | | | 10 35 25.5 | -51 08 07 | 282.67 | 6.13 | 214 | 46.6 | -59.4 | 13x 8 | 19.2 | .50 | | | | | |
| 2758 | | | 10 35 30.0 | -67 14 36 | 290.59 | -7.91 | 92 | 99.3 | -121.2 | 15x 5 | 19.1 | .38 | | | | | |
| 2759 | | I | 10 35 33.8 | -52 18 37 | 283.27 | 5.12 | 168 | 69.8 | 143.5 | 28x 11 | 17.0 | .39 | | S | | | |
| 2760 | | | 10 35 48.8 | -49 06 10 | 281.72 | 7.93 | 214 | 49.6 | 49.6 | 13x 7 | 19.5 | .41 | ? | | | | vLSB |
| 2761 | | | 10 35 54.5 | -52 35 19 | 283.46 | 4.90 | 168 | 72.2 | 128.5 | 16x 16 | 16.8 | .36 | | E ? | | | |
| 2762 | | | 10 36 02.7 | -52 47 42 | 283.58 | 4.73 | 168 | 73.0 | 117.4 | 23x 11 | 18.1 | .37 | | S 5 | E | | |
| 2763 | L214-016 | | 10 36 04.0 | -49 53 56 | 282.15 | 7.26 | 214 | 51.9 | 6.9 | 148x 15 | 15.8 | .53 | | S 4 | E | S | dist.abs.lane |
| 2764 | | | 10 36 12.9 | -53 59 59 | 284.19 | 3.69 | 168 | 72.5 | 52.8 | 17x 5 | 18.9 | .50 | | S | | | pair w.2765 |
| 2765 | | | 10 36 14.7 | -54 00 09 | 284.19 | 3.69 | 168 | 72.7 | 52.7 | 17x 5 | 18.9 | .50 | | S | | | pair w.2764 |
| 2766 | | | 10 36 15.6 | -65 01 33 | 289.57 | -5.94 | 92 | 112.5 | -3.0 | 27x 12 | 17.6 | .42 | ? | S L: | | S | |
| 2767 | | | 10 36 15.7 | -52 09 00 | 283.29 | 5.31 | 214 | 53.7 | -113.7 | 19x 5 | 19.3 | .47 | | S ? | | | v.obs. |
| 2768 | | | 10 36 19.2 | -52 45 39 | 283.59 | 4.78 | 168 | 75.3 | 119.2 | 34x 12 | 16.3 | .37 | | S E | | | |
| 2769 | | | 10 36 25.9 | -49 36 46 | 282.06 | 7.54 | 214 | 55.0 | 22.3 | 20x 17 | 18.0 | .51 | | S | F | | small nucl, vLSBspiral |
| 2770 | | | 10 36 27.9 | -66 36 31 | 290.36 | -7.31 | 93 | -127.6 | -91.3 | 16x 7 | 19.3 | .36 | | | | | LSB |
| 2771 | | | 10 36 29.1 | -65 50 35 | 289.99 | -6.64 | 92 | 110.3 | -46.8 | 47x 43 | 16.2 | .38 | ? | S | | | vLSBdisk |
| 2772 | | | 10 36 34.0 | -52 44 39 | 283.62 | 4.81 | 168 | 77.3 | 120.0 | 13x 9 | 18.1 | .41 | | E ? | | | or * |
| 2773 | | | 10 36 38.1 | -50 16 43 | 282.42 | 6.97 | 214 | 56.8 | -13.4 | 24x 16 | 17.9 | .55 | ? | S | | | vLSB |
| 2774 | | I | 10 36 55.3 | -50 09 39 | 282.40 | 7.10 | 214 | 59.2 | -7.1 | 43x 31 | 15.7 | .55 | | SB 3 | | | |
| 2775 | | | 10 36 58.4 | -49 18 47 | 281.99 | 7.84 | 214 | 59.8 | 38.3 | 19x 9 | 18.1 | .44 | | | | | w.comp? |
| 2776 | | I | 10 37 04.2 | -53 26 08 | 284.02 | 4.25 | 168 | 80.2 | 82.8 | 16x 7 | 18.6 | .52 | | | | | smooth |
| 2777 | | | 10 37 07.0 | -50 01 41 | 282.37 | 7.23 | 214 | 60.9 | .0 | 22x 8 | 18.0 | .52 | | S E | | | w.sev.comp. |
| 2778 | | | 10 37 12.1 | -52 10 37 | 283.43 | 5.36 | 214 | 61.4 | -115.2 | 40x 20 | 16.6 | .44 | | S | | S | v.obs. |
| 2779 | | | 10 37 12.2 | -49 39 27 | 282.20 | 7.56 | 214 | 61.7 | 19.9 | 13x 13 | 18.0 | .49 | | E ? | | | |
| 2780 | | | 10 37 21.6 | -53 11 40 | 283.94 | 4.48 | 168 | 83.0 | 95.7 | 24x 13 | 17.4 | .45 | ? | S | | 2 | |
| 2781 | | | 10 37 28.4 | -49 23 45 | 282.11 | 7.81 | 214 | 64.1 | 33.9 | 17x 5 | 19.4 | .44 | | S | | | vLSB |
| 2782 | | | 10 37 31.8 | -49 10 12 | 282.00 | 8.01 | 214 | 64.6 | 46.0 | 15x 5 | 18.8 | .41 | ? | S | E | | |
| 2783 | | | 10 37 34.5 | -65 49 55 | 290.08 | -6.58 | 93 | -125.3 | -49.2 | 20x 7 | 18.9 | .38 | ? | S M | | | LSB |
| 2784 | | | 10 37 38.5 | -50 16 26 | 282.56 | 7.05 | 214 | 65.4 | -13.2 | 43x 40 | 16.2 | .53 | | S RM | F | S | br.blg, large LSBdisk |
| 2785 | | | 10 37 39.0 | -65 51 35 | 290.10 | -6.60 | 92 | 116.6 | -48.2 | 13x 4 | 19.6 | .37 | | | | | |
| 2786 | | | 10 37 39.2 | -53 21 45 | 284.06 | 4.35 | 168 | 85.0 | 86.6 | 20x 12 | 18.2 | .52 | | S 3 | | | |
| 2787 | | | 10 37 50.3 | -51 15 13 | 283.06 | 6.21 | 214 | 66.9 | -65.7 | 20x 3 | 19.7 | .49 | | S | | | v.obs. |
| 2788 | | | 10 37 51.0 | -51 10 40 | 283.03 | 6.28 | 214 | 67.0 | -61.7 | 13x 8 | 19.2 | .51 | | S M | | | dist.nucl, vvLSBdisk |
| 2789 | | | 10 37 54.3 | -54 12 08 | 284.50 | 3.63 | 168 | 85.4 | 41.5 | 16x 11 | 18.5 | .64 | ? | | | | mult.syst. |
| 2790 | | | 10 37 54.6 | -49 40 01 | 282.30 | 7.61 | 214 | 67.9 | 19.3 | 22x 15 | 17.7 | .46 | | S M | | | LSB spiral |
| 2791 | | | 10 37 55.4 | -54 16 54 | 284.54 | 3.57 | 168 | 85.4 | 37.3 | 20x 5 | 18.9 | .67 | | S 5 | E | | |
| 2792 | | | 10 37 56.3 | -49 47 43 | 282.37 | 7.50 | 214 | 68.1 | 12.4 | 11x 8 | 18.5 | .51 | | E | | | or blg |
| 2793 | | | 10 37 56.7 | -50 04 24 | 282.50 | 7.25 | 214 | 68.1 | -2.5 | 20x 7 | 18.9 | .52 | | S ? | E | | vLSB |
| 2794 | | | 10 38 00.9 | -50 26 24 | 282.69 | 6.94 | 214 | 68.6 | -22.1 | 13x 11 | 18.7 | .49 | ? | | | | |
| 2795 | | | 10 38 10.1 | -67 08 25 | 290.77 | -7.69 | 92 | 113.6 | -116.8 | 13x 13 | 18.4 | .40 | | S | F | | LSB |
| 2796 | | | 10 38 13.3 | -49 16 06 | 282.15 | 7.98 | 214 | 70.7 | 40.7 | 15x 4 | 19.0 | .38 | | | | | HSB |
| 2797 | | | 10 38 27.4 | -49 09 20 | 282.13 | 8.10 | 214 | 72.8 | 46.7 | 42x 31 | 16.1 | .38 | | I 9 ? | | S | LSB, smooth |
| 2798 | | | 10 38 33.7 | -64 30 40 | 289.53 | -5.37 | 93 | -125.6 | 21.8 | 27x 7 | 18.3 | .58 | | I ? | | 1 | |
| 2799 | | | 10 38 34.8 | -64 55 28 | 289.73 | -5.73 | 92 | 126.0 | 1.4 | 13x 9 | 18.8 | .43 | ? | | | | |
| 2800 | | | 10 38 35.2 | -52 34 50 | 283.81 | 5.10 | 168 | 94.0 | 128.2 | 13x 13 | 18.8 | .42 | | S 3 | | | pair w.2801 |

**Table 1.** Galaxies in the southern ZOA – I. The Hydra/Antlia Extension (continued)

| RKK Ident (1) | Other Ident (2) | IR (3) | R.A. (h m s) (4) | Dec. (° ′ ″) (5) | gal $\ell$ (°) (6) | gal $b$ (°) (7) | SRC (8) | X (mm) (9) | Y (mm) (10) | D x d (″) (11) | $B_J$ ($^m$) (12) | $N_{HI}$ (13) | u | Type class. (14) | o * | Remarks (15) |
|---|---|---|---|---|---|---|---|---|---|---|---|---|---|---|---|---|
| 2801 | | | 10 38 39.6 | -52 35 35 | 283.83 | 5.10 | 168 | 94.6 | 127.5 | 13x 13 | 18.8 | .42 | | | | pair w.2800 |
| 2802 | | | 10 38 39.7 | -67 17 14 | 290.88 | -7.80 | 92 | 115.5 | -124.9 | 13x 3 | 19.9 | .40 | ? | S | E1 | |
| 2803 | | | 10 38 54.1 | -65 59 38 | 290.28 | -6.65 | 92 | 122.8 | -55.9 | 16x 4 | 19.4 | .36 | | S | E | |
| 2804 | | | 10 39 01.7 | -49 59 53 | 282.62 | 7.40 | 214 | 77.4 | 1.5 | 11x 9 | 18.5 | .49 | | E | | or blg |
| 2805 | | | 10 39 05.1 | -49 41 32 | 282.48 | 7.67 | 214 | 78.0 | 17.9 | 23x 9 | 18.0 | .43 | | S E? | | |
| 2806 | | | 10 39 15.8 | -49 15 02 | 282.29 | 8.08 | 214 | 79.8 | 41.6 | 24x 9 | 17.9 | .35 | ? | S L? | 1 | |
| 2807 | | | 10 39 26.4 | -51 16 49 | 283.30 | 6.31 | 214 | 80.3 | -67.2 | 43x 9 | 17.5 | .44 | | S | | |
| 2808 | | | 10 39 42.8 | -51 38 29 | 283.51 | 6.01 | 214 | 82.3 | -86.6 | 17x 7 | 18.5 | .50 | | S | 1 | |
| 2809 | | | 10 39 47.0 | -49 25 05 | 282.45 | 7.97 | 214 | 84.3 | 32.5 | 20x 11 | 17.7 | .36 | | F | | |
| 2810 | | | 10 39 52.6 | -49 57 21 | 282.72 | 7.50 | 214 | 84.7 | 3.7 | 13x 5 | 19.2 | .50 | | | | |
| 2811 | | | 10 39 59.9 | -50 22 33 | 282.94 | 7.15 | 214 | 85.5 | -18.8 | 24x 24 | 17.2 | .58 | ? | | | vLSB |
| 2812 | | | 10 40 17.0 | -52 04 11 | 283.79 | 5.68 | 214 | 86.8 | -109.6 | 34x 13 | 17.6 | .48 | | S | S | v.obs. |
| 2813 | | | 10 40 53.7 | -50 27 04 | 283.10 | 7.15 | 214 | 93.1 | -23.0 | 12x 9 | 18.8 | .49 | | | | between 2 ** |
| 2814 | | | 10 40 54.2 | -52 25 14 | 284.04 | 5.41 | 214 | 91.6 | -128.5 | 17x 5 | 18.9 | .47 | | S E? | | |
| 2815 | | | 10 41 47.0 | -52 07 01 | 284.02 | 5.74 | 214 | 99.1 | -112.4 | 16x 5 | 19.5 | .45 | | S | | v.obs. |
| 2816 | | | 10 41 49.8 | -52 12 06 | 284.07 | 5.67 | 214 | 99.4 | -116.9 | 13x 5 | 19.2 | .44 | | S ? | | |
| 2817 | | | 10 41 52.3 | -51 28 17 | 283.73 | 6.32 | 214 | 100.5 | -77.8 | 20x 9 | 18.7 | .47 | ? | | S | v.obs. |
| 2818 | | | 10 41 58.5 | -51 26 18 | 283.72 | 6.36 | 214 | 101.4 | -76.0 | 16x 8 | 18.9 | .46 | ? | | | v.obs. |
| 2819 | | | 10 42 02.8 | -52 12 58 | 284.10 | 5.68 | 214 | 101.2 | -117.7 | 13x 9 | 18.4 | .44 | | L ? | | or blg |
| 2820 | | P | 10 42 14.2 | -50 46 24 | 283.45 | 6.96 | 214 | 104.2 | -40.4 | 30x 7 | 18.3 | .43 | | | | vLSB |
| 2821 | | | 10 42 34.1 | -52 25 52 | 284.27 | 5.52 | 168 | 126.9 | 134.8 | 16x 5 | 19.6 | .48 | ? | S M | E1 | |
| 2822 | | | 10 42 39.2 | -52 11 34 | 284.17 | 5.74 | 214 | 106.2 | -116.6 | 27x 27 | 17.1 | .44 | | S 5 | F S | v.obs. |
| 2823 | | | 10 42 40.5 | -51 03 04 | 283.64 | 6.75 | 214 | 107.6 | -55.4 | 13x 3 | 19.5 | .44 | ? | S ? | | |
| 2824 | | | 10 42 46.9 | -49 37 40 | 282.98 | 8.02 | 214 | 110.2 | 20.9 | 19x 13 | 17.8 | .39 | | S ? | | p.cov.by *, w.comp. |
| 2825 | | | 10 42 50.7 | -49 57 52 | 283.15 | 7.72 | 214 | 110.3 | 2.8 | 16x 5 | 19.3 | .47 | ? | | | LSB Im? |
| 2826 | | | 10 43 39.6 | -49 55 38 | 283.25 | 7.82 | 214 | 117.4 | 4.7 | 34x 13 | 17.3 | .41 | | S | NS | br.blg |
| 2827 | | | 10 43 42.2 | -53 40 42 | 285.01 | 4.50 | 169 | -129.0 | 67.2 | 23x 7 | 18.1 | .34 | | | S | |
| 2828 | | | 10 44 03.8 | -49 48 40 | 283.25 | 7.95 | 214 | 121.0 | 10.8 | 20x 16 | 17.6 | .38 | | S M? | V | |
| 2829 | | | 10 44 04.1 | -52 01 21 | 284.29 | 5.99 | 214 | 118.0 | -107.7 | 34x 13 | 17.8 | .48 | | S | | v.obs. |
| 2830 | | | 10 44 04.4 | -49 52 54 | 283.29 | 7.89 | 214 | 121.0 | 7.0 | 20x 7 | 18.4 | .39 | | S M: | | br.blg |
| 2831 | | | 10 44 07.3 | -51 40 51 | 284.14 | 6.30 | 214 | 119.0 | -89.4 | 12x 8 | 18.7 | .47 | ? | | | |
| 2832 | | | 10 44 19.5 | -51 10 10 | 283.93 | 6.77 | 214 | 121.4 | -62.1 | 13x 7 | 18.6 | .46 | ? | | | or blended ** |
| 2833 | | | 10 44 25.9 | -53 51 32 | 285.19 | 4.39 | 168 | 137.6 | 57.5 | 27x 5 | 18.8 | .32 | | S 5 | E | |
| 2834 | | | 10 44 26.2 | -51 09 16 | 283.93 | 6.79 | 214 | 122.3 | -61.3 | 32x 5 | 18.9 | .46 | | | S | v.obs. |
| 2835 | | | 10 44 29.9 | -50 02 06 | 283.42 | 7.78 | 214 | 124.5 | -1.3 | 32x 8 | 17.9 | .40 | | S M | NS | |
| 2836 | | | 10 44 32.3 | -53 31 16 | 285.05 | 4.70 | 168 | 139.5 | 75.5 | 13x 8 | 17.8 | .30 | ? | S E | | |
| 2837 | | | 10 44 33.9 | -53 17 04 | 284.94 | 4.91 | 168 | 140.5 | 88.2 | 24x 8 | 18.0 | .34 | | S 3 | | w.LSB comp.(face-on) |
| 2838 | | | 10 44 42.7 | -51 55 39 | 284.33 | 6.12 | 214 | 123.5 | -102.8 | 13x 7 | 19.0 | .49 | | S | | br.blg, poss.larger |
| 2839 | | | 10 45 01.9 | -50 18 46 | 283.63 | 7.58 | 214 | 128.6 | -16.3 | 28x 20 | 17.3 | .43 | | S M | V | dist.blg, LSBdisk |
| 2840 | | | 10 45 05.0 | -67 22 42 | 291.48 | -7.59 | 93 | -79.6 | -129.2 | 13x 11 | 18.3 | .34 | ? | E 2 | | or br.* w.halo |
| 2841 | | | 10 45 11.3 | -50 13 05 | 283.60 | 7.67 | 214 | 130.1 | -11.3 | 47x 15 | 16.6 | .42 | | S 2 | | large br.blg |
| 2842 | | | 10 45 19.0 | -53 37 40 | 285.20 | 4.65 | 168 | 145.4 | 69.4 | 20x 13 | 17.4 | .33 | | S | | |
| 2843 | | | 10 45 28.6 | -66 22 59 | 291.05 | -6.68 | 93 | -80.5 | -75.8 | 13x 9 | 19.0 | .37 | ? | | | w.comp. |
| 2844 | | | 10 45 37.8 | -50 07 56 | 283.63 | 7.78 | 214 | 134.1 | -6.8 | 20x 16 | 17.7 | .42 | ? | | | br.nucl, LSBdisk |
| 2845 | | | 10 45 59.2 | -50 13 55 | 283.73 | 7.72 | 214 | 136.9 | -12.2 | 47x 17 | 16.5 | .43 | | S L | | or Im |
| 2846 | | | 10 46 01.7 | -51 13 25 | 284.19 | 6.84 | 214 | 135.6 | -65.4 | 20x 5 | 19.3 | .49 | | S | | v.obs. |
| 2847 | L169-002 | I | 10 46 09.0 | -53 02 18 | 285.04 | 5.23 | 169 | -114.5 | 104.5 | 94x 27 | 15.2 | .50 | | S L | | |
| 2848 | | | 10 46 14.2 | -51 02 00 | 284.13 | 7.03 | 214 | 137.7 | -55.2 | 12x 7 | 18.8 | .46 | | S ? | | |
| 2849 | | | 10 46 15.8 | -66 26 00 | 291.15 | -6.69 | 93 | -76.1 | -78.2 | 16x 5 | 19.3 | .41 | ? | | 1 | |
| 2850 | | | 10 46 16.0 | -49 56 59 | 283.64 | 7.99 | 214 | 139.9 | 2.8 | 15x 5 | 19.1 | .41 | | S | S | |
| 2851 | | | 10 46 18.0 | -50 11 31 | 283.75 | 7.78 | 214 | 139.7 | -10.2 | 13x 8 | 18.7 | .42 | | S E: | | br.blg, LSBdisk |
| 2852 | | I | 10 46 22.0 | -65 03 34 | 290.53 | -5.47 | 93 | -79.2 | -4.6 | 27x 27 | 16.6 | .53 | | | S | |
| 2853 | | | 10 46 23.3 | -66 52 38 | 291.36 | -7.08 | 93 | -74.2 | -102.0 | 20x 11 | 18.2 | .50 | ? | | | dbl syst. of E |
| 2854 | | | 10 46 47.0 | -53 33 46 | 285.36 | 4.81 | 169 | -104.8 | 74.6 | 22x 16 | 17.2 | .40 | ? | S 5 : | | dist.blg |
| 2855 | | | 10 46 49.4 | -51 41 38 | 284.52 | 6.48 | 214 | 141.4 | -90.8 | 27x 15 | 17.0 | .50 | | L ? | S | |
| 2856 | | | 10 46 59.2 | -54 04 40 | 285.62 | 4.36 | 169 | -102.0 | 47.0 | 27x 20 | 16.9 | .53 | ? | | S | smooth |
| 2857 | | | 10 47 43.9 | -52 49 57 | 285.16 | 5.53 | 169 | -98.9 | 114.0 | 13x 13 | 18.1 | .43 | ? | | | br.* s.p. |
| 2858 | | | 10 48 06.1 | -67 32 01 | 291.81 | -7.59 | 93 | -63.7 | -136.7 | 17x 15 | 18.1 | .34 | | S ? | | dist.nucl. |
| 2859 | | I | 10 49 08.0 | -66 55 34 | 291.62 | -7.00 | 93 | -59.7 | -104.0 | 27x 17 | 17.6 | .41 | | S | | LSB |
| 2860 | | NG | 10 49 16.5 | -67 07 22 | 291.73 | -7.17 | 93 | -58.5 | -114.5 | 13x 7 | 18.9 | .39 | ? | | 2 | |
| 2861 | | | 10 49 32.6 | -64 48 49 | 290.72 | -5.09 | 93 | -61.7 | 9.3 | 34x 27 | 16.8 | .56 | ? | S ? | S | LSB |
| 2862 | | I | 10 49 34.5 | -67 07 53 | 291.76 | -7.17 | 93 | -57.0 | -114.9 | 17x 7 | 18.8 | .39 | ? | S ? | S | |
| 2863 | | NG | 10 49 50.0 | -67 08 04 | 291.78 | -7.16 | 93 | -55.6 | -115.0 | 22x 11 | 17.7 | .39 | | S M | S | |
| 2864 | | | 10 51 11.2 | -54 06 03 | 286.19 | 4.62 | 169 | -69.0 | 47.0 | 13x 8 | 18.6 | .40 | | S ? | | |
| 2865 | | | 10 51 32.3 | -54 04 28 | 286.22 | 4.66 | 169 | -66.2 | 48.5 | 20x 7 | 18.3 | .39 | | S ? | | |
| 2866 | PMNJ1054 | | 10 51 54.3 | -53 50 01 | 286.16 | 4.90 | 169 | -63.7 | 61.5 | 16x 7 | 18.6 | .39 | | | | br.* s.p. |
| 2867 | | | 10 52 28.0 | -66 23 13 | 291.68 | -6.38 | 93 | -42.9 | -74.5 | 23x 4 | 19.3 | .46 | | S M | E | |
| 2868 | | | 10 56 35.1 | -53 57 12 | 286.84 | 5.09 | 169 | -26.6 | 55.8 | 20x 16 | 17.5 | .36 | ? | S ? | | |
| 2869 | | | 10 57 21.0 | -67 18 29 | 292.52 | -7.00 | 93 | -16.4 | -123.5 | 15x 8 | 18.7 | .37 | ? | S ? | | |
| 2870 | L093-003 | I | 10 57 32.6 | -66 03 51 | 292.01 | -5.87 | 93 | -15.8 | -56.8 | 94x 47 | 14.7 | .55 | | S 3 | S | |
| 2871 | | | 10 57 54.8 | -64 33 45 | 291.42 | -4.48 | 93 | -14.1 | 23.7 | 13x 8 | 18.4 | .69 | ? | | | |
| 2872 | | | 10 58 03.4 | -64 32 38 | 291.43 | -4.46 | 93 | -13.2 | 24.8 | 13x 12 | 18.0 | .70 | ? | | 1 | |
| 2873 | | P | 10 58 39.4 | -53 41 31 | 287.01 | 5.46 | 169 | -10.3 | 69.9 | 20x 13 | 17.6 | .36 | | S M | | |
| 2874 | | | 10 58 49.1 | -55 52 48 | 287.94 | 3.47 | 169 | -8.8 | -47.4 | 19x 16 | 17.5 | .42 | ? | SB | S | |
| 2875 | | | 10 59 05. | -56 10.1 : | 288.09 | 3.22 | 169 | -6.5 | -62.8 | 27x 8 | 17.8 | .38 | | S | S | smooth |
| 2876 | | | 10 59 46.2 | -53 38 10 | 287.14 | 5.58 | 169 | -1.5 | 72.9 | 20x 16 | 17.2 | .27 | | S ? | | |
| 2877 | | | 11 00 03.3 | -52 41 48 | 286.79 | 6.45 | 169 | .9 | 123.3 | 27x 16 | 16.8 | .36 | | E 3 | | |
| 2878 | | | 11 00 06. | -56 26.5 : | 288.33 | 3.03 | 169 | -10.0 | -30.2 | 20x 7 | 18.3 | .43 | | S L | E | |
| 2879 | L169-005 | | 11 00 20.4 | -53 22 26 | 287.11 | 5.85 | 169 | 3.1 | 86.9 | 67x 47 | 14.8 | .34 | | SB 1 | | |
| 2880 | | | 11 00 30.6 | -53 07 51 | 287.03 | 6.08 | 169 | 4.5 | 100.0 | 20x 4 | 19.1 | .39 | ? | | | larger vLSBdisk? |
| 2881 | | | 11 01 23.8 | -52 19 13 | 286.83 | 6.88 | 169 | 12.0 | 143.4 | 13x 7 | 19.0 | .30 | | | | fuzzy |
| 2882 | | | 11 01 41.4 | -53 07 17 | 287.19 | 6.16 | 169 | 14.0 | 100.4 | 13x 5 | 19.2 | .41 | | S | | dist.blg |
| 2883 | | I | 11 01 45.7 | -67 32 46 | 293.00 | -7.05 | 93 | 6.3 | -136.3 | 24x 8 | 18.2 | .31 | ? | P | S | 1 dist.sp.arm |
| 2884 | | | 11 01 53.6 | -66 31 03 | 292.60 | -6.10 | 93 | 7.6 | -81.2 | 27x 17 | 17.5 | .45 | ? | S ? | | dist.nucl, LSBdisk |
| 2885 | | | 11 02 24. | -56 10.7 : | 288.52 | 3.40 | 169 | -27.0 | -16.5 | 17x 17 | 18.0 | .39 | | S | F | dist.nucl. |
| 2886 | | | 11 02 44.0 | -52 35 30 | 287.12 | 6.71 | 169 | 22.8 | 128.7 | 19x 13 | 17.6 | .34 | | S | V | dist.blg, LSBdisk |
| 2887 | | | 11 03 00.3 | -52 17 38 | 287.04 | 7.00 | 169 | 25.2 | 144.7 | 20x 7 | 18.3 | .37 | | S 3 : | | |
| 2888 | | | 11 03 04.4 | -54 43 24 | 288.02 | 4.78 | 169 | 24.0 | 14.4 | 13x 8 | 18.4 | .29 | | | | |
| 2889 | | | 11 04 58.1 | -52 18 27 | 287.32 | 7.11 | 169 | 41.2 | 143.7 | 16x 16 | 17.4 | .32 | ? | S M | | br.blg or s.p.* |
| 2890 | | | 11 05 03.3 | -67 05 05 | 293.11 | -6.50 | 93 | 23.8 | -111.9 | 23x 8 | 18.3 | .37 | | S M | | |
| 2891 | | | 11 05 44.1 | -65 53 02 | 292.70 | -5.36 | 93 | 29.0 | -47.6 | 17x 16 | 18.0 | .60 | ? | | F | |
| 2892 | | | 11 06 28.8 | -67 04 51 | 293.24 | -6.44 | 93 | 31.2 | -111.9 | 15x 12 | 18.3 | .35 | ? | S ? | | |
| 2893 | | | 11 07 31.7 | -52 19 30 | 287.70 | 7.24 | 169 | 62.2 | 142.2 | 27x 8 | 17.7 | .28 | | S L | | |
| 2894 | | | 11 07 35.4 | -66 36 31 | 293.16 | -5.96 | 93 | 37.9 | -86.7 | 17x 12 | 18.0 | .46 | ? | S ? | S | |
| 2895 | | I | 11 08 07.3 | -67 07 22 | 293.40 | -6.42 | 93 | 39.7 | -114.4 | 74x 40 | 15.6 | .35 | | SB 3 | S | LSB |
| 2896 | | Q | 11 09 58.5 | -57 16 51 | 289.90 | 2.79 | 169 | 72.2 | -124.0 | 30x 3 | 18.8 | .86 | ? | S | E | |
| 2897 | | | 11 10 08.4 | -53 00 06 | 288.32 | 6.77 | 169 | 82.3 | 105.2 | 13x 11 | 18.6 | .25 | | | | |
| 2898 | | | 11 10 08.6 | -67 04 07 | 293.57 | -6.29 | 93 | 50.3 | -111.9 | 34x 7 | 17.5 | .37 | | S 1 | | |
| 2899 | | | 11 10 54.5 | -56 30 31 | 289.73 | 3.55 | 169 | 80.7 | -82.9 | 17x 8 | 18.6 | .75 | ? | S | | LSBdisk |
| 2900 | | | 11 11 03.9 | -52 38 06 | 288.32 | 7.16 | 169 | 90.5 | 124.6 | 16x 11 | 18.1 | .34 | | | 1 | |

**Table 1.** Galaxies in the southern ZOA – I. The Hydra/Antlia Extension (continued)

| RKK Ident (1) | Other Ident (2) | IR (3) | R.A. (h m s) (4) | Dec. (° ′ ″) (5) | gal $\ell$ (°) (6) | gal $b$ (°) (7) | SRC (8) | X (mm) (9) | Y (mm) (10) | D x d (″) (11) | $B_J$ ($^m$) (12) | $N_{HI}$ (13) | u | Type class. (14) | o | * | Remarks (15) |
|---|---|---|---|---|---|---|---|---|---|---|---|---|---|---|---|---|---|
| 2901 | | I | 11 11 06.1 | -66 50 52 | 293.57 | -6.05 | 93 | 55.9 | -100.3 | 26x 9 | 17.6 | .41 | ? | | | | cl.to br.* |
| 2902 | | I | 11 11 12.6 | -67 15 45 | 293.74 | -6.43 | 93 | 55.4 | -122.6 | 60x 34 | 15.7 | .33 | | SB 5 | | S | br.blg, LSBdisk |
| 2903 | | | 11 11 28.6 | -52 48 58 | 288.44 | 7.01 | 169 | 93.4 | 114.7 | 20x 8 | 18.2 | .25 | | S 5 | | | |
| 2904 | | | 11 11 39.9 | -54 17 30 | 289.01 | 5.65 | 169 | 91.5 | 35.6 | 15x 8 | 18.7 | .26 | | S ? | | | |
| 2905 | | | 11 11 54.0 | -52 19 09 | 288.32 | 7.50 | 169 | 98.0 | 141.2 | 16x 7 | 18.5 | .22 | | S | | | |
| 2906 | | | 11 12 22.4 | -60 43 41 | 291.46 | -.31 | 129 | -78.4 | -40.1 | 13x 9 | 18.4 | .51 | | S | | | |
| 2907 | L215-036 | | 11 12 27.6 | -52 25 39 | 288.44 | 7.43 | 169 | 102.4 | 135.2 | 47x 38 | 15.6 | .22 | ? | | | | neb? |
| 2908 | | | 11 12 48.5 | -53 14 08 | 288.78 | 6.70 | 169 | 103.2 | 91.8 | 16x 16 | 17.6 | .33 | | | | | |
| 2909 | | | 11 13 02.0 | -53 39 33 | 288.97 | 6.31 | 169 | 103.9 | 69.0 | 23x 11 | 17.2 | .25 | | SB 1 | | | br.blg |
| 2910 | | | 11 13 29.3 | -52 58 31 | 288.78 | 6.98 | 169 | 109.3 | 105.5 | 15x 9 | 18.4 | .27 | ? | | | | |
| 2911 | | | 11 13 34.9 | -54 56 08 | 289.51 | 5.15 | 169 | 104.8 | .4 | 27x 13 | 17.3 | .49 | ? | | | S | smooth, cl.to br.* |
| 2912 | | | 11 13 36.1 | -67 24 12 | 294.00 | -6.48 | 93 | 67.3 | -130.8 | 17x 5 | 19.2 | .31 | | S M | | | |
| 2913 | | | 11 13 38.1 | -52 41 01 | 288.70 | 7.25 | 169 | 111.3 | 121.0 | 30x 11 | 17.4 | .26 | | S E | | | |
| 2914 | | | 11 15 20.5 | -54 50 49 | 289.72 | 5.33 | 169 | 118.6 | 4.4 | 23x 5 | 18.4 | .46 | | S | | 1 | |
| 2915 | | | 11 15 50.0 | -67 14 07 | 294.15 | -6.25 | 93 | 79.4 | -122.5 | 20x 4 | 19.5 | .39 | | S | | | LSB |
| 2916 | | | 11 15 56.6 | -53 21 20 | 289.27 | 6.75 | 169 | 127.9 | 84.0 | 27x 7 | 17.9 | .21 | | S | E | | |
| 2917 | | | 11 16 14.2 | -54 43 26 | 289.79 | 5.49 | 169 | 125.9 | 10.6 | 19x 11 | 17.8 | .39 | | S 0 | | | |
| 2918 | | | 11 16 33.4 | -56 35 23 | 290.49 | 3.75 | 169 | 122.2 | -89.4 | 16x 4 | 19.4 | .59 | | S | E | | |
| 2919 | | | 11 17 01.5 | -54 40 30 | 289.88 | 5.57 | 169 | 132.1 | 12.9 | 27x 11 | 17.1 | .39 | | S E: | | | dist.nucl, dist.sp.arms |
| 2920 | | | 11 17 05.3 | -54 29 49 | 289.83 | 5.74 | 170 | -124.7 | 25.2 | 13x 4 | 19.8 | .34 | | S ? | | | |
| 2921 | | | 11 17 07.8 | -53 27 39 | 289.47 | 6.71 | 169 | 137.0 | 77.8 | 27x 13 | 17.0 | .20 | | S M? | | | smooth |
| 2922 | | | 11 17 19.3 | -56 42 58 | 290.63 | 3.67 | 170 | -116.5 | -93.5 | 24x 5 | 19.1 | .54 | ? | S | | | |
| 2923 | | | 11 17 20.1 | -53 34 54 | 289.54 | 6.61 | 170 | -125.4 | 74.3 | 13x 3 | 19.5 | .21 | | S | E | | |
| 2924 | | | 11 17 22.2 | -56 16 53 | 290.49 | 4.08 | 170 | -117.4 | -70.2 | 19x 5 | 18.9 | .54 | | S ? | | | |
| 2925 | | | 11 17 34.4 | -53 55 27 | 289.70 | 6.30 | 169 | 139.0 | 52.8 | 16x 7 | 18.5 | .21 | | S M? | | | pair w.2926 |
| 2926 | | | 11 17 37.0 | -53 55 04 | 289.70 | 6.31 | 169 | 139.3 | 53.1 | 13x 7 | 18.9 | .21 | | S | | | pair w.2925 |
| 2927 | | | 11 17 42.1 | -54 12 17 | 289.81 | 6.05 | 170 | -120.8 | 41.1 | 23x 7 | 18.4 | .28 | | | | | in st.diffr.pat. |
| 2928 | | | 11 17 52.6 | -53 51 17 | 289.71 | 6.38 | 170 | -120.3 | 59.9 | 17x 8 | 18.4 | .20 | ? | S | | S | blg or s.p.* |
| 2929 | | | 11 17 57.3 | -53 07 25 | 289.47 | 7.07 | 170 | -121.7 | 99.1 | 13x 7 | 19.1 | .18 | | S | | | |
| 2930 | | | 11 18 05.6 | -53 05 39 | 289.48 | 7.11 | 170 | -120.7 | 100.7 | 20x 7 | 18.5 | .19 | | S ? | | | |
| 2931 | | | 11 18 07.9 | -58 54 00 | 291.49 | 1.66 | 129 | -43.4 | 59.3 | 40x 11 | 17.0 | .79 | | S ? | | S | |
| 2932 | | | 11 18 09.8 | -53 25 59 | 289.61 | 6.79 | 170 | -119.2 | 82.6 | 23x 7 | 18.1 | .17 | | S | | | |
| 2933 | | | 11 18 18.4 | -56 54 29 | 290.83 | 3.54 | 170 | -108.7 | -103.4 | 13x 13 | 18.6 | .50 | ? | | | | |
| 2934 | | | 11 18 21.9 | -52 26 33 | 289.29 | 7.73 | 170 | -120.2 | 135.8 | 24x 7 | 18.6 | .19 | | S | | | |
| 2935 | | | 11 18 26.7 | -52 24 12 | 289.29 | 7.78 | 170 | -119.7 | 137.9 | 24x 8 | 17.9 | .18 | | S M | | | |
| 2936 | | | 11 18 27.5 | -54 18 41 | 289.95 | 5.98 | 170 | -114.5 | 35.7 | 13x 5 | 19.4 | .27 | | S ? | | | cl.to br.* |
| 2937 | FGCE0884 | | 11 18 29.6 | -53 24 01 | 289.64 | 6.84 | 170 | -116.7 | 84.5 | 54x 9 | 16.9 | .16 | | S | E | S | |
| 2938 | | | 11 18 54.4 | -53 31 48 | 289.75 | 6.74 | 170 | -113.0 | 77.7 | 16x 11 | 17.8 | .17 | ? | S E? | | | or blg |
| 2939 | | | 11 18 56.8 | -56 02 43 | 290.61 | 4.38 | 170 | -106.3 | -57.0 | 15x 8 | 18.8 | .54 | | | | 1 | |
| 2940 | | | 11 18 58.5 | -55 53 18 | 290.56 | 4.53 | 170 | -106.5 | -48.6 | 15x 9 | 18.8 | .52 | ? | | | S | |
| 2941 | | I | 11 19 04.5 | -54 32 50 | 290.12 | 5.79 | 170 | -109.1 | 23.3 | 17x 8 | 18.3 | .41 | | S | | | w.LSB comp. |
| 2942 | | | 11 19 05.3 | -56 53 21 | 290.92 | 3.59 | 170 | -103.0 | -102.1 | 13x 3 | 19.9 | .44 | | | | | |
| 2943 | | | 11 19 05.4 | -56 51 54 | 290.91 | 3.62 | 170 | -103.1 | -100.9 | 20x 11 | 17.8 | .44 | ? | | | | |
| 2944 | | | 11 19 08.5 | -52 36 31 | 289.46 | 7.62 | 170 | -113.5 | 127.2 | 16x 8 | 18.5 | .22 | | | | | |
| 2945 | | | 11 19 09.1 | -52 39 14 | 289.48 | 7.58 | 170 | -113.3 | 124.7 | 16x 3 | 19.3 | .22 | | S | | | |
| 2946 | | | 11 19 18.6 | -52 50 36 | 289.57 | 7.41 | 170 | -111.5 | 114.6 | 34x 20 | 17.1 | .19 | | | | S | |
| 2947 | | | 11 19 23.1 | -52 58 31 | 289.62 | 7.29 | 170 | -110.6 | 107.6 | 20x 11 | 18.0 | .19 | | | | S | |
| 2948 | | | 11 19 24.0 | -53 45 14 | 289.89 | 6.56 | 170 | -108.6 | 65.9 | 27x 8 | 18.2 | .20 | | S M | | S | |
| 2949 | | | 11 19 29.5 | -54 07 39 | 290.03 | 6.21 | 170 | -106.9 | 45.9 | 13x 7 | 19.1 | .24 | | | | | smooth |
| 2950 | | | 11 19 36.1 | -55 37 28 | 290.56 | 4.81 | 170 | -102.4 | -34.2 | 22x 11 | 18.1 | .46 | | | | | |
| 2951 | | | 11 19 50.0 | -54 20 44 | 290.15 | 6.02 | 170 | -103.7 | 34.4 | 17x 5 | 19.2 | .28 | | S M | | | w.sev.comp. |
| 2952 | | | 11 20 01.0 | -60 39 18 | 292.31 | .09 | 129 | -28.5 | -34.5 | 16x 3 | 19.3 | .24 | ? | S | | 1 | |
| 2953 | | | 11 20 22.5 | -53 12 31 | 289.84 | 7.12 | 170 | -102.1 | 95.5 | 17x 8 | 18.2 | .20 | | | | | |
| 2954 | | | 11 20 30.4 | -55 34 11 | 290.66 | 4.90 | 170 | -95.7 | -31.0 | 17x 17 | 18.0 | .47 | ? | S ? | F | | |
| 2955 | | | 11 20 31.2 | -53 57 12 | 290.12 | 6.42 | 170 | -99.3 | 55.6 | 13x 5 | 19.4 | .22 | | S | | | |
| 2956 | | | 11 20 40.0 | -56 01 19 | 290.83 | 4.48 | 170 | -93.5 | -55.2 | 32x 27 | 16.8 | .39 | | S M | F | S | |
| 2957 | | | 11 21 07.2 | -56 28 46 | 291.05 | 4.07 | 170 | -89.1 | -79.5 | 17x 11 | 18.1 | .41 | ? | | | | cl.to br.* |
| 2958 | | | 11 21 07.9 | -56 04 52 | 290.91 | 4.45 | 170 | -89.9 | -58.2 | 20x 16 | 17.7 | .37 | ? | S | | | cl.to br.* |
| 2959 | | | 11 21 11.2 | -52 22 16 | 289.68 | 7.95 | 170 | -97.3 | 140.6 | 20x 7 | 18.3 | .24 | | S | | | |
| 2960 | | | 11 21 30.6 | -56 46 25 | 291.19 | 3.81 | 170 | -85.6 | -95.2 | 22x 9 | 17.9 | .37 | | | | | |
| 2961 | | | 11 21 40.3 | -52 45 48 | 289.88 | 7.60 | 170 | -92.6 | 119.7 | 15x 7 | 18.4 | .24 | | | | | |
| 2962 | | | 11 21 43.4 | -55 39 05 | 290.85 | 4.88 | 170 | -86.3 | -35.0 | 13x 5 | 19.4 | .42 | | S | | | w.sev.comp. |
| 2963 | | | 11 21 44.4 | -53 01 19 | 289.97 | 7.36 | 170 | -91.5 | 105.9 | 20x 17 | 17.9 | .24 | | | | | |
| 2964 | | Q2 | 11 21 50.6 | -53 40 25 | 290.21 | 6.75 | 170 | -89.4 | 71.0 | 40x 11 | 17.0 | .19 | | S L | E | | pair w.2966 |
| 2965 | | | 11 21 51.4 | -52 21 48 | 289.77 | 7.99 | 170 | -91.9 | 141.2 | 30x 7 | 18.2 | .24 | | S | | | |
| 2966 | | Q2 | 11 21 52.9 | -53 40 06 | 290.21 | 6.76 | 170 | -89.1 | 71.3 | 15x 7 | 18.6 | .19 | | | | | pair w.2964 |
| 2967 | | | 11 21 59.6 | -56 00 37 | 291.00 | 4.56 | 170 | -83.6 | -54.1 | 74x 40 | 15.3 | .41 | | S M | | S | vLSBdisk |
| 2968 | | | 11 22 00.5 | -53 34 09 | 290.20 | 6.86 | 170 | -88.3 | 76.7 | 27x 8 | 18.2 | .21 | ? | S | | | dist.nucl. or s.p.* |
| 2969 | | | 11 22 03.5 | -52 23 47 | 289.81 | 7.97 | 170 | -90.1 | 139.5 | 17x 5 | 19.6 | .24 | | S ? | | | LSB |
| 2970 | | | 11 22 04.0 | -55 26 43 | 290.83 | 5.09 | 170 | -84.1 | -23.9 | 20x 8 | 18.7 | .49 | | S | | | |
| 2971 | | | 11 22 08.1 | -54 10 58 | 290.42 | 6.29 | 170 | -86.1 | 43.8 | 47x 12 | 17.0 | .29 | | S | E | S | |
| 2972 | | | 11 22 08.3 | -53 08 14 | 290.07 | 7.27 | 170 | -88.1 | 99.8 | 17x 16 | 18.2 | .24 | | | | | LSB |
| 2973 | | | 11 22 11.8 | -67 27 52 | 294.80 | -6.25 | 93 | 111.2 | -137.4 | 16x 4 | 19.5 | .45 | ? | S M | E | | |
| 2974 | | | 11 22 23.0 | -53 43 29 | 290.30 | 6.73 | 170 | -85.0 | 68.4 | 27x 20 | 17.2 | .22 | | S 5 | | S | br.* s.p. |
| 2975 | | | 11 22 33.1 | -53 07 23 | 290.12 | 7.31 | 170 | -84.8 | 100.7 | 27x 17 | 16.9 | .25 | ? | S ? | | | |
| 2976 | | | 11 22 34.0 | -56 54 45 | 291.38 | 3.73 | 170 | -77.6 | -102.3 | 74x 60 | 15.1 | .38 | | S | | S | vLSB |
| 2977 | | | 11 22 35.4 | -56 07 36 | 291.12 | 4.47 | 170 | -78.9 | -60.2 | 19x 4 | 19.3 | .43 | | S ? | | | |
| 2978 | | | 11 22 44.0 | -54 06 30 | 290.48 | 6.39 | 170 | -81.5 | 48.0 | 12x 5 | 19.3 | .29 | | | | | neighb.of 2980 |
| 2979 | | | 11 22 46.0 | -56 01 39 | 291.11 | 4.57 | 170 | -77.7 | -54.9 | 24x 8 | 18.1 | .43 | | | | | |
| 2980 | | | 11 22 48.6 | -54 06 21 | 290.49 | 6.39 | 170 | -80.9 | 48.1 | 24x 13 | 17.5 | .29 | | S L: | | S | |
| 2981 | | | 11 22 53.2 | -52 59 59 | 290.13 | 7.44 | 170 | -82.3 | 107.4 | 17x 5 | 19.1 | .27 | | S ? | | | |
| 2982 | | | 11 22 58.4 | -56 22 13 | 291.25 | 4.26 | 170 | -75.6 | -73.2 | 34x 20 | 16.9 | .45 | | | | S | |
| 2983 | | | 11 23 01.0 | -55 28 28 | 290.96 | 5.11 | 170 | -76.9 | -25.2 | 20x 5 | 19.1 | .46 | | S ? | | | |
| 2984 | | | 11 23 02.1 | -55 47 27 | 291.07 | 4.81 | 170 | -76.2 | -42.1 | 28x 17 | 17.4 | .43 | | | | | |
| 2985 | | | 11 23 06.5 | -53 01 14 | 290.17 | 7.43 | 170 | -80.5 | 106.4 | 40x 12 | 16.8 | .26 | | S E | | | |
| 2986 | | | 11 23 07.8 | -57 43 21 | 291.71 | 2.99 | 129 | -9.4 | 123.1 | 13x 11 | 18.4 | .48 | | S | | | w.comp. |
| 2987 | | | 11 23 11.4 | -53 53 20 | 290.47 | 6.62 | 170 | -78.3 | 59.9 | 19x 7 | 18.5 | .24 | | S | | S | |
| 2988 | | | 11 23 11.9 | -53 45 57 | 290.43 | 6.73 | 170 | -78.5 | 66.4 | 31x 8 | 17.7 | .23 | | S E: | | | |
| 2989 | | | 11 23 23.4 | -53 37 55 | 290.41 | 6.87 | 170 | -77.2 | 73.7 | 20x 20 | 17.3 | .23 | | L ? | | 1 | |
| 2990 | | | 11 23 40.7 | -56 11 19 | 291.29 | 4.46 | 170 | -70.6 | -63.3 | 15x 12 | 18.1 | .46 | | | | | |
| 2991 | L170-002 | I | 11 23 43.3 | -52 30 16 | 290.09 | 7.95 | 170 | -76.4 | 134.2 | 81x 81 | 14.5 | .26 | | SBR5 | | S | |
| 2992 | L170-003 | | 11 23 48.9 | -53 57 37 | 290.58 | 6.58 | 170 | -73.3 | 56.2 | 94x 40 | 14.6 | .26 | | S 2 | | S | |
| 2993 | | | 11 23 50.1 | -54 08 56 | 290.64 | 6.40 | 170 | -72.8 | 46.1 | 30x 20 | 17.1 | .30 | | SB 1 | | | |
| 2994 | | | 11 24 16.0 | -52 45 58 | 290.25 | 7.73 | 170 | -71.5 | 120.3 | 34x 13 | 17.2 | .26 | | S E | | | |
| 2995 | L170-004 | | 11 24 21.2 | -53 58 51 | 290.66 | 6.58 | 170 | -69.0 | 55.2 | 112x 17 | 15.8 | .27 | | S M | E | S | |
| 2996 | | | 11 24 23.9 | -53 27 02 | 290.49 | 7.09 | 170 | -69.4 | 83.6 | 20x 7 | 18.5 | .25 | | | | | |
| 2997 | | | 11 24 27.4 | -52 35 27 | 290.22 | 7.90 | 170 | -70.2 | 129.7 | 13x 12 | 18.6 | .27 | ? | | | | |
| 2998 | | | 11 24 32.2 | -53 21 26 | 290.48 | 7.18 | 170 | -68.5 | 88.7 | 16x 11 | 18.4 | .26 | ? | | | | gas? |
| 2999 | | | 11 24 37.8 | -54 46 41 | 290.96 | 5.84 | 170 | -65.6 | 12.5 | 20x 17 | 17.8 | .50 | | S | F | | |
| 3000 | | | 11 24 38.0 | -54 09 46 | 290.76 | 6.42 | 170 | -66.5 | 45.5 | 27x 17 | 17.4 | .29 | | | | | |

**Table 1.** Galaxies in the southern ZOA – I. The Hydra/Antlia Extension (continued)

| RKK Ident (1) | Other Ident (2) | IR (3) | R.A. (h m s) (4) | Dec. (° ′ ″) (5) | gal $\ell$ (°) (6) | gal $b$ (°) (7) | SRC (8) | X (mm) (9) | Y (mm) (10) | D x d (″) (11) | $B_J$ ($^m$) (12) | $N_{HI}$ (13) | u | Type class. (14) | o | * | Remarks (15) |
|---|---|---|---|---|---|---|---|---|---|---|---|---|---|---|---|---|---|
| 3001 | | | 11 24 43.8 | -55 18 38 | 291.14 | 5.34 | 170 | -64.1 | -16.0 | 13x 11 | 18.7 | .50 | ? | | | | |
| 3002 | | | 11 24 44.7 | -54 32 10 | 290.90 | 6.08 | 170 | -65.1 | 25.5 | 13x 7 | 19.4 | .37 | ? | | | | |
| 3003 | | | 11 24 50.2 | -57 56 04 | 292.00 | 2.86 | 129 | 2.9 | 111.8 | 22x 16 | 17.0 | .58 | | | | | |
| 3004 | | | 11 24 50.9 | -53 53 09 | 290.70 | 6.70 | 170 | -65.2 | 60.4 | 16x 8 | 18.1 | .26 | | S E | | | cl.to br.* |
| 3005 | | | 11 24 58.2 | -54 07 37 | 290.79 | 6.47 | 170 | -63.9 | 47.5 | 16x 9 | 18.3 | .29 | | | | | |
| 3006 | | | 11 25 03.7 | -56 00 54 | 291.41 | 4.69 | 170 | -60.6 | -53.7 | 34x 34 | 16.0 | .46 | | SBR4 | | S | w.sev.comp. |
| 3007 | | | 11 25 04.2 | -52 47 06 | 290.38 | 7.75 | 170 | -65.0 | 119.5 | 30x 16 | 17.1 | .23 | | | | | |
| 3008 | | | 11 25 13.3 | -53 03 55 | 290.49 | 7.49 | 170 | -63.4 | 104.5 | 13x 13 | 18.0 | .31 | ? | E ? | | | |
| 3009 | | | 11 25 15.9 | -52 42 07 | 290.38 | 7.84 | 170 | -63.5 | 123.9 | 30x 17 | 17.0 | .25 | | S ? | | S | |
| 3010 | | | 11 25 30.1 | -54 05 33 | 290.86 | 6.53 | 170 | -59.8 | 49.5 | 13x 11 | 18.4 | .29 | ? | | | | |
| 3011 | | | 11 25 30.5 | -55 16 38 | 291.24 | 5.41 | 170 | -58.2 | -14.0 | 16x 4 | 19.7 | .52 | ? | | | | |
| 3012 | | | 11 25 33.8 | -54 07 36 | 290.88 | 6.50 | 170 | -59.3 | 47.6 | 13x 5 | 19.2 | .30 | | S ? | | | |
| 3013 | | | 11 26 03.5 | -54 09 07 | 290.95 | 6.50 | 170 | -55.4 | 46.4 | 20x 13 | 18.1 | .30 | | S | | | dist.nucl. |
| 3014 | | | 11 26 15.1 | -53 53 10 | 290.90 | 6.76 | 170 | -54.1 | 60.7 | 12x 5 | 19.3 | .34 | | | | | cl.to br.* |
| 3015 | | | 11 26 25.0 | -54 59 16 | 291.27 | 5.73 | 170 | -51.6 | 1.6 | 16x 9 | 18.4 | .41 | ? | | | | |
| 3016 | | | 11 26 32.8 | -53 38 18 | 290.86 | 7.01 | 170 | -52.1 | 74.0 | 13x 3 | 20.0 | .34 | | S | E | | |
| 3017 | | | 11 26 41.1 | -58 08 49 | 292.30 | 2.74 | 129 | 16.0 | 100.4 | 74x 34 | 15.3 | .63 | | SX 3 | | S | |
| 3018 | | | 11 26 44.5 | -52 28 32 | 290.52 | 8.12 | 170 | -51.7 | 136.4 | 23x 20 | 17.2 | .20 | | S ? | | S | |
| 3019 | | | 11 26 55.0 | -54 34 35 | 291.21 | 6.14 | 170 | -48.2 | 23.8 | 15x 15 | 18.4 | .35 | ? | S | | | dist.nucl. |
| 3020 | | | 11 27 10.2 | -55 46 01 | 291.62 | 5.02 | 170 | -45.0 | -40.0 | 30x 24 | 17.1 | .39 | | S | F | S | LSB |
| 3021 | | | 11 27 22.7 | -53 33 10 | 290.95 | 7.13 | 170 | -45.5 | 78.7 | 20x 8 | 18.1 | .31 | | S ? | | S | cl.to br.* |
| 3022 | | I | 11 27 24.5 | -55 48 06 | 291.66 | 5.00 | 170 | -43.2 | -41.9 | 16x 9 | 18.3 | .36 | | | | | |
| 3023 | | | 11 27 32.6 | -52 52 11 | 290.76 | 7.79 | 170 | -44.8 | 115.3 | 31x 8 | 17.9 | .27 | | S | E | | |
| 3024 | | | 11 27 33.8 | -55 11 37 | 291.49 | 5.58 | 170 | -42.6 | -9.2 | 38x 35 | 15.8 | .47 | | E 0 | | S | |
| 3025 | | | 11 27 36.2 | -53 28 27 | 290.96 | 7.22 | 170 | -43.8 | 82.9 | 13x 4 | 19.8 | .29 | | S ? | | | |
| 3026 | | | 11 27 45.4 | -54 53 53 | 291.42 | 5.87 | 170 | -41.4 | 6.6 | 20x 8 | 18.7 | .42 | | | | | |
| 3027 | | | 11 27 53.6 | -54 17 05 | 291.25 | 6.46 | 170 | -40.8 | 39.5 | 17x 5 | 18.8 | .25 | | S | | | |
| 3028 | | | 11 27 58.6 | -52 48 08 | 290.80 | 7.87 | 170 | -41.4 | 119.0 | 15x 7 | 18.7 | .23 | | S ? | | | |
| 3029 | | | 11 28 08.2 | -56 11 15 | 291.88 | 4.66 | 170 | -37.4 | -62.4 | 27x 8 | 18.1 | .51 | | | | | p.cov.by * |
| 3030 | | | 11 28 08.6 | -53 42 11 | 291.11 | 7.02 | 170 | -39.3 | 70.7 | 17x 9 | 18.1 | .29 | | | | 1 | |
| 3031 | | | 11 28 12.1 | -53 35 18 | 291.08 | 7.14 | 170 | -39.0 | 76.9 | 13x 8 | 18.4 | .29 | ? | | | 1 | |
| 3032 | | | 11 28 21.7 | -53 39 37 | 291.12 | 7.07 | 170 | -37.6 | 73.1 | 11x 4 | 19.3 | .30 | | S ? | | | |
| 3033 | | | 11 28 23.3 | -53 40 33 | 291.13 | 7.06 | 170 | -37.4 | 72.2 | 13x 5 | 19.5 | .30 | | | | | |
| 3034 | | | 11 28 24.7 | -52 45 15 | 290.85 | 7.94 | 170 | -37.9 | 121.6 | 16x 13 | 17.9 | .23 | | SB 5 | | | |
| 3035 | | | 11 28 25.5 | -53 40 04 | 291.14 | 7.07 | 170 | -37.1 | 72.7 | 23x 12 | 17.7 | .30 | | S ? | | | |
| 3036 | | | 11 28 28.3 | -53 41 29 | 291.15 | 7.05 | 170 | -36.7 | 71.4 | 11x 9 | 18.7 | .30 | | E ? | | | |
| 3037 | | | 11 28 33.0 | -55 35 07 | 291.75 | 5.25 | 170 | -34.8 | -30.1 | 17x 12 | 18.2 | .51 | ? | | | | |
| 3038 | | | 11 28 42.5 | -53 40 21 | 291.18 | 7.08 | 170 | -34.9 | 72.5 | 15x 15 | 18.3 | .30 | ? | | | | |
| 3039 | | | 11 28 43.7 | -53 00 42 | 290.98 | 7.71 | 170 | -35.2 | 107.9 | 17x 5 | 19.3 | .26 | | S | | | |
| 3040 | | | 11 28 43.7 | -53 49 02 | 291.22 | 6.94 | 170 | -34.6 | 64.7 | 20x 13 | 18.0 | .28 | ? | S | | | |
| 3041 | | | 11 28 48.9 | -52 46 28 | 290.91 | 7.94 | 170 | -34.6 | 120.6 | 20x 8 | 18.6 | .25 | | S | | | |
| 3042 | | | 11 28 59.6 | -52 27 27 | 290.84 | 8.25 | 170 | -33.4 | 137.6 | 20x 7 | 18.6 | .24 | | S ? | | | br.* s.p. |
| 3043 | | | 11 28 59.9 | -52 48 02 | 290.95 | 7.92 | 170 | -33.1 | 119.2 | 30x 7 | 18.2 | .27 | | S | | | |
| 3044 | | | 11 29 14.0 | -54 15 21 | 291.43 | 6.55 | 170 | -30.4 | 41.2 | 12x 7 | 19.0 | .29 | | | | | |
| 3045 | | | 11 29 25.7 | -54 12 31 | 291.44 | 6.60 | 170 | -28.9 | 43.8 | 15x 4 | 19.7 | .31 | | S | E | | |
| 3046 | | | 11 29 35.0 | -55 21 08 | 291.81 | 5.52 | 170 | -27.0 | -17.5 | 12x 7 | 19.3 | .40 | | S | | | |
| 3047 | | | 11 29 35.5 | -53 19 24 | 291.20 | 7.45 | 170 | -28.0 | 91.2 | 20x 17 | 17.9 | .22 | | S | F | | dist.nucl. |
| 3048 | | | 11 29 41.8 | -53 10 00 | 291.16 | 7.61 | 170 | -27.3 | 99.7 | 13x 8 | 19.1 | .24 | | | | 1 | |
| 3049 | | | 11 29 47.5 | -53 24 05 | 291.25 | 7.39 | 170 | -26.4 | 87.1 | 13x 5 | 19.5 | .23 | | S | | | |
| 3050 | | | 11 29 49.5 | -53 22 36 | 291.25 | 7.41 | 170 | -26.2 | 88.4 | 9x 5 | 19.6 | .23 | | | | | |
| 3051 | | | 11 29 50.9 | -53 20 49 | 291.24 | 7.44 | 170 | -26.0 | 90.0 | 27x 20 | 16.9 | .23 | | E | | | |
| 3052 | | | 11 29 58.2 | -52 25 02 | 290.97 | 8.33 | 170 | -25.4 | 139.8 | 22x 7 | 18.3 | .28 | | S | | | |
| 3053 | | | 11 29 58.5 | -55 16 43 | 291.85 | 5.60 | 170 | -24.1 | -13.6 | 23x 7 | 18.6 | .43 | | S | E | | |
| 3054 | | P | 11 30 06.8 | -55 39 59 | 291.98 | 5.24 | 170 | -22.9 | -34.3 | 13x 9 | 18.6 | .44 | | | | | |
| 3055 | | | 11 30 07.3 | -53 10 03 | 291.22 | 7.62 | 170 | -23.9 | 99.6 | 47x 27 | 16.3 | .24 | | S M | V | S | |
| 3056 | | | 11 30 08.8 | -55 19 25 | 291.88 | 5.57 | 170 | -22.8 | -16.0 | 27x 7 | 18.2 | .43 | | I | | | w.comp. |
| 3057 | | | 11 30 09.1 | -52 19 32 | 290.97 | 8.43 | 170 | -24.0 | 144.8 | 11x 9 | 18.3 | .28 | | E | | | |
| 3058 | | | 11 30 13.6 | -52 18 30 | 290.98 | 8.45 | 170 | -23.4 | 145.7 | 15x 8 | 18.3 | .27 | | | | | |
| 3059 | | | 11 30 15.9 | -52 34 21 | 291.06 | 8.20 | 170 | -22.9 | 131.5 | 13x 7 | 18.7 | .26 | | S | | | |
| 3060 | | | 11 30 21.6 | -52 32 08 | 291.07 | 8.24 | 170 | -22.2 | 133.5 | 47x 8 | 17.5 | .25 | | S 5 | E | | |
| 3061 | | | 11 30 28.5 | -56 35 56 | 292.31 | 4.37 | 170 | -19.8 | -84.3 | 35x 31 | 16.5 | .31 | | S ? | F | S | |
| 3062 | | | 11 30 29.8 | -55 17 39 | 291.92 | 5.61 | 170 | -20.1 | -14.4 | 34x 7 | 18.1 | .43 | | S | E | | |
| 3063 | | P | 11 30 29.9 | -52 50 28 | 291.18 | 7.95 | 170 | -20.9 | 117.2 | 22x 11 | 18.2 | .29 | | | | S | |
| 3064 | | | 11 30 30.8 | -52 20 28 | 291.03 | 8.43 | 170 | -21.0 | 144.0 | 15x 5 | 19.2 | .27 | | | | | |
| 3065 | | | 11 30 36.5 | -53 03 06 | 291.26 | 7.76 | 170 | -20.0 | 105.9 | 13x 7 | 18.9 | .28 | ? | S ? | | | |
| 3066 | | | 11 30 38.1 | -53 10 56 | 291.30 | 7.63 | 170 | -19.7 | 98.9 | 15x 9 | 18.9 | .23 | | | | | |
| 3067 | | I | 11 30 41.8 | -55 16 26 | 291.94 | 5.64 | 170 | -18.6 | -13.3 | 34x 27 | 16.8 | .41 | | S 5 | | S | larger LSBdisk? |
| 3068 | | NG | 11 30 42.2 | -52 28 20 | 291.10 | 8.31 | 170 | -19.4 | 136.9 | 24x 5 | 18.8 | .25 | | S ? | | | |
| 3069 | | I | 11 30 46.8 | -52 28 17 | 291.11 | 8.32 | 170 | -18.8 | 137.0 | 27x 24 | 16.5 | .25 | | S ? | | | |
| 3070 | | NG | 11 30 49.8 | -52 28 36 | 291.12 | 8.32 | 170 | -18.4 | 136.7 | 12x 5 | 19.0 | .25 | | S | | | |
| 3071 | | | 11 30 51.8 | -52 25 36 | 291.11 | 8.36 | 170 | -18.1 | 139.4 | 9x 8 | 19.2 | .25 | | | | | neighb.of 3072 |
| 3072 | | | 11 30 52.6 | -52 25 30 | 291.11 | 8.37 | 170 | -18.0 | 139.5 | 20x 13 | 17.7 | .25 | | S E: | | | |
| 3073 | | NG | 11 30 54.1 | -52 28 36 | 291.13 | 8.32 | 170 | -17.8 | 136.7 | 15x 5 | 19.4 | .25 | | S | | | |
| 3074 | | | 11 30 55.7 | -52 25 44 | 291.12 | 8.37 | 170 | -17.6 | 139.3 | 16x 4 | 19.6 | .25 | | S | | | |
| 3075 | | | 11 30 58.7 | -52 23 14 | 291.11 | 8.41 | 170 | -17.2 | 141.5 | 27x 17 | 16.9 | .25 | | S E: | | | |
| 3076 | | | 11 31 01.2 | -56 15 55 | 292.28 | 4.71 | 170 | -15.9 | -66.4 | 17x 12 | 18.0 | .32 | | | | 1 | |
| 3077 | | | 11 31 02.6 | -54 00 07 | 291.61 | 6.87 | 170 | -16.3 | 54.9 | 13x 7 | 19.1 | .26 | | S ? | | | |
| 3078 | | | 11 31 03.5 | -54 49 45 | 291.86 | 6.08 | 170 | -15.9 | 10.6 | 20x 5 | 19.1 | .36 | | S M | | | |
| 3079 | | | 11 31 07.5 | -54 45 39 | 291.85 | 6.15 | 170 | -15.4 | 14.3 | 13x 8 | 18.8 | .41 | | | | | |
| 3080 | | I | 11 31 08.8 | -52 45 45 | 291.25 | 8.06 | 170 | -15.7 | 121.4 | 40x 40 | 16.1 | .34 | | SX 5 | | | vLSBdisk, cl.to br.* |
| 3081 | | I | 11 31 10.9 | -55 26 28 | 292.06 | 5.50 | 170 | -14.9 | -22.2 | 17x 8 | 18.4 | .35 | | | | | |
| 3082 | | | 11 31 11.6 | -52 28 25 | 291.17 | 8.33 | 170 | -15.4 | 136.9 | 15x 7 | 18.5 | .24 | | S | | | |
| 3083 | | | 11 31 21.3 | -54 46 04 | 291.88 | 6.15 | 170 | -13.7 | 13.9 | 15x 7 | 19.2 | .42 | | S | | | dist.nucl, LSBdisk |
| 3084 | | | 11 31 21.5 | -53 01 40 | 291.36 | 7.81 | 170 | -13.9 | 107.2 | 27x 9 | 17.9 | .25 | | | | | |
| 3085 | | | 11 31 32.1 | -52 25 23 | 291.20 | 8.40 | 170 | -12.6 | 139.6 | 13x 11 | 18.3 | .25 | | | | | |
| 3086 | | | 11 31 37.2 | -53 25 22 | 291.52 | 7.45 | 170 | -11.8 | 86.0 | 34x 34 | 16.4 | .23 | | S M | | S | larger LSBdisk? |
| 3087 | | | 11 31 48.5 | -55 32 40 | 292.17 | 5.43 | 170 | -10.1 | -27.7 | 22x 7 | 18.7 | .36 | | S | | | |
| 3088 | | | 11 31 51.6 | -54 07 01 | 291.76 | 6.79 | 170 | -9.8 | 48.8 | 16x 11 | 18.4 | .25 | | | | | in st.diffr.pat. |
| 3089 | | | 11 31 56.1 | -54 06 35 | 291.77 | 6.81 | 170 | -9.2 | 49.2 | 24x 20 | 17.1 | .24 | | S | | | in st.diffr.pat. |
| 3090 | | | 11 31 56.2 | -52 19 36 | 291.24 | 8.51 | 170 | -9.3 | 144.8 | 13x 9 | 18.7 | .28 | ? | | | | |
| 3091 | | | 11 32 07.8 | -52 34 05 | 291.34 | 8.29 | 170 | -7.7 | 131.8 | 13x 8 | 18.6 | .26 | | | | | |
| 3092 | | | 11 32 18.0 | -53 31 36 | 291.65 | 7.38 | 170 | -6.4 | 80.5 | 13x 7 | 19.1 | .23 | | S ? | | | |
| 3093 | | I | 11 32 20.7 | -52 20 18 | 291.30 | 8.52 | 170 | -6.0 | 144.2 | 13x 5 | 19.5 | .28 | | | | | |
| 3094 | | | 11 32 22.6 | -53 02 47 | 291.51 | 7.84 | 170 | -5.7 | 106.2 | 19x 11 | 17.8 | .24 | | L ? | | | |
| 3095 | | | 11 32 23.6 | -52 36 16 | 291.39 | 8.26 | 170 | -5.6 | 129.9 | 20x 7 | 18.6 | .27 | | S | | | |
| 3096 | | | 11 32 24.3 | -56 49 55 | 292.63 | 4.22 | 170 | -5.6 | -96.8 | 17x 7 | 18.7 | .41 | ? | | | | or blended ** |
| 3097 | | I | 11 32 28.1 | -57 06 18 | 292.72 | 3.96 | 170 | -5.1 | -111.4 | 20x 13 | 18.1 | .46 | | S ? | | | |
| 3098 | | | 11 32 32.1 | -52 37 04 | 291.41 | 8.26 | 170 | -4.4 | 129.2 | 27x 9 | 17.8 | .27 | | S M | | S | |
| 3099 | | | 11 32 41.2 | -52 52 54 | 291.51 | 8.01 | 170 | -3.2 | 115.0 | 16x 11 | 18.4 | .30 | | | | S | |
| 3100 | | | 11 32 46.6 | -52 44 43 | 291.48 | 8.15 | 170 | -2.5 | 122.3 | 34x 27 | 16.7 | .30 | | SB | F | S | vLSBdisk |

**Table 1.** Galaxies in the southern ZOA – I. The Hydra/Antlia Extension (continued)

| RKK Ident (1) | Other Ident (2) | IR (3) | R.A. (h m s) (4) | Dec. (° ′ ″) (5) | gal ℓ (°) (6) | gal b (°) (7) | SRC (8) | X (mm) (9) | Y (mm) (10) | D x d (″) (11) | $B_J$ ($^m$) (12) | $N_{HI}$ (13) | u | Type class. (14) | o * | Remarks (15) |
|---|---|---|---|---|---|---|---|---|---|---|---|---|---|---|---|---|
| 3101 | | | 11 32 55.4 | -52 22 25 | 291.39 | 8.51 | 170 | -1.3 | 142.3 | 17x 8 | 18.2 | .25 | | S | | |
| 3102 | | | 11 32 58.5 | -55 01 28 | 292.18 | 5.97 | 170 | -1.2 | .1 | 34x 27 | 16.7 | .37 | | | S | smooth |
| 3103 | | | 11 33 04.1 | -52 53 07 | 291.57 | 8.03 | 170 | -.1 | 114.8 | 13x 13 | 18.0 | .28 | ? | | | |
| 3104 | | | 11 33 05.9 | -52 59 43 | 291.60 | 7.92 | 170 | .1 | 108.9 | 13x 7 | 18.7 | .25 | | S E | | |
| 3105 | | | 11 33 13.9 | -54 47 12 | 292.15 | 6.21 | 170 | .9 | 12.9 | 20x 13 | 17.9 | .36 | ? | S ? | | |
| 3106 | | | 11 33 22.2 | -56 04 27 | 292.54 | 4.99 | 170 | 1.7 | -56.2 | 20x 13 | 18.0 | .47 | | S ? | | or dbl syst. |
| 3107 | | | 11 33 24.8 | -56 25 07 | 292.65 | 4.66 | 170 | 1.9 | -74.6 | 16x 3 | 20.1 | .30 | | S | E S | |
| 3108 | | | 11 33 30.4 | -56 28 51 | 292.68 | 4.60 | 170 | 2.6 | -78.0 | 27x 27 | 16.5 | .31 | ? | | | br.** s.p. |
| 3109 | | | 11 33 31.9 | -52 37 44 | 291.56 | 8.29 | 170 | 3.7 | 128.6 | 13x 9 | 18.6 | .28 | | | | |
| 3110 | | | 11 33 32.0 | -52 46 25 | 291.60 | 8.15 | 170 | 3.6 | 120.8 | 19x 4 | 19.7 | .30 | | S | | LSB |
| 3111 | | | 11 33 34.9 | -54 42 06 | 292.17 | 6.31 | 170 | 3.6 | 17.4 | 32x 13 | 17.5 | .31 | | | S | |
| 3112 | | | 11 33 35.2 | -53 50 08 | 291.92 | 7.14 | 170 | 3.8 | 63.9 | 22x 7 | 18.2 | .19 | | S | | |
| 3113 | | | 11 33 35.9 | -56 09 23 | 292.59 | 4.92 | 170 | 3.4 | -60.6 | 23x 5 | 19.0 | .36 | | S | E | |
| 3114 | | | 11 33 37.7 | -53 02 35 | 291.69 | 7.90 | 170 | 4.3 | 106.4 | 20x 20 | 17.3 | .23 | | | | |
| 3115 | | | 11 33 40.5 | -52 42 00 | 291.60 | 8.23 | 170 | 4.8 | 124.7 | 16x 13 | 17.9 | .30 | | | | |
| 3116 | | Q | 11 33 41.3 | -53 37 25 | 291.87 | 7.35 | 170 | 4.7 | 75.2 | 13x 9 | 18.7 | .22 | ? | | | |
| 3117 | | | 11 33 43.4 | -56 20 28 | 292.66 | 4.74 | 170 | 4.2 | -70.5 | 15x 13 | 18.3 | .30 | | S ? | | |
| 3118 | | | 11 33 43.7 | -53 04 17 | 291.72 | 7.88 | 170 | 5.1 | 104.8 | 22x 16 | 17.4 | .22 | | S L | | |
| 3119 | | | 11 33 46.5 | -52 57 39 | 291.69 | 7.98 | 170 | 5.6 | 110.8 | 16x 9 | 18.0 | .23 | | S E? | | |
| 3120 | | | 11 33 50.6 | -55 14 25 | 292.36 | 5.80 | 170 | 5.4 | -11.5 | 20x 17 | 17.3 | .35 | | E | | or S..E |
| 3121 | | | 11 33 55.1 | -52 33 24 | 291.60 | 8.38 | 170 | 6.8 | 132.4 | 30x 5 | 18.6 | .26 | | S 5 | | |
| 3122 | | I | 11 33 57.0 | -55 09 03 | 292.35 | 5.89 | 170 | 6.3 | -6.7 | 15x 5 | 19.1 | .35 | | | | |
| 3123 | | | 11 34 02.0 | -52 38 59 | 291.64 | 8.29 | 170 | 7.7 | 127.4 | 20x 20 | 17.3 | .28 | | E ? | | |
| 3124 | | | 11 34 10.5 | -56 25 26 | 292.75 | 4.68 | 170 | 7.6 | -74.9 | 19x 12 | 18.4 | .33 | | S | F | |
| 3125 | | | 11 34 13.9 | -52 35 32 | 291.65 | 8.36 | 170 | 9.4 | 130.5 | 27x 9 | 17.9 | .26 | | S ? | S | p.cov.by br.* |
| 3126 | | | 11 34 17.0 | -52 33 27 | 291.65 | 8.39 | 170 | 9.8 | 132.4 | 40x 7 | 17.9 | .25 | | S L | E | |
| 3127 | | | 11 34 17.1 | -52 43 17 | 291.70 | 8.24 | 170 | 9.8 | 123.6 | 40x 8 | 17.5 | .28 | | S 2 | S | |
| 3128 | | | 11 34 18.1 | -52 47 00 | 291.72 | 8.18 | 170 | 9.9 | 120.3 | 22x 17 | 17.0 | .28 | | E 2 ? | | |
| 3129 | | | 11 34 18.7 | -52 31 35 | 291.64 | 8.42 | 170 | 10.1 | 134.0 | 12x 5 | 19.1 | .24 | | S M | | neighb.of 3131 |
| 3130 | | | 11 34 20.2 | -53 00 09 | 291.79 | 7.97 | 170 | 10.1 | 108.5 | 17x 15 | 17.5 | .21 | ? | E | | |
| 3131 | | | 11 34 20.7 | -52 31 21 | 291.65 | 8.43 | 170 | 10.3 | 134.2 | 16x 13 | 18.1 | .24 | | S | F | w.comp. |
| 3132 | | | 11 34 24.2 | -52 48 25 | 291.74 | 8.16 | 170 | 10.7 | 119.0 | 13x 13 | 18.7 | .27 | | S ? | F | |
| 3133 | | | 11 34 24.9 | -52 50 17 | 291.75 | 8.13 | 170 | 10.8 | 117.3 | 13x 4 | 19.7 | .26 | | | | |
| 3134 | | | 11 34 25.7 | -53 57 56 | 292.08 | 7.05 | 170 | 10.4 | 56.9 | 20x 11 | 17.7 | .20 | | E | | or S..E |
| 3135 | | | 11 34 29.0 | -56 29 02 | 292.81 | 4.64 | 170 | 9.8 | -78.2 | 22x 8 | 18.3 | .33 | | S ? | S | |
| 3136 | | | 11 34 32.0 | -52 46 50 | 291.75 | 8.19 | 170 | 11.8 | 120.4 | 13x 8 | 19.1 | .26 | | | | |
| 3137 | | | 11 34 36.9 | -53 24 10 | 291.94 | 7.60 | 170 | 12.1 | 87.0 | 13x 7 | 18.4 | .21 | | S E | | or blg |
| 3138 | | | 11 34 40.1 | -52 49 08 | 291.78 | 8.16 | 170 | 12.8 | 118.3 | 20x 13 | 18.0 | .25 | | | | |
| 3139 | | | 11 35 03.9 | -52 46 56 | 291.83 | 8.21 | 170 | 16.1 | 120.3 | 13x 4 | 19.7 | .22 | | | | |
| 3140 | | | 11 35 16.0 | -52 38 01 | 291.82 | 8.36 | 170 | 17.8 | 128.2 | 13x 4 | 19.8 | .20 | | | | pair w.3141 |
| 3141 | | | 11 35 16.4 | -52 37 53 | 291.82 | 8.36 | 170 | 17.8 | 128.3 | 12x 7 | 19.1 | .20 | | | | pair w.3140 |
| 3142 | | | 11 35 16.7 | -54 39 13 | 292.39 | 6.42 | 170 | 16.7 | 19.9 | 13x 9 | 19.1 | .30 | | | | LSB |
| 3143 | | | 11 35 20.1 | -52 39 44 | 291.83 | 8.34 | 170 | 18.3 | 126.7 | 16x 13 | 18.3 | .20 | | | | |
| 3144 | | | 11 35 21.7 | -54 02 24 | 292.23 | 7.02 | 170 | 17.7 | 52.8 | 22x 4 | 19.1 | .23 | ? | S | S | |
| 3145 | | | 11 35 27.1 | -54 07 24 | 292.27 | 6.94 | 170 | 18.4 | 48.3 | 16x 13 | 18.3 | .24 | | S M: | | |
| 3146 | | | 11 35 27.1 | -54 10 15 | 292.28 | 6.90 | 170 | 18.4 | 45.8 | 20x 13 | 17.9 | .25 | ? | | S | |
| 3147 | | | 11 35 28.4 | -53 01 54 | 291.96 | 7.99 | 170 | 19.2 | 106.8 | 24x 5 | 18.8 | .19 | | S | | |
| 3148 | | Q2 | 11 35 31.1 | -52 27 53 | 291.80 | 8.53 | 170 | 19.9 | 137.2 | 13x 4 | 20.0 | .20 | | S | | pair w.3149 |
| 3149 | | Q2 | 11 35 35.3 | -52 28 08 | 291.82 | 8.53 | 170 | 20.5 | 137.0 | 13x 4 | 20.2 | .19 | | S | | pair w.3148 |
| 3150 | | I2 | 11 35 40.4 | -52 32 41 | 291.85 | 8.46 | 170 | 21.1 | 132.9 | 20x 7 | 18.3 | .19 | | | | pair w.3152 |
| 3151 | | | 11 35 41.3 | -53 13 45 | 292.05 | 7.81 | 170 | 20.8 | 96.2 | 16x 7 | 19.0 | .19 | | | | LSB |
| 3152 | | I2 | 11 35 47.1 | -52 31 59 | 291.86 | 8.48 | 170 | 22.1 | 133.6 | 15x 13 | 17.7 | .19 | | E ? | | pair w.3150 |
| 3153 | | | 11 35 49.4 | -55 13 27 | 292.63 | 5.90 | 170 | 20.6 | -10.7 | 15x 7 | 19.0 | .30 | ? | S ? | | |
| 3154 | L170-008 | | 11 36 00.8 | -52 47 03 | 291.97 | 8.25 | 170 | 23.8 | 120.1 | 47x 16 | 16.6 | .19 | | L | | |
| 3155 | | | 11 36 01.1 | -54 27 40 | 292.44 | 6.64 | 170 | 22.6 | 30.2 | 17x 15 | 18.1 | .27 | | S ? | | LSB |
| 3156 | | | 11 36 02.8 | -52 48 47 | 291.98 | 8.22 | 170 | 24.0 | 118.5 | 16x 15 | 17.8 | .19 | ? | S ? | | |
| 3157 | | | 11 36 13.6 | -54 33 32 | 292.50 | 6.55 | 170 | 24.2 | 24.9 | 40x 31 | 16.2 | .28 | | S M | S | LSB |
| 3158 | | | 11 36 16.3 | -53 11 39 | 292.12 | 7.87 | 170 | 25.5 | 98.1 | 27x 8 | 17.9 | .20 | | S 0 | | |
| 3159 | | | 11 36 17.4 | -52 30 05 | 291.93 | 8.53 | 170 | 26.2 | 135.2 | 20x 16 | 17.7 | .19 | ? | S | F | vLSBdisk |
| 3160 | | | 11 36 23.4 | -59 15 59 | 293.83 | 2.04 | 129 | 82.2 | 38.8 | 13x 11 | 18.4 | .78 | | S | | dist.sp.arms |
| 3161 | AM1136 | | 11 36 28.0 | -52 28 11 | 291.95 | 8.57 | 170 | 27.7 | 136.9 | 47x 27 | 15.8 | .19 | | S 1 | | |
| 3162 | | | 11 36 28.4 | -53 27 39 | 292.23 | 7.62 | 170 | 27.0 | 83.7 | 22x 16 | 17.6 | .23 | | S M | S | |
| 3163 | | | 11 36 41.5 | -52 40 05 | 292.04 | 8.39 | 170 | 29.4 | 126.2 | 16x 13 | 18.5 | .20 | | S L | | |
| 3164 | | | 11 36 50.0 | -55 20 59 | 292.80 | 5.82 | 170 | 28.2 | -17.6 | 16x 9 | 18.7 | .30 | | | | LSB |
| 3165 | L170-009 | I | 11 36 51.3 | -53 16 34 | 292.23 | 7.81 | 170 | 30.2 | 93.6 | 74x 27 | 15.8 | .22 | | S 2 | S | |
| 3166 | | | 11 36 53.7 | -54 25 10 | 292.55 | 6.72 | 170 | 29.5 | 32.3 | 16x 8 | 18.4 | .28 | | | | |
| 3167 | | | 11 36 54.1 | -56 51 01 | 293.23 | 4.38 | 170 | 27.4 | -98.0 | 20x 16 | 18.2 | .38 | ? | | | |
| 3168 | | | 11 37 09.7 | -58 55 01 | 293.83 | 2.40 | 129 | 88.3 | 57.3 | 13x 5 | 19.2 | .66 | ? | | | |
| 3169 | | | 11 37 11.7 | -54 34 28 | 292.64 | 6.58 | 170 | 31.7 | 23.9 | 20x 8 | 18.0 | .29 | | S E | | |
| 3170 | | | 11 37 23.9 | -53 45 14 | 292.44 | 7.38 | 170 | 34.0 | 67.9 | 15x 5 | 19.3 | .23 | | S ? | | |
| 3171 | | | 11 37 25.1 | -54 40 48 | 292.70 | 6.49 | 170 | 33.3 | 18.3 | 20x 16 | 17.7 | .31 | | | | LSB, cl.to br.* |
| 3172 | | Q | 11 37 33.2 | -53 20 39 | 292.35 | 7.78 | 170 | 35.7 | 89.8 | 20x 13 | 17.7 | .22 | | | | |
| 3173 | | | 11 37 37.2 | -55 29 57 | 292.95 | 5.71 | 170 | 34.0 | -25.7 | 17x 5 | 19.2 | .41 | | S | | |
| 3174 | | | 11 37 37.5 | -54 13 07 | 292.60 | 6.94 | 170 | 35.4 | 43.0 | 22x 22 | 17.3 | .32 | | S M: | F | dist.nucl. |
| 3175 | | | 11 37 37.9 | -52 33 59 | 292.15 | 8.53 | 170 | 37.1 | 131.5 | 19x 16 | 17.9 | .23 | | S M | | LSB |
| 3176 | L170-010 | I | 11 37 39.5 | -54 07 13 | 292.58 | 7.03 | 170 | 35.7 | 48.2 | 60x 54 | 15.1 | .29 | | SB 4 | S | LSB |
| 3177 | | | 11 37 45.1 | -58 39 57 | 293.84 | 2.66 | 129 | 93.0 | 70.6 | 22x 7 | 18.2 | .59 | | S M: | S | |
| 3178 | | | 11 37 46.5 | -60 08 14 | 294.24 | 1.25 | 129 | 89.5 | -8.2 | 15x 3 | 19.6 | .22 | | S | E | |
| 3179 | | | 11 37 53.2 | -53 00 59 | 292.31 | 8.10 | 170 | 38.7 | 107.4 | 27x 9 | 17.9 | .24 | | SB M | | br.* s.p, faint sp.arms |
| 3180 | | I | 11 37 54.1 | -56 03 37 | 293.15 | 5.18 | 170 | 35.6 | -55.8 | 20x 15 | 17.7 | .38 | | S M | F | |
| 3181 | | | 11 37 56.7 | -52 29 15 | 292.17 | 8.62 | 170 | 39.7 | 135.7 | 23x 13 | 17.6 | .24 | | S 3 | | |
| 3182 | | | 11 38 00.4 | -53 59 38 | 292.59 | 7.17 | 170 | 38.6 | 54.9 | 13x 11 | 18.6 | .26 | | | S | |
| 3183 | | | 11 38 05.3 | -57 03 03 | 293.44 | 4.23 | 170 | 35.9 | -108.9 | 13x 13 | 18.8 | .39 | | SB ? | | LSB |
| 3184 | | | 11 38 20.7 | -56 35 54 | 293.35 | 4.68 | 170 | 38.3 | -84.7 | 13x 7 | 19.1 | .34 | | | | |
| 3185 | | I | 11 38 26.8 | -52 27 21 | 292.24 | 8.67 | 170 | 43.9 | 137.3 | 20x 20 | 17.0 | .24 | ? | E | 1 | |
| 3186 | | | 11 38 33.4 | -52 59 46 | 292.40 | 8.15 | 170 | 44.1 | 108.3 | 32x 8 | 17.6 | .23 | | S 3 | | |
| 3187 | | | 11 38 39.4 | -58 40 36 | 293.95 | 2.69 | 129 | 99.2 | 69.7 | 20x 7 | 18.4 | .59 | | S | | |
| 3188 | | | 11 38 54.6 | -52 37 27 | 292.35 | 8.52 | 170 | 47.4 | 128.2 | 13x 11 | 18.0 | .24 | | E ? | 1 | |
| 3189 | | | 11 39 11.5 | -56 14 33 | 293.37 | 5.05 | 170 | 45.0 | -65.8 | 24x 12 | 17.5 | .37 | | | | |
| 3190 | | | 11 39 23.7 | -52 21 52 | 292.35 | 8.79 | 170 | 51.7 | 142.0 | 20x 13 | 18.3 | .21 | | S | | dist.nucl, LSBdisk |
| 3191 | | | 11 39 24.8 | -52 33 51 | 292.41 | 8.60 | 170 | 51.6 | 131.3 | 17x 16 | 17.3 | .26 | | E 2 | | |
| 3192 | | | 11 39 35.3 | -53 44 31 | 292.75 | 7.47 | 170 | 51.4 | 68.2 | 19x 4 | 19.2 | .23 | | S ? | | |
| 3193 | | | 11 39 49.2 | -53 04 45 | 292.61 | 8.12 | 170 | 54.2 | 103.6 | 17x 7 | 18.6 | .22 | | S ? | | |
| 3194 | | | 11 40 06.3 | -53 05 17 | 292.65 | 8.13 | 170 | 56.5 | 103.1 | 12x 7 | 19.3 | .22 | | | | smooth |
| 3195 | | | 11 40 10.5 | -52 54 09 | 292.61 | 8.31 | 170 | 57.3 | 113.0 | 20x 4 | 18.8 | .22 | | S | | |
| 3196 | | | 11 40 20.7 | -56 47 00 | 293.67 | 4.57 | 170 | 52.8 | -95.0 | 30x 4 | 19.1 | .37 | | S | E | |
| 3197 | | | 11 40 26.8 | -53 01 48 | 292.69 | 8.20 | 170 | 59.3 | 106.1 | 17x 12 | 17.6 | .22 | | E ? | | |
| 3198 | | I | 11 40 41.0 | -53 03 11 | 292.73 | 8.18 | 170 | 61.2 | 104.8 | 15x 12 | 18.1 | .22 | | E ? | | |
| 3199 | | | 11 40 49.9 | -53 31 28 | 292.87 | 7.73 | 170 | 61.6 | 79.6 | 16x 13 | 18.0 | .19 | | S | | |
| 3200 | | | 11 40 58.9 | -55 33 09 | 293.43 | 5.78 | 170 | 59.5 | -29.2 | 23x 7 | 18.5 | .33 | | S | | |

**Table 1.** Galaxies in the southern ZOA – I. The Hydra/Antlia Extension (continued)

| RKK Ident (1) | Other Ident (2) | IR (3) | R.A. (h m s) (4) | Dec. (° ′ ″) (5) | gal ℓ (°) (6) | gal b (°) (7) | SRC (8) | X (mm) (9) | Y (mm) (10) | D x d (″) (11) | $B_J$ ($^m$) (12) | $N_{HI}$ (13) | u | Type class. (14) | o * | Remarks (15) |
|---|---|---|---|---|---|---|---|---|---|---|---|---|---|---|---|---|
| 3201 | | | 11 41 10.9 | -53 13 42 | 292.85 | 8.03 | 170 | 64.9 | 95.3 | 23x 5 | 18.9 | .23 | | S | E | br.* s.p. |
| 3202 | | | 11 41 11.7 | -55 39 42 | 293.49 | 5.68 | 170 | 60.9 | -35.1 | 13x 12 | 18.3 | .35 | | | | cl.to br.* |
| 3203 | | | 11 41 12.8 | -57 35 33 | 293.99 | 3.82 | 129 | 120.4 | 126.8 | 13x 4 | 19.0 | .54 | | S | | cl.to br.* |
| 3204 | L216-035 | | 11 41 37.1 | -52 20 00 | 292.67 | 8.91 | 170 | 70.0 | 143.2 | 74x 67 | 14.9 | .20 | | S S5 | F | |
| 3205 | | | 11 41 38.6 | -55 55 27 | 293.62 | 5.45 | 170 | 63.8 | -49.3 | 8x 8 | 18.9 | .36 | | E 0 ? | | |
| 3206 | | | 11 41 44.6 | -56 09 55 | 293.69 | 5.22 | 170 | 64.1 | -62.2 | 13x 13 | 18.2 | .41 | | | | |
| 3207 | | | 11 41 44.8 | -53 46 59 | 293.07 | 7.52 | 170 | 68.5 | 65.5 | 24x 8 | 17.8 | .24 | | S | | |
| 3208 | | | 11 41 46.6 | -56 11 02 | 293.70 | 5.20 | 170 | 64.4 | -63.2 | 19x 16 | 17.7 | .41 | | S ? | F | |
| 3209 | | | 11 41 47.3 | -53 03 26 | 292.89 | 8.22 | 170 | 70.1 | 104.3 | 17x 16 | 17.9 | .21 | | SB M: | | |
| 3210 | | | 11 41 47.3 | -55 57 39 | 293.65 | 5.42 | 170 | 64.8 | -51.3 | 13x 8 | 18.4 | .36 | | S E | | |
| 3211 | | | 11 42 03.8 | -54 39 57 | 293.35 | 6.68 | 170 | 69.3 | 18.1 | 16x 3 | 19.7 | .38 | | S ? | 1 | |
| 3212 | | | 11 42 07.0 | -54 12 00 | 293.23 | 7.13 | 170 | 70.6 | 43.0 | 24x 12 | 18.2 | .32 | | S | 1 | LSB |
| 3213 | | | 11 42 30.1 | -56 16 18 | 293.82 | 5.14 | 170 | 69.6 | -68.1 | 24x 11 | 17.7 | .41 | | | | |
| 3214 | | | 11 42 31.6 | -53 16 22 | 293.05 | 8.04 | 170 | 75.6 | 92.6 | 34x 16 | 17.3 | .19 | ? | S ? | S | |
| 3215 | | | 11 42 31.9 | -52 58 54 | 292.98 | 8.32 | 170 | 76.2 | 108.2 | 26x 8 | 17.7 | .19 | | S ? | | |
| 3216 | | | 11 42 33.7 | -53 34 27 | 293.14 | 7.75 | 170 | 75.3 | 76.4 | 12x 9 | 18.4 | .22 | ? | | | |
| 3217 | | | 11 42 35.2 | -53 46 36 | 293.19 | 7.56 | 170 | 75.1 | 65.6 | 12x 9 | 18.4 | .23 | ? | E ? | | |
| 3218 | | | 11 42 54.6 | -53 54 59 | 293.28 | 7.43 | 170 | 77.4 | 58.0 | 17x 16 | 17.4 | .28 | | E 1 | | cl.to br.* |
| 3219 | | | 11 42 55.2 | -57 01 21 | 294.07 | 4.43 | 170 | 71.2 | -108.4 | 27x 4 | 19.2 | .41 | | S | E | w.LSB comp. |
| 3220 | | | 11 42 55.5 | -55 16 57 | 293.63 | 6.11 | 170 | 74.8 | -15.2 | 26x 7 | 18.6 | .37 | | S M | | dist.nucl. |
| 3221 | | | 11 42 58.6 | -52 34 41 | 292.94 | 8.73 | 170 | 80.6 | 129.7 | 13x 3 | 20.0 | .25 | | S ? | | |
| 3222 | | | 11 43 05.1 | -52 54 06 | 293.04 | 8.42 | 170 | 80.8 | 112.3 | 20x 15 | 17.7 | .23 | | S | F | |
| 3223 | | | 11 43 10.6 | -52 43 10 | 293.01 | 8.60 | 170 | 82.0 | 122.0 | 27x 11 | 17.7 | .26 | | S 4 | | |
| 3224 | | | 11 43 16.0 | -53 51 09 | 293.31 | 7.51 | 170 | 80.3 | 61.3 | 17x 13 | 18.4 | .26 | ? | S | | dist.nucl, LSBdisk |
| 3225 | | | 11 43 18.2 | -54 09 32 | 293.39 | 7.21 | 170 | 80.0 | 44.9 | 15x 11 | 18.0 | .31 | | E 2 : | | |
| 3226 | | | 11 43 36.4 | -53 15 22 | 293.21 | 8.10 | 170 | 84.3 | 93.2 | 16x 11 | 18.3 | .20 | | E ? | | |
| 3227 | | | 11 43 37.6 | -56 15 40 | 293.97 | 5.19 | 170 | 78.0 | -67.9 | 24x 9 | 17.9 | .38 | | | | |
| 3228 | L170-011 | I | 11 43 39.4 | -56 06 37 | 293.94 | 5.34 | 170 | 78.5 | -59.8 | 202x108 | 12.8 | .39 | | S 4 ? | S | no dist.blg |
| 3229 | | | 11 43 56.6 | -54 45 33 | 293.64 | 6.66 | 170 | 83.7 | 12.5 | 24x 13 | 17.3 | .38 | | L | | |
| 3230 | | | 11 44 10.5 | -54 15 33 | 293.54 | 7.15 | 170 | 86.6 | 39.2 | 20x 7 | 18.4 | .30 | | S E | | |
| 3231 | | | 11 44 19.2 | -54 33 44 | 293.64 | 6.86 | 170 | 87.0 | 22.9 | 13x 5 | 19.6 | .28 | | S ? | | |
| 3232 | | | 11 44 26.4 | -53 58 21 | 293.51 | 7.44 | 170 | 89.3 | 54.5 | 22x 11 | 18.1 | .29 | | S ? | | |
| 3233 | | | 11 44 32.1 | -53 54 47 | 293.51 | 7.50 | 170 | 90.2 | 57.6 | 27x 4 | 18.9 | .30 | | S | E | |
| 3234 | | | 11 44 33.4 | -54 40 24 | 293.70 | 6.76 | 170 | 88.6 | 16.9 | 13x 7 | 18.7 | .31 | | S ? | | |
| 3235 | | | 11 44 33.9 | -52 51 47 | 293.25 | 8.52 | 170 | 92.9 | 113.9 | 20x 4 | 18.9 | .23 | | S M | | p.cov.by br.* |
| 3236 | | | 11 44 38.0 | -53 05 22 | 293.32 | 8.30 | 170 | 92.9 | 101.7 | 20x 11 | 17.5 | .19 | ? | E | | br.* s.p. |
| 3237 | | | 11 44 43.9 | -52 32 34 | 293.19 | 8.83 | 170 | 95.0 | 131.0 | 17x 3 | 19.6 | .20 | | S M | E | |
| 3238 | | | 11 44 44.1 | -52 24 16 | 293.16 | 8.97 | 170 | 95.3 | 138.4 | 20x 16 | 17.3 | .20 | | E 2 | | |
| 3239 | | | 11 44 49.2 | -52 57 14 | 293.31 | 8.44 | 170 | 94.7 | 108.9 | 44x 30 | 16.0 | .22 | | SBS4 | S | |
| 3240 | | | 11 45 07.6 | -55 37 04 | 294.02 | 5.87 | 170 | 90.7 | -33.9 | 13x 5 | 19.0 | .43 | | S E: | | |
| 3241 | | | 11 45 08.6 | -54 03 19 | 293.63 | 7.38 | 170 | 94.7 | 49.8 | 17x 8 | 18.3 | .29 | | S | | * on edge of gal. |
| 3242 | | | 11 45 10.1 | -54 22 05 | 293.71 | 7.08 | 170 | 94.1 | 33.1 | 22x 5 | 19.1 | .31 | | S | E | dist.nucl. |
| 3243 | | | 11 45 14.1 | -52 29 27 | 293.25 | 8.90 | 170 | 99.2 | 133.6 | 15x 4 | 19.6 | .19 | | S M | | dist.nucl. |
| 3244 | | | 11 45 27.1 | -54 15 27 | 293.72 | 7.20 | 170 | 96.6 | 38.9 | 15x 7 | 18.6 | .31 | | S ? | | w.comp. |
| 3245 | L170-012 | | 11 45 29.4 | -53 37 32 | 293.57 | 7.81 | 170 | 98.5 | 72.7 | 47x 34 | 16.0 | .22 | | S 5 | | br.* on sp.arm |
| 3246 | | | 11 45 29.7 | -53 54 02 | 293.64 | 7.54 | 170 | 97.8 | 58.0 | 13x 7 | 19.1 | .29 | | | | w.comp. |
| 3247 | | | 11 45 40.4 | -52 48 58 | 293.40 | 8.60 | 170 | 102.0 | 116.0 | 13x 13 | 18.6 | .19 | | | | LSB |
| 3248 | | | 11 45 41.9 | -54 04 47 | 293.72 | 7.38 | 170 | 99.0 | 48.3 | 16x 7 | 18.3 | .28 | | L ? | | |
| 3249 | | | 11 45 48.4 | -53 47 35 | 293.66 | 7.66 | 170 | 100.6 | 63.6 | 17x 9 | 18.4 | .29 | | | | |
| 3250 | | | 11 45 59.5 | -53 36 35 | 293.64 | 7.84 | 170 | 102.5 | 73.4 | 24x 11 | 17.7 | .23 | | | | dbl nucl? |
| 3251 | | | 11 46 18.0 | -53 49 02 | 293.74 | 7.65 | 170 | 104.4 | 62.1 | 19x 16 | 17.3 | .29 | | S L | S | |
| 3252 | | | 11 46 27.6 | -53 12 01 | 293.61 | 8.26 | 170 | 107.3 | 95.1 | 16x 9 | 18.3 | .20 | | S 1 : | | w.comp. |
| 3253 | | | 11 46 40.7 | -53 54 30 | 293.82 | 7.58 | 170 | 107.1 | 57.1 | 12x 7 | 18.6 | .29 | | | | w.comp. |
| 3254 | | | 11 46 44.9 | -52 55 45 | 293.59 | 8.53 | 170 | 110.4 | 109.5 | 27x 16 | 17.2 | .19 | | | | |
| 3255 | | | 11 46 48.7 | -52 34 08 | 293.51 | 8.88 | 170 | 111.9 | 128.8 | 13x 11 | 18.1 | .17 | ? | | | p.cov.by br.* |
| 3256 | | | 11 46 50.9 | -54 38 32 | 294.02 | 6.87 | 170 | 106.5 | 17.7 | 20x 7 | 18.4 | .27 | | S | | |
| 3257 | | I | 11 47 07.3 | -52 35 19 | 293.56 | 8.88 | 170 | 114.3 | 127.6 | 47x 19 | 16.1 | .17 | | E 6 | | |
| 3258 | | | 11 47 14.5 | -53 01 17 | 293.68 | 8.46 | 170 | 114.1 | 104.4 | 20x 5 | 18.5 | .19 | | S ? | | |
| 3259 | | | 11 47 17.4 | -54 22 03 | 294.01 | 7.15 | 170 | 110.7 | 32.3 | 23x 20 | 16.9 | .33 | | | S | |
| 3260 | | | 11 47 27.1 | -57 22 53 | 294.75 | 4.23 | 170 | 103.2 | -129.2 | 15x 9 | 18.4 | .51 | | | 1 | |
| 3261 | | | 11 47 30.3 | -53 41 14 | 293.88 | 7.82 | 170 | 114.3 | 68.6 | 24x 7 | 18.4 | .28 | | | | |
| 3262 | | | 11 47 30.9 | -54 26 10 | 294.06 | 7.10 | 170 | 112.2 | 28.5 | 27x 7 | 18.1 | .31 | | S | | |
| 3263 | | | 11 47 35.5 | -52 54 54 | 293.71 | 8.58 | 170 | 117.2 | 109.9 | 13x 7 | 19.0 | .18 | | S M | S | |
| 3264 | | | 11 47 39.5 | -56 06 03 | 294.48 | 5.48 | 170 | 108.4 | -60.7 | 13x 5 | 19.2 | .43 | ? | | | |
| 3265 | | | 11 47 45.7 | -54 26 04 | 294.10 | 7.11 | 170 | 114.1 | 28.5 | 17x 11 | 17.7 | .32 | ? | L ? | | |
| 3266 | | | 11 48 03.7 | -52 46 08 | 293.74 | 8.74 | 170 | 121.4 | 117.5 | 20x 16 | 17.4 | .18 | | E 3 : | | |
| 3267 | | | 11 48 04.0 | -52 21 03 | 293.64 | 9.14 | 170 | 122.7 | 139.9 | 17x 16 | 17.9 | .19 | ? | S 5 | | LSB |
| 3268 | | | 11 48 04.2 | -54 20 28 | 294.12 | 7.21 | 170 | 116.8 | 33.3 | 40x 24 | 16.6 | .34 | | S M: | | |
| 3269 | | | 11 48 04.9 | -54 10 24 | 294.08 | 7.37 | 170 | 117.4 | 42.3 | 27x 8 | 17.9 | .28 | | S M | S | |
| 3270 | | | 11 48 05.2 | -57 16 31 | 294.81 | 4.36 | 170 | 108.1 | -123.8 | 19x 12 | 18.1 | .47 | | | | |
| 3271 | | | 11 48 10.6 | -53 18 49 | 293.89 | 8.21 | 170 | 120.7 | 88.3 | 16x 8 | 18.5 | .22 | | S M | | cl.to br.** |
| 3272 | | | 11 48 16.2 | -52 28 06 | 293.70 | 9.03 | 170 | 124.0 | 133.5 | 13x 5 | 19.5 | .16 | | | | w.comp. |
| 3273 | | | 11 48 24.1 | -52 57 41 | 293.84 | 8.56 | 170 | 123.6 | 107.1 | 16x 15 | 17.7 | .17 | | E 1 | | |
| 3274 | | | 11 48 43.2 | -54 43 23 | 294.30 | 6.86 | 170 | 120.7 | 12.6 | 35x 16 | 16.9 | .27 | | S M | S | br.* s.p. |
| 3275 | | | 11 49 29.6 | -57 04 49 | 294.95 | 4.59 | 170 | 118.9 | -113.9 | 23x 7 | 18.4 | .42 | | | S | |
| 3276 | | | 11 49 33.6 | -57 02 15 | 294.95 | 4.63 | 170 | 119.5 | -111.7 | 15x 7 | 18.6 | .41 | | S ? | | |
| 3277 | | | 11 49 38.4 | -55 32 11 | 294.62 | 6.10 | 170 | 125.1 | -31.4 | 23x 8 | 18.3 | .42 | | S | E | |
| 3278 | | | 11 49 52.0 | -57 03 37 | 295.00 | 4.62 | 170 | 121.7 | -113.0 | 27x 13 | 17.6 | .42 | | | | br.blg |
| 3279 | | | 11 50 06.7 | -57 00 33 | 295.02 | 4.68 | 170 | 123.6 | -110.4 | 81x 54 | 14.7 | .41 | | S E | F S | |

**Table 2.** IRAS Galaxies in the Hydra/Antlia ZOA Region

| IRAS PSC Ident. (1) | | IR (2) | RKK (3) | | R.A. (h m s) (4) | Dec. (° ′ ″) (5) | gal ℓ (°) (6) | gal b (°) (7) | D x d (″) (8) | $B_J$ ($^m$) (9) | Type class. (10) | Sep (″) (11) | $f_{12}$ (12) | $f_{25}$ (13) | $f_{60}$ (14) | $f_{100}$ (15) | Qual. (16) | $col_1$ (17) | $col_2$ (18) |
|---|---|---|---|---|---|---|---|---|---|---|---|---|---|---|---|---|---|---|---|
| | | | | | optical | | | | | | | IRAS | | | | | | | |
| I08327-5536 | | P | 38 | | 08 32 43.2 | -55 35 50 | 272.02 | -9.2 | 23x 7 | 18.3 | S M | 50 | 0.42 | 0.25 | 0.40 | 10.90 | 3 1 1 1 | 0.66 | 27.25 |
| I08336-5553 | Y | I | 60 | | 08 33 37.1 | -55 53 08 | 272.33 | -9.3 | 16x 16 | 17.4 | E | 8 | 0.28 | 0.25 | 1.38 | 1.97 | 1 1 3 2 | 0.04 | 1.43 |
| I08338-5500 | | I | 67 | | 08 33 46.0 | -55 00 22 | 271.62 | -8.7 | 50x 34 | 15.7 | S 0 | 18 | 0.26 | 0.77 | 0.82 | 4.87 | 1 1 3 1 | 0.30 | 5.94 |
| I08343-5432 | | I | 87 | | 08 34 19.6 | -54 32 39 | 271.29 | -8.4 | 47x 23 | 16.2 | F | 5 | 0.25 | 0.25 | 0.98 | 11.50 | 1 1 3 1 | 0.07 | 11.73 |
| I08344-5708 | | I | 97 | | 08 34 29.2 | -57 08 18 | 273.43 | -9.9 | 38x 23 | 16.1 | | 39 | 0.39 | 0.46 | 0.42 | 2.49 | 1 1 3 1 | 1.02 | 5.93 |
| I08349-5531 | | Q | 107 | | 08 34 49.9 | -55 30 38 | 272.12 | -8.9 | 19x 9 | 18.1 | | 92 | 0.59 | 0.22 | 0.40 | 1.65 | 3 3 1 1 | 0.81 | 4.12 |
| I08360-5456 | Y | I | 162 | L | 08 36 05.0 | -54 56 51 | 271.77 | -8.4 | 242x215 | 12.0 | L | 1 | 0.29 | 0.35 | 4.37 | 10.60 | 3 3 3 3 | 0.01 | 2.43 |
| I08360-5723 | | Q | 166 | L | 08 36 09.7 | -57 22 41 | 273.76 | -9.9 | 74x 16 | 16.1 | S 3 | 65 | 8.41 | 6.74 | 0.72 | 1.15 | 3 3 3 1 | 109.34 | 1.60 |
| I08363-5656 | | P | 171 | | 08 36 15.0 | -56 56 14 | 273.41 | -9.6 | 13x 7 | 18.9 | S | 43 | 0.46 | 0.25 | 0.40 | 1.53 | 3 1 1 1 | 0.72 | 3.82 |
| I08393-5522 | | P | 292 | | 08 39 29.3 | -55 22 16 | 272.41 | -8.3 | 13x 4 | 19.6 | | 63 | 0.74 | 0.43 | 0.40 | 2.17 | 1 1 3 2 | 1.99 | 5.42 |
| I08401-5440 | | I | 310 | | 08 40 07.9 | -54 39 35 | 271.90 | -7.8 | 27x 9 | 17.7 | S L | 27 | 0.27 | 0.25 | 0.42 | 3.38 | 1 1 3 1 | 0.38 | 8.05 |
| I08407-5721 | | Q | 334 | | 08 40 48.8 | -57 19 38 | 274.10 | -9.3 | 13x 8 | 18.4 | E | 105 | 0.25 | 0.25 | 0.40 | 2.05 | 1 1 1 3 | 0.39 | 5.12 |
| I08408-5529 | | Q | 336 | | 08 40 51.2 | -55 30 08 | 272.64 | -8.2 | 16x 11 | 18.3 | | 30 | 3.20 | 2.18 | 0.50 | 2.01 | 3 3 3 1 | 27.90 | 4.02 |
| I08423-5451 | | P | 382 | | 08 42 25.8 | -54 51 15 | 272.26 | -7.6 | 13x 8 | 18.6 | S | 31 | 0.27 | 0.25 | 0.40 | 8.49 | 3 1 1 1 | 0.42 | 21.22 |
| I08434-5308 | | P | 402 | | 08 43 27.0 | -53 08 29 | 270.99 | -6.5 | 13x 7 | 18.9 | | 13 | 0.37 | 0.25 | 0.38 | 17.60 | 1 1 3 1 | 0.64 | 46.32 |
| I08488-5805 | | I2 | 545 | | 08 48 48.6 | -58 05 43 | 275.38 | -9.0 | 31x 7 | 17.8 | S | 7 | 0.38 | 0.25 | 0.57 | 5.42 | 1 1 3 1 | 0.29 | 9.51 |
| I08488-5805 | | I2 | 547 | | 08 48 50.6 | -58 05 41 | 275.38 | -8.9 | 16x 8 | 18.4 | | 9 | 0.38 | 0.25 | 0.57 | 5.42 | 1 1 3 1 | 0.29 | 9.51 |
| I08491-5709 | | Q | 555 | | 08 49 08.0 | -57 07 49 | 274.65 | -8.3 | 17x 8 | 18.9 | S | 117 | 0.25 | 0.25 | 0.27 | 2.65 | 1 1 3 2 | 0.86 | 9.81 |
| I08504-5510 | | I | 606 | | 08 50 32.6 | -55 11 07 | 273.25 | -6.9 | 16x 7 | 18.6 | | 53 | 0.25 | 0.25 | 0.59 | 2.16 | 1 1 1 3 | 0.18 | 3.66 |
| I08508-5731 | | Q | 618 | | 08 50 56.9 | -57 32 37 | 275.13 | -8.4 | 13x 3 | 19.9 | S | 86 | 0.42 | 0.25 | 0.40 | 8.35 | 3 1 1 1 | 0.66 | 20.87 |
| I08510-5421 | | I | 624 | | 08 51 04.8 | -54 22 03 | 272.67 | -6.3 | 16x 11 | 18.3 | S | 55 | 0.33 | 0.14 | 0.40 | 6.82 | 3 3 1 1 | 0.29 | 17.05 |
| I08514-5427 | | I | 631 | | 08 51 22.8 | -54 27 24 | 272.77 | -6.4 | 13x 7 | 18.6 | S E | 26 | 0.81 | 0.25 | 0.28 | 8.92 | 1 1 3 1 | 2.58 | 31.86 |
| I08531-5536 | | P | 686 | | 08 53 13.4 | -55 36 33 | 273.83 | -6.9 | 58x 40 | 15.4 | S L | 31 | 1.19 | 0.58 | 0.29 | 12.40 | 1 1 3 1 | 8.21 | 42.76 |
| I08535-5541 | | I | 694 | L | 08 53 32.8 | -55 41 58 | 273.93 | -6.9 | 67x 13 | 16.3 | S | 2 | 0.43 | 0.25 | 0.31 | 9.11 | 1 1 3 1 | 1.12 | 29.39 |
| I08561-5400 | | I | 757 | | 08 56 05.8 | -54 00 13 | 272.86 | -5.6 | 27x 7 | 18.3 | S | 15 | 0.35 | 0.25 | 0.40 | 18.80 | 1 1 3 1 | 0.55 | 47.00 |
| I08575-5512 | Y | I | 799 | | 08 57 31.4 | -55 12 26 | 273.92 | -6.2 | 24x 19 | 17.0 | | 4 | 0.25 | 0.44 | 2.03 | 2.24 | 1 3 3 3 | 0.03 | 1.10 |
| I08575-5303 | | P2 | 801 | | 08 57 32.5 | -53 02 58 | 272.28 | -4.8 | 9x 8 | 19.2 | | 52 | 0.93 | 0.44 | 0.59 | 21.10 | 3 3 2 1 | 1.18 | 35.76 |
| I08575-5303 | | P2 | 802 | | 08 57 33.0 | -53 03 07 | 272.28 | -4.8 | 13x 7 | 19.0 | | 45 | 0.93 | 0.44 | 0.59 | 21.10 | 3 3 2 1 | 1.18 | 35.76 |
| I08579-5239 | | I | 808 | | 08 57 58. | -52 39.7 : | 272.03 | -4.5 | 17x 12 | 17.8 | S 3 | 10 | 0.27 | 0.25 | 0.63 | 5.69 | 1 1 3 1 | 0.177 | 9.03 |
| I08581-6050 | Y | I | 824 | | 08 58 13.7 | -60 51 24 | 278.31 | -9.8 | 16x 8 | 18.4 | S E | 55 | 0.25 | 0.25 | 1.29 | 2.29 | 1 1 3 3 | 0.04 | 1.78 |
| I09009-5544 | Y | I | 887 | | 09 00 54.1 | -55 45 04 | 274.65 | -6.2 | 17x 15 | 17.8 | | 17 | 0.29 | 0.25 | 0.74 | 2.03 | 1 1 3 1 | 0.13 | 2.74 |
| I09009-5733 | | I | 888 | | 09 00 56.0 | -57 33 11 | 276.01 | -7.4 | 16x 7 | 18.5 | S L | 41 | 0.33 | 0.57 | 0.35 | 1.34 | 1 1 3 1 | 1.54 | 3.83 |
| I09027-5757 | Y | I | 943 | | 09 02 47.4 | -57 57 38 | 276.48 | -7.5 | 17x 9 | 18.3 | S M | 2 | 0.25 | 0.25 | 2.90 | 6.04 | 1 1 3 3 | 0.01 | 2.08 |
| I09034-5932 | | Q | 959 | | 09 03 32.9 | -59 31 50 | 277.73 | -8.4 | 17x 7 | 18.7 | | 69 | 3.59 | 3.99 | 1.04 | 1.97 | 3 3 3 1 | 13.24 | 1.89 |
| I09037-5414 | Y | I | 966 | | 09 03 43.8 | -54 14 28 | 273.79 | -4.9 | 40x 7 | 17.5 | S L | 22 | 0.29 | 0.25 | 0.82 | 2.40 | 1 1 3 1 | 0.11 | 2.93 |
| I09066-5357 | | Q | 1032 | | 09 06 42.4 | -53 58 58 | 273.89 | -4.4 | 24x 5 | 18.5 | S M | 109 | 12.40 | 8.46 | 1.32 | 3.22 | 3 3 3 1 | 60.21 | 2.44 |
| I09083-5911 | | P | 1060 | | 09 08 25.3 | -59 12 14 | 277.91 | -7.7 | 13x 9 | 18.8 | S | 80 | 0.50 | 0.15 | 0.40 | 1.62 | 3 2 1 1 | 0.47 | 4.05 |
| I09096-5541 | | I | 1097 | | 09 09 37.1 | -55 41 13 | 275.43 | -5.2 | 27x 7 | 17.9 | S | 6 | 0.45 | 0.25 | 0.47 | 13.70 | 1 1 3 1 | 0.51 | 29.15 |
| I09112-5838 | Y | I | 1125 | L | 09 11 17.4 | -58 38 06 | 277.74 | -7.1 | 56x 34 | 15.2 | S 3 | 2 | 0.25 | 0.26 | 2.80 | 4.68 | 2 3 3 3 | 0.01 | 1.67 |
| I09118-5931 | | P | 1131 | | 09 11 44.5 | -59 30 08 | 278.42 | -7.6 | 19x 3 | 19.9 | S | 86 | 0.33 | 0.25 | 0.40 | 2.16 | 3 1 1 1 | 0.52 | 5.40 |
| I09122-6034 | Y | I | 1146 | L | 09 12 16.1 | -60 34 56 | 279.26 | -8.3 | 67x 60 | 14.2 | SB 2 | 2 | 0.50 | 2.02 | 12.10 | 15.60 | 3 3 3 3 | 0.01 | 1.29 |
| I09123-6112 | | P | 1147 | | 09 12 20.3 | -61 11 40 | 279.72 | -8.7 | 27x 4 | 18.5 | S | 52 | 0.34 | 0.14 | 0.49 | 5.46 | 3 2 1 1 | 0.20 | 11.14 |
| I09133-6013 | Y | I | 1159 | L | 09 13 23.0 | -60 13 30 | 279.09 | -8.0 | 114x 67 | 13.5 | SB 4 | 7 | 0.32 | 0.65 | 4.39 | 10.50 | 3 3 3 3 | 0.01 | 2.39 |
| I09136-5916 | * | I | 1163 | | 09 13 36.3 | -59 16 51 | 278.42 | -7.3 | 16x 12 | 17.1 | | 6 | 0.42 | 0.25 | 0.68 | 1.71 | 1 1 3 1 | 0.23 | 2.51 |
| I09144-6251 | Y | I | 1170 | L | 09 14 29.1 | -62 51 37 | 281.12 | -9.7 | 121x 94 | 13.5 | S 1 | 0 | 0.22 | 0.34 | 3.33 | 7.93 | 2 3 3 3 | 0.01 | 2.38 |
| I09162-5932 | Y | I | 1191 | | 09 16 13.5 | -59 32 58 | 278.85 | -7.3 | 16x 9 | 18.0 | | 5 | 0.35 | 0.25 | 0.72 | 2.19 | 1 1 3 1 | 0.17 | 3.04 |
| I09163-6240 | Y | I | 1196 | L | 09 16 21.9 | -62 40 19 | 281.13 | -9.4 | 121x 81 | 13.4 | S 5 | 5 | 0.25 | 0.25 | 0.79 | 1.77 | 1 1 3 3 | 0.10 | 2.24 |
| I09168-6141 | Y | I | 1205 | | 09 16 52.2 | -61 41 30 | 280.46 | -8.7 | 22x 5 | 18.1 | | 5 | 0.41 | 0.25 | 0.63 | 1.66 | 1 1 3 1 | 0.26 | 2.63 |
| I09174-6210 | | P | 1215 | | 09 17 31.2 | -62 12 15 | 280.89 | -9.0 | 27x 20 | 16.9 | | 88 | 0.18 | 0.25 | 0.40 | 2.29 | 3 1 1 1 | 0.28 | 5.72 |
| I09196-6113 | | I | 1244 | | 09 19 39.0 | -61 12 57 | 280.35 | -8.1 | 20x 4 | 18.9 | S | 42 | 0.41 | 0.25 | 0.40 | 1.40 | 3 1 1 1 | 0.64 | 3.50 |
| I09211-6050 | Y | I | 1265 | L | 09 21 08.3 | -60 50 05 | 280.21 | -7.7 | 121x 40 | 14.0 | S 3 | 5 | 0.26 | 0.25 | 1.97 | 7.47 | 3 3 3 3 | 0.02 | 3.79 |
| I09211-6021 | | P | 1267 | | 09 21 10.8 | -60 22 46 | 279.89 | -7.4 | 24x 5 | 18.4 | S | 73 | 0.42 | 0.25 | 0.54 | 12.80 | 3 1 1 1 | 0.36 | 23.70 |
| I09214-5655 | Y | I | 1272 | | 09 21 28.3 | -56 55 44 | 277.47 | -4.9 | 24x 15 | 17.1 | S L | 3 | 0.25 | 0.25 | 1.75 | 5.79 | 1 1 3 1 | 0.02 | 3.31 |
| I09217-5648 | Y | I | 1275 | | 09 21 43.2 | -56 48 22 | 277.41 | -4.8 | 23x 7 | 18.4 | S | 5 | 0.33 | 0.27 | 1.41 | 4.60 | 1 1 3 2 | 0.04 | 3.26 |
| I09221-6250 | | P | 1283 | | 09 22 10.7 | -62 51 55 | 281.74 | -9.1 | 27x 19 | 17.0 | S | 87 | 0.31 | 0.25 | 0.40 | 2.34 | 3 1 1 1 | 0.48 | 5.85 |
| I09222-6327 | Y | I | 1288 | L | 09 22 18.2 | -63 27 50 | 282.18 | -9.5 | 101x 12 | 15.8 | S 3 | 5 | 0.25 | 0.25 | 0.78 | 1.97 | 1 1 3 3 | 0.10 | 2.53 |
| I09253-6318 | | I | 1335 | | 09 25 27.1 | -63 18 50 | 282.33 | -9.1 | 15x 8 | 19.1 | | 43 | 0.35 | 0.25 | 0.40 | 1.13 | 1 1 1 3 | 0.55 | 2.82 |
| I09262-6033 | Y | I | 1353 | L | 09 26 12.6 | -60 33 07 | 280.46 | -7.1 | 128x 47 | 14.2 | S 1 | 2 | 0.44 | 0.23 | 1.64 | 4.14 | 2 3 3 3 | 0.04 | 2.52 |
| I09264-5853 | Y | I | 1356 | | 09 26 25.7 | -58 53 16 | 279.32 | -5.9 | 16x 9 | 17.8 | E | 10 | 0.26 | 0.25 | 1.09 | 2.25 | 1 1 3 2 | 0.05 | 2.06 |
| I09264-5706 | | Q | 1358 | | 09 26 30.6 | -57 07 47 | 278.11 | -4.6 | 11x 11 | 18.1 | | 76 | 4.84 | 1.27 | 0.40 | 17.60 | 3 3 1 1 | 38.42 | 44.00 |
| I09286-6157 | | I | 1397 | L | 09 28 42.0 | -61 57 44 | 281.66 | -7.9 | 108x 60 | 13.8 | S 0 | 9 | 0.29 | 0.25 | 0.48 | 3.36 | 1 1 3 1 | 0.31 | 7.00 |
| I09293-6448 | | I | 1408 | | 09 29 18.8 | -64 48 49 | 283.69 | -9.9 | 38x 11 | 16.9 | S E | 17 | 0.38 | 0.25 | 0.57 | 3.40 | 1 1 3 1 | 0.29 | 5.96 |
| I09293-5750 | | I | 1409 | | 09 29 20.6 | -57 50 15 | 278.87 | -4.8 | 34x 5 | 18.0 | S M | 12 | 0.26 | 0.25 | 0.56 | 4.14 | 1 1 3 1 | 0.21 | 7.39 |
| I09294-6304 | | I3 | 1410 | | 09 29 25.3 | -63 04 32 | 282.49 | -8.7 | 8x 4 | 20.0 | S | 22 | 0.26 | 0.25 | 0.51 | 2.36 | 1 1 3 1 | 0.25 | 4.63 |
| I09294-6304 | | I3 | 1411 | | 09 29 25.4 | -57 23 36 | 278.58 | -4.5 | 34x 27 | 16.4 | S | 10 | 0.26 | 0.25 | 0.51 | 2.36 | 1 1 3 1 | 0.25 | 4.63 |
| I09294-6304 | | I3 | 1412 | | 09 29 25.4 | -63 04 20 | 282.49 | -8.7 | 24x 7 | 18.0 | S | 16 | 0.26 | 0.25 | 0.51 | 2.36 | 1 1 3 1 | 0.25 | 4.63 |
| I09294-5723 | | I | 1413 | | 09 29 27.3 | -63 04 08 | 282.49 | -8.7 | 11x 5 | 19.1 | S | 13 | 0.71 | 0.25 | 0.58 | 25.80 | 1 1 3 1 | 0.53 | 44.48 |
| I09302-4724 | | I | 1426 | | 09 30 13.3 | -47 24 18 | 271.88 | 2.8 | 16x 13 | 17.8 | S | 17 | 0.46 | 0.25 | 0.60 | 3.40 | 1 1 3 3 | 0.32 | 5.67 |
| I09328-6103 | Y | I | 1464 | L | 09 32 52.2 | -61 03 34 | 281.40 | -6.9 | 108x 81 | 14.2 | S 6 | 4 | 0.26 | 0.25 | 1.13 | 3.41 | 1 1 3 3 | 0.05 | 3.02 |
| I09365-6155 | Y | I | 1506 | L | 09 36 31.6 | -61 55 31 | 282.31 | -7.3 | 114x 47 | 14.1 | S 3 | 11 | 0.25 | 0.28 | 2.06 | 6.50 | 1 1 3 3 | 0.02 | 3.16 |
| I09369-6315 | Y | I | 1514 | L | 09 36 57.6 | -63 15 38 | 283.25 | -8.2 | 108x 27 | 14.9 | S 0 | 1 | 0.28 | 0.77 | 6.63 | 11.20 | 3 3 3 3 | 0.00 | 1.69 |
| I09369-6454 | | I2 | 1515 | | 09 36 59.1 | -64 54 33 | 284.37 | -9.4 | 23x 5 | 18.7 | S | 15 | 0.26 | 0.25 | 0.39 | 2.11 | 1 1 3 1 | 0.43 | 5.41 |
| I09369-6454 | | I2 | 1516 | | 09 36 59.9 | -64 55 53 | 284.38 | -9.5 | 23x 16 | 17.4 | L | 68 | 0.26 | 0.25 | 0.39 | 2.11 | 1 1 3 1 | 0.43 | 5.41 |
| I09371-6136 | Y | I | 1517 | L | 09 37 08.0 | -61 36 09 | 282.15 | -7.0 | 108x 74 | 13.7 | S 4 | 6 | 0.25 | 0.25 | 1.66 | 6.23 | 1 1 3 3 | 0.02 | 3.75 |
| I09371-6130 | * | I | 1519 | L | 09 37 12.5 | -61 30 38 | 282.10 | -6.9 | 54x 40 | 14.9 | L | 8 | 0.30 | 0.25 | 0.73 | 3.47 | 1 1 3 1 | 0.14 | 4.75 |
| I09373-6344 | Y | I | 1524 | | 09 37 19.5 | -63 44 46 | 283.61 | -8.6 | 54x 40 | 15.4 | SB 1 | 2 | 0.25 | 0.42 | 2.60 | 3.22 | 1 3 3 3 | 0.02 | 1.24 |
| I09380-6159 | | I | 1534 | | 09 38 08.9 | -61 58 06 | 282.49 | -7.2 | 34x 4 | 18.6 | S | 66 | 0.25 | 0.25 | 0.63 | 5.35 | 1 1 3 3 | 0.16 | 8.49 |
| I09403-6034 | | I | 1561 | | 09 40 18.1 | -60 34 10 | 281.75 | -5.9 | 34x 20 | 16.2 | E 4 | 1 | 0.28 | 0.25 | 0.80 | 35.50 | 1 1 3 1 | 0.11 | 44.38 |
| I09408-4838 | | Q | 1569 | | 09 40 48.5 | -48 40 20 | 274.05 | 3.0 | 20x 7 | 18.1 | S M | 98 | 0.36 | 0.25 | 0.40 | 8.62 | 3 1 1 1 | 0.56 | 21.55 |
| I09421-6606 | | I | 1583 | | 09 42 09.6 | -66 06 27 | 285.58 | -10.0 | 16x 7 | 18.8 | S M | 7 | 0.67 | 0.25 | 0.43 | 0.95 | 1 1 3 3 | 0.91 | 2.21 |

**Table 2.** IRAS Galaxies in the Hydra/Antlia ZOA Region – continued

| | | | optical | | | | | | | IRAS | | | | | | | |
|---|---|---|---|---|---|---|---|---|---|---|---|---|---|---|---|---|---|
| IRAS PSC Ident. | IR | RKK | R.A. (h m s) | Dec. (° ′ ″) | gal ℓ (°) | gal b (°) | D x d (″) | $B_J$ ($^m$) | Type class. | Sep (″) | Flux Density $f_{12}$ | $f_{25}$ | $f_{60}$ | $f_{100}$ | Qual. | Color $col_1$ | $col_2$ |
| (1) | (2) | (3) | (4) | (5) | (6) | (7) | (8) | (9) | (10) | (11) | (12) | (13) | (14) | (15) | (16) | (17) | (18) |
| I09447-6302 Y | I | 1628 L | 09 44 48.0 | -63 02 24 | 283.78 | -7.5 | 141x 16 | 15.1 | S 2 | 3 | 0.25 | 0.25 | 0.76 | 1.36 | 1 1 3 2 | 0.11 | 1.79 |
| I09473-4741 Y | I | 1664 L | 09 47 22.9 | -47 41 09 | 274.26 | 4.5 | 101x 94 | 13.2 | L | 9 | 0.48 | 0.27 | 1.27 | 3.11 | 1 1 3 3 | 0.08 | 2.45 |
| I09475-4746 | P | 1666 | 09 47 31.4 | -47 47 33 | 274.34 | 4.4 | 43x 16 | 16.4 | E 7 | 75 | 0.54 | 0.63 | 0.46 | 3.17 | 1 1 3 1 | 1.61 | 6.89 |
| I09485-4831 | I | 1682 | 09 48 32.9 | -48 31 40 | 274.94 | 4.0 | 17x 13 | 18.0 | | 13 | 0.60 | 0.20 | 0.40 | 1.96 | 3 2 1 1 | 0.75 | 4.90 |
| I09487-4928 Y | I | 1685 L | 09 48 43.1 | -49 28 40 | 275.56 | 3.3 | 54x 40 | 15.2 | S 5 | 14 | 0.42 | 0.25 | 0.63 | 2.76 | 1 1 3 1 | 0.26 | 4.38 |
| I09506-4846 | I | 1720 | 09 50 43.3 | -48 46 16 | 275.38 | 4.0 | 19x 8 | 18.0 | | 20 | 0.29 | 0.25 | 0.73 | 8.78 | 1 1 3 1 | 0.14 | 12.03 |
| I09512-4841 | Q | 1728 | 09 51 08.3 | -48 42 09 | 275.39 | 4.1 | 34x 16 | 16.5 | SX 5 | 87 | 0.43 | 0.25 | 0.49 | 2.48 | 3 1 1 1 | 0.45 | 5.06 |
| I09515-6117 | P | 1733 | 09 51 28. | -61 18.1 : | 283.27 | -5.6 | 20x 13 | 17.9 | E 3 | 48 | 0.41 | 0.25 | 0.60 | 15.10 | 1 1 3 1 | 0.28 | 25.17 |
| I09532-4906 | I | 1757 | 09 53 16.6 | -49 06 51 | 275.92 | 4.0 | 19x 27 | 16.8 | SB | 5 | 0.30 | 0.25 | 0.48 | 7.94 | 1 1 3 1 | 0.33 | 16.54 |
| I09532-4911 Y | I | 1758 | 09 53 16.8 | -49 11 57 | 275.97 | 3.9 | 24x 13 | 17.0 | S 2 | 3 | 0.25 | 0.26 | 2.59 | 4.01 | 1 3 3 3 | 0.01 | 1.55 |
| I09594-4916 Y | I | 1876 | 09 59 28.0 | -49 16 22 | 276.82 | 4.5 | 27x 7 | 17.8 | S 0 | 31 | 0.39 | 0.25 | 0.77 | 1.88 | 1 1 3 1 | 0.16 | 2.44 |
| I10017-6508 | I | 1902 | 10 01 41.8 | -65 08 29 | 286.55 | -8.0 | 16x 4 | 19.3 | S | 42 | 0.28 | 0.25 | 0.40 | 1.76 | 3 1 1 1 | 0.44 | 4.40 |
| I10019-6443 Y | I | 1907 L | 10 01 54.6 | -64 43 26 | 286.32 | -7.6 | 148x 87 | 13.5 | SXR3 | 5 | 0.25 | 0.25 | 1.24 | 4.68 | 1 1 3 3 | 0.04 | 3.77 |
| I10039-6730 Y | I | 1938 L | 10 03 59.3 | -67 30 14 | 288.17 | -9.7 | 60x 54 | 15.1 | SX 3 | 5 | 0.25 | 0.25 | 1.07 | 1.91 | 1 1 3 3 | 0.05 | 1.79 |
| I10044-4748 | I | 1942 | 10 04 22.4 | -47 48 28 | 276.60 | 6.2 | 13x 8 | 18.6 | I | 65 | 0.39 | 0.25 | 0.43 | 2.44 | 1 1 1 3 | 0.53 | 5.67 |
| I10047-6502 | Q | 1946 | 10 04 41.6 | -65 03 33 | 286.76 | -7.7 | 13x 4 | 19.1 | S | 94 | 9.11 | 5.33 | 0.89 | 1.44 | 3 3 3 1 | 61.30 | 1.62 |
| I10048-4728 | Q2 | 1949 L | 10 04 47.4 | -47 27 05 | 276.45 | 6.5 | 74x 47 | 14.5 | SB 3 | 98 | 3.06 | 2.27 | 2.49 | 3.94 | 3 3 3 3 | 1.12 | 1.58 |
| I10048-4728 | Q2 | 1954 | 10 04 59.4 | -47 28 58 | 276.49 | 6.5 | 20x 13 | 17.0 | S | 71 | 3.06 | 2.27 | 2.49 | 3.94 | 3 3 3 3 | 1.12 | 1.58 |
| I10050-4941 | Q | 1955 | 10 05 01.5 | -49 42 05 | 277.80 | 4.7 | 13x 5 | 19.2 | | 52 | 4.56 | 2.72 | 0.40 | 1.60 | 3 3 1 1 | 77.52 | 4.00 |
| I10055-6407 * | I | 1963 L | 10 05 32.4 | -64 07 06 | 286.28 | -6.9 | 74x 74 | 14.3 | S 3 | 3 | 0.42 | 0.25 | 1.18 | 2.20 | 1 1 3 1 | 0.08 | 1.86 |
| I10059-6249 | I | 1970 | 10 05 54.0 | -62 49 34 | 285.55 | -5.8 | 20x 5 | 18.5 | S | 26 | 0.75 | 0.25 | 0.47 | 2.70 | 1 1 3 1 | 0.85 | 5.74 |
| I10063-4858 | I | 1977 | 10 06 19.2 | -48 58 24 | 277.55 | 5.4 | 20x 13 | 17.2 | | 10 | 0.31 | 0.44 | 0.71 | 8.98 | 1 1 3 1 | 0.27 | 12.65 |
| I10068-6443 | I | 1988 | 10 06 49.2 | -64 43 03 | 286.74 | -7.3 | 15x 13 | 17.8 | S E | 18 | 0.28 | 0.25 | 0.50 | 2.27 | 1 1 3 1 | 0.28 | 4.54 |
| I10075-6647 Y | I | 2000 L | 10 07 30.2 | -66 47 02 | 288.02 | -8.9 | 195x 74 | 13.1 | SB 3 | 6 | 0.99 | 2.77 | 16.00 | 31.30 | 3 3 3 3 | 0.01 | 1.96 |
| I10082-5647 * | I | 2017 | 10 08 16.4 | -56 47 07 | 282.30 | -.7 | 67x 27 | 15.8 | S | 20 | 8.70 | 22.70 | 56.80 | 143.00 | 3 3 3 1 | 0.06 | 2.52 |
| I10094-4815 | Q | 2037 | 10 09 17.5 | -48 14 11 | 277.53 | 6.3 | 30x 11 | 17.3 | E | 115 | 0.25 | 0.25 | 0.40 | 2.32 | 1 1 1 3 | 0.39 | 5.80 |
| I10096-6653 Y | I | 2043 L | 10 09 39.4 | -66 53 50 | 288.26 | -8.9 | 60x 47 | 15.0 | SX 3 | 2 | 0.40 | 0.25 | 0.89 | 1.92 | 1 1 3 3 | 0.13 | 2.16 |
| I10106-6217 Y | I | 2063 L | 10 10 36.5 | -62 17 11 | 285.69 | -5.1 | 67x 27 | 15.1 | S 3 | 5 | 1.00 | 5.10 | 27.80 | 30.50 | 3 3 3 3 | 0.01 | 1.10 |
| I10111-6346 Y | I | 2081 | 10 11 08.4 | -63 46 59 | 286.59 | -6.3 | 34x 11 | 16.9 | S 0 | 2 | 0.25 | 0.25 | 1.75 | 2.63 | 1 1 3 1 | 0.02 | 1.50 |
| I10114-5216 | I | 2095 | 10 11 31.0 | -52 16 27 | 280.12 | 3.2 | 40x 20 | 16.0 | S 3 | 14 | 0.25 | 0.26 | 1.24 | 8.83 | 1 1 3 1 | 0.04 | 7.12 |
| I10116-5043 | Q | 2104 | 10 11 45.4 | -50 41 33 | 279.26 | 4.5 | 20x 16 | 17.2 | L | 106 | 1.11 | 0.30 | 0.40 | 7.31 | 3 3 1 1 | 2.08 | 18.27 |
| I10118-5202 | I | 2109 L | 10 11 50.9 | -52 02 16 | 280.03 | 3.4 | 47x 40 | 15.3 | SB 3 | 18 | 1.26 | 0.25 | 0.62 | 8.58 | 1 1 3 1 | 0.82 | 13.84 |
| I10126-5222 | Q | 2138 | 10 12 45.8 | -52 21 42 | 280.33 | 3.2 | 13x 5 | 19.2 | S | 90 | 0.35 | 0.25 | 0.43 | 10.10 | 3 1 1 1 | 0.47 | 23.49 |
| I10133-4954 | P | 2156 | 10 13 23.6 | -49 53 22 | 279.02 | 5.3 | 20x 5 | 18.6 | S L | 79 | 0.27 | 0.25 | 0.40 | 2.30 | 3 1 1 1 | 0.42 | 5.75 |
| I10148-6436 | I | 2192 L | 10 14 44.9 | -64 37 00 | 287.39 | -6.7 | 67x 8 | 16.9 | S | 39 | 0.67 | 0.25 | 0.40 | 2.27 | 3 1 1 1 | 1.05 | 5.67 |
| I10149-4837 Y | I | 2200 L | 10 14 55.2 | -48 37 47 | 278.52 | 6.5 | 269x175 | 11.9 | SB 4 | 9 | 0.37 | 0.74 | 2.85 | 12.10 | 1 1 3 3 | 0.03 | 4.25 |
| I10150-5031 | I | 2205 | 10 15 01.7 | -50 31 55 | 279.60 | 4.9 | 13x 8 | 18.5 | S L | 50 | 0.23 | 0.25 | 0.40 | 2.90 | 3 1 1 1 | 0.36 | 7.25 |
| I10152-5136 Y | I | 2213 | 10 15 13.8 | -51 36 10 | 280.22 | 4.0 | 17x 8 | 17.9 | S | 13 | 0.25 | 0.16 | 1.58 | 6.44 | 1 2 3 1 | 0.02 | 4.08 |
| I10166-6600 | I | 2266 | 10 16 44.5 | -66 01 09 | 288.35 | -7.8 | 13x 11 | 18.0 | E | 43 | 0.25 | 0.25 | 0.57 | 1.40 | 1 1 3 3 | 0.19 | 2.46 |
| I10173-4931 Y | I3 | 2287 | 10 17 13.4 | -49 31 55 | 279.34 | 5.9 | 12x 3 | 19.8 | S | 77 | 0.29 | 0.25 | 1.07 | 1.71 | 1 1 3 2 | 0.06 | 1.60 |
| I10173-4931 Y | I3 | 2291 | 10 17 19.8 | -49 31 26 | 279.35 | 6.0 | 23x 9 | 17.7 | S M | 17 | 0.29 | 0.25 | 1.07 | 1.71 | 1 1 3 2 | 0.06 | 1.60 |
| I10173-4931 Y | I3 | 2294 | 10 17 21.8 | -49 32 05 | 279.36 | 6.0 | 12x 7 | 18.6 | | 28 | 0.29 | 0.25 | 1.07 | 1.71 | 1 1 3 2 | 0.06 | 1.60 |
| I10177-6043 | I2 | 2310 | 10 17 43.4 | -60 44 12 | 285.53 | -3.3 | 54x 24 | 15.2 | E | 17 | 0.30 | 0.50 | 1.57 | 16.50 | 1 3 1 1 | 0.06 | 10.51 |
| I10177-6043 | I2 | 2314 | 10 17 47.1 | -60 44 08 | 285.53 | -3.3 | 34x 24 | 15.7 | E | 17 | 0.30 | 0.50 | 1.57 | 16.50 | 1 3 1 1 | 0.06 | 10.51 |
| I10178-4732 Y | I | 2317 | 10 17 50.9 | -47 32 22 | 278.33 | 7.7 | 40x 22 | 16.4 | S M | 12 | 0.25 | 0.28 | 0.75 | 1.87 | 1 1 3 2 | 0.12 | 2.49 |
| I10185-6345 | I | 2341 | 10 18 35.5 | -63 45 52 | 287.27 | -5.8 | 47x 31 | 15.7 | L | 6 | 0.33 | 0.25 | 0.76 | 5.62 | 1 1 3 1 | 0.14 | 7.39 |
| I10196-4921 Y | I | 2380 | 10 19 39.9 | -49 21 43 | 279.58 | 6.3 | 40x 30 | 16.3 | | 13 | 0.42 | 0.25 | 0.80 | 2.39 | 1 1 3 1 | 0.16 | 2.99 |
| I10203-5007 | Q | 2407 | 10 20 30.1 | -50 08 40 | 280.12 | 5.7 | 16x 4 | 19.4 | S | 85 | 0.25 | 0.54 | 0.63 | 1.64 | 1 1 1 3 | 0.34 | 2.60 |
| I10211-4913 Y | I | 2426 L | 10 21 10.1 | -49 13 02 | 279.71 | 6.6 | 77x 44 | 14.8 | S 2 | 3 | 0.26 | 0.21 | 1.66 | 4.61 | 1 2 3 3 | 0.02 | 2.78 |
| I10221-5144 | I2 | 2456 | 10 22 09.8 | -51 43 50 | 281.20 | 4.5 | 24x 5 | 18.7 | | 19 | 0.52 | 0.25 | 0.48 | 2.77 | 1 1 3 1 | 0.56 | 5.77 |
| I10221-5144 | I2 | 2457 | 10 22 14.2 | -51 44 02 | 281.21 | 4.5 | 22x 4 | 19.1 | SY M | 30 | 0.52 | 0.25 | 0.48 | 2.77 | 1 1 3 1 | 0.56 | 5.77 |
| I10234-4957 Y | I2 | 2485 L | 10 23 24.8 | -49 57 59 | 280.42 | 6.1 | 17x 8 | 18.3 | | 28 | 0.27 | 0.32 | 1.37 | 2.03 | 1 1 3 1 | 0.05 | 1.48 |
| I10234-4957 * | I2 | 2486 L | 10 23 25.9 | -49 57 24 | 280.42 | 6.1 | 34x 16 | 16.8 | S 5 | 8 | 0.27 | 0.32 | 1.37 | 2.03 | 1 1 3 1 | 0.05 | 1.48 |
| I10240-5337 | I | 2507 | 10 24 01.8 | -53 37 04 | 282.44 | 3.1 | 51x 34 | 15.4 | SB 5 | 22 | 0.25 | 0.25 | 1.42 | 5.56 | 1 1 1 3 | 0.03 | 3.92 |
| I10240-6729 Y | I | 2508 L | 10 24 04.9 | -67 29 26 | 289.77 | -8.6 | 60x 60 | 14.9 | SX 7 | 0 | 0.25 | 0.25 | 1.02 | 2.91 | 1 1 3 3 | 0.06 | 2.85 |
| I10240-5122 | Q | 2510 | 10 24 10.9 | -51 22 47 | 281.28 | 5.0 | 30x 5 | 18.1 | S | 71 | 0.70 | 0.18 | 0.40 | 2.10 | 3 2 1 1 | 0.79 | 5.25 |
| I10241-5330 | Q | 2512 | 10 24 13.1 | -53 31 37 | 282.42 | 3.2 | 27x 27 | 16.1 | F | 113 | 0.96 | 0.46 | 0.47 | 13.90 | 3 3 1 1 | 2.00 | 29.57 |
| I10245-4811 Y | I | 2518 | 10 24 34.6 | -48 10 55 | 279.64 | 7.7 | 15x 7 | 18.6 | S | 15 | 0.34 | 0.25 | 1.59 | 2.55 | 1 1 3 3 | 0.03 | 1.60 |
| I10248-4853 | I | 2525 L | 10 24 51.7 | -48 53 34 | 280.06 | 7.2 | 47x 20 | 16.2 | S 3 | 8 | 0.25 | 0.25 | 0.64 | 5.75 | 1 1 3 1 | 0.15 | 8.98 |
| I10252-6711 | I | 2545 | 10 25 15.5 | -67 11 34 | 289.71 | -8.3 | 13x 4 | 19.7 | S | 7 | 0.47 | 0.25 | 0.41 | 2.18 | 1 1 3 1 | 0.70 | 5.32 |
| I10262-5106 | Q | 2580 | 10 26 15.8 | -51 05 15 | 281.40 | 5.4 | 23x 12 | 17.7 | S | 69 | 1.39 | 0.88 | 0.34 | 1.68 | 3 3 2 1 | 10.58 | 4.94 |
| I10263-5026 Y | I | 2585 L | 10 26 23.3 | -50 26 07 | 281.08 | 6.0 | 65x 47 | 14.9 | S M | 3 | 0.44 | 0.25 | 0.89 | 2.74 | 1 1 3 2 | 0.14 | 3.08 |
| I10264-6413 Y | I | 2588 | 10 26 26.4 | -64 13 54 | 288.25 | -5.7 | 39x 20 | 16.9 | S | 17 | 0.26 | 0.28 | 1.02 | 4.32 | 1 1 3 1 | 0.07 | 4.24 |
| I10267-5042 Y | I | 2595 L | 10 26 42.5 | -50 42 53 | 281.27 | 5.8 | 81x 22 | 15.6 | S 1 | 4 | 0.25 | 0.44 | 0.73 | 2.92 | 1 1 3 1 | 0.21 | 4.00 |
| I10270-5125 | I | 2604 | 10 27 02.5 | -51 25 58 | 281.69 | 5.2 | 17x 12 | 17.6 | | 27 | 0.31 | 0.25 | 0.41 | 2.15 | 1 1 3 1 | 0.46 | 5.24 |
| I10274-6715 | I | 2608 | 10 27 27.1 | -67 15 28 | 289.92 | -8.3 | 40x 13 | 16.6 | S 0 | 28 | 0.47 | 0.25 | 0.43 | 1.74 | 1 1 3 1 | 0.64 | 4.05 |
| I10280-4950 | Q | 2625 | 10 28 05.0 | -49 48 58 | 280.99 | 6.6 | 12x 9 | 18.9 | | 80 | 0.79 | 0.41 | 0.40 | 1.53 | 3 3 1 1 | 2.02 | 3.82 |
| I10282-5141 | I | 2631 | 10 28 13.4 | -51 41 43 | 281.98 | 5.0 | 30x 9 | 17.6 | S E | 12 | 0.47 | 0.25 | 0.50 | 2.00 | 1 1 3 1 | 0.47 | 4.00 |
| I10289-5107 | P | 2649 | 10 28 52.6 | -51 06 54 | 281.77 | 5.6 | 36x 27 | 16.4 | S M | 68 | 0.40 | 0.25 | 0.43 | 2.66 | 1 1 3 1 | 0.54 | 6.19 |
| I10289-4847 Y | I | 2652 L | 10 28 57.0 | -48 47 25 | 280.58 | 7.6 | 87x 48 | 14.5 | S 3 | 9 | 0.25 | 0.34 | 1.85 | 4.17 | 1 3 3 3 | 0.02 | 2.25 |
| I10302-6327 | I | 2684 L | 10 30 14.5 | -63 27 02 | 288.21 | -4.9 | 128x 47 | 14.6 | S 7 | 49 | 0.36 | 0.32 | 0.76 | 6.72 | 1 1 3 1 | 0.20 | 8.84 |
| I10333-6727 Y | I | 2733 | 10 33 22.1 | -67 27 48 | 290.52 | -8.2 | 40x 35 | 15.7 | S 1 | 5 | 0.25 | 0.28 | 1.24 | 2.31 | 1 3 3 3 | 0.05 | 1.86 |
| I10338-5350 Y | I | 2740 | 10 33 52.1 | -53 50 54 | 283.81 | 3.6 | 54x 5 | 17.6 | S 5 | 13 | 0.25 | 0.26 | 2.89 | 4.62 | 1 1 3 3 | 0.01 | 1.60 |
| I10353-5440 | I | 2754 L | 10 35 18.1 | -54 40 31 | 284.40 | 3.0 | 54x 24 | 16.1 | S 5 | 32 | 1.65 | 0.88 | 0.38 | 19.00 | 3 3 2 1 | 10.06 | 50.00 |
| I10353-4850 * | I | 2756 | 10 35 22.5 | -48 50 22 | 281.53 | 8.1 | 13x 5 | 19.2 | | 24 | 0.29 | 0.25 | 0.78 | 2.65 | 1 1 3 1 | 0.12 | 3.40 |
| I10355-5218 Y | I | 2759 | 10 35 33.8 | -52 18 37 | 283.27 | 5.1 | 28x 11 | 17.0 | S | 8 | 0.25 | 0.28 | 2.15 | 3.60 | 1 3 3 3 | 0.02 | 1.67 |
| I10369-5009 Y | I | 2774 | 10 36 55.3 | -50 09 39 | 282.40 | 7.1 | 43x 31 | 15.7 | SB 3 | 3 | 0.25 | 0.31 | 1.31 | 3.36 | 1 1 3 3 | 0.05 | 2.56 |
| I10371-5326 | I | 2776 | 10 37 04.2 | -53 26 08 | 284.02 | 4.2 | 16x 7 | 18.6 | | 21 | 0.26 | 0.25 | 0.54 | 3.46 | 1 1 1 3 | 0.22 | 6.41 |
| I10421-5045 | P | 2820 | 10 42 14.2 | -50 46 24 | 283.45 | 6.9 | 30x 7 | 18.3 | | 83 | 0.25 | 0.25 | 0.40 | 10.90 | 3 1 1 1 | 0.39 | 27.25 |
| I10461-5302 | I | 2847 L | 10 46 09.0 | -53 02 18 | 285.04 | 5.2 | 94x 27 | 15.2 | S L | 12 | 0.42 | 0.25 | 0.93 | 16.80 | 1 1 3 1 | 0.12 | 18.06 |

**Table 2.** IRAS Galaxies in the Hydra/Antlia ZOA Region – continued

| | | | optical | | | | | | | IRAS | | | | | | | |
|---|---|---|---|---|---|---|---|---|---|---|---|---|---|---|---|---|---|
| IRAS PSC Ident. | IR | RKK | R.A. (h m s) | Dec. (° ′ ″) | gal ℓ (°) | gal b (°) | D x d (″) | $B_J$ ($^m$) | Type class. | Sep (″) | Flux Density $f_{12}$ | $f_{25}$ | $f_{60}$ | $f_{100}$ | Qual. | Color $col_1$ | $col_2$ |
| (1) | (2) | (3) | (4) | (5) | (6) | (7) | (8) | (9) | (10) | (11) | (12) | (13) | (14) | (15) | (16) | (17) | (18) |
| I10463-6503 Y | I | 2852 | 10 46 22.0 | -65 03 34 | 290.53 | -5.4 | 27x 27 | 16.6 | | 4 | 0.25 | 0.40 | 2.44 | 3.82 | 1 3 3 1 | 0.02 | 1.57 |
| I10491-6655 | I | 2859 | 10 49 08.0 | -66 55 34 | 291.62 | -7.0 | 27x 17 | 17.6 | S | 14 | 0.42 | 0.25 | 0.56 | 2.50 | 1 1 3 1 | 0.33 | 4.46 |
| I10495-6707 Y | I | 2862 | 10 49 34.5 | -67 07 53 | 291.76 | -7.1 | 17x 7 | 18.8 | S | 16 | 0.27 | 0.25 | 1.05 | 1.79 | 1 1 3 2 | 0.06 | 1.70 |
| I10574-6603 Y | I | 2870 L | 10 57 32.6 | -66 03 51 | 292.01 | -5.8 | 94x 47 | 14.7 | S 3 | 40 | 0.55 | 0.89 | 10.30 | 19.10 | 3 3 3 3 | 0.00 | 1.85 |
| I10587-5342 | P | 2873 | 10 58 39.4 | -53 41 31 | 287.01 | 5.4 | 20x 13 | 17.6 | S M | 50 | 1.16 | 0.63 | 0.40 | 10.20 | 3 3 1 1 | 4.57 | 25.50 |
| I11017-6732 | I | 2883 | 11 01 45.7 | -67 32 46 | 293.00 | -7.0 | 24x 8 | 18.2 | P | 12 | 0.42 | 0.25 | 0.56 | 1.86 | 1 1 3 1 | 0.33 | 3.32 |
| I11081-6707 | I | 2895 | 11 08 07.3 | -67 07 22 | 293.40 | -6.4 | 74x 40 | 15.6 | SB 3 | 9 | 0.43 | 0.25 | 0.48 | 2.74 | 1 1 3 1 | 0.47 | 5.71 |
| I11098-5718 | Q | 2896 | 11 09 58.5 | -57 16 51 | 289.90 | 2.7 | 30x 3 | 18.8 | S | 104 | 0.30 | 0.25 | 1.32 | 8.33 | 1 1 1 3 | 0.04 | 6.31 |
| I11110-6651 | I | 2901 | 11 11 06.1 | -66 50 52 | 293.57 | -6.0 | 26x 9 | 17.6 | | 33 | 0.67 | 0.25 | 0.46 | 3.87 | 1 1 3 1 | 0.79 | 8.41 |
| I11112-6715 Y | I | 2902 | 11 11 12.6 | -67 15 45 | 293.74 | -6.4 | 60x 34 | 15.7 | SB 5 | 1 | 0.25 | 0.45 | 2.67 | 3.77 | 1 3 3 3 | 0.02 | 1.41 |
| I11190-5432 | I | 2941 | 11 19 04.5 | -54 32 50 | 290.12 | 5.7 | 17x 8 | 18.3 | S | 47 | 0.38 | 0.25 | 0.59 | 2.81 | 1 1 3 2 | 0.27 | 4.76 |
| I11217-5341 | Q2 | 2964 | 11 21 50.6 | -53 40 25 | 290.21 | 6.7 | 40x 11 | 17.0 | S L | 63 | 3.04 | 0.78 | 0.40 | 1.89 | 3 3 1 1 | 14.82 | 4.72 |
| I11217-5341 | Q2 | 2966 | 11 21 52.9 | -53 40 06 | 290.21 | 6.7 | 15x 7 | 18.6 | | 91 | 3.04 | 0.78 | 0.40 | 1.89 | 3 3 1 1 | 14.82 | 4.72 |
| I11237-5230 Y | I | 2991 L | 11 23 43.3 | -52 30 16 | 290.09 | 7.9 | 81x 81 | 14.5 | SBR5 | 17 | 0.25 | 0.25 | 0.70 | 2.83 | 1 1 3 2 | 0.13 | 4.04 |
| I11274-5548 Y | I | 3022 | 11 27 24.5 | -55 48 06 | 291.66 | 5.0 | 16x 9 | 18.3 | | 11 | 0.38 | 0.25 | 0.75 | 3.11 | 1 1 3 3 | 0.17 | 4.15 |
| I11300-5539 | P | 3054 | 11 30 06.8 | -55 39 59 | 291.98 | 5.2 | 13x 9 | 18.6 | | 72 | 0.35 | 0.25 | 1.63 | 14.20 | 3 1 1 1 | 0.03 | 8.71 |
| I11305-5251 | P | 3063 | 11 30 29.9 | -52 50 28 | 291.18 | 7.9 | 22x 11 | 18.2 | | 61 | 0.34 | 0.25 | 0.40 | 7.37 | 3 1 1 1 | 0.53 | 18.43 |
| I11307-5516 Y | I | 3067 | 11 30 41.8 | -55 16 26 | 291.94 | 5.6 | 34x 27 | 16.8 | S 5 | 13 | 0.25 | 0.25 | 2.03 | 7.93 | 1 1 3 1 | 0.02 | 3.91 |
| I11307-5228 * | I | 3069 | 11 30 46.8 | -52 28 17 | 291.11 | 8.3 | 27x 24 | 16.5 | S | 12 | 0.25 | 0.25 | 0.70 | 2.34 | 1 1 3 3 | 0.13 | 3.34 |
| I11311-5245 | I | 3080 | 11 31 08.8 | -52 45 45 | 291.25 | 8.0 | 40x 40 | 16.1 | SX 5 | 20 | 0.31 | 0.25 | 1.09 | 12.20 | 3 1 1 1 | 0.07 | 11.19 |
| I11311-5526 | I | 3081 | 11 31 10.9 | -55 26 28 | 292.06 | 5.5 | 17x 8 | 18.4 | | 32 | 1.24 | 0.32 | 1.43 | 9.25 | 3 3 1 1 | 0.19 | 6.47 |
| I11322-5220 | I | 3093 | 11 32 20.7 | -52 20 18 | 291.30 | 8.5 | 13x 5 | 19.5 | | 59 | 0.25 | 0.25 | 0.40 | 2.60 | 1 1 1 3 | 0.39 | 6.50 |
| I11325-5705 | I | 3097 | 11 32 28.1 | -57 06 18 | 292.72 | 3.9 | 20x 13 | 18.1 | S | 30 | 0.36 | 0.25 | 0.48 | 4.57 | 1 1 3 1 | 0.39 | 9.52 |
| I11337-5339 | Q | 3116 | 11 33 41.3 | -53 37 25 | 291.87 | 7.3 | 13x 9 | 18.7 | | 118 | 0.35 | 0.47 | 0.40 | 1.44 | 3 1 1 1 | 1.03 | 3.60 |
| I11339-5509 | I | 3122 | 11 33 57.0 | -55 09 03 | 292.35 | 5.8 | 15x 5 | 19.1 | | 32 | 0.43 | 0.39 | 0.64 | 3.41 | 1 1 3 2 | 0.41 | 5.33 |
| I11356-5227 | Q2 | 3148 | 11 35 31.1 | -52 27 53 | 291.80 | 8.5 | 13x 4 | 20.0 | S | 99 | 0.42 | 0.25 | 0.40 | 6.26 | 3 1 1 1 | 0.66 | 15.65 |
| I11356-5227 | Q2 | 3149 | 11 35 35.3 | -52 28 08 | 291.82 | 8.5 | 13x 4 | 20.2 | S | 67 | 0.42 | 0.25 | 0.40 | 6.26 | 3 1 1 1 | 0.66 | 15.65 |
| I11358-5232 | I2 | 3150 | 11 35 40.4 | -52 32 41 | 291.85 | 8.4 | 20x 7 | 18.3 | | 76 | 0.41 | 0.25 | 0.69 | 5.22 | 1 1 3 1 | 0.22 | 7.57 |
| I11358-5232 | I2 | 3152 | 11 35 47.1 | -52 31 59 | 291.86 | 8.4 | 15x 13 | 17.7 | E | 18 | 0.41 | 0.25 | 0.69 | 5.22 | 1 1 3 1 | 0.22 | 7.57 |
| I11368-5316 Y | I | 3165 L | 11 36 51.3 | -53 16 34 | 292.23 | 7.8 | 74x 27 | 15.8 | S 2 | 6 | 0.25 | 0.25 | 1.00 | 2.75 | 1 1 3 2 | 0.06 | 2.75 |
| I11376-5321 | Q | 3172 | 11 37 33.2 | -53 20 39 | 292.35 | 7.7 | 20x 13 | 17.7 | | 80 | 0.47 | 0.25 | 0.40 | 5.88 | 3 1 1 1 | 0.73 | 14.70 |
| I11376-5406 Y | I | 3176 L | 11 37 39.5 | -54 07 13 | 292.58 | 7.0 | 60x 54 | 15.1 | SB 4 | 38 | 0.25 | 0.25 | 0.62 | 2.34 | 1 1 3 3 | 0.16 | 3.77 |
| I11379-5604 | I | 3180 | 11 37 54.1 | -56 03 37 | 293.15 | 5.1 | 20x 15 | 17.7 | S M | 60 | 0.34 | 0.25 | 0.51 | 14.70 | 1 1 3 1 | 0.33 | 28.82 |
| I11384-5227 | I | 3185 | 11 38 26.8 | -52 27 21 | 292.24 | 8.6 | 20x 20 | 17.0 | E | 32 | 0.45 | 0.25 | 0.45 | 5.57 | 1 1 3 1 | 0.56 | 12.38 |
| I11407-5303 Y | I | 3198 | 11 40 41.0 | -53 03 11 | 292.73 | 8.1 | 15x 12 | 18.1 | E | 15 | 0.25 | 0.25 | 0.67 | 2.07 | 1 1 3 3 | 0.14 | 3.09 |
| I11436-5606 Y | I | 3228 L | 11 43 39.4 | -56 06 37 | 293.94 | 5.3 | 202x108 | 12.8 | S 4 | 21 | 0.90 | 1.68 | 17.50 | 34.40 | 3 3 3 3 | 0.00 | 1.97 |
| I11471-5235 Y | I | 3257 | 11 47 07.3 | -52 35 19 | 293.56 | 8.8 | 47x 19 | 16.1 | E 6 | 11 | 0.25 | 0.25 | 1.43 | 2.88 | 1 1 3 3 | 0.03 | 2.01 |